# VLBI20-30: a scientific roadmap for the next decade

## The future of the European VLBI Network

Editors: Tiziana Venturi, Zsolt Paragi & Michael Lindqvist

JIVE
Joint Institute for VLBI
ERIC

EUROPEAN VLBI NETWORK

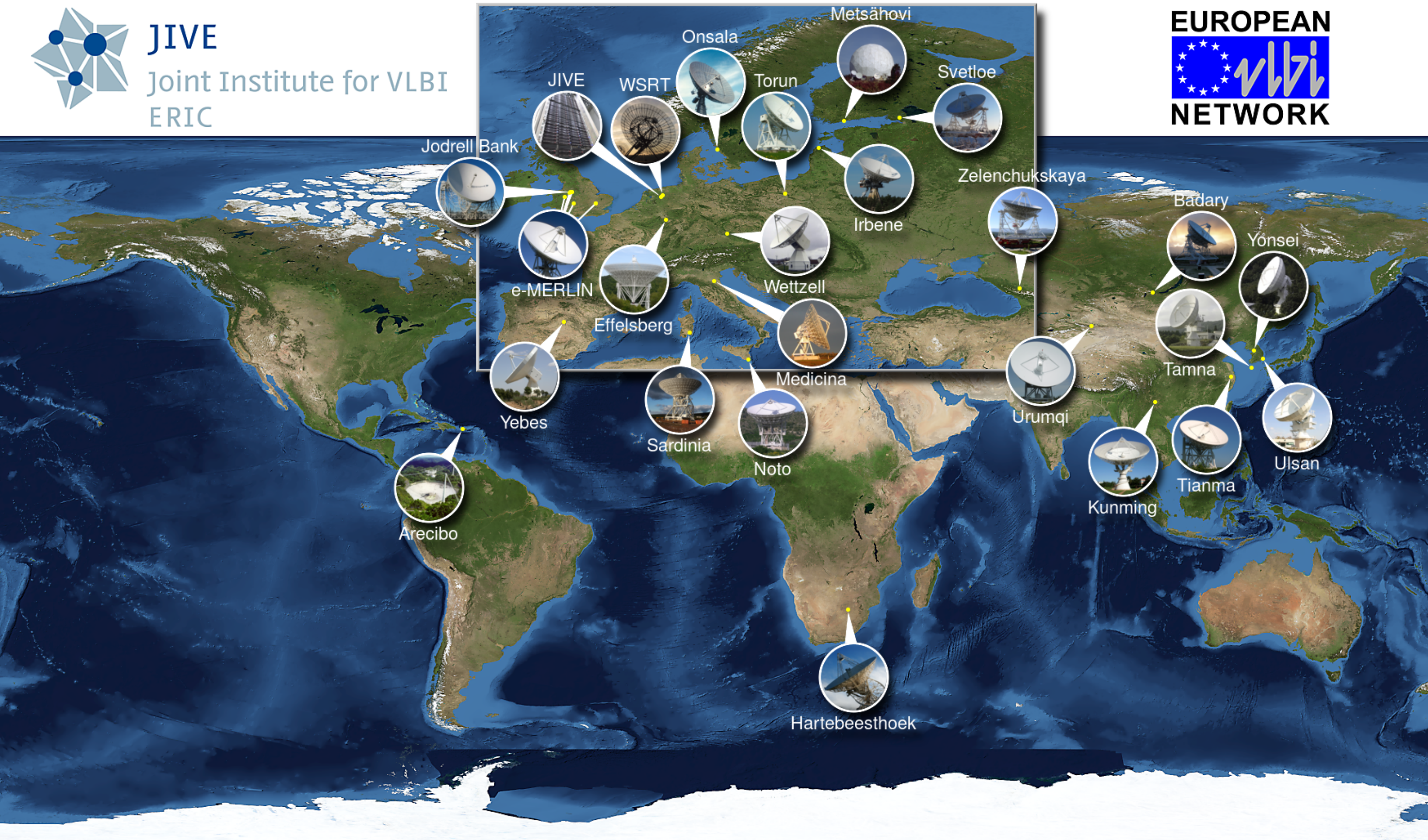

Image by Paul Boven (boven@jive.eu). Satellite image: Blue Marble Next Generation, courtesy of Nasa Visible Earth (visibleearth.nasa.gov).

Endorsed by the EVN Consortium Board of Directors

EUROPEAN VLBI NETWORK

JIVE
Joint Institute for VLBI
ERIC

JUMPING JIVE

European Commission

*Acknowledgements*

Following on from the EVN Directors' decision to commission a new science vision document for the EVN, the EC-H2020 project JUMPING JIVE (grant agreement No. 730884) dedicated a Work Package (WP7, "VLBI Future") to provide the organisational and financial support for its development. A core team formed by R. Beswick, T. Bogdanović, W. Brisken, P. Charlot, M. Lindqvist, A. Lobanov, Z. Paragi, A. Szomoru and T. Venturi started the process and led the discussion and coordination effort. While the main focus of this report is the EVN instrument, VLBI as a technique is a global effort, and indeed some science goals are best addressed by the EVN making joint observations with other VLBI networks around the world. Most of the science goals identified in this report apply to VLBI in general, even if the specific recommendations of the report apply mainly to the EVN; hence this document's title *VLBI20-30, a scientific roadmap for the next decade: The future of the European VLBI Network*.

The first face-to-face meeting of the project was held in Zaandam, Netherlands on February 28th – March 1st, 2018 and involved invited experts covering a wide variety of VLBI science. Further progress was made during the European Week of Astronomy and Space Science in Liverpool, at the Special Session 11, "Exploring the Universe: A European vision for the future of VLBI". At the 14th EVN Symposium, held in Granada on October 8-11, 2018 and attended by 170 people, an early draft of this science vision document was presented and received further input from the participants.

The VLBI community of users has provided enthusiastic and constructive input to this science roadmap document, which is updated as of November 30th, 2019. We thank all who have contributed as chapter authors and co-authors (see below) and all the many others from the community who have provided comments and advice on specific scientific questions. We warmly thank Prof. R. Schilizzi and Prof. P. Wilkinson for insightful suggestions. Thanks are due to Emma van der Wateren for careful proofreading.

Tiziana Venturi (on behalf of the VLBI20-30 Editorial team)
Director IRA, INAF, Italy
Vice-Chair EVN Consortium Board of Directors

Coordinators:

- Main coordinators and editors: Tiziana Venturi, Michael Lindqvist, Zsolt Paragi
- Executive summary: Tiziana Venturi, John Conway, Huib van Langevelde, Zsolt Paragi
- Chapter 1: John McKean
- Chapter 2: Rob Beswick, Tom Muxlow, Raffaella Morganti, Robert Schulz
- Chapter 3: Andrei Lobanov, Sándor Frey
- Chapter 4: Miguel Pérez-Torres, Zsolt Paragi
- Chapter 5: José Carlos Guirado, Anna Bartkiewicz, Kazi Rygl
- Chapter 6: Patrick Charlot, Leonid Gurvits
- Chapter 7: Tiziana Venturi, Francisco Colomer, John Conway, Huib van Langevelde, Michael Lindqvist, Zsolt Paragi
- Appendix 1: Walter Brisken, Zsolt Paragi, Michael Lindqvist, Tiziana Venturi
- Appendix 2: Walter Brisken, Arpad Szomoru, Pablo de Vicente
- External advisor: Tamara Bogdanović

Contributors:

- Tao An
- Guillem Anglada
- Megan Argo
- Rebecca Azulay
- Anna Bartkiewicz
- Ilse van Bemmel
- Rob Beswick
- Tatiana Bocanegra
- Biagina Boccardi
- Tamara Bogdanović
- Walter Brisken
- Paola Castangia
- Patrick Charlot
- James Chibueze
- Giuseppe Cimò
- Juan-Bautista Climent
- Francisco Colomer
- John Conway
- Roger Deane
- Adam Deller
- Richard Dodson
- Dmitry Duev
- Sandra Etoka
- Danielle Fenech
- Sándor Frey
- Krisztina Gabányi
- Denise Gabuzda
- Michael Garrett
- Marcin Gawroński
- Giancarlo Ghirlanda
- Marcello Giroletti
- Ciriaco Goddi
- José Luis Gómez
- Malcolm Gray
- Jane Greaves
- José Carlos Guirado
- Leonid Gurvits
- Jason Hessels
- Alexander van der Horst
- Todd Hunter
- Robert Laing
- Dharam Vir Lal
- Sébastien Lambert
- Huib van Langevelde
- Michael Lindqvist
- Andrei Lobanov
- Laurent Loinard
- Benito Marcote
- John McKean
- Andrea Merloni
- James Miller-Jones
- Guifré Molera Calvés
- Raffaella Morganti
- Luca Moscadelli
- Tom Muxlow
- Hans Olofsson
- Zsolt Paragi
- Leonid Petrov
- Miguel Pérez-Torres
- Roberto Pizzo
- Andrea Possenti
- Luis Henry Quiroga-Nuñez
- Cormac Reynolds
- Anita Richards
- Maria Rioja
- Kazi Rygl
- Alberto Sanna
- Tuomas Savolainen
- Tullia Sbarrato
- Robert Schulz
- Cristiana Spingola
- Gabriele Surcis
- Arpad Szomoru
- Corrado Trigilio
- Eskil Varenius
- Tiziana Venturi
- Pablo de Vicente
- Wouter Vlemmings
- Sjoert van Velzen
- Johan van der Walt

# EXECUTIVE SUMMARY

The European VLBI Network (EVN) Consortium Board of Directors commissioned in Autumn 2016 the EVN community to produce a Science Vision document covering the period till 2030. Since the last such document (EVN2015) was issued over a decade ago, the technical capabilities and scientific potential of Very Long Baseline Interferometry (VLBI) have considerably expanded. New technical developments on the horizon promise even greater future capabilities and scientific possibilities for VLBI in the future. Additionally, since the last science vision report many new exciting questions have emerged within astrophysics, and VLBI observations have already proved to be invaluable. Finally, the coming decade will see the beginning of operations of the Square Kilometre Array (SKA), alongside major new telescopes at other wavebands, which will significantly change the overall radio astronomy landscape. The above considerations make it timely to consider EVN and VLBI science priorities for the period 2020-2030.

The EVN is a distributed long-baseline radio interferometric array, that operates at the very forefront of astronomical research. Recent results, together with the new science possibilities outlined in this vision document, demonstrate the EVN's potential to generate new and exciting results that will transform our view of the cosmos. Together with *e*-MERLIN, the EVN already provides a range of baseline lengths that permit unique studies of faint radio sources to be made over a wide range of spatial scales. A major role for the EVN now and in the future relates to making initial or follow-up observations of transient sources, which are increasingly important areas of modern astronomy. In order to fully capitalise on its unique transient source capabilities, the EVN should develop new operational procedures that will permit it to respond to events triggered by other instruments. This may require more regular sessions, new advanced interfaces to react on triggers (with at least a flexible sub-array), and developments in the correlator and the EVN pipeline to aid searching for a broad parameter space for transient detection.

Anticipated technical upgrades to all aspects of the EVN's telescope and correlator facilities in the years ahead are expected to revolutionise the EVN's technical capabilities. New receiver developments will furnish low-noise systems that incorporate large instantaneous bandwidths and/or simultaneous multi-band capability. These new wide band receivers will improve continuum sensitivity by factors greater than 4 and also give broad band spectral and polarisation information for every continuum source. By facilitating simultaneous observations of multiple lines, similar increases in sensitivity can be realised for spectral line observations. Future EVN technical upgrades will support observing modes that allow joint observations with the SKA1-MID telescope. This will allow the phased-up SKA1-MID to be used as an element of EVN/global-VLBI giving an additional factor in sensitivity improvement larger than a factor of 2 to VLBI. In summary, the overall increase in sensitivity compared to today's VLBI is expected to be larger than a factor of 8. In the future, the EVN, either acting alone or in combination with SKA1-MID, will provide huge increases in performance relative to its present day capabilities. Such improvements will open a whole new set of scientific opportunities that can benefit from the exquisite spatial and spectral resolution of VLBI.

As the community prepares for the era of SKA science operations, the way in which the European radio astronomy facilities collaborate and organise themselves is likely to change. Already, there is significant discussion of these issues among the radio astronomy institutes in Europe. For the EVN a more centralised approach, in which JIVE plays an increasingly prominent role, is likely to emerge. In the future, close engagement with the SKA, its regional centres and other complementary telescopes, such as the International LOFAR Telescope (ILT), will be mandatory. Within the coming decade extending the application of SKA technologies to VLBI can also be expected. It is noteworthy that the e-EVN is already an SKA pathfinder, and crucial SKA technologies (such as real-time data transport and correlation over 1000s of kilometres, as well as central distribution of highly accurate time and frequency signals) have already been developed by our community. Developments of ultra-wideband receivers going significantly beyond the present SKA1-MID specifications could be a future example where EVN develops crucial technology for an SKA upgrade.

## Role of EVN/VLBI in the future astronomy landscape

Complementary to other radio astronomy facilities, VLBI provides (sub-)milliarcsecond angular resolution and ∼10 microarcseconds relative astrometric precision at cm wavelengths, and event-horizon scales in supermassive black holes (SMBHs) at the extreme and challenging mm and sub-mm wavelengths. The current role VLBI is playing in the astrophysical landscape has been assessed very favourably in the past, and over the past few years VLBI science has considerably broadened in the scientific topics covered. While parsec-scale imaging of the inner radio jets of Active Galactic Nuclei AGN as well as masers in early stages of star formation, are still important areas of research, a burst in the successful applications of VLBI within new emerging topics has recently taken place. The current level of image sensitivity, reaching $\mu$Jy levels for VLBI arrays at 18 cm and 6 cm has allowed remarkable, novel science applications, which were unforeseen in previous VLBI science cases. Most noticeably, the EVN has responded very skillfully to a range of astrophysical transients, from the counterparts of gravitational waves, to the still elusive Fast Radio Bursts, Tidal Disruption Events, Gamma-ray bursts and supernova explosions. For all of these transients the combination of astrometric precision, angular resolution and imaging capability which only VLBI can deliver, provides unique and essential information for unlocking the astrophysics of these sources and using them as tools for cosmology. A niche application only possible by arranging an ad-hoc network is imaging the event-horizon scales in selected super-massive black holes, as provided by the Event Horizon Telescope project with the recent success of the first image of a black hole in Messier 87 at 1.3 mm wavelengths. Figure 1 clearly shows how the science distribution of VLBI papers has broadened from 1980-1999 to 2000-2019.

At the same time radio astronomy is starting to unveil the faint slowly varying Universe at milliarsecond resolution. The (sub-)$\mu$Jy sensitivity of the EVN will throw a light on the formation of supermassive black holes at high redshift and cosmic star formation history of the Universe; in particular only VLBI can provide the angular resolution needed to discriminate between the activity of the nuclear and stellar components in distant galaxies during their most significant evolutionary stage. Much closer, in our home Galaxy, we have the chance to study in detail the life cycle of baryons through the formation and death of stars. VLBI offers unique tools to zoom in on the formation of stars through molecular masers that probe kinematics, physical (and chemical) states and even magnetic field strengths. VLBI astrometry provides a unique contribution to mapping the structure of our Galaxy by providing distances to stars behind interstellar and circumstellar dust. Pulsar proper motions and parallax distances can be refined, which, together with accurate pulsar timing will let us probe neutron star interiors, test fundamental physics, and detect the low-frequency gravitational wave background. Finally, absolute VLBI astrometry has been used to define the terrestrial and celestial reference frames, and it provides the only way to tie those to dynamical reference frames. By applying similar tools to spacecraft signals it is also possible to study the Solar System by measuring the gravitational fields of planets and moons with extreme accuracy.

The challenge for the future is how to optimise this unique role of VLBI in the context of a rapidly evolving set of global astronomical facilities as planned for the next few decades. Clearly, astronomy is set to go through another revolution, as leaps forward in sensitivity, survey speed, agility, frequency range and post-processing capabilities change how observations are made. Fundamental science questions will be addressed both by very detailed observations and extremely large surveys. Major impacts are expected from the ELT and other 30 m-class optical telescopes, the LSST, LIGO/Virgo, CTA, *Euclid*, the *JWST*, *Athena*, opening exciting new frontiers in our knowledge and understanding of the Universe and the fundamental laws governing it. Many of these observatories will provide massive datasets with sub-arcsecond angular resolution, available for statistical analysis.

Radio astronomy will be a vital part of the above revolution. With ALMA we now have an unprecedented view on the cold Universe, to be complemented by NOEMA soon in the Northern hemisphere. In the m- and cm-range the SKA is designed to survey and monitor the radio sky, providing more sensitive coverage than we have ever seen before. Additionally, dedicated arrays such as CHIME, and HERA are focusing on new phenomena. The current plans for the ngVLA are particularly relevant, as the integration of very long baseline capabilities in this unique project reinforces the role of very high angular resolution in the landscape of future radio astronomical facilities.

At present a number of facilities, such as LOFAR, uGMRT, APERTIF at the WSRT, VLA and MeerKAT are bridging the gap between the past and the future in terms of sensitivity, already challenging data transfer, storage, calibration and analysis methods. It is foreseen that VLBI in general, and the EVN in particular, will be a vital part of the above data revolution. Expected increases in correlator and data imaging capacity, possibly combined with Phased Array Feeds on the larger EVN telescopes, will make imaging over large fields of view a standard capability. This will for instance allow complementary very high spatial resolution centimetre-wave simultaneous imaging of large numbers of weak extragalactic sources. In this way the EVN/*e*-MERLIN can follow up sources first detected by radio survey instruments such as LOFAR and SKA with sub-arcsecond to millarcsecond resolution. Moreover, the EVN will also be flexible to follow up transient events detected either in large field of view radio surveys, or by instruments operating in other parts of the electromagnetic spectrum.

## Key science goals for VLBI in the next decade

This document provides a comprehensive roadmap of science cases whose progress relies on VLBI, and represents the starting point for the technological and operational priorities of the European VLBI Network. Based on the scientific content of the document the following are the main questions, which VLBI can provide a unique contribution to during the coming decade.

WHAT IS THE NATURE OF DARK MATTER AND DARK ENERGY?
Dark matter and dark energy are the key ingredients to understand the evolution of our Universe. Primary goals of VLBI are to determine the nature of dark matter and its distribution, and probe the equation-of-state of dark energy in the Universe. The former will be achieved by high dynamic range VLBI imaging of gravitational lenses, which can reveal the distribution of dark matter halos in a range of mass scales around galaxies. The latter requires measuring the expansion history of the Universe at different epochs. There are a number of ways VLBI can determine $H_0$, the rate of expansion: by spatially resolving $H_2O$ megamasers in nearby galaxies, by observing variability time delays between gravitationally lensed components of high-redshift quasars, and by high-resolution imaging of the electromagnetic counterparts to gravitational wave “standard sirens”, mergers of compact objects (neutron stars and black holes).

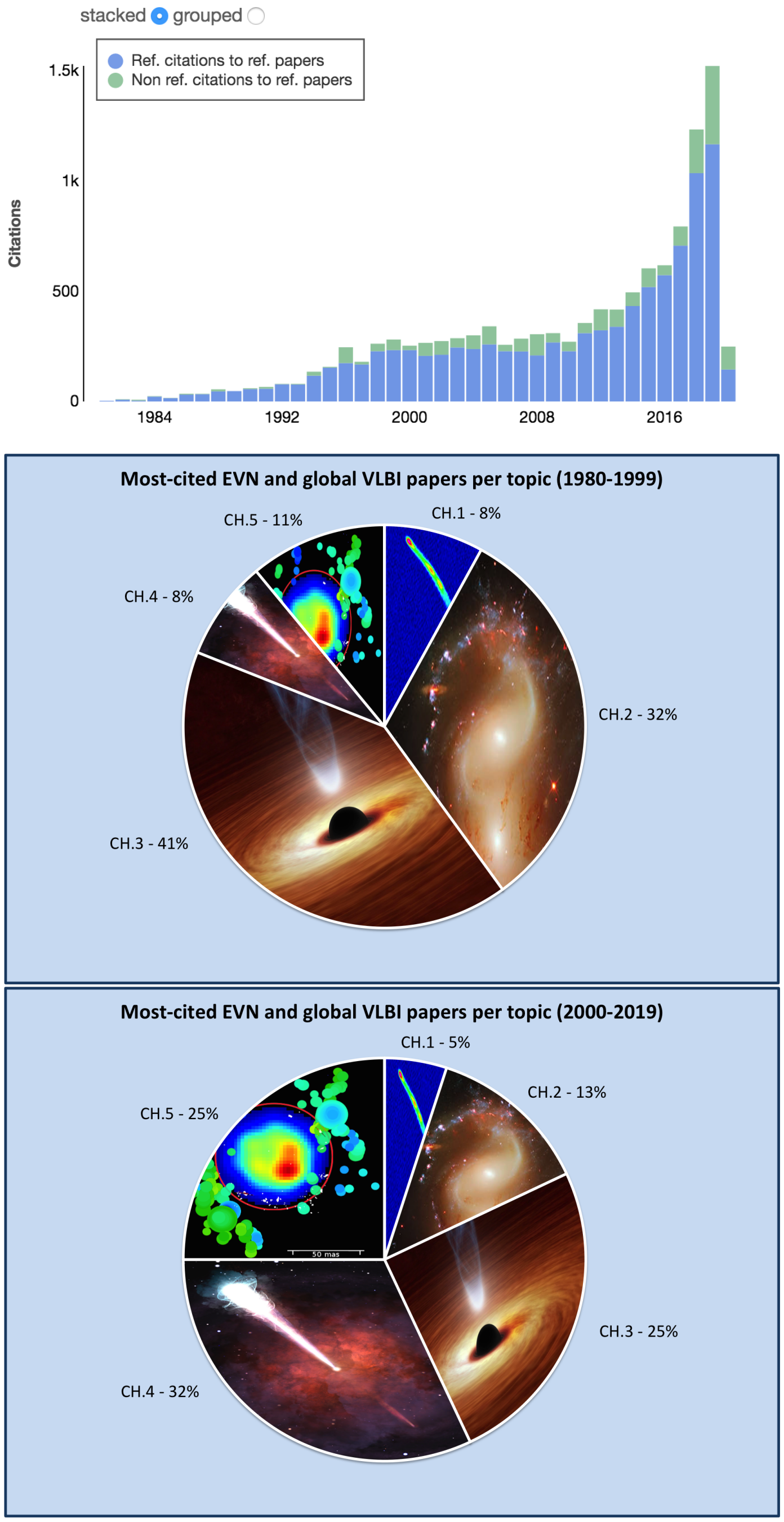

Figure 1: How the science done with the EVN and global VLBI has changed in the past 40 years. (upper panel) The citation history to papers with EVN data. (lower panel) The distribution of science topics addressed by the EVN before and after 2000. Cosmology (CH.1), Galaxies and AGN feedback (CH.2), Blazars and peculiar massive black hole systems (CH.3), Transient phenomena and stellar compact objects (CH.4), Stars/ISM (CH.5), Earth and Space applications (CH.6). Based on the top-100 most cited papers with EVN data, using NASA ADS. See also Chapter 7.

WHEN AND HOW DID THE FIRST BLACK HOLES FORM?
Our knowledge of the co-evolution of supermassive black holes and their host galaxies from the early Universe till today is still limited. At the same time this is a fundamental question for our understanding of the Universe. VLBI provides a very powerful tool to detect nuclear activity from massive black holes, especially in regions of space where emission from other parts of the electromagnetic spectrum is absorbed by dense nuclear dust and gas. This allows us to separate emission from AGN powered by black holes and radio emission from star-formation in the early Universe. It also lets us detect peculiar objects like the elusive intermediate-mass black holes (possible seeds of supermassive black hole formation), and small-separation binary AGN undetectable by other techniques, that are potential sources for low-frequency gravitational waves that will be studied by future space interferometers such as *LISA*.

HOW DO RELATIVISTIC JETS FORM? WHAT IS THEIR IMPACT ON THE HOST GALAXY?
The jet-launching mechanism by supermassive black holes as well as their exact physical conditions are still a mystery. This can be probed by very high resolution total intensity and polarisation VLBI imaging, as well as joint multi-messenger observations, including in particular those with neutrino observatories and high energy particle detectors. Measuring the jet polarisation over a range of angular scales and a very broad bandwidth is crucial to determine the strength and 3-D structure of magnetic fields, but also provides invaluable information on the surrounding ionised medium in which the jets propagate. VLBI also has a unique power to reveal the impact of AGN jets on their large-scale galactic environment ("AGN feedback") by high angular and velocity resolution imaging of the 21 cm spin-flip spectral line of atomic hydrogen, known to be associated with AGN-driven massive outflows.

WHAT IS THE PHYSICS OF EXPLOSIONS FOLLOWING GRAVITATIONAL WAVE EVENTS?
The detection of gravitational waves has opened the door to what is now referred to as multi-messenger astrophysics. VLBI plays an invaluable role in the detection and classification of the radio counterparts of gravitational waves at medium and high redshifts. Particularly relevant are the determination of the morphology, velocity and orientation of any relativistic jets produced and of the accurate positions needed to locate these sources within their host galaxies.

WHAT ARE THE ELUSIVE FAST RADIO BURSTS?
The recent developments in time-domain radio astronomy have led to the discovery of fast radio bursts, whose nature, origin and environment are still unknown. Only VLBI can locate such transients to high astrometric accuracy within their parent galaxies, and has the potential to address their relation, if any, to other non-transient radio emitting sources in these galaxies. This is important to reveal their immediate environment, and eventually the nature of their progenitors.

ARE WE ALONE?
The detection of signatures of extraterrestrial intelligence is a key goal of all the new forthcoming radio facilities, which will provide an enormous increase of the observables parameter space. At that point it will be critical to "clean" the detected signal. VLBI will be crucial to SETI to separate detections from spurious signals, as it is much less sensitive to Radio Frequency Interference RFI than other radio detection methods and exploits consistent sky positions to filter signals. Follow-up of candidate positive signals can successfully be carried out with very sensitive VLBI arrays, which can uniquely locate the origin of any detection, including determining the originating stellar system and the parallax, orbit or other motion of the transmitter.

HOW WAS THE MILKY WAY BORN?
The accurate morphological classification of our Milky Way, the knowledge of the kinematics in its inner parts and the structure of the molecular clouds in its spiral arms are essential to understand and constrain the nature and history of the Milky Way. VLBI uniquely contributes to the above through astrometry of young and evolved stars in the spiral arms or inner regions of the Milky Way. Such observations are highly complementary to those from the *Gaia* satellite which in the galactic plane is limited to stars within a few kpc due to interstellar extinction.

HOW DO STARS FORM? HOW DO THEY IMPACT THE ENVIRONMENT AT THEIR DEATH?
As of today, after almost a century of stellar evolution studies, the processes giving birth to stars are still largely unknown. VLBI is essential to our knowledge and understanding of stellar evolution, particularly in the pre-main sequence and last evolutionary stages. The unique study of several species of masers allows us to disentangle the kinematics of the star-forming clouds. At the same time the monitoring of supernova explosions, which can be followed up for several years, can uniquely discriminate among different models of stellar evolution, and derive information on the environment in which the material ejected by the supernovae propagates.

## Technological priorities for the next decade

The European VLBI Network, as a consortium of individual observatories plus JIVE - which contributes central correlator and central support functions - operates by dedicating observing time, consumables and expertise to the common goals of the network. The Memorandum of Understanding establishing the EVN consortium commits its members to maintain common inter-operable observing capabilities. The development of the array compliant with the scientific requirements implies the agreement on a common strategic technical roadmap. This document is the inventory of selected scientific areas of investigation, as suggested by the user community and constrained by the expected future technical capabilities, which should be taken into account in the development of the future technological roadmap for the EVN. Based on the science cases presented here, the EVN Consortium Board of Directors has prioritised the following changes and upgrades of the network:

- Develop broad-band EVN antenna/receiver systems that are compatible with SKA1-MID on the short term, and can be further upgraded to larger bandwidths and higher frequencies in the long term. This may include both C/X/U (SKA Band5a,b) and possibly the L- ($\sim$SKA Band 2) and UHF bands ($\sim$SKA Band 1).
- Increase the recording bit-rate to at least 32 Gbps to ensure the requested forward leap in sensitivity.
- Increase the number of telescopes operating at frequencies above 22 GHz.
- Consider the VLBI antennas distributed in the African Continent as an attractive option to enhance the European distributed radio astronomy facilities with more optimal frequency and *uv*-coordinates-coverage.
- Develop an EVN correlator platform that is capable of handling a large number of telescopes (N > 30), some of which supply multiple beams, and that has modes optimised for processing a) wide fields of view, b) short transients and c) high spectral resolution modes (for masers, SETI and Spacecraft applications).
- Develop a large field of view archive that provides advanced access to VLBI and other radio astronomy data products in support of the science goals of other instruments such as SKA and LOFAR in the radio and survey data at other wavebands, also complementing efforts at shorter wavelengths like the Global mm-VLBI Array and the Event Horizon Telescope (EHT).

# THE CONSORTIUM BOARD OF DIRECTORS

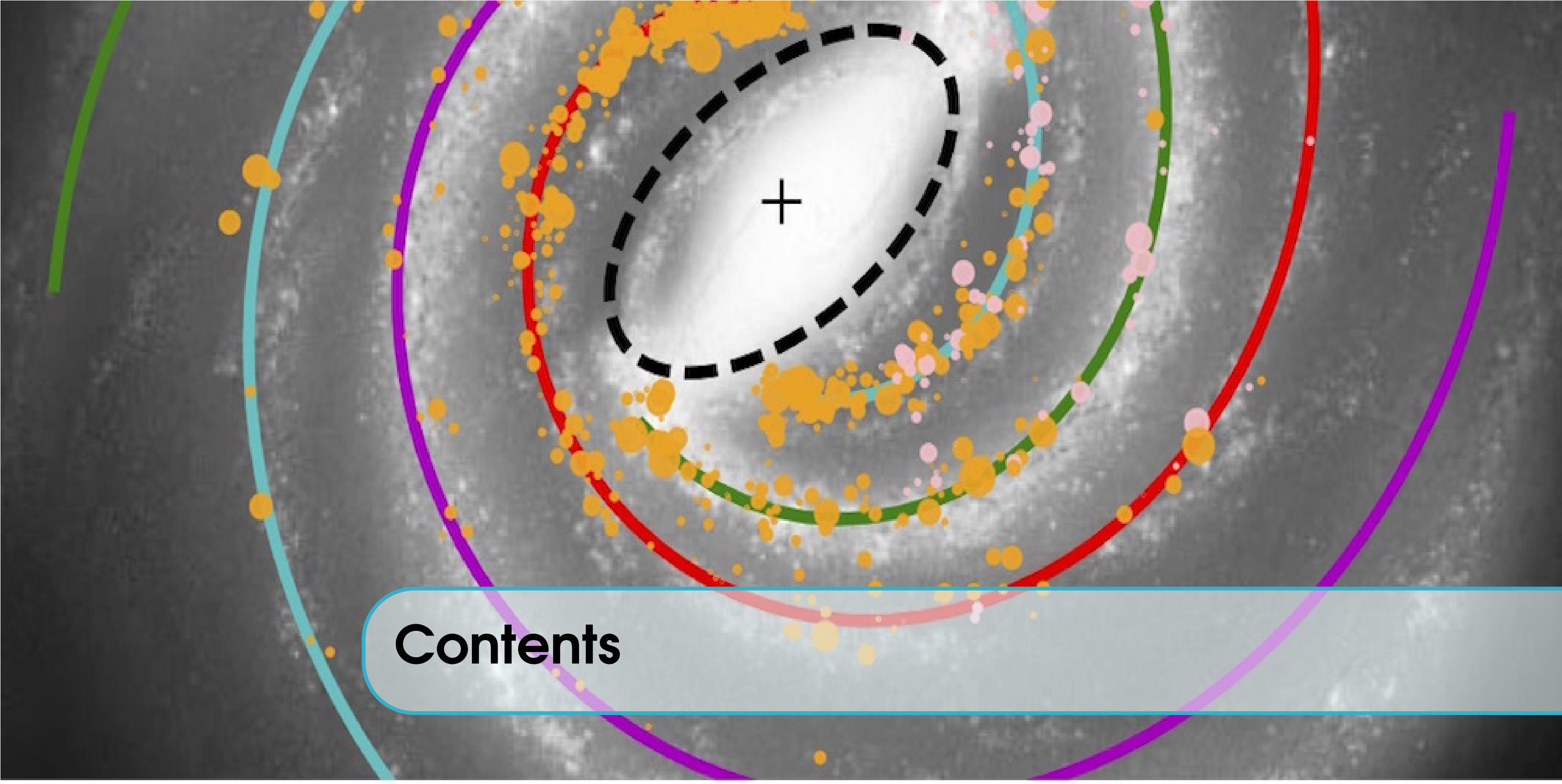

# Contents

Image: Simulated distribution of High Mass Star Forming Regions in the Milky Way. The star forming regions that could be detected by the VLBA and SKA are represented with pink and orange dots, respectively. Credit: background image adapted from artistic impression of the Milky Way (R. Hurt: NASA/JPL/Caltech, J. Green: CSIRO, J. Urquhart: U. of Kent). Galactic simulations implemented by L.H. Quiroga-Nuñez (NRAO/UNM), H. van Langevelde (JIVE/Leiden) and M. Reid (CfA). Quiroga-Nuñez et al., 2017, *A&A* **604**, A72.

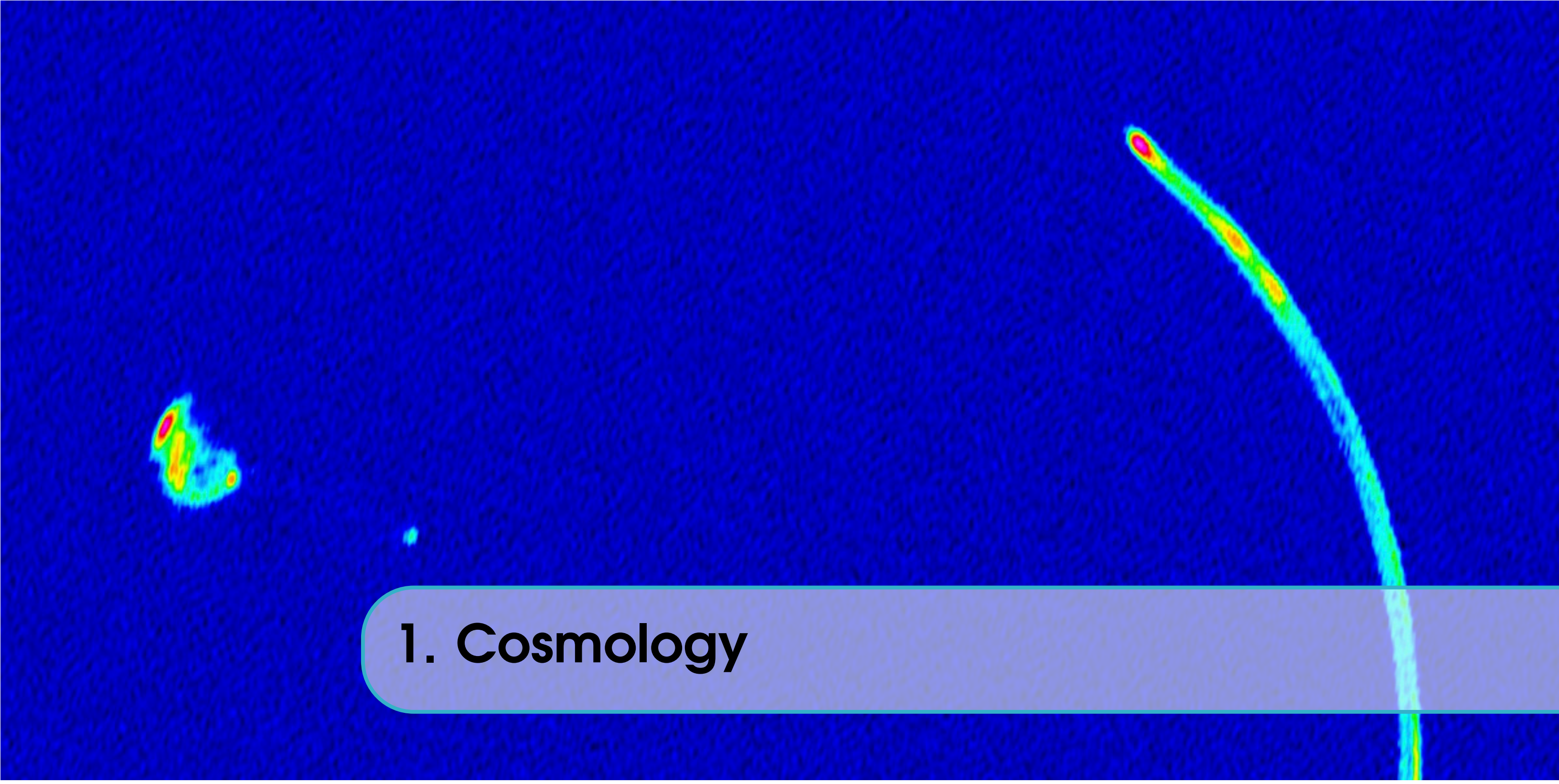

# 1. Cosmology

Understanding the origin and fate of our Universe has been a major theme within the fields of physics and astronomy for over a century, and although tremendous progress has been made through the discovery of the cosmic microwave background (CMB; Penzias & Wilson 1965) and the accelerating Universe (Riess et al. 1998), there are still controversies and inconsistencies that need to be resolved over the next decade. Typically, the structure of our Universe is defined via the cosmological model, which can be further parameterised into a set of fundamental constants that describe the matter, energy and curvature density of our Universe ($\Omega_{\rm M}$, $\Omega_\Lambda$ and $\Omega_{\rm k}$, respectively), the rate of expansion ($H_0$) and the dark energy equation-of-state ($w$). Through observations of the CMB power spectrum (only), most recently by the European Space Agency (ESA) *Planck* satellite (Planck Collaboration 2019), this model has become highly constrained, confirming the concordance of $\Lambda$CDM (cold dark matter) model for cosmology and galaxy formation. In this model, the Universe is close to flat ($\Omega_{\rm k} = -0.056^{+0.044}_{-0.050}$) and is dominated by the energy density of the Universe ($\Omega_\Lambda = 0.679 \pm 0.013$). In addition, the overall matter density ($\Omega_{\rm M} = 0.321 \pm 0.013$) is much higher than the overall baryon density ($\Omega_{\rm b}h^2 = 0.02212 \pm 0.00022$), leading to the need for an extensive dark matter component within our Universe.

Determining the nature of both dark matter and dark energy is now the focus of several on-going and future large-scale ground-based surveys at optical and infrared wavelengths (e.g. Dark Energy Survey, DES; KIlo-Degree Survey, KIDS; Large Synoptic Survey Telescope, LSST) and future space-based missions (e.g. ESA *Euclid* space mission). When the data from CMB power-spectra measurements are combined with the results from these galaxy surveys, which include information from weak gravitational lensing tomography (e.g. Massey et al. 2007), baryonic acoustic oscillations (e.g. Percival et al. 2010) and high redshift supernovae (e.g. Riess et al., 2004), degeneracies between the different parameters can be broken and the overall precision in the measurements is increased (e.g. Hildebrandt et al. 2017). For example, extracting information on the Hubble constant requires some knowledge about the baryon density of the Universe (Planck Collaboration 2019). These

Chapter image credit: Global VLBI (EVN, VLBA and GBT) observations of the gravitationally lensed radio source MG J0751+2716 taken at 1.7 GHz (Fig. 1, Spingola et al. 2018).

next generation surveys are predicted to further tightly constrain the dark energy equation-of-state (e.g. see Fig. 1.1). However, even though such observations are predicted to have a high level of precision, it is likely that the measurements of the cosmological parameters will be dominated by systematics. For example, there is currently tension between the measurement of the Hubble constant made using the CMB and with other probes at the 2.9-$\sigma$ level (e.g. see Fig. 1.1; see also Riess et al. 2018). Therefore, it is important to provide independent, but still competitive constraints to the cosmological model with as many different methods as possible.

Although most cosmological probes involve dedicated wide-field surveys covering a significant fraction of the sky, detailed studies of single objects (or samples of objects) with VLBI have been shown to provide competitive tests for cosmology. Here, the compact radio sources are either standard rulers or standard candles, which can be used to determine the cosmological model via the dependence on the luminosity or angular diameter distances. The best used methods involve imaging on mas-scales the emission from strong gravitational lenses with a variable background radio source (e.g. Wucknitz et al. 2004), the kinematics of water masers within the circum-nuclear accretion disk of a super massive black hole (e.g. Herrnstein et al. 1999) or, as has been shown recently, measuring the expansion velocity of radio-jets associated with gravitational wave events at cosmological distances (e.g. Mooley et al. 2018). A brief summary of these methods and the technical requirements for obtaining competitive constraints to the cosmological model with VLBI are presented in this chapter.

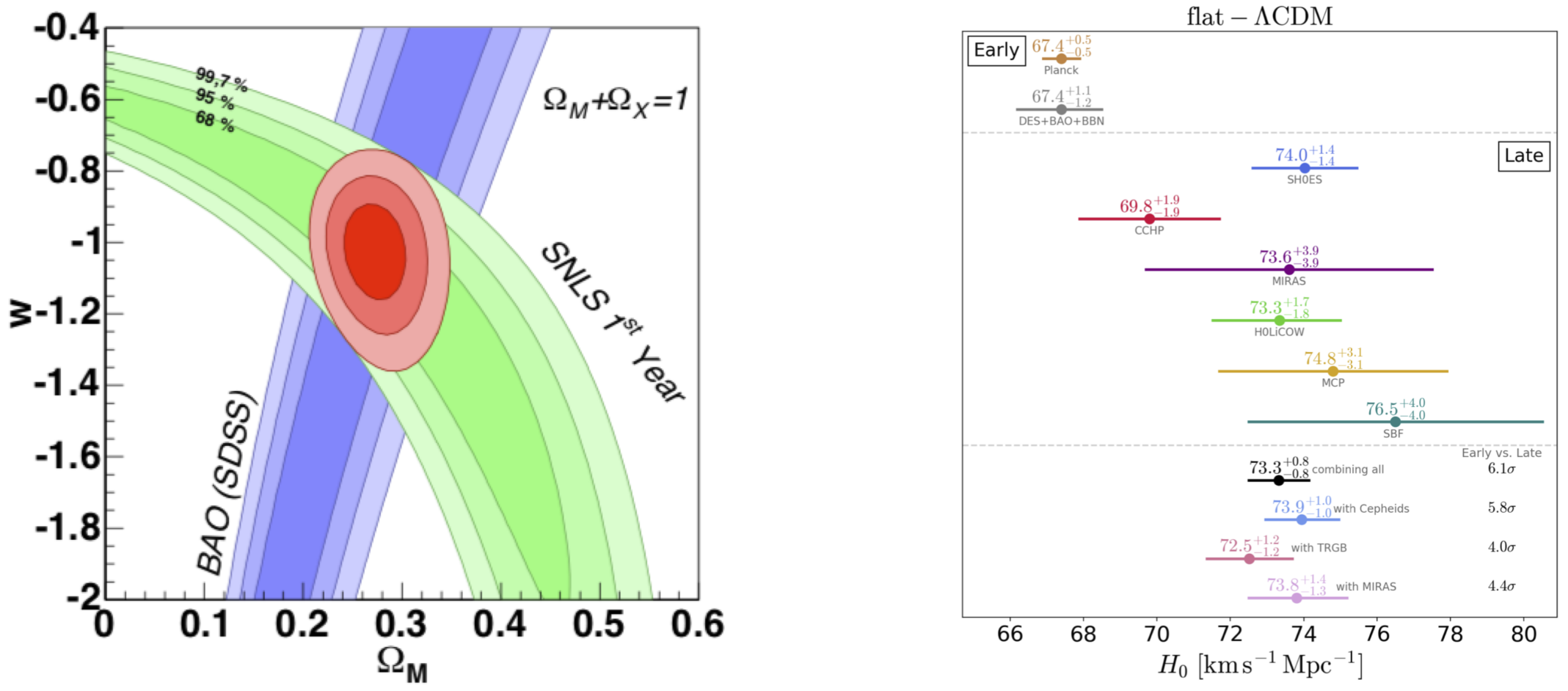

Figure 1.1: (left) The predictions for the dark energy equation-of-state ($w$) as a function of the matter density ($\Omega_\mathrm{M}$), when the results from the *Euclid* supernovae survey (SNLS) are combined with the results from BAO observations from SDSS. Taken from the *Euclid* Science Book (2010). (right) The various current measurements of the Hubble constant from different observational probes (Verde, Treu & Riess 2019).

In addition to constraining dark energy, understanding the nature of dark matter is also a key goal of the galaxy survey experiments described above. Here, the aim is to constrain the parameter space of the different candidate particles for dark matter, which is typically defined as a mass (or energy). In the hierarchical model of galaxy formation, the structure we observe today is predicted to form through the mergers of lower mass haloes, which are dominated by dark matter (e.g. Frenk & White 2012). However, whether dark matter is warm or cold can have a profound effect on the mass structure in our Universe on the smallest-scales, which is best represented by the halo mass function. Simply put, if dark matter is warm, then the formation of structure is suppressed, leading to less low mass haloes in the early Universe that can merge to form larger structures. Also, since warm dark matter haloes are less centrally concentrated, they are more easily destroyed during the merger process. These two effects result in an expected several orders of magnitude difference in the number of dark matter haloes for different models of dark matter on the lowest mass scales (Lovell et al. 2012; see Fig. 1.2). This small-scale crisis in the ΛCDM galaxy formation model has persisted for almost two decades (e.g. Moore et al. 1999). However, observations with VLBI provide a unique probe of galaxy formation on parsec-scales, due to the extremely high angular resolutions that can be achieved. This is typically done through studies of strong gravitational lenses, as the light deflections produced by a $10^6$ $M_\odot$ dark matter halo is of the order of 1 to 5 mas, and is therefore, well matched to global VLBI experiments at GHz frequencies (e.g. McKean et al. 2015). This chapter also provides a brief outline to this method, and the requirements needed to test various models for dark matter over the next decade.

In this chapter, each section describes the various observational probes of cosmology that can be carried out with VLBI and also gives a review of what is predicted to be achievable with the advancements in the sensitivity, field of view and frequency coverage of the EVN over the next decade.

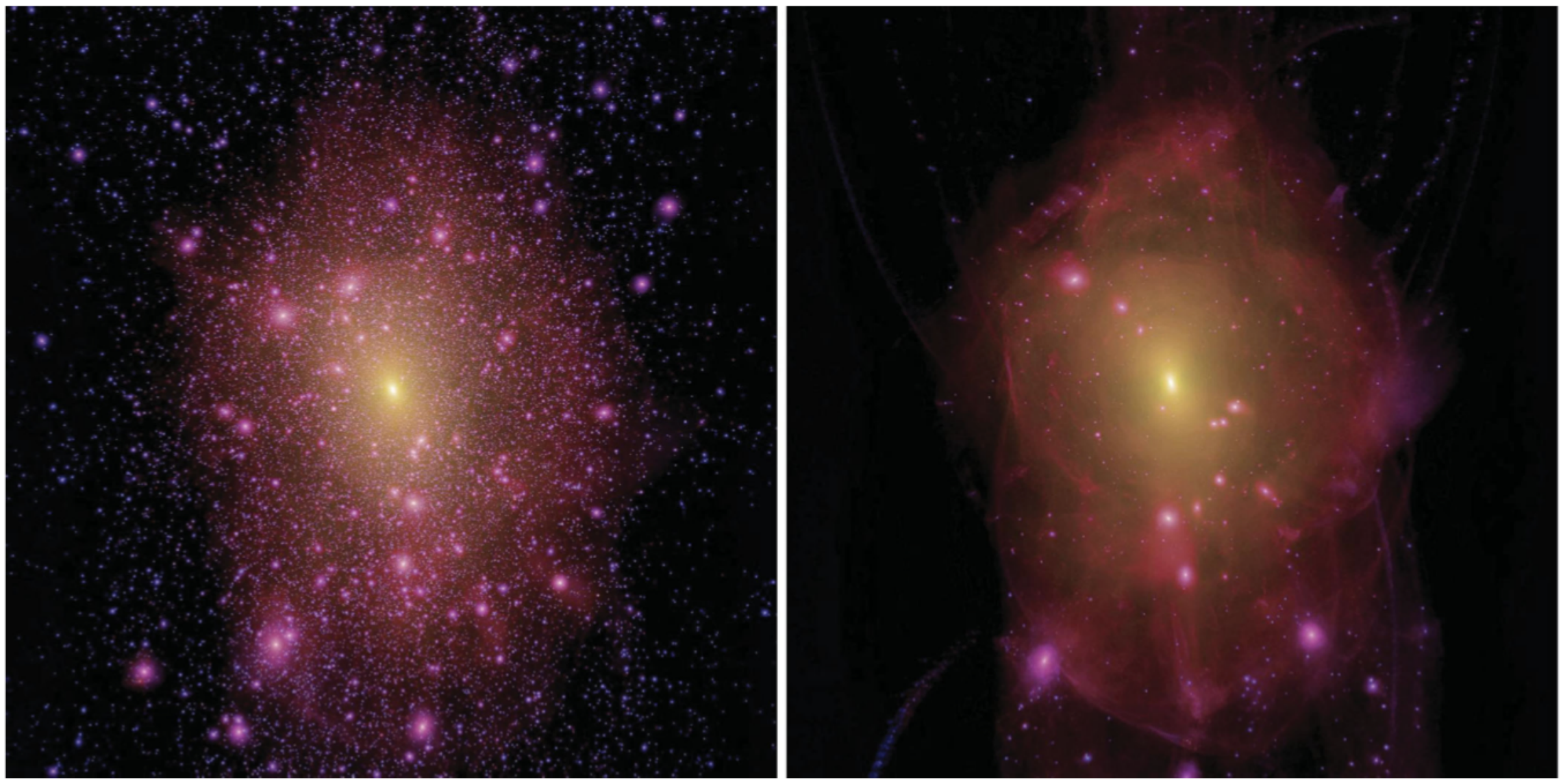

Figure 1.2: The dark matter (only) distribution for a Milky Way sized halo formed in a cold dark matter (CDM; left) and a warm dark matter (WDM; right) Universe (Fig. 3, Lovell et al. 2012).

## 1.1 Revealing the nature of dark matter

Strong gravitational lensing occurs when a distant background object and a foreground massive galaxy (with a sufficiently high surface-mass density) are suitably aligned, the result of which is the forming of multiple images of the background object. By measuring the distortions of the surface brightness distribution of the gravitationally lensed images, it is possible to place tight constraints on the lensing mass distribution, which can be used to test models for the global (baryonic and dark) matter distribution (e.g. Koopmans et al. 2009), the slope and normalisation of the halo mass function (e.g. Vegetti et al. 2012), and the slope of the stellar initial mass function (e.g. Spiniello et al. 2012). Each of these applications uniquely test the cold dark matter model for structure formation on (sub-)galactic scales. In addition, competitive tests of dark energy (w/ BAO, SN1a) can be made when the background object is variable and the (gravitational + geometrical) time-delay is measured between the different gravitationally lensed images (e.g. Bonvin et al. 2017).

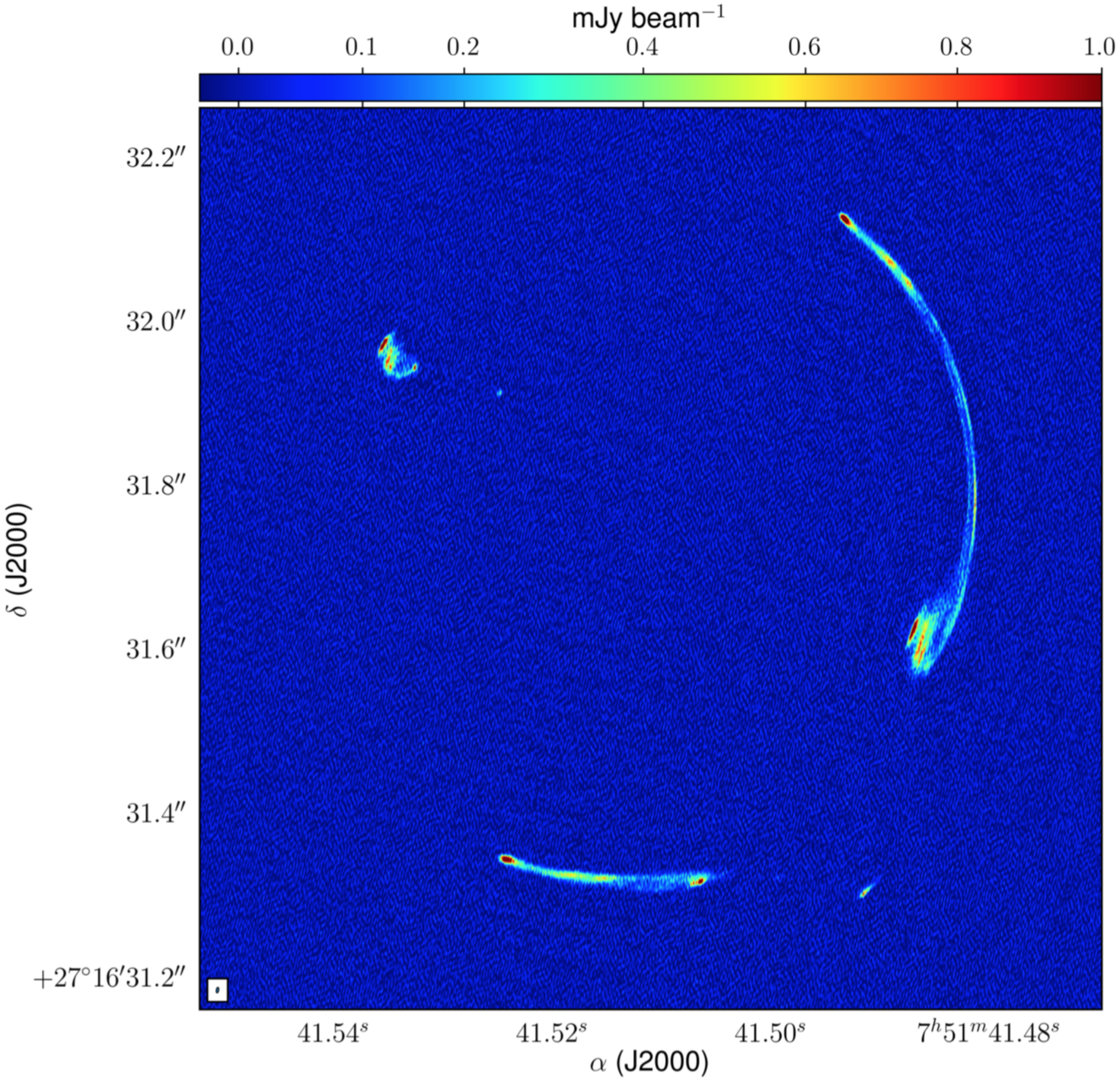

Figure 1.3: Global VLBI (EVN, VLBA and GBT) observations of the gravitationally lensed radio source MG J0751+2716 taken at 1.7 GHz (Fig. 1, Spingola et al. 2018).

Strong gravitational lensing provides a very accurate measurement of the mass within the Einstein radius of the lens, which for galaxy-scale systems is of the order of 0.15 to 2 arcsec, equivalent to a physical projected radius of around 1 to 12 kpc. As lensing is sensitive to the total mass within this radius, it can be used to investigate the contribution of dark and baryonic matter at the inner part of the lensing galaxies. The main astrophysical applications are to i) test hierarchical galaxy formation models by measuring the inner slope of the dark matter halo, ii) test models for dark matter by determining the slope of the dark matter halo mass function, and iii) test models of black hole–host galaxy growth through the detection (or non-detection) of central images. We discuss each of these applications in turn.

### 1.1.1 Probing the inner mass distribution of dark matter haloes

One of the principle applications of strong gravitational lensing is to test models for the mass distribution of galaxies, that is, determining the overall shape (ellipticity) and radial density profile ($\rho(r) \propto r^{\gamma}$, where $r$ is the radius) of massive dark matter haloes. However, as the lensing mass distribution is only probed where the images are formed, only a few constraints (positions and flux-densities of the images) can be used to constrain the overall mass model when the data are at a low-angular resolution and the lensed images are compact. However, when the lensed images are extended in the form of large (tangential) gravitational arcs, or have structure in the radial direction, then the shape and the radial density profile of the lensing mass distribution can be precisely measured (e.g. in the cases of JVAS B0218+357; Wucknitz et al. 2004, HS 0810+2554; Hartley et al. 2019 and CLASS B1933+503; Suyu et al. 2012). A spectacular example of mas-scale observations tightly constraining the mass distribution of a lensing galaxy was demonstrated from 1.7 GHz global VLBI imaging of the extended radio source MG J0751+2716, where the gravitational lens produces large 200 to 600 mas arcs that are highly resolved in both the radial and tangential direction (see Fig. 1.3; Spingola et al. 2018). Here, the data quality is so high that the parameters of the global mass model can be recovered with sub-percent precision; in fact, the limiting factor in the mass modelling is our knowledge of the complex structure of galaxies on a few tens of parsec-scales (see below). In particular, these data required the gravitational lensing galaxy to have a radial mass-density profile with $\gamma = 2.08 \pm 0.02$, which is significantly steeper than for an isothermal mass distribution ($\gamma_{\rm iso} = 2$). This was taken as evidence for the two-phase galaxy formation model predicted from the hierarchical scheme for structure formation; these measurements, such as for MG J0751+2716 and the other lensed radio sources mentioned above, are only possible due to the unique mas-scale angular resolution provided by VLBI.

### 1.1.2 Probing low-mass structure in the Universe

In the case of MG J0751+2716, it was also shown that the positions of the lensed images could not be reproduced by a smooth mass distribution within the measurement errors ($\sim 40\ \mu$as), with rms residuals between the observed and predicted positions of around 3 mas; given the sensitivity and angular resolution of the VLBI imaging, this corresponded to a mis-match between the data and the model at up to the $700\sigma$-level. Another example where the image positions cannot be well-fit by a globally smooth mass model is CLASS B0128+437 (Phillips et al. 2000; see Fig. 1.4), which has four images of a distant background radio source at redshift 3.124 (McKean et al. 2004). When observed at 50 mas angular resolution with MERLIN, the source is found to have four unresolved images, which can be well-fit with a simple elliptical mass model with an external shear component (to account for any complexity in the environment); note that this angular-resolution scale is what is also typically obtained with optical imaging with either the *HST* or with adaptive-optics corrections on 8 to 10-m class ground-based telescopes. However, when observed with VLBI at mas-scale angular resolution, the lensed images are resolved into multiple components, providing many more observational constraints, and now, the different image surface brightness distributions can no longer be explained by a simple smooth mass density distribution (Biggs et al. 2004). The reasons for this mis-match between the data and the model are not clear, but could be due to baryonic structures like massive companion galaxies (e.g. in the case of MG B2016+112; More et al. 2009) or galactic-scale disks (e.g. in the cases of CLASS B1555+375 and CLASS B0712+472; Hsueh et al. 2016, 2017) associated with the main lensing galaxy. The high angular resolution provided by VLBI has allowed the lensing effect of these baryonic structures to be quantified and compared with hydrodynamical simulations of galaxy formation (e.g. Hsueh et al. 2018). Also, the multi-frequency capability of VLBI allows the frequency dependent structure of the lensed images to be measured on parsec-scales at different lines-of-sight through the lensing galaxy. This has typically been used to determine the (differential) electron density of large-scale baryonic components like galactic disks, through measurements of interstellar scattering (e.g. Biggs et al. 2003, 2004) and free-free absorption (e.g. Winn et al. 2003; Mittal et al. 2007), but can also be potentially measured through studies of differential Faraday rotation between the lensed images, as has been done at a lower angular resolution (e.g. Mao et al. 2017).

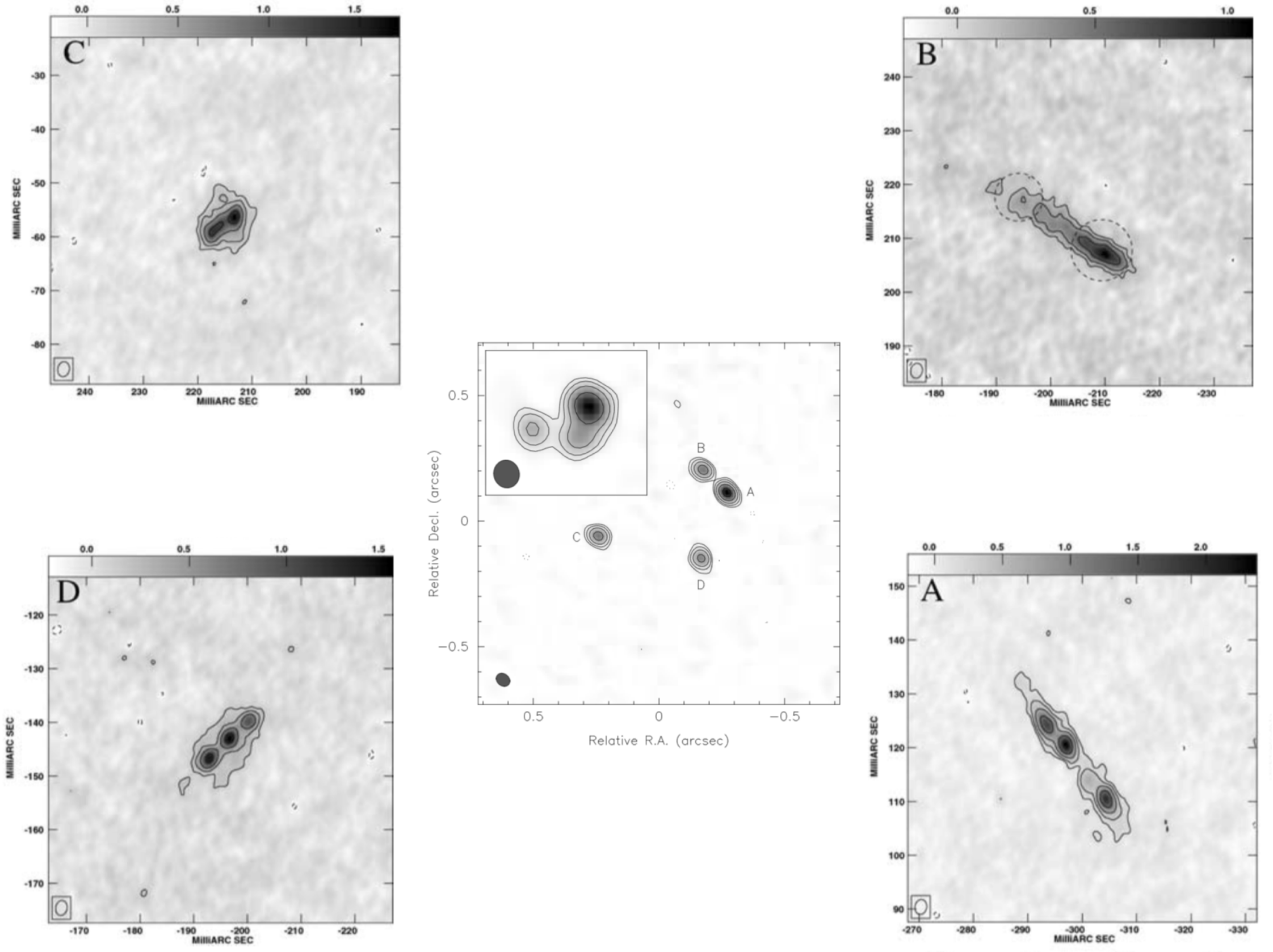

Figure 1.4: Example of the improved angular resolution and information obtained over VLA and MERLIN (centre and inset; Fig. 1, Phillips et al. 2000) when observations are made with VLBI (Fig. 4, Biggs et al. 2004), for the case of CLASS B0128+437.

Understanding the contribution of these baryonic effects to the observed surface brightness distribution (flux-densities and positions) of the lensed images is important if the contribution from clumpy models for dark matter are also to be determined. As shown in Fig. 1.2, different dark matter models predict a varying abundance for low mass haloes both within the lensing galaxy, but also along the line-of-sight to the background object (e.g. Despali et al. 2018). By searching for differences in the structure of the multiple lensed images, deviations from a smooth dark matter model for the lensing potential can be observed. For example, observations at mas-scales with VLBI uniquely probe the deviations caused by a population of $\sim 10^6$ $M_\odot$ haloes, where the discrepancy between warm and cold dark matter models is most significant (see Fig. 1.5 for an example of the sub-halo mass function), and potentially, different prescriptions for warm dark matter can be ruled-out. Again, this is a science application where the unique angular resolution provided by VLBI can be used to confirm that there is a mass perturbation in the lens model by detecting the localised change in the surface brightness distribution of the lensed images, and quantify it through modelling to determine the location and the mass of the perturbing halo (see Fig. 1.5 for a simulation of the perturbing effect a sub-halo, with a varying mass, has on the extended lensed jet-structure from an AGN observed with VLBI). This is also demonstrated in Fig. 1.4, where the relative magnifications of the lensed images, which equates to the relative flux-ratios when the images are unresolved, disagree from what is expected from a smooth mass model. However, when observed at high angular resolution with VLBI, the magnifications of the different images (their relative sizes) are directly measured, which allows the contribution from low-mass haloes to be directly measured. Of course, our ability to detect the perturbing effect of dark matter clumps depends on the signal-to-noise ratio of the data, and how extended the lensed images are. However, detections have been made using both compact lensed images (extended by 5 to 10 mas; e.g. JVAS B1422+231; Bradač et al. 2002) and large-scale lensed radio-jets (extended by 200 mas; e.g. MG J0414+0534; MacLeod et al. 2013). The current limiting factor in this astrophysical application of gravitational lensing is the relatively low number of objects that can be used in the analysis (see below for further discussion).

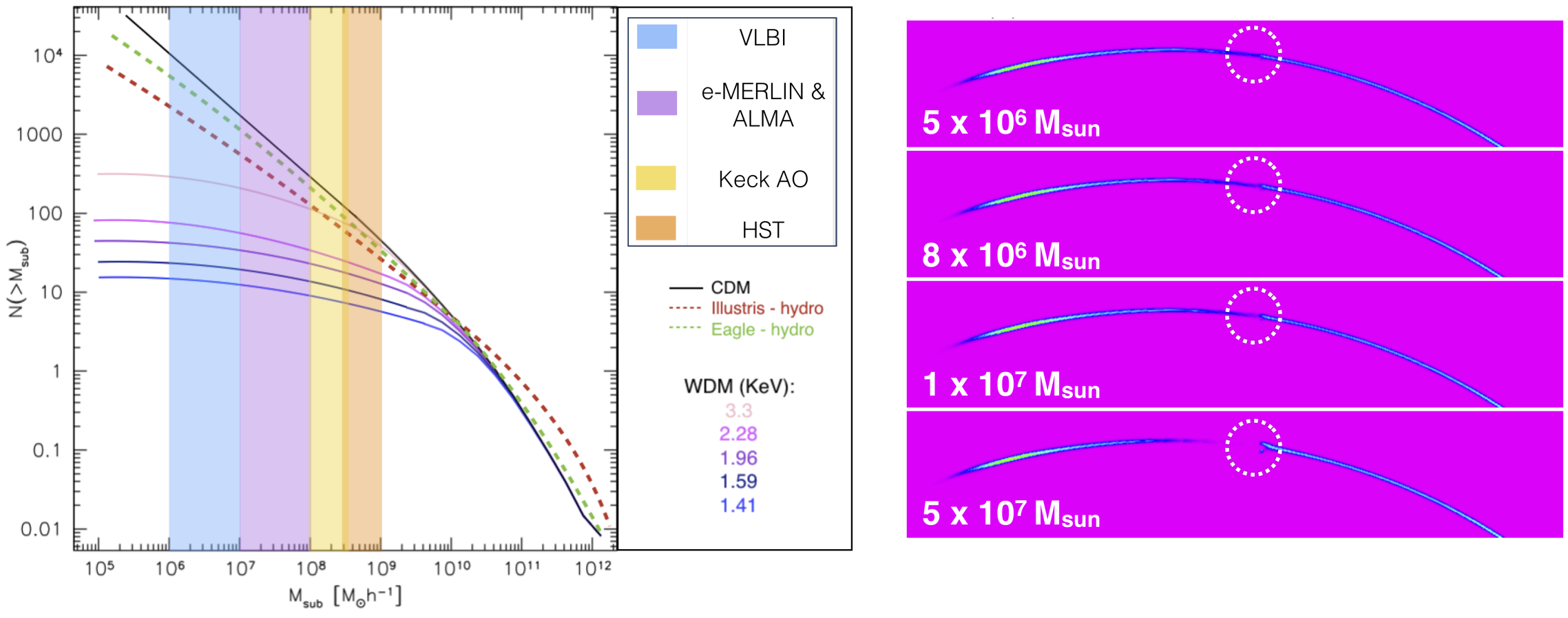

Figure 1.5: (left) The sub-halo cumulative mass function from numerical simulations with different dark matter models (CDM and variants of WDM), and for simulations that include the effect of baryons (e.g. Despali & Vegetti 2017). The coloured regions indicate the varying sub-halo detection sensitivities, given the different angular resolutions of the various instruments. (right) Noiseless simulation of different dark matter halo masses seen against a gravitational arc observed at mas angular resolution (courtesy of Gulia Despali; see also McKean et al. 2015).

### 1.1.3 Measuring the mass function of black holes throughout the Universe

In addition to the baryonic and dark matter structures discussed above, massive black holes that are either free-floating within a galaxy (e.g. Banik et al. 2019) or located at the centre of a galaxy (Mao et al. 2001) can also have a detectable gravitational lensing signature, which provides an additional unique test of cosmology and galaxy formation with VLBI. The former is expected to be extremely rare, given the abundance of free-floating black holes and the cross-section of the lensed images being quite small. However, detecting super massive black holes at the centres of quiescent lensing galaxies is expected to be possible through the measurement of central lensed images. This application uses the property that all non-singular mass density distributions should produce an odd number of images (3 or 5), as opposed to the even number (2 or 4) that are typically observed. The odd image is predicted to be located very close to the centre of the lensing potential, giving valuable information on the inner mass density slope (which probes dark matter; see above), but also on the presence of a super massive black hole. In this case, the central lensed image is further strongly gravitationally lensed by the super massive black hole, producing another set of lensed images that are separated by 5 to 100 mas, for black hole masses between $10^6$ to $10^9$ $M_\odot$, which is a resolution scale that is well-matched by VLBI at cm-wavelengths. To date, only one system has been found to produce a central lensed image (see Fig. 1.6, PMN) J1632−0033; Winn et al. 2004), which demonstrates that such detections can be made. However, given that the relative magnification between highest flux-density lensed image and the central lensed image is predicted to be around $10^{-3}$, the dynamic range of current facilities has made detecting such central lensed images challenging in other systems (e.g. Zhang et al. 2017; Quinn et al. 2016). In addition, multi-frequency data are required to discriminated between a low-luminosity AGN within the lensing galaxy and a genuine central lensed image (see Fig. 1.6).

### 1.1.4 Requirements for probing dark matter with VLBI

To realise the scientific potential of investigating dark matter with gravitational lensing will require increasing the number of lensed objects detected on VLBI-scales from around $\sim 35$ by several orders of magnitude. This is needed to improve on the statistics from studies of individual samples, but also so that rare, special systems can be found for specific science cases. Given the very large sky areas that have already been surveyed for lensed compact objects by the Cosmic Lens All-Sky Survey (CLASS; e.g., Browne et al. 2003; Myers et al. 2003) and the Parkes-MIT-NRAO (PMN; e.g., Winn et al. 2000), this will require more sensitive VLBI arrays in the future. The general requirements for such surveys and follow-up are now discussed.

- **Wide-area surveys with VLBI:** Given the shallow (differential) number counts of compact radio sources (e.g. $n(s) \propto S^{-2}$; McKean et al. 2007), surveys for gravitational lenses are always most efficient when the searches are undertaken over a larger area, as opposed to a greater depth. For this reason, having a wide-field VLBI facility can potentially provide a one-stop survey for lensed objects. This is because the VLBI observations can directly resolve the candidate lensed images, allowing their surface brightness and morphologies to be tested against lens models, as was recently demonstrated by Spingola et al. (2019a), who found two gravitationally lensed radio sources in the mJIVE–20 wide-field VLBI survey. The approximate sky density of faint radio sources at L-band (1–2 GHz; $S_{1.4} > 15$ $\mu$Jy) is around $\sim 750$ deg$^{-2}$. Such objects would be easily detectable within the field of view of the SKA-MID array (phase 1) in a few minutes of integration (e.g., McKean et al. 2015). However, for such objects to be also imaged on VLBI-scales would require the ability to process around $\sim 10^3$ phase centres from each observation. This example is for a low-frequency survey, however, observations at higher frequencies are also advantageous as they can potentially provide a larger instantaneous bandwidth, which is useful for identifying lensed objects through comparing their spectral energy distributions and polarisation as a function of frequency, but also for achieving various science goals (see above). However, observations at higher frequencies (or at lower frequencies with the large antennas of the EVN) are less efficient for surveys given the order of magnitude decrease in the primary beam field of view. This could be mitigated by rolling out phased array feeds on the EVN antennas, which can increase the effective field of view and also provide a more uniform response as a function of position on the sky.

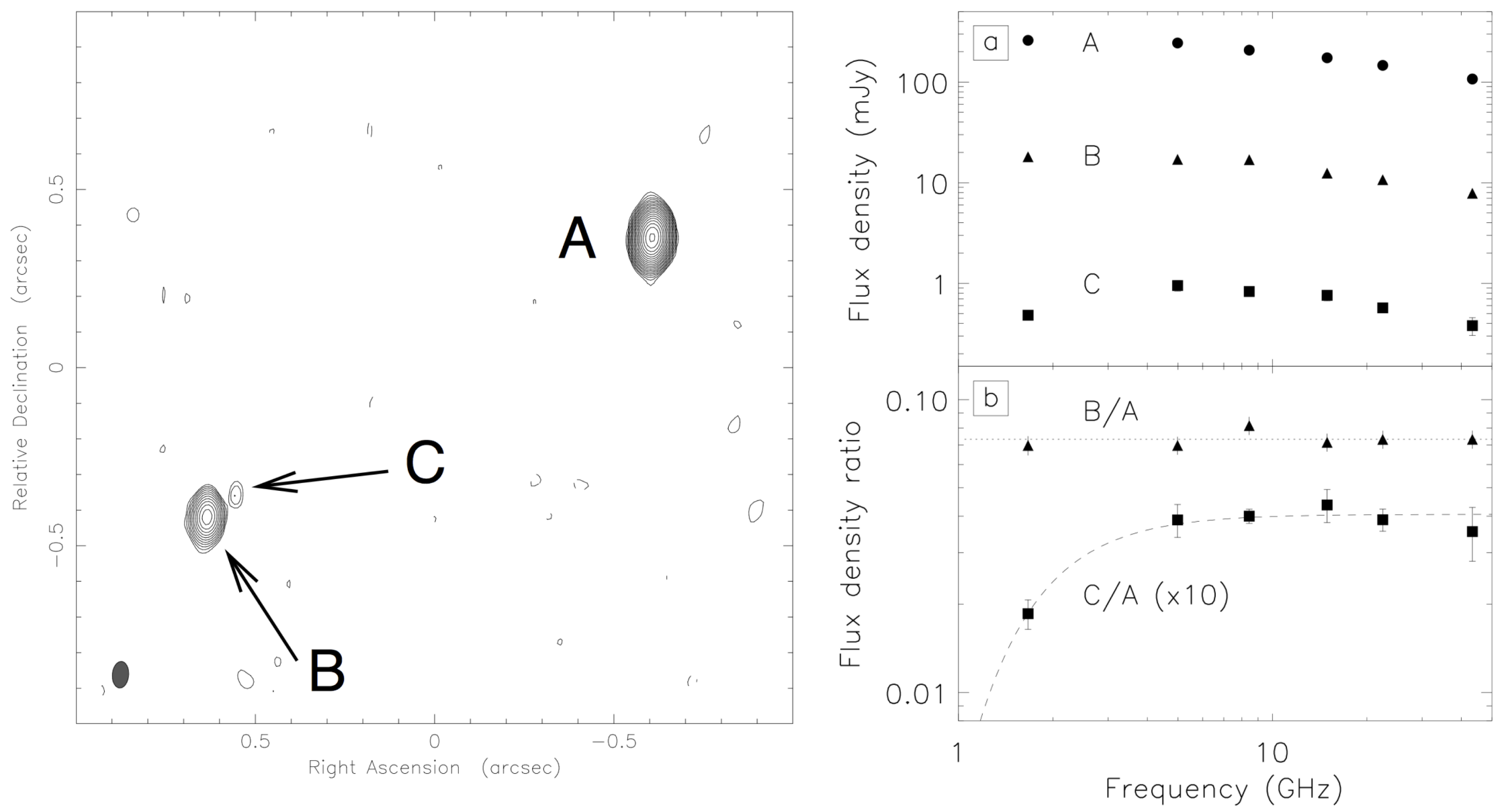

Figure 1.6: (left) A MERLIN 5 GHz image of the gravitationally lensed radio source PMN) J1632−0033, which shows two lensed images (A and B), and an elusive central lensed image (C), which is both highly demagnified relative to the other images and located close to the lensing galaxy position. (right) The radio spectral energy distribution of the three lensed images, showing that at low radio frequencies, image C also shows free-free absorption due to passing through the centre of the lensing galaxy (Winn et al. 2004).

- **Improved sensitivity (thermal noise and dynamic range):** Although wide-area surveys are needed, an improved sensitivity is also required for such surveys to be efficient. This is because even though there will be a large parent population of sources to be surveyed, sufficiently deep observations are also needed to resolve the different lensed images, which can have flux-ratios of the order of unity to around 40. Also, once cases of multiple imaging have been identified, further long-track observations will be needed to increase the sensitivity to extended structure by improving the *uv*-coverage and the surface brightness sensitivity. For example, to image a lensed object similar to that shown in Fig. 1.3, but with a total flux-density at the few mJy-level would require a thermal noise limited dataset with an rms of 0.5 $\mu$Jy beam$^{-1}$. This would require the current set of EVN telescopes to operate with a contiguous bandwidth of around 4 GHz between 1 to 5 GHz, which requires a recording-rate of 32 Gbit s$^{-1}$ (assuming 2 bit-sampling, although increasing this to help mitigate radio frequency interference would also be needed). Such a large bandwidth would also be important for polarisation studies, such as plasma lensing effects, but would improve the overall dynamic range as the *uv*-coverage is better sampled through multi-frequency synthesis.

## 1.2 Measuring the expansion-rate of the Universe over cosmic time

There are several unique measurements of cosmic expansion that can be carried out with VLBI observations. These are mainly related to measuring the Hubble constant, $H_0$, but through measuring the change in the rate of expansion, it is also possible to constrain the dark energy equation-of-state, $w$. There are three main observational channels to measure $H_0$ with VLBI; these are gravitational lenses, water megamasers and from measuring the jet properties of gravitational wave events.

### 1.2.1 Probing dark energy with gravitational lenses

Even before the first discovery of gravitational lensing by Walsh et al. (1979), it was already suggested that if the background source is variable, then there would be a time-delay in the fluctuations observed between the different lensed images (Refsdal 1964). This time-delay is related to general relativistic effects as the light rays pass through different parts of the lensing potential, which requires some knowledge of the gravitational lensing mass model, but is also due to the different path lengths that the light from the different images take towards the observer. The latter is dependent on the Hubble constant, and as such, allows a one-step determination of the cosmological model. Although the first precise time-delays were measured at radio-wavelengths (e.g. Biggs et al., 1999; Fassnacht et al. 2002), high cadence monitoring at optical wavelengths with a dedicated network of small to medium sized telescopes now provides accurate light-curves for a large sample of lensed quasars, allowing the effects of micro-lensing (by stars within the lensing galaxy) and the intrinsic variability of the distant quasar to be disentangled (e.g. Bonvin et al. 2018). This, coupled with sophisticated lens modelling techniques, has provided tests of dark energy that are competitive with those from BAO and SN1a observations (see Fig. 1.1). The SKA is predicted to detect $10^5$ gravitationally lensed radio sources, from which around 1–2% are expected to be variable, providing a sample of $\sim 10^3$ gravitationally lensed radio sources that can be monitored for variability in both total intensity and polarisation. Although VLBI will be required to also provide the precise lens models to determine the gravitational time-delay (see above), there is also the exciting possibility that through regular monitoring at high angular resolution, variability of individual jet components or proper motion of the lensed radio jets can be measured, which would provide a new avenue for determining time-delays (e.g. Spingola et al. 2019b). This would require a dedicated VLBI facility to monitor gravitationally lensed radio sources at a high cadence of a few hours for up to several weeks.

### 1.2.2 Probing cosmology and black holes with water vapour emission at high redshift

The luminous emission ($> 10$ L$_\odot$) from the $6_{16}$–$5_{23}$ water maser line (rest-frequency 22.23508 GHz) is exclusively associated with AGN activity, where the large particle densities ($10^7$ to $10^{11}$ cm$^{-3}$) and temperatures ($> 300$ K) of the molecular gas allow for collisional-excitation to occur (e.g. see Lo 2005 for a review). Also, the strong continuum emission from the powerful AGN jets provide the seed photons needed to stimulate the megamaser emission. Typically, the water megamasers are coincident with the extended radio jets ($<$ 30-pc from the black hole) or the circumnuclear accretion disk ($<$ 1-pc from the black hole), and given the small angular-scales involved, VLBI observations have been vital for mapping the water megamaser regions within nearby AGN (e.g. Miyoshi et al. 1995). In the case of water megamasers related to the radio-jets, it is thought that the conditions for masing are driven by a radiative-shock as the bulk motion of the jet passes through the interstellar medium (ISM) (e.g. Peck et al. 2003). However, water megamasers that are associated with accretion disks are significantly more common (mainly due to observational biases), and these are primarily used to probe cosmology (e.g. Herrnstein et al. 1999), although they also constrain the shape of the accretion disk and provide an accurate measurement of the mass of the central super massive black hole (e.g. Miyoshi et al. 1995). This is done by monitoring the water megamaser lines as a function of time with a large single dish radio telescope to determine the change in velocity as the maser lines orbit the central super massive black hole, from which the centripetal acceleration can be determined. Then, by mapping the positions of the red- and blue-shifted water megamaser regions at the extremity of the disk with VLBI and fitting a (thin) disk model to the resulting position–velocity diagram, the size and structure of the disk can be determined. By comparing the angular- and physical-size of the disk, the water megamaser system becomes a standard ruler, from which the Hubble constant can be derived to a precision of about 10% for an individual system (e.g. Braatz et al. 2010). Thus far, the largest survey for water megamasers has been carried out by the Megamaser Cosmology Project (MCP; Reid et al. 2009) with the VLBA, who have reported measurements of $H_0$ from four galaxies in the Hubble flow (out to 150 Mpc) that have a weighted mean value of $69.3 \pm 4.2$ km s$^{-1}$ (Braatz et al. 2018), which is in good agreement with the value obtained from the CMB (see Fig. 1.1). The final precision is expected to be around 4% when the MCP is completed. However, in order to provide a robust and competitive test of the cosmological model to the 1% level, an improved sensitivity of current VLBI facilities at 22 GHz is required. This increase in sensitivity is also needed to test for systematics associated with un-modelled peculiar motions of the central super massive black hole, and uncertainties in the systemic velocity of the galaxies (e.g. Pesce et al. 2018).

As $H_0$ has thus far been measured using only water megamaser galaxies in the local Universe, this methodology provides only a weak constraint on the dark energy equation-of-state $w$ (see Fig. 1.8). To date, there are only two confirmed detections of AGN hosting water megamasers at cosmological distances; from a type 2 quasar at redshift 0.66 (Barvainis & Antonucci 2005) and from a gravitationally lensed quasar at redshift 2.64 (Impellizzeri et al. 2008). Both of these water megamaser systems are extremely powerful, with intrinsic isotropic luminosities $> 10^4$ $L_\odot$, although whether they are associated with the AGN accretion disk or jets has yet to be determined. However, these detections demonstrate that powerful water megamaser systems can be found at high redshift. By assuming that the slope of the local water megamaser luminosity function does not change with redshift and by using the observed isotropic luminosity of the two detected systems as a normalisation of the luminosity function, McKean et al. (2011) estimated that there are potentially 7 600 water megamaser galaxies per steradian between redshift 1.2 and 4.5 that are detectable with the SKA (Phase-2); an increase of four orders of magnitude over the current sample of known water megamaser systems at high redshift. Such systems can potentially be used to constrain cosmology if they are associated with an accretion disk, and if the high-velocity features can be imaged with VLBI. They would also provide a unique measurement of the black hole mass function within massive galaxies at cosmological distances.

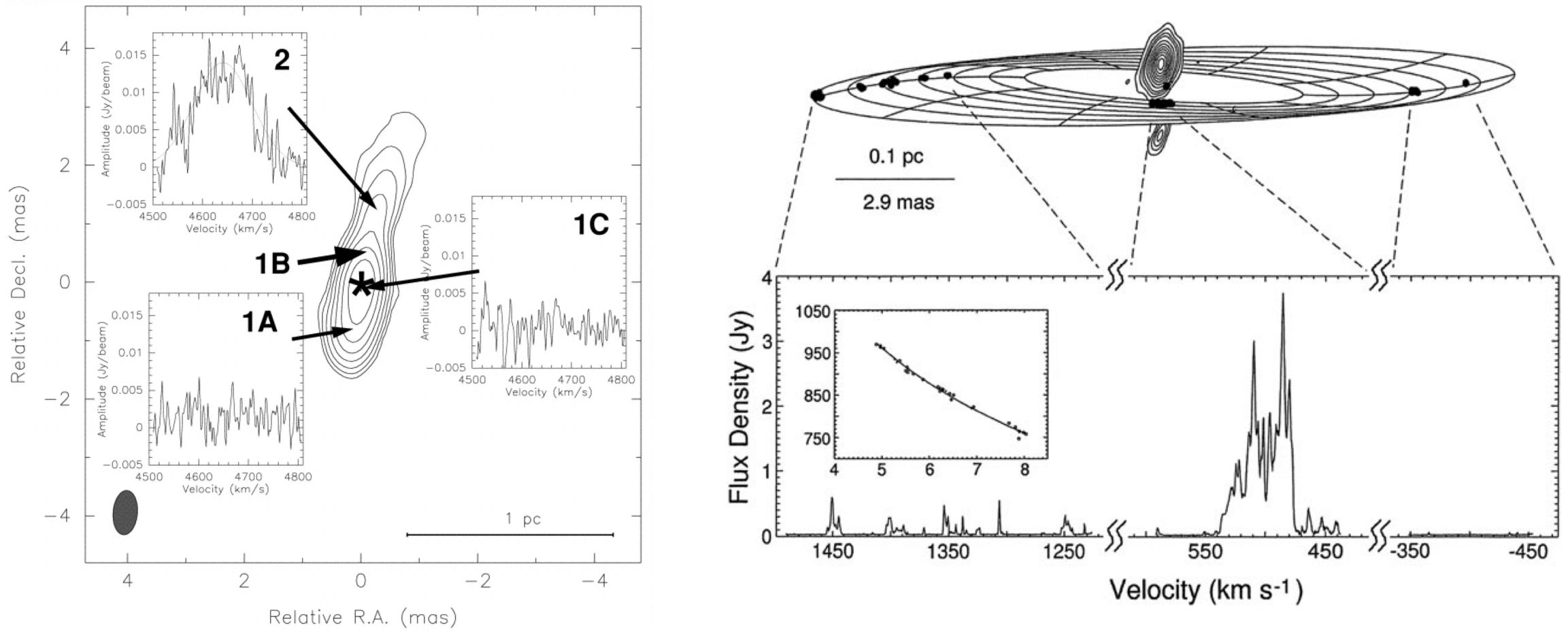

Figure 1.7: (left) Example of the water megamasers associated with the radio jets of Markarian 348, which are thought to be due to radiative-shocks from the jets passing through the ISM of the galaxy (Peck et al. 2003) ©AAS. Reproduced with permission. (right) Example of the water megamasers associated with the accretion disk of NGC 4258, where the systemic lines provide the centripetal accelerations, while the high velocity lines probe the size and structure of the accretion disk (Herrnstein et al. 1999).

### 1.2.3 Gravitational wave events

Recently, a new avenue for determining the Hubble constant has emerged from gravitational wave events, which can be used as cosmic sirens to provide an independent test of cosmology. This is done though modelling the gravitational wave data from neutron star–black hole mergers, under the assumption of general relativity, which provides a number of source parameters, including the luminosity distance and the merger orbital inclination (e.g. Schutz 1986). By combining this measurement with the electro-magnetic data (to localise the galaxy, obtain its recessional velocity and determine its proper motion), the Hubble constant can be derived, for example, in the case of GW 170817, where the Hubble constant was found to be $H_0 = 74.0^{+16.0}_{-8.0}$ km s$^{-1}$ Mpc$^{-1}$ from the Virgo/LIGO dataset (Abbott et al. 2017). However, high resolution imaging of the radio emission from GW 170817 with VLBI also revealed evidence for a narrow, relativistic jet from the source (Mooley et al. 2018). Further observations with VLBI determined the apparent velocity of the jet, which constrained the jet-opening angle from this source. By combining this with the model for the gravitational-wave data, which predicts the distance and the observing angle to the source, and the after-glow light-curve, the parameter space is further limited, and a value of $H_0 = 68.9^{+4.7}_{-4.6}$ km s$^{-1}$ Mpc$^{-1}$ was obtained (see Fig. 1.9; Hotokezaka et al. 2019). By adding the data from the VLBI observations to the gravitational wave observations, the degeneracy between the distance and the observing angle in both methods can be broken. Currently, the precision of the measurement is at the 7% level, but it is estimated that with around $\sim 15$ neutron star–black hole merger events, with similar narrow-jets and small inclination angles to that of GW 170817, where the gravitational wave signal can be measured and the radio emission is resolved, then the overall precision will be reduced to around $\sim 1\%$. As the sensitivity of the gravitational wave detectors increases, and more detectors join the global array, the number of gravitational wave events is expected to increase significantly over the next decade. Having the required sensitivity and resolution from VLBI arrays will also have to be maintained during this period. This science case highlights an important synergy between high quality VLBI imaging and the emerging field of gravitational wave astronomy.

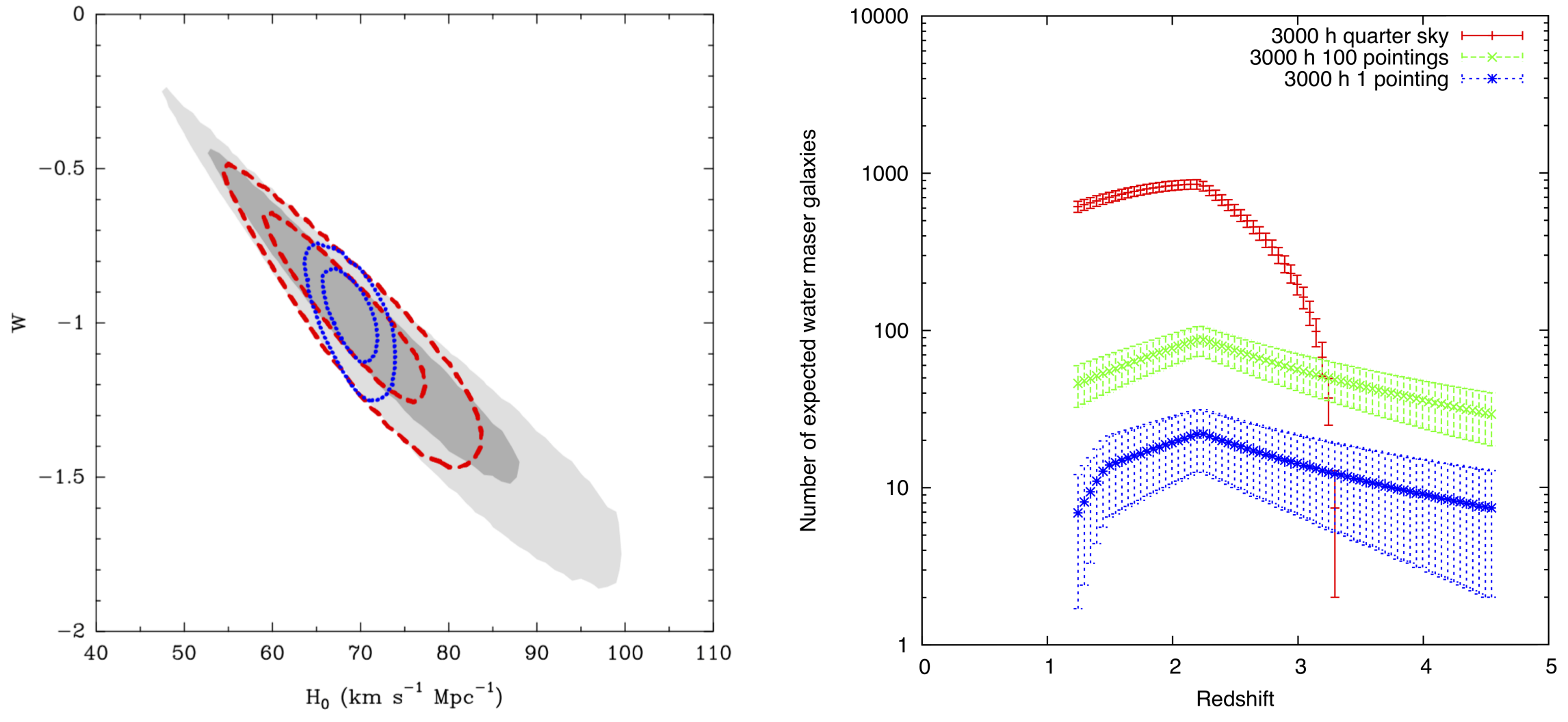

Figure 1.8: (left) The constraints on $w$ and $H_0$ from WMAP (grey region), and the combined constraints from WMAP and the MCP for one galaxy (red), and those expected when combined with MCP data for ten galaxies (blue) (Reid et al. 2013). ©AAS. Reproduced with permission. (right) Estimated number of water megamaser galaxies that are detectable with the SKA (Phase-2) in 3000 h of integration, for a wide-field, medium deep and deep pointings (Fig. 4, McKean et al. 2011).

### 1.2.4 Cosmological tests with precision astrometry

Stochastic gravitational waves deflect light rays in a quadrupolar pattern. The gravitational waves that will produce extragalactic proper motions lie in the frequency range $10^{-18}$ Hz $< f <$ $10^{-8}$ Hz, which overlaps the CMB polarisation and the pulsar timing techniques, but uniquely covers about seven orders of magnitude of frequency space between the two methods (e.g. Gwinn et al. 1997). Measuring or constraining the proper motion quadrupole power can therefore detect or place limits on primordial gravitational waves in a unique portion of the gravitational wave spectrum. Measurement of precise proper motions of super massive black holes residing within AGN enables tests of another cosmological parameter: the observed over-density of galaxies on the scale of 150 Mpc in co-moving coordinates from BAO. At redshift 0.5, the proper motions associated with BAO are of order 1 $\mu$as yr$^{-1}$ (Darling et al. 2018a). Anisotropic expansion of the Universe will cause a pattern with proper motions of $\Delta H/H_0$ 15 $\mu$as yr$^{-1}$ (Darling et al. 2018b). A detection of a signal at the level of 1 $\mu$as yr$^{-1}$ with that pattern will correspond to variations in the expansion rate at a level of 7%.

The temperature dipole of the CMB radiation that is caused by the motion of our Solar system barycentre with respect to the CMB rest coordinate (e.g. Hinshaw et al. 2009) is 370 km s$^{-1}$, which induces a maximum secular parallax of 78 $\mu$as yr$^{-1}$ Mpc$^{-1}$. Proper motions measured with an accuracy of 1 $\mu$as yr$^{-1}$ will allow the detection of a secular parallax within 78 Mpc. To date, there are 88 RFC[1] objects within that distance. Although peculiar velocities of individual objects will cause a bias and variance in the measurement of the secular parallax, the signal has a distinctive correlation structure with quadrupolar, octupolar, and higher-order angular structure due to correlations in the peculiar velocity field. Hall (2019) explores the measurability of the secular parallax in detail.

[1]http://astrogeo.org/rfc/

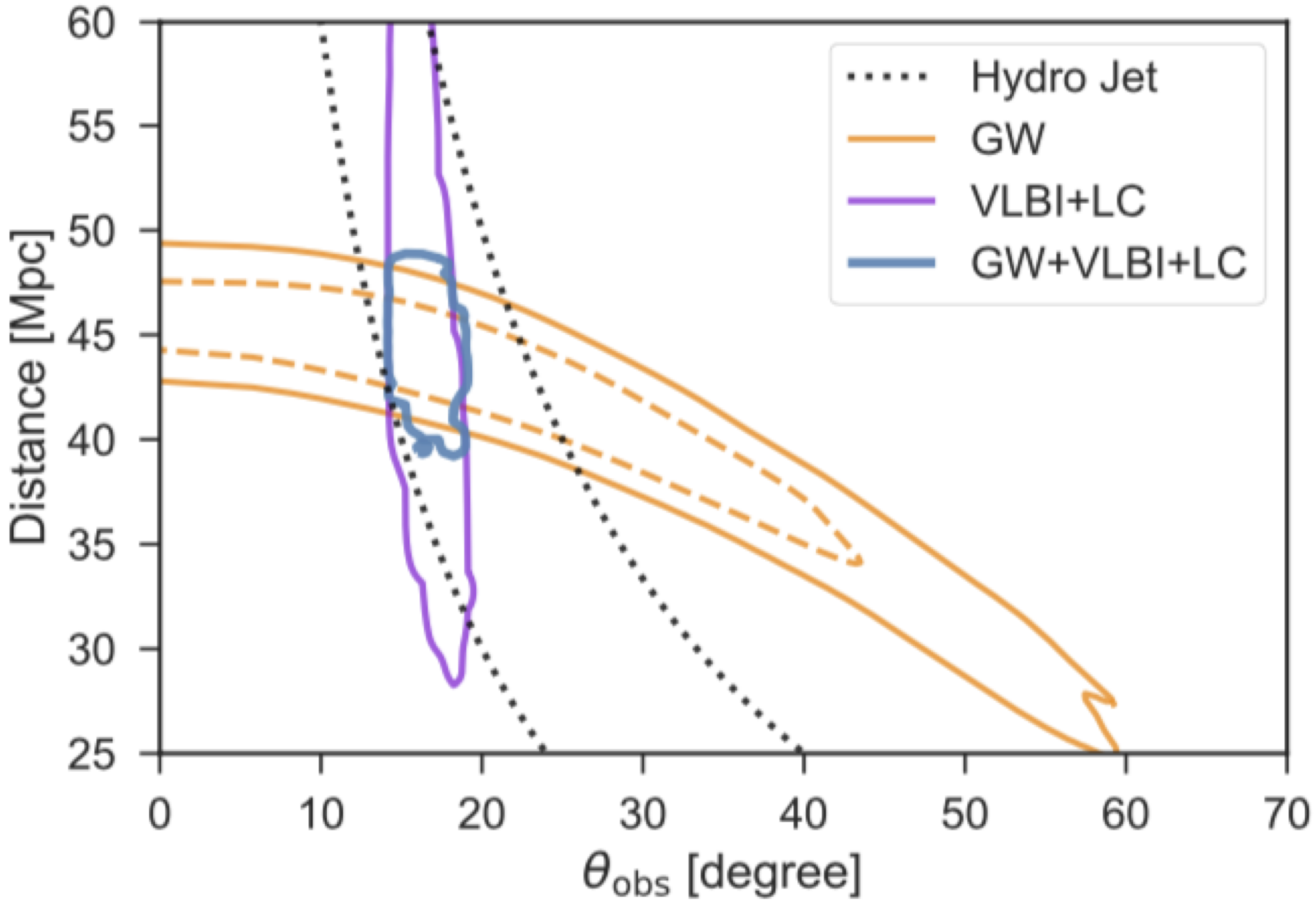

Figure 1.9: The parameter space of the distance and the observing angle for GW 170817 from the gravitational wave data (GW), and from the VLBI and the light-curve data (VLBI-LC). Also shown is the constraint when a hydro-jet model is used. The joint GW+VLBI+LC constraints break the degeneracies in both methods, tightly constraining the distance to GW 170818 (Hotokezaka et al. 2019).

In order to make these cosmology tests with precision astrometry, it is crucial to provide a long time series of VLBI astrometric observations, obtain high fidelity images, and trace variability of the core-shift.

### 1.2.5 Requirements for probing dark energy with VLBI

Similar to the discussion for studies of dark matter above, our ability to test models for dark energy is currently limited by the number of suitable gravitationally lensed sources that can be used for this science case. However, as with the other tests described in this section, an improved imaging sensitivity and dynamic range, and monitoring cadence is needed in the future. We discuss these points here.

- **Imaging cadence:** Observing radio sources at high angular resolution over short and long time-scales is needed for measuring variability in the source surface brightness distribution. In the case of gravitational lensing, this may constrain either the lensing time-delay or the mass distribution on very small spatial-scales, testing models for both dark energy and dark matter. Here, the jet-emission would be strongest a low radio frequencies. As the time-delays in the close merging images of suitable gravitational lenses are of order hours to days, a cadence of hours may be needed to fully exploit this science case in the future. In the case of water megamasers and precision astrometric measurements, high cadence at higher radio frequencies is needed due to the frequency of the water line (even for high redshift sources, see below) and the need to have both precise measurements of the source positions and to determine any core-shift. Therefore, having a continuously operational VLBI facility, with the required frequency coverage between 1 to 22 GHz on all elements of the array, will be needed to enable studies of cosmology with variable radio sources in the future.

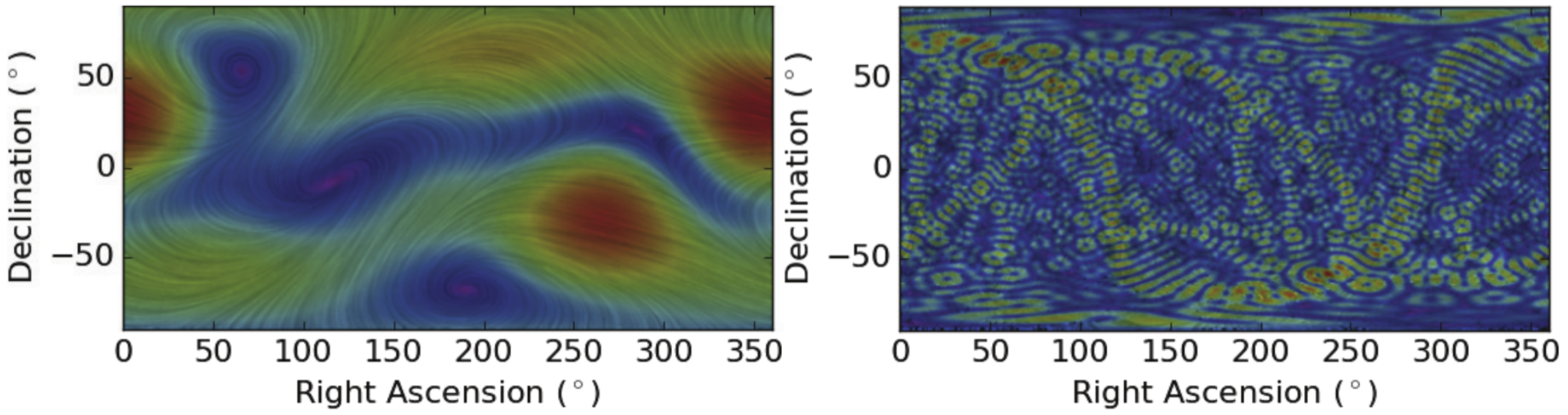

Figure 1.10: All-sky stream plots. Streamlines indicate the vector field direction, and the colours indicate the vector amplitude, from violet (zero) to red (maximum). (left) Randomly generated gravitational wave stream plot. (right) Randomly generated baryon acoustic oscillation streamlines (Darling et al. 2018b). ©AAS. Reproduced with permission.

- **Frequency coverage:** As discussed above, using water megamaser galaxies at high redshift to probe cosmology will require a VLBI array that has continuous frequency coverage from around 10 to 22 GHz, corresponding to water megamaser galaxies within the local volume to redshift 1.2. Currently, the EVN does not have this frequency flexibility and the recent development of broad band receivers (e.g. BRAND) can provide this up to around 15 GHz, which would be sensitive to water megamaser galaxies at redshift $> 0.48$. Also, not all EVN antennas are equipped with 22.2 GHz receiver systems, mainly due to the surface accuracy of those antennas. By increasing the number of antennas in the array that operate at 22 GHz, the overall sensitivity to water megamaser systems in the local Universe will be improved (this is the limiting factor in current studies). Also, to measure the core-shift in radio sources so that their positions can be precisely measured requires simultaneous multi-frequency imaging, for example with a BRAND-like system.

- **Angular resolution:** Observing radio sources at higher redshift, such as in the cases of water megamaser galaxies, has the advantage of better probing the dark energy equation-of-state and also limiting the systematics associated with peculiar motions of the target source. However, observing at higher redshift also results in a smaller physical scale being observed. For example, in the case of UGC 3789 at $49.9 \pm 7$ Mpc, the accretion disk is about 2 mas in diameter, which is equivalent to around 0.5 pc (i.e. about 20 synthesised beams). For the same galaxy at redshifts 0.5 and 1, the angular-scale changes by a factor of between 25 and 35, meaning that the water megamaser components will no longer be resolved (disk sizes of about 0.06 to 0.08 mas). However, the larger volume of the Universe that is probed will likely find systems that are associated with more massive black holes, which may also result in supporting larger accretion disks. In these cases, an increase in the black hole mass by two orders of magnitude would also increase the accretion disk size in a similar way. Such systems would be resolvable with a global VLBI array operating between 10 and 22 GHz. For lower mass black holes, space-based VLBI would be needed to achieve the required angular resolutions. Alternatively, there may be a few special cases at very high redshift where the water megamasers are also gravitationally lensed. In such cases, the lensing magnification may help overcome the lower angular-scales at such large distances, but would also provide two routes to constraining cosmology, one via the lensing time-delay and the other via the water megamaser geometric distances.

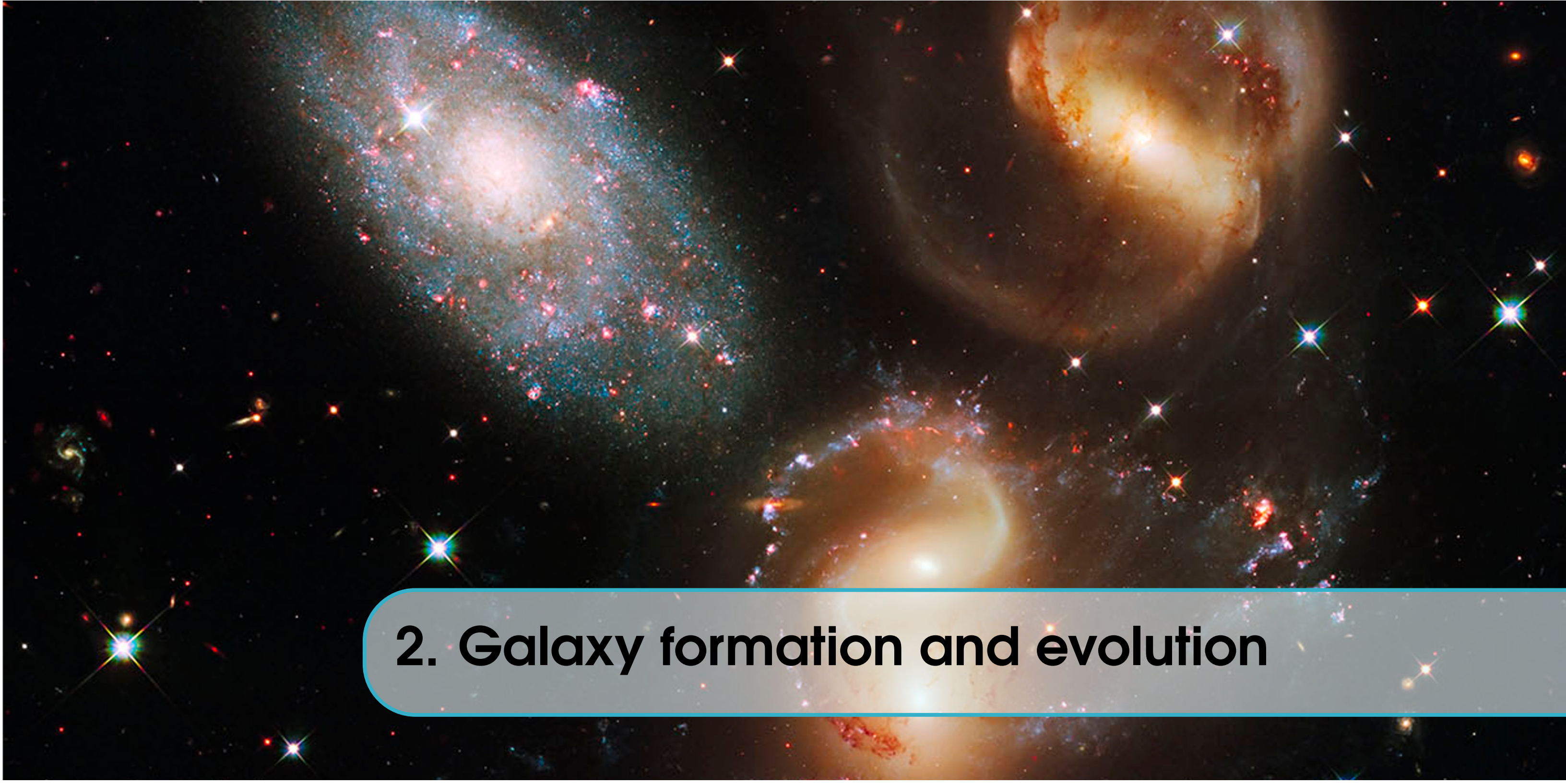

# 2. Galaxy formation and evolution

## 2.1 Star-formation, accretion and feedback over cosmic time

### 2.1.1 Galaxy formation

One of the most important questions regarding the evolution of our Universe is how the rate of star-formation has changed throughout cosmic time – from which the history of the assembly of galaxies can be derived. Star-formation rates within individual galaxies can be traced over a large range of wavelengths in both continuum - for example ultra-violet and infrared wavelengths, and line emission such as H$\alpha$. However, in the cases of many of these tracers, the dust obscuration which is so prevalent in the most massive star-forming galaxies often results in reduced estimates of the star-formation rate, which then require significant correction terms. In addition, at high redshifts line tracers such as H$\alpha$, are transformed to parts of the electromagnetic spectrum that are not easily observed with ground-based telescopes. A recent excellent derivation of the evolution of cosmic star-formation rate density with time from Madau & Dickinson (2014), incorporating our best obscuration correction terms is shown in Figure 2.1 (left & centre).

There is a continued debate regarding both the dust obscuration in high redshift galaxies and derived corrections applied to observed ultraviolet UV emission at redshifts above 2, in addition to the contribution to the star-formation rate in the infrared (IR) from low luminosity star-forming objects (e.g. Katsianis et al., 2017). The resulting slope of the star-formation rate density (SFRD) distribution at high redshift, which is dominated by rest-frame UV data, appears somewhat flatter than that derived from IR data at redshifts above the peak of the distribution which lies close to z=2. In addition, the extinction-corrected optical/UV SFRD compilation by Hopkins & Beacom (2006), and SFRD estimates derived from high-redshift $\gamma$-ray bursts (GRBs) by Kistler et al. (2009) both suggest that the slope of the SFRD distribution above z$\sim$2 may be flatter still. This range of possible SFRD distributions is illustrated in Figure 2.1 (Right) which is taken from Gruppioni et al. (2017).

Both thermal and non-thermal radio emission can be related to the ongoing star-formation rate in galaxies at all redshifts and uniquely offers an un-obscured tracer of the current star-formation. This relationship is underlined by the observed correlation over 4 orders of magnitude in luminosity

Chapter image credit: Stephan's Quintet – HUBBLESITE.org.

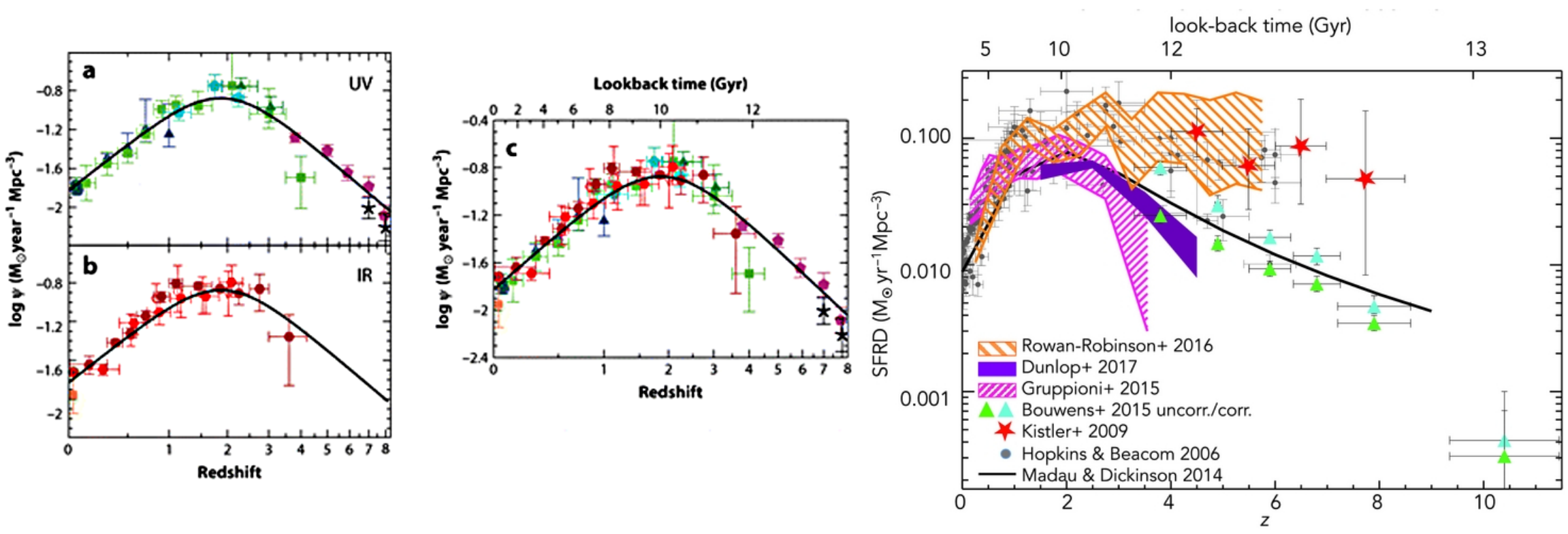

Figure 2.1: (left) The history of cosmic star-formation from (a) FUV, (b) IR, & (c) FUV+IR rest-frame measurementsc(Madau & Dickinson 2014). (right) Redshift evolution of co-moving SFRD. Different derivations of the obscured and unobscured SFRD are compared (Gruppioni et al. 2017)

between the radio and infrared emission, both of which are related to star-formation; the infrared emission emanating from dust heated by photons from young stars, whilst the radio emission arises from synchrotron radiation due to the acceleration of charged particles produced in supernova explosions. Deep radio surveys of the sky are able to detect many thousands of faint, star-forming galaxies out to several redshifts and allow a definitive radio-based, and hence obscuration-free, census of the star-formation rate history of the Universe – provided that radio emission from AGN-jet activity can be separated from that associated with star-formation.

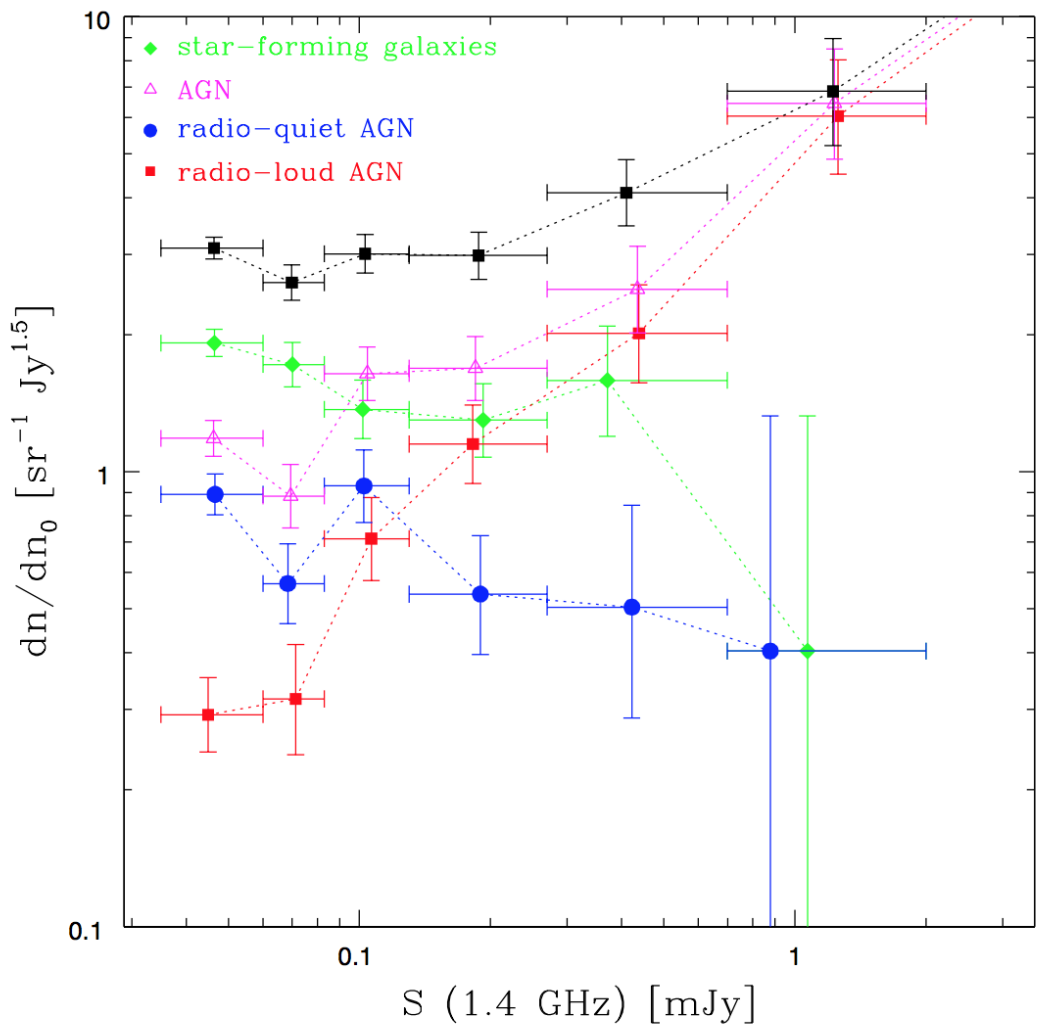

Figure 2.2: Euclidean normalised 1.4 GHz E-CDFS source counts for the complete sample (black filled squares) & various classes of radio source delineated by colour (Fig. 1, Padovani et al. 2015)

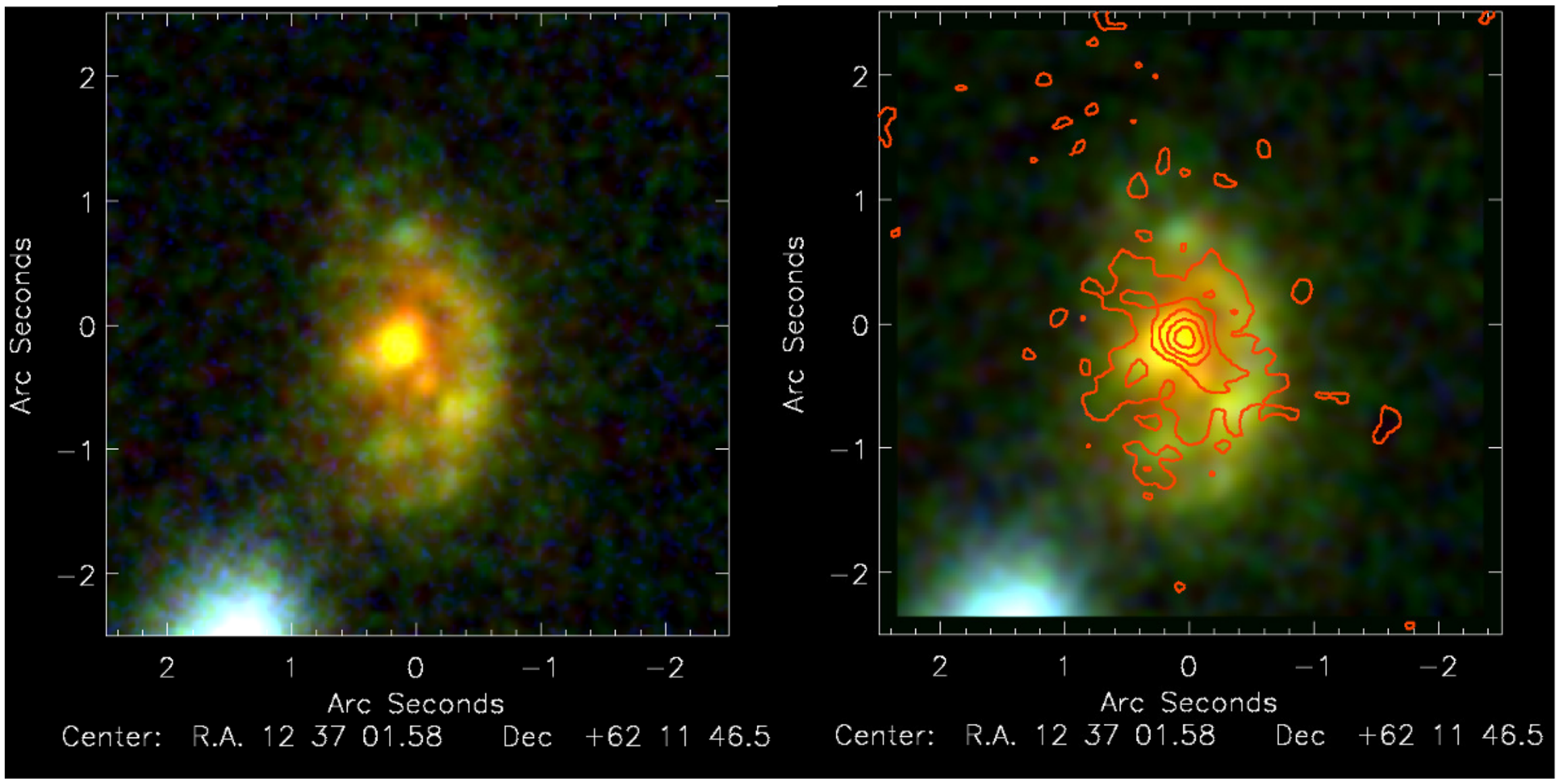

Figure 2.3: (left) Sub-mm very red dusty Seyfert galaxy GN17 at $z = 1.76$. S-F rate $\sim$720 $M_\odot$/yr. Possible extended SFG with a nuclear starburst. (right) *e*-MERLIN+JVLA 1.4 GHz image overlaid CI=4.5 $\mu$Jy/bm (Beam 280 $\times$ 262 mas). Courtesy of Tom Muxlow.

### 2.1.2 Nature of the Faint Radio Population

At low 1.4 GHz flux densities the radio source population evolves significantly with star-forming galaxies starting to dominate the source counts over radio-loud AGN at flux densities below 100$\mu$Jy. In addition, radio-quiet AGN systems begin to dominate the AGN population detected as shown in Figure 2.2 from Padovani et al. (2015) from deep VLA observations of a sample of radio sources in the Extended *Chandra* Deep Field-South. Padovani et al. identify the radio-loud AGN population as associated with 'jet mode' accretion with bright compact cores and extended radio jet structures on sub-galactic scales; and the radio-quiet AGN with 'radiative mode' accretion, the local luminosity function of the latter being not inconsistent with Seyfert galaxies. Star-forming galaxies clearly dominate the radio populations at $\mu$Jy flux density level and contain steep-spectrum extended radio components on sub-galactic scales. Similar extended radio emission is also seen in many of the radio-quiet sources indicating that the radio emission in such systems may also be dominated by that generated in star-formation processes with AGN activity mainly detected in wavebands other than the radio. However, the high resolution radio imaging to date shows a variety of radio structures for the radio-faint AGN population with the implication that the population is heterogeneous and that higher angular resolution still is required to reliably separate any embedded AGN-jet structures from extended radio emission from star-formation regions across the host galaxy.

#### Separating AGN and star-formation related radio emission

VLBI with mas-scale angular resolution has historically been used as a definitive tracer of radio-loud AGN activity with its ability to identify very high brightness temperature compact core radio components associated with AGN 'jet-mode' accretion activity, together with the inner extended jet components ejected from such activity. Sub-arcsecond resolution imaging from compact VLBI arrays such as *e*-MERLIN, and the JVLA at higher frequencies, have been used to resolve and separate AGN-jet structures from extended regions of star-formation in a number of multi-band deep fields (e.g. Richards et al. 1998, Chapman et al. 2003, Morrison et al. 2008, Smolčić et al. 2017) allowing obscuration-free estimates of star-formation rate (SFR) to be derived. The high resolution imaging has also shown that luminous star-forming galaxies (SFGs) are likely to contain compact nuclear starbursts in addition to extended regions of star-formation. These nuclear starbursts are resolved by *e*-MERLIN at L-Band (Muxlow et al. 2005) and the JVLA at 10 GHz (Murphy et al. 2017), but not detected by VLBI at mas resolution to present sensitivity levels. Murphy et al. find that the median FWHM size of the nuclear starbursts from a sample of 32 SFGs at 10 GHz in GOODS-N is $\sim$170 mas, whereas *e*-MERLIN+JVLA 1.5 GHz imaging shows both the nuclear starbursts in addition to extended steep-spectrum star-formation related emission with resulting median angular size values of $\sim$1.2 arcsecond. Some such systems are also associated with AGN activity in other wavebands (e.g. X-rays, optical broad-lines), complicating the radio source classification within the scheme shown in Figure 2.2, blurring the division between the SFG and radio-faint AGN populations. This issue is illustrated in Figure 2.3 where the z=1.76 dusty sub-mm galaxy GN17 identified with the *Scuba* source HDF 850.6 has historically been interpreted as a system undergoing a burst of intense star-formation through SED fitting from modified local ULIRG profiles (e.g. Pope et al. 2006). The faint *Chandra* X-ray detection also favours star-formation rather than an AGN interpretation. The sub-mm and IR emission from GN 17 indicate very significant reprocessing of UV radiation by dust; however, in spite of the large amount of dust obscuration, GN 17 possesses a bright optical nucleus and the 270mas resolution radio image shows a morphology which could be interpreted as either an extended SFG with a resolved nuclear starburst or a SFG with an embedded AGN-jet system. The relatively compact radio component overlies the bright optical nucleus, is not detected by the EVN at mas resolutions, and possesses a brightness temperature ($T_B$<$10^3$K) - well below that expected for a radio-loud AGN core - so the SFG interpretation remains the most likely, but it is unclear what contribution UV radiation from the bright optical nucleus makes to the IR flux density of GN 17.

Intermediate resolution imaging between full VLBI and *e*-MERLIN angular resolutions is required to investigate in detail the radio structure of GN 17 and many other potential SFG/AGN hybrid systems seen in deep radio fields - and thus estimate any potential AGN contamination to the SFG radio flux densities. The EVN has pioneered combination imaging with *e*-MERLIN, and recent developments have allowed full integration of *e*-MERLIN antennas into the EVN, permitting radio imaging over such intermediate angular scales between 30 - 150 mas. Such imaging will also investigate AGN feedback on the detailed structure of the high luminosity SFGs nuclear starbursts in the presence of AGN active in other wavebands - in addition to resolving and imaging any embedded faint AGN-jet radio components which may also be present and contributing to the feedback processes.

#### The Faint Radio-Loud AGN Population

The radio structures of the $\mu$Jy radio-loud AGN population as imaged with *e*-MERLIN+JVLA at L-Band (Muxlow et al. (2005), *e*-MERGE Consortium - private communication) show that both one- and two-sided extended radio structures are seen on sub-galactic scales associated with a central active compact AGN radio core (See Figure 2.4). This is in contrast to the classical double structures found on either side of the host galaxy which are typically associated with radio-loud AGN sources with 1.5 GHz flux densities $\gtrapprox$ 1mJy. Such systems may be young emerging AGN, or perhaps sources where the radio jets are disrupted by entrained circum-nuclear material in the nuclei of host galaxies that have yet to be swept clear of star-forming gas and dust by AGN activity. In GOODS-N the *e*-MERGE sub-arcsecond study of radio-loud AGN sources identifies the brighter $\mu$Jy examples of the population from both spectral properties and EVN detections at full mas resolution of bright core components with brightness temperatures $>10^5$K (Chi, Barthel & Garrett 2013, Radcliffe et al. 2018). This selection favours systems aligned to the line-of-sight and thus initial one-sided jet structures would be expected if the jets are moving with relativistic velocities. Any jet entrainment from circum-nuclear material would result in disruption of the jet flow and deceleration of the jet plasma resulting in the appearance of two-sided jets. Intermediate resolution EVN+*e*-MERLIN imaging on scales of 10s of mas at L-Band are required to investigate the scales on which jets are one- or two-sided and the implied scales on which the initially one-sided jets are slowed. In Figure 2.4 (left) the insert image from archival VLA and MERLIN L-Band data demonstrates that for J123642+621331 the jet starts off one-sided and becomes two sided on scales between ~100 - 250 mas (~0.8 - 2 kpc).

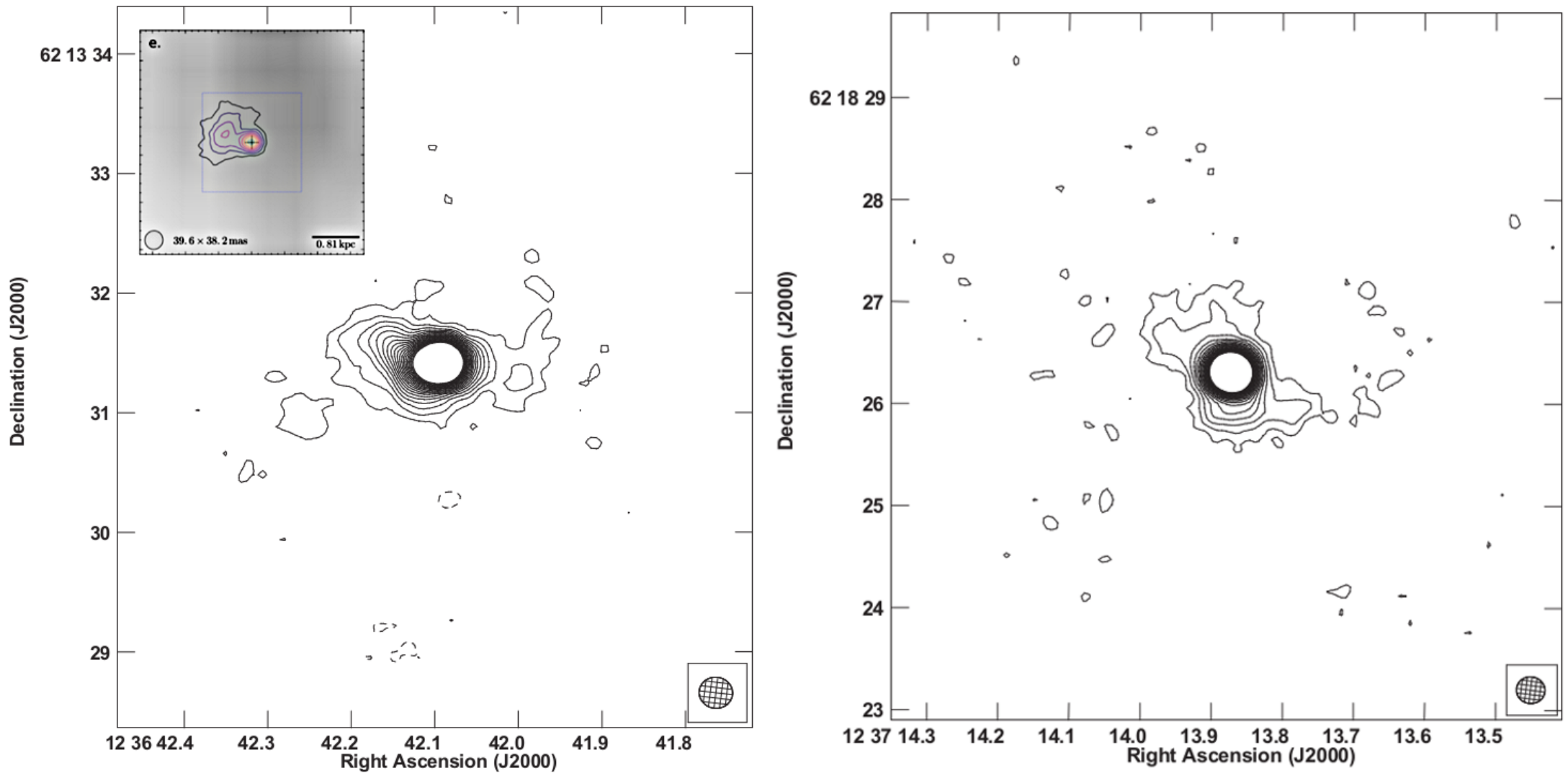

Figure 2.4: *e*-MERLIN+JVLA L-Band images (*e*-MERGE Consortium) of the faint $\mu$Jy radio-loud AGN systems J123642+621331 (left) & J123713+621826 (right) showing 2-sided jet emission on arcsecond scales. Insert for J1213642+621331 displays an EVN+MERLIN combination image showing initial one-sided core-jet structure on ~100 mas scales, imaged with a beam of $40 \times 38$ mas (Jack Radcliffe, private communication).

In recent years VLBI instruments have been employed to investigate a high redshift population of infra-red faint radio sources (IFRS) ( e.g. Middelberg et al. 2008, Herzog et al. 2015). IFRS objects are thought to be radio-loud AGN systems at redshifts >2, and to be important probes into the co-evolution history of SMBH growth and galaxy assembly. Although VLBI observations have confirmed the presence of active AGN in many IFRS objects it remains unclear as to the nature of the extended radio emission also present in these sources. Intermediate resolution imaging on 10s of mas resolution scales is required to characterise the true nature of this class of object and to confirm that the radio emission is wholly associated with AGN activity.

### 2.1.3 Star-formation and accretion in the local Universe

At high redshifts, VLBI observations of galaxies can pinpoint the location of compact AGN related emission; however in the local Universe sensitive VLBI experiments can be used to investigate the detailed physics of star-formation and accretion in individual objects on linear scales matching the physical processes, e.g. sub-parsec scales. In such a scenario each galaxy provides an individual laboratory of accretion powered sources such as AGN, as well as rapidly evolving stars, enshrouded by gas and dust, all of which are essentially at the same distance. The high angular resolution and sensitivity of interferometers are ideally suited to detailed studies of these objects - providing a unique probe to the physics of compact objects such as AGN. Supernovae and their evolving remnants probe the content and dynamics of the cool, neutral, and molecular ISM via molecular masers and absorption experiments at the highest angular resolution.

Supermassive black-holes exist at the centres of all galaxies, but only a fraction of these are in active states of accretion at any one epoch. This accretion is one of the most significant energy sources in the Universe, with the potential to clear star-forming gas from the galactic bulge area and regulate the growth of whole galaxies in clusters (e.g. di Matteo et al. 2015). This mechanical feedback mechanism through jets and outflows (see Section 3.1.3) is powered by the central accretion processes. Whilst the role and influence of accretion in luminous AGN sources has been well studied (see Chapter 3), comparatively little is known about the accretion activity of supermassive black-holes at low radiative luminosities, despite both their ubiquitous nature and importance to galaxy growth and evolution.

The low radiative efficiency of low-luminosity AGN, which are also commonly embedded in nuclear star-formation regions, means that the use of optical or X-ray diagnostics becomes increasingly challenging. However, sensitive, high resolution (sub- to milli-arcsecond angular resolution) radio observations can provide a direct probe of the jet power (see Fig. 2.5) hidden to other wavebands, and hence its mechanical influence on feedback (Nagar, Falcke & Wilson 2005 ). VLBI facilities have a critical role in such studies (e.g. Baldi et al. 2018; Park et al. 2017; Pérez-Torres et al. 2010) by reaching sensitivities of a few 10s of $\mu$Jy they can probe the equivalent X-ray flux (applying the 'fundamental plane' relation for accreting black-holes, e.g. Falcke, Körding & Markoff 2004) of $10^{38}\,\mathrm{erg\,s^{-1}}$ for a $10^{7}\,\mathrm{M_{\odot}}$ black-hole at 20 Mpc, which is a factor of 20 lower than the typical equivalent [O III] detection limit of traditional optical spectroscopic surveys. Over the coming decade sensitive, milliarcsecond resolution combined VLBI and *e*-MERLIN surveys of the compact core emission, and importantly the imaging of the inner jet structure, from samples of low-luminosity AGN will provide a unique characterisation of the SMBH accretion activity in the local Universe, allowing the links to SMBH activity with other galaxy properties and putting SMBH activity in the wider context of its implication on black-hole growth, and the regulation of galaxy evolution via feedback.

Beyond the most massive systems, accretion remains a dominant process across a range of black-hole system masses down to intermediate-mass and stellar-mass black-hole systems. Outside our own Galaxy such accreting systems are commonly seen as off-nuclear Ultra Luminous X-ray (ULXs), or microquasars/X-ray binary (XRBs) sources in many local galaxies, but are typically faint at radio wavelengths and only detectable in the most extreme cases and in the deepest VLBI observations. By their nature these are associated with the endpoints of massive star-formation.

In the case of extragalactic microquasars an extrapolation of the radio characteristics from galactic equivalents (see Chapter 4) are expected to be compact, highly variable objects which may show superluminal motion. Within nearby star-forming galaxies ($<$100 Mpc) VLBI has thus shown tentative evidence for persistent and variable emission from such sources in a few cases, notably in Arp 220, Arp 299 and M82 (Batejat et al. 2012; Pérez-Torres et al. 2009; Muxlow et al. 2010) with detections of variable and superluminous sub-mJy radio sources. Persistent radio emission from an ULX source, NGC 2276 3c, was tentatively reported on milliarcsecond scales (ULX NGC 2276 3c, Mezcua et al. 2015) revealing compact jets 1.8 pc within its larger, 650 pc radio lobe with the EVN.

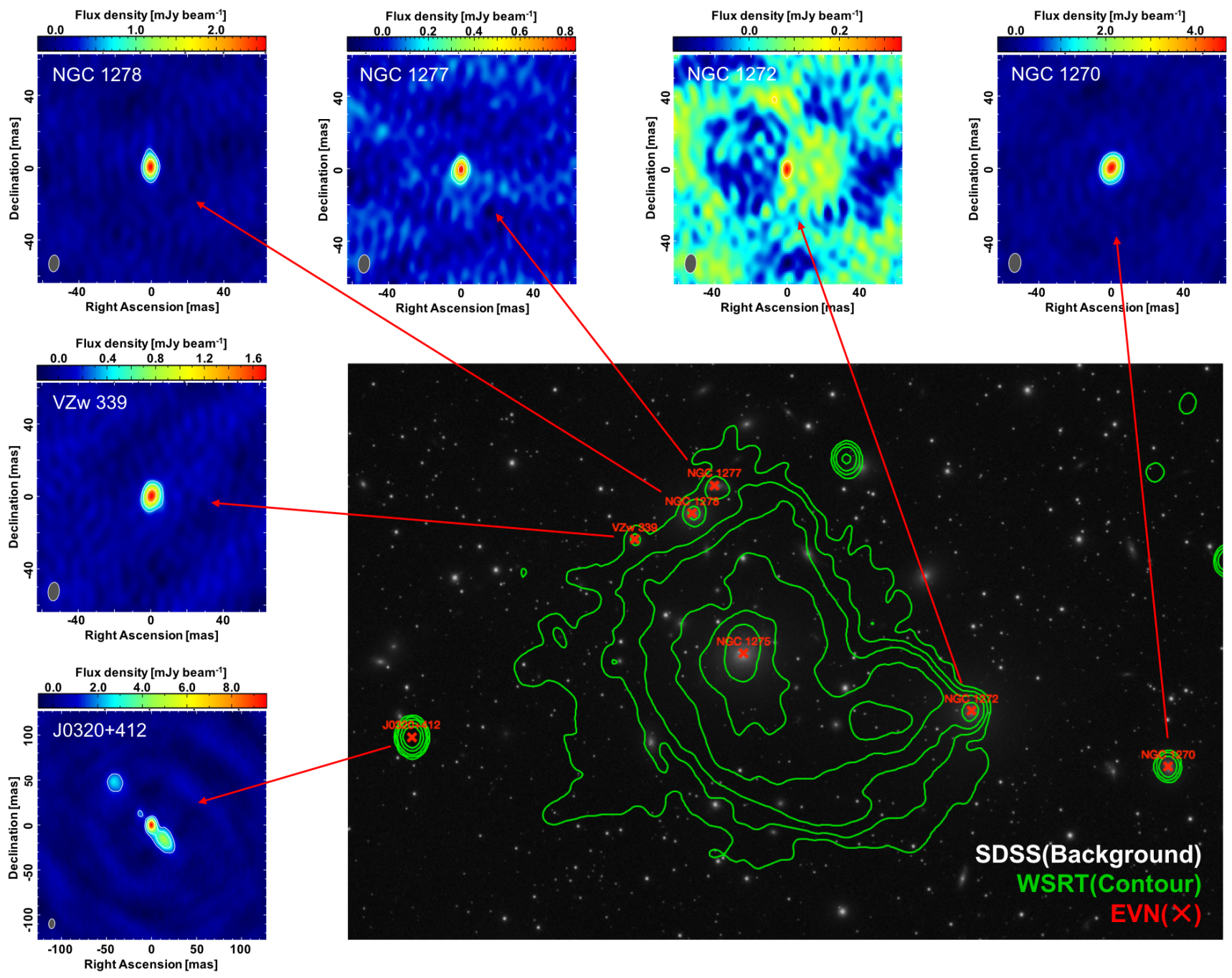

Figure 2.5: Combined image of the radio-optical images with compact radio sources detected from the EVN observation on a mas scale. The green contours represent the 1.4-GHz WSRT radio continuum that is overlaid on an SDSS g-band image (Fig. 1, Park et al. 2017).

This result remains at the cusp of current capabilities (Yang et al. 2017). Such studies are at the limit of existing VLBI arrays and limited to the most luminous examples in their class. However, advances in the sensitivity and capabilities of VLBI arrays during the coming years will open up the possibility of imaging many of these targets which will provide vital constraints on radio emission from the full mass range of black-hole systems across a range of galaxy types.

### 2.1.4 Star-formation processes

Characterising the star-formation history, along with the current level of star-formation, within individual and ensembles of galaxies remains a crucial physical parameter in understanding galaxy evolution. Equally important to our understanding is the role of massive stars and their subsequent stellar end points, on the surrounding ISM in galaxies and the part they play in self-regulation of the triggering and fuelling of both star-formation and accretion. Both within the local and distant Universe, much of the most intense star-formation activity occurs in the dust-enshrouded centres of star-forming galaxies, which are essentially invisible to traditional observational tracers of star-formation such as optical and even, in the most extreme cases, the infrared and X-rays. It is well recognised that the levels of both radio synchrotron and thermal Bremsstrahlung radio emission are excellent tracers of the large scale star-formation rates in galaxies (Beswick et al. 2015). However, the link between synchrotron emission and star-formation is via a complex physical mechanism, and the isolation of the fainter thermal free-free emission component requires detailed characterisation of the radio spectral energy distribution and/or the use of other a priori information such as optical recombination data (e.g. Westcott et al. 2018; Galvin et al. 2016).

Radio observations, and in particular high resolution (VLBI), provide a clean and fundamental tracer of the products of star-formation, such as supernovae (see Chapter 4) and supernova remnants, which is unbiased by obscuration. Whereas low resolution radio observations of star-forming galaxies trace the diffuse radio emission primarily originating from charged particles that have escaped old supernova remnants, high sensitivity VLBI+*e*-MERLIN observations can systematically detect, characterise and resolve individual radio supernova and their remnants on a galaxy by galaxy basis, critically resolving away the 'confusion' of the diffuse galaxy emission. By the characterisation of this population the highly obscured massive supernovae/star-formation rates of starburst and ultra-luminous galaxies can be directly inferred.

To date a number of such studies have been undertaken in individual galaxies (e.g. Varenius et al. 2019; Pérez-Torres et al. 2009; Bondi et al. 2012; Fenech et al. 2010). These have identified large populations 'supernovae factories' of compact radio sources with each galaxy effectively acting as a laboratory containing multiple discrete radio sources at different stages in their evolution, in different physical environments which all lie at essentially the same distance and can be studied in a systematic manner. One of the best studied example is the ULIRG Arp 220 (see Fig. 2.6, Varenius et al. 2019 and references therein) for which VLBI observations over the last two decades have detected nearly 100 compact radio sources including multiple new supernovae, supernova remnants and variable sources. The ultimate resolution that VLBI arrays provide has allowed a number of these sources to be resolved allowing new insights into the evolution of supernova remnants, as well as preferentially probing the top-end of the stellar initial mass function within the dense environments of these extreme starformers.

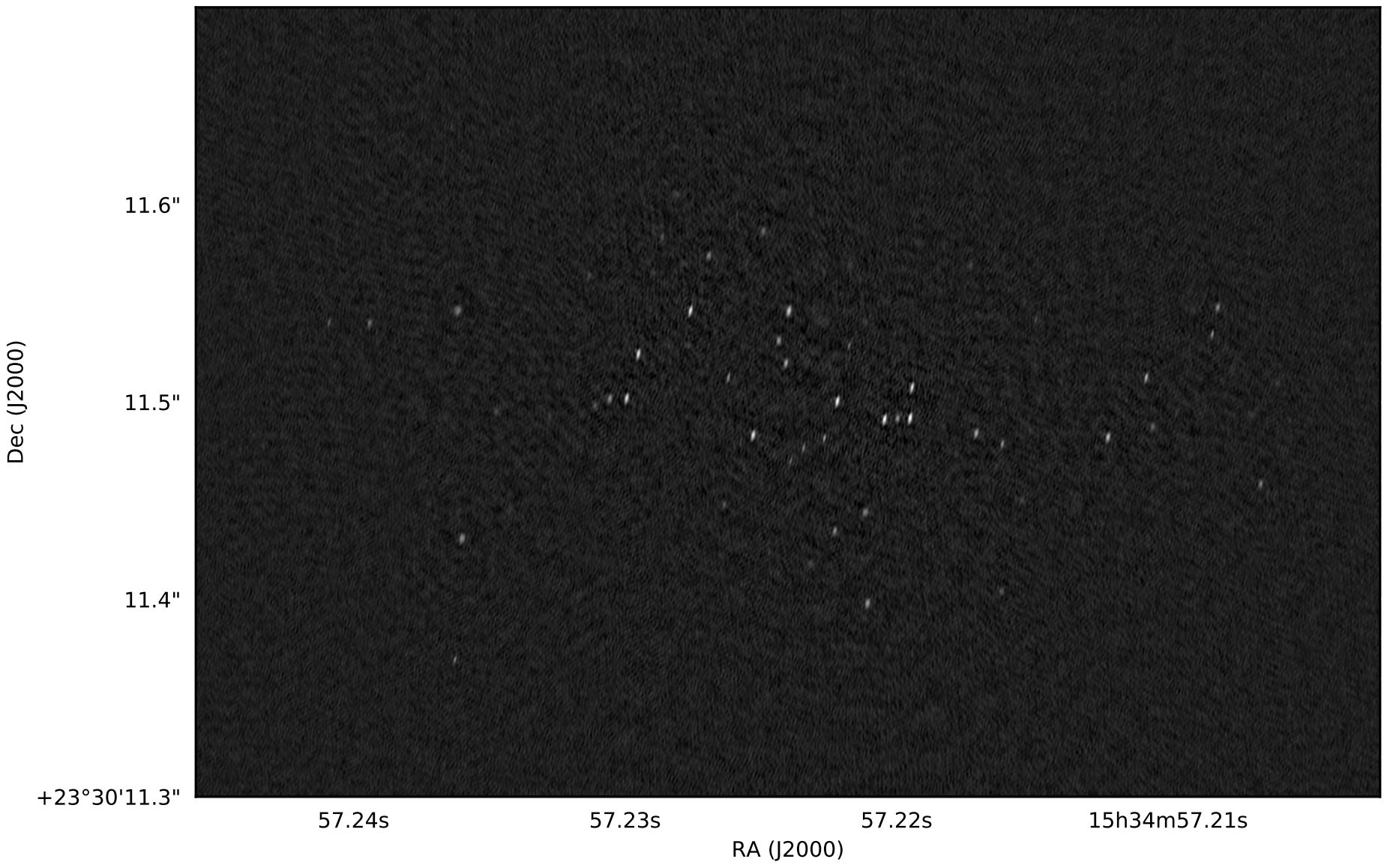

Figure 2.6: Deep (rms 4 $\mu$Jy/b) 6 cm VLBI imaging of the western nucleus of Arp 220. Credit: Varenius et al. (2019), reproduced with permission ©ESO.

However, studies of these extreme star-forming local galaxies (e.g. Arp 220, Arp 299 and others) have implications beyond their important contributions to the physics of star-formation and stellar evolution within these individual galaxies. These extreme star-forming environments, with star-formation rates exceeding 1000s $M_{\odot}\,yr^{-1}$, provide the best local Universe analogues available for the sites of extreme star-formation in the optically-faint, dusty star-forming galaxies (e.g. sub-mm galaxies) in the distant Universe (see above). In the distant Universe observations do not have the spatial resolving power to unravel the individual compact source properties, whereas VLBI studies of these local analogues do. Such local galaxy studies are crucial in providing the key anchor point for our wider understanding and interpretation of cosmic star-formation evolution in galaxies at all redshifts.

### 2.1.5 Requirements and synergies (continuum)

Unique VLBI combination $\mu$Jy sensitivity wide-field L-Band imaging on intermediate angular resolutions between 20 and 200mas is required to investigate the detailed radio structures of the SFG nuclear starbursts to determine how they differ from their low-redshift equivalents; to investigate any feedback from AGN activity in other wavebands; to characterise direct mechanical feedback from any faint embedded radio AGN-jet activity that may be detected in both the SFGs and diverse radio-quiet source population. Such studies are fundamental to accurately separate star-formation associated radio emission from AGN-jet contamination in deep-field investigations of the evolution of star-formation rate density with cosmic time - and demonstrate a new research area for wide-field VLBI in intermediate resolution studies of high redshift star-formation radio structures. They are also vital in assessing the role of AGN in promoting and controlling star-formation activity at high redshifts, and provide a link to understanding the co-evolution of SMBH growth with the assembly of galaxies in the early Universe. Intermediate resolution combination imaging of the faint $\mu$Jy radio-loud sources at both L- and C-band will also characterise the jet properties of this new population, and investigate the role of the circum-nuclear environment in disrupting jet flow on sub-Galactic scales. Similar requirements apply for studying star formation and accretion in the local Universe. In addition, isolation of thermal free-free and synchrotron processes would benefit from very wide bands. Detailed studies of star formation product tracers (SNe) require (sub-)mas imaging at GHz frequencies.

## 2.2 Active Galactic Nuclei and their impact on galaxy evolution

After almost three decades of observations, more and more pieces of evidence call for a revision of the Unified Model of active galactic nuclei (AGN). The unification scheme proposed by Antonucci (1993) postulates that all AGN are intrinsically the same physical object and that the observed dichotomy between broad line (*type 1*) and narrow line (*type 2*) AGN is only due to the orientation relative to a dusty toroidal structure surrounding the nuclear engine (the “torus”) (Antonucci 1993; Urry & Padovani 1995). However, X–ray and IR studies have changed our view of the classical dusty torus, from a uniform isolated entity to a clumpy structure connected with the host galaxy via gas inflows and outflows (for a recent review see Ramos Almeida & Ricci 2017). Mid-IR interferometry has revealed that the emitting dust is concentrated on scales of 0.1–10 pc and, in most cases, can be modelled with two nuclear components, instead of a single disk or toroidal structure (Burtscher et al. 2013). The most recent models for IR emission in AGN imply an equatorial inflowing disk and a polar-extended structure, which may originate from a dusty outflowing wind (e.g. Hoenig & Kishimoto 2017). Despite the great progress made, whether the torus is geometrically thick, if the polar elongation is always present, and whether the broad line region (BLR) and the torus are produced by accretion disk winds are still a matter of debate.

The study of the physical properties, structure, and kinematics of the gas surrounding SMBHs is fundamental in answering these open-ended questions and in building detailed models of AGN.

These studies, however, are complicated by the extremely small spatial scales (less than 10 pc) and by the complex structures of the nuclear components. Moreover, particularly in type 2 AGN, the inner regions are often obscured at optical and UV wavelengths. Current X-ray and IR instruments can access these regions but are not able to resolve them, and information on their structure have to be inferred by modelling the emitted radiation. The radio emission can penetrate the large column densities of gas and dust that often obscure the line of sight to the nucleus. In particular, at radio wavelengths, studies of luminous water and hydroxil masers (traditionally referred to as "megamasers") permit to directly map the molecular gas at tens of parsec or even at sub-parsec distances from the SMBH. Indeed, the high brightness temperature and small size of the maser spots make them perfect targets for Very Long Baseline Interferometry (VLBI) observations, through which angular resolutions of the order of 0.1 mas can be reached.

### 2.2.1 Extragalactic studies with masers and megamasers

Extragalactic OH masers are typically observed at the rest frequencies of 1665 and 1667 MHz, corresponding to the first two of the four hyperfine transitions of the ground rotational levels of the molecule (for recent reviews see e.g. Lo 2005; Tarchi 2012). These masers are typically associated with starburst regions within a few hundred pc from the nucleus and have luminosities intermediate between Galactic stellar masers and extragalactic megamasers which are associated with AGN activity. In only two cases (Mrk 231 and IIIZw 35) OH emission, mapped with VLBI, appears to be related to AGN activity, tracing a rotating dusty structure located between 30 and 100 pc from the nuclear engine and a parsec-scale ring, respectively (Kloeckner, Baan, & Garret 2003; Pihlstroem et al. 2001). Hydroxyl masers are used to investigate the gas dynamics in star-forming regions.

VLA studies of hydroxyl maser emission within the central kiloparsec region of the nearby (D=3.7 Mpc) starburst galaxy M 82 (Argo et al. 2010) detected 13 masers distributed across the central 30 arcseconds. Whilst remaining spatially unresolved, several of these masers show significant velocity structure indicating that such masers are associated with multiple masing regions within the beam along the line of sight. The masers can be used to investigate the distribution of the molecular and atomic gas present in the nuclear region of M 82 by overlying a position-velocity ($p-v$) plot of the molecular hydroxyl masers on the atomic H I absorption. The comparison plot is shown in figure 2.7 which is aligned along the major axis of M 82 where the 1667 MHz masers are contoured over the H I absorption seen in the atomic gas in grayscale; the H I absorption being derived from 1420 MHz data from the VLA archive (programme AW444; Wills et al. 2000).

The distribution of the masers is well correlated with the H I absorption with the rotation of the nuclear region of M 82 shown running $\sim$linearly from top left to bottom right. However, it is clear that there is an additional blue velocity arc running from the galaxy centre which contains both atomic and molecular gas with several of the brightest masers located on the arc. This feature has been suggested to be either an expanding super-bubble or the $x_2$-orbits of an inner bar (Wills et al. 2000).

EVN observations of two masers within the blue arc (Argo 2018), spatially resolve the masing regions and show a significant difference in line ratios for the different spatial components within the masing regions. These main-line OH masers are radiatively pumped and collisional de-excited to the rotational ground state with the line ratios being dependent on the physical conditions within the masing cloud (Gray 2007) – see figure 2.8. Thus high resolution imaging of hydroxyl masers can act as a detailed probe of the physical conditions within the masing regions in nearby star-forming galaxies.

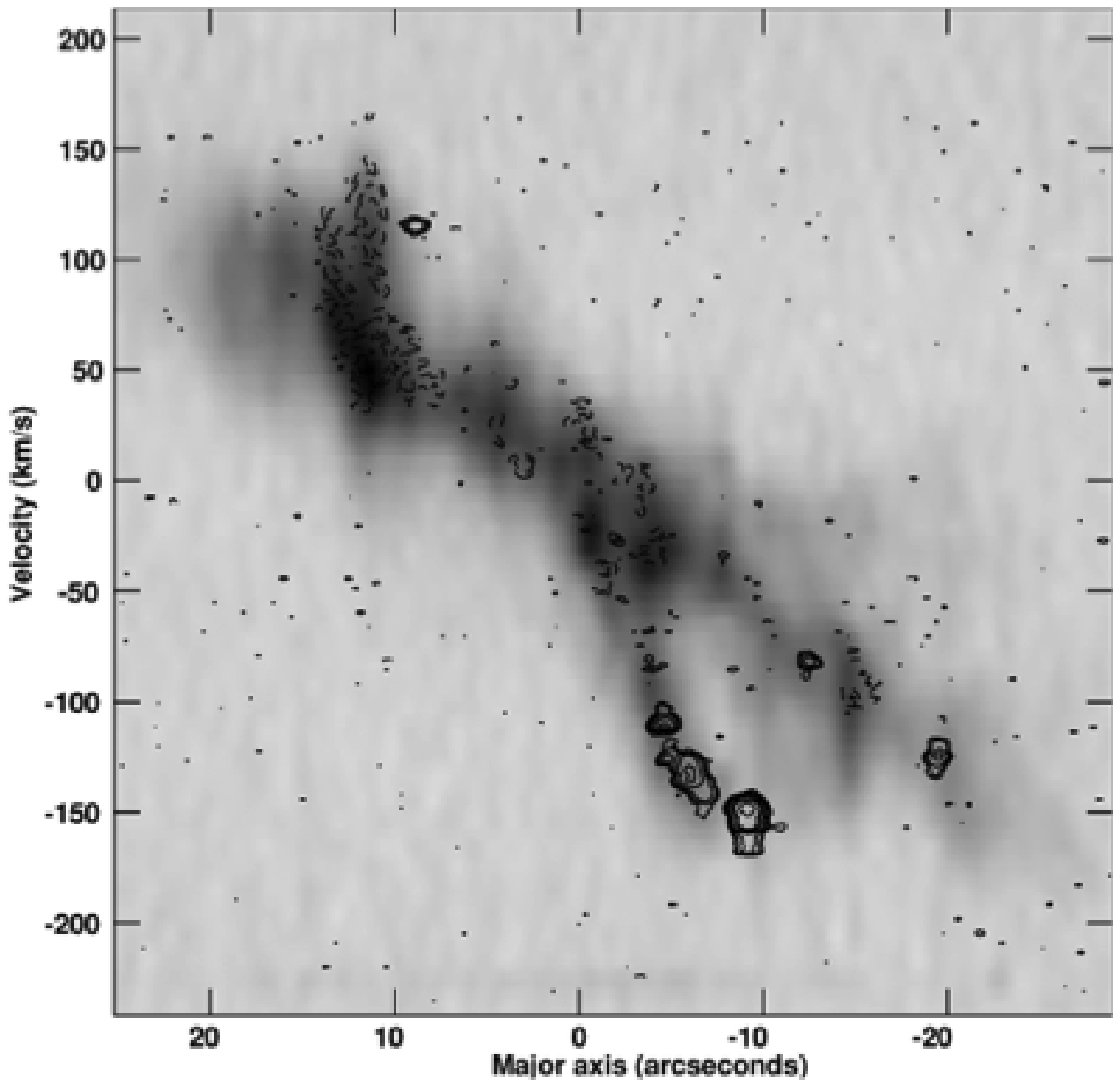

Figure 2.7: 1667 MHz masers contoured over H I absorption (grayscale). Offsets from dynamical centre & velocity relative to systemic (225 km s$^{-1}$). From Argo (2018).

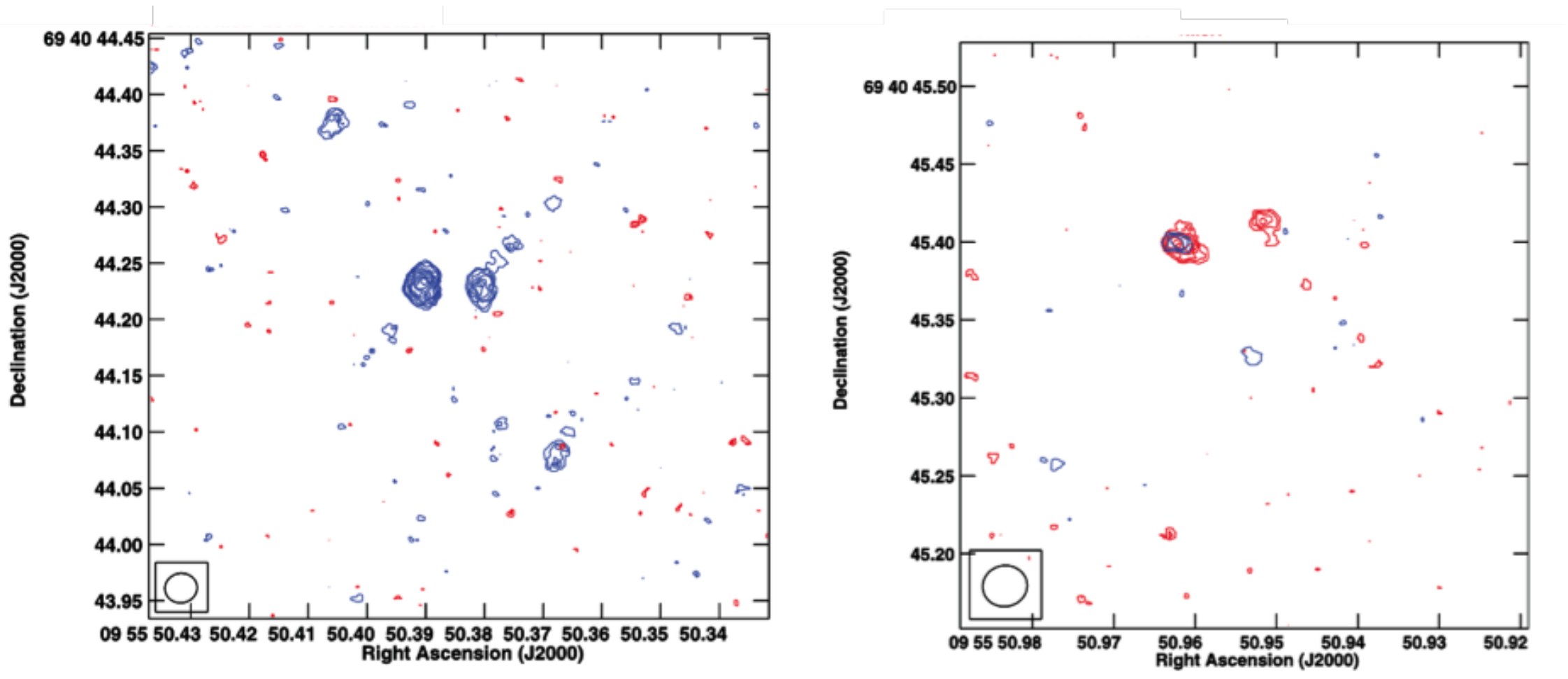

Figure 2.8: Comparison of the 1667 MHz (blue) and 1665 MHz (red) and emission for the masers 50.37+44.3 (left panel) and 50.95+45.4 (right panel). From Argo (2018).

In the remainder of this section, we will focus on water megamasers that, differently from hydroxil ones, have always been found to be associated with nuclear activity. The 22 GHz maser line of ortho-$H_2O$ may trace three distinct phenomena related to AGN. Typical triple-peak line systems are associated with edge-on accretion disks. VLBI and single-dish monitoring studies of disk-masers allow us to map accretion disks and to determine the enclosed dynamical masses (e.g. Kuo et al. 2011; Gao et al. 2017; Zhao et al. 2018). In addition, radio continuum observations of disk-maser galaxies have recently been used to test some aspects of the AGN paradigm, that is, the alignment between the radio jet and the rotation axis of the accretion disk (Kamali et al. 2019, and references therein). $H_2O$ megamasers may also trace nuclear ejection processes in the form of jets or winds. Indeed, they have been found to be associated with the interaction of the jets with molecular clouds in the host galaxy (Gallimore et al. 2001; Peck et al. 2003; Castangia et al. 2019) or with wide-angle outflows at less than 1 pc from the nuclear engine (Greenhill et al. 2003). Observations of the best-studied jet-maser, in Mrk 348, using the reverberation mapping technique, provided estimates of relevant physical quantities such as the jet velocity, the shock speed, and the densities of the material in the jet and in the ambient medium (Peck et al. 2003). Water maser observations in Circinus (Greenhill et al. 2003) and NGC 3079 (Kondratko et al. 2005), instead, seem to have resolved individual outflowing torus clouds at <1 pc from the nuclear engine, probing the geometry and kinematics of nuclear winds. Therefore, each megamaser source provides a plethora of information on the (sub-)parsec-scale environment around AGN, making the discovery of new sources and their interferometric follow-up extremely important for AGN studies.

Very bright masers (with peak flux densities of the order of 1 Jy) as, for example, the ones in NGC 4258, NGC 3079, and NGC 1068 can be observed with any existing VLBI array. On the contrary, high angular resolution observations of weaker but still relatively bright masers (with peak flux densities from tens to hundreds of mJy) require sensitive instruments like the VLBA and the EVN, capable of reaching rms of a few mJy per 0.2 km/s channel within a few hours on-source. This is essential to image the largest possible number of maser spots and Doppler components, whose spectral linewidths can be as narrow as 1 km/s or less. The weakest masers, with peak flux density up to 10 mJy, instead, can be observed only with the most sensitive arrays like the High Sensitivity Array (HSA), or Global VLBI. The EVN may play an important role in extragalactic maser studies, since with the inclusion of large antennas (e.g. the new 64-m Sardinia Radio Telescope) the sensitivity of the full array at 22 GHz is much higher than that of other VLBI arrays. However, in order to fully exploit this enhanced feature, the frequency coverage of the receivers available at each station has to be as homogeneous as possible. For example, at K band, the water maser line (rest freq. 22.23508 GHz) in a galaxy with recessional velocity of 10000 km/s ($z\sim0.03$) would be detectable at $\sim$21.5 GHz. If the K-band receivers of a number of stations do not cover this frequency, the VLBI array usable for observing such a transition would turn out to be incomplete, with an impact (significant, in some cases) on the final sensitivity and *uv*-coverage of the measurements.

### 2.2.2 Outflows and AGN feedback

Hydrogen is the most abundant element in the Universe occurring in different phases in various structures that cover a range of spatial scales. The radio regime is particularly well suited for the study of neutral atomic hydrogen (H I), because of the hyperfine transition at 1420 MHz (or 21 cm). H I provides a wealth of information on the nature of gaseous matter in the Universe which is why it has been identified as one of the key science drivers for the SKA (Staveley-Smith & Oosterloo 2015; Morganti et al. 2015) and its precursors.

In general, H I can be observed in *emission* and *absorption*. However, only connected interferometers or single-dish telescopes have the brightness temperature sensitivity to detect H I emission, mostly in the nearby Universe. The situation is different for H I absorption, which only requires a radio background source (often a radio AGN) and a sufficient optical depth[1] of the gas.

It is common to distinguish two types of H I absorption: *intervening* and *associated* absorption. In the former case, the observed gas and background source belong to two independent systems. This allows the study of H I gas in a galaxy (even our own) which is in front of a more distant active galaxy along the observer's line of sight. As such, intervening absorption can be used to trace the gas content over a range of redshifts which gives insight into evolution of galaxies in the Universe (Curran 2017). In case of *associated* absorption, the observed gas and background source are located in the same system, e.g., the same galaxy and the gas has to be in front of the background source. This constraint is a great benefit, because it allows to deduce the kinematic structure of the gas, e.g., whether it is regularly rotating, infalling or outflowing gas. This is particularly interesting for the understanding of feeding and feedback in radio AGN. Several studies have also shown a higher detection rate of H I absorption in young and restarted AGN which could be related to a denser ISM and may provide insight into the evolution of AGN (e.g. Morganti & Oosterloo 2018). A vital step in understanding the gas dynamics is to resolve the distribution and kinematics of the gas close to the AGN. In most cases, this requires reaching sub-kiloparsec scales and sub-arcsec angular resolution which can be provided by VLBI. Almost no other observing technique can acquire information on the gas content of a galaxy on these small spatial scales.

Systematic searches for H I absorption are generally conducted with lower-resolution interconnected interferometers or single-dish instruments. The upcoming large surveys planned with SKA pathfinders and precursors promise a major increase in the number of H I absorption using "blind searches" (see Table 2.1), thanks to the large instantaneous band. This will open up the possibility of statistical studies of the properties of the H I gas in larger and new groups of sources. The follow-up with VLBI will be a key addition in order to maximise the science, in particular in the SKA era when VLBI with the SKA promises a significant boost in sensitivity.

Other sources of associated H I absorption are star forming galaxies and supernova remnants (e.g. Muxlow et al. 2006; Leahy & Tian 2010), but the largest contribution of H I VLBI continues to be AGN studies. Thus, the majority of this chapter will focus on this topic. Intervening absorption studies with VLBI may become more prominent in the future (e.g. Biggs et al. 2016; Gupta et al. 2018).

### 2.2.3 Tracing neutral atomic hydrogen on parsec scales

H I VLBI has a long tradition and a range of studies have shown how tracing H I on parsec-scales reveals a variety of gas structures in the close surroundings of the central supermassive black hole. Here, we highlight only some recent results. Particularly important are the studies of the presence of H I in tori and circumnuclear disks. These structures are considered a key ingredient in AGN, affecting the view of the central regions and providing a reservoir of gas for the fueling of the central engine. Recent theories predict them to be clumpy and perhaps even warped (Ramos Almeida & Ricci 2017). In particular in radio galaxies, VLBI has provided important information (see e.g. Araya et al. 2010; Morganti & Oosterloo 2018 for an overview). Struve & Conway (2010) confirmed the presence of a nuclear disk in Cygnus A using the VLBA (see Fig. 2.9). The disk is located at a radius of about 80 pc, has a scale height of about 20 pc and a density of $> 10^4\,\mathrm{cm}^{-3}$. Espada et al. (2010) detected absorption in Centaurus A about 0.4 pc away from the nucleus which could be part of a circumnuclear disk, but this is not yet confirmed. In the restarted, giant radio galaxy 3C 236 a thick disk of H I gas is observed at radius of about 880 pc and a scale height of at least 400 pc (Schilizzi et al. 2001; Struve & Conway 2012; Schulz et al. 2018). The orientation of the disk coincides with a dust-lane and a disk of ionised and molecular gas. Struve & Conway (2012) also suggested the existence of an almost edge-on disk or torus with a radius of $< 200$ pc in 4C 31.04. H I gas has also been detected in absorption in the centre of cool-core galaxy clusters. The limited available data have shown that the gas may be concentrated in tori-like structures.

[1]The optical depth $\tau$ is defined as $\tau = \log(1 - \Delta S_{\rm abs}/(c_f S_{\rm cont}))$ where $\Delta S_{\rm abs}$ is the absorbed flux density, $S_{\rm cont}$ the continuum flux density, and $c_f$ the covering factor.

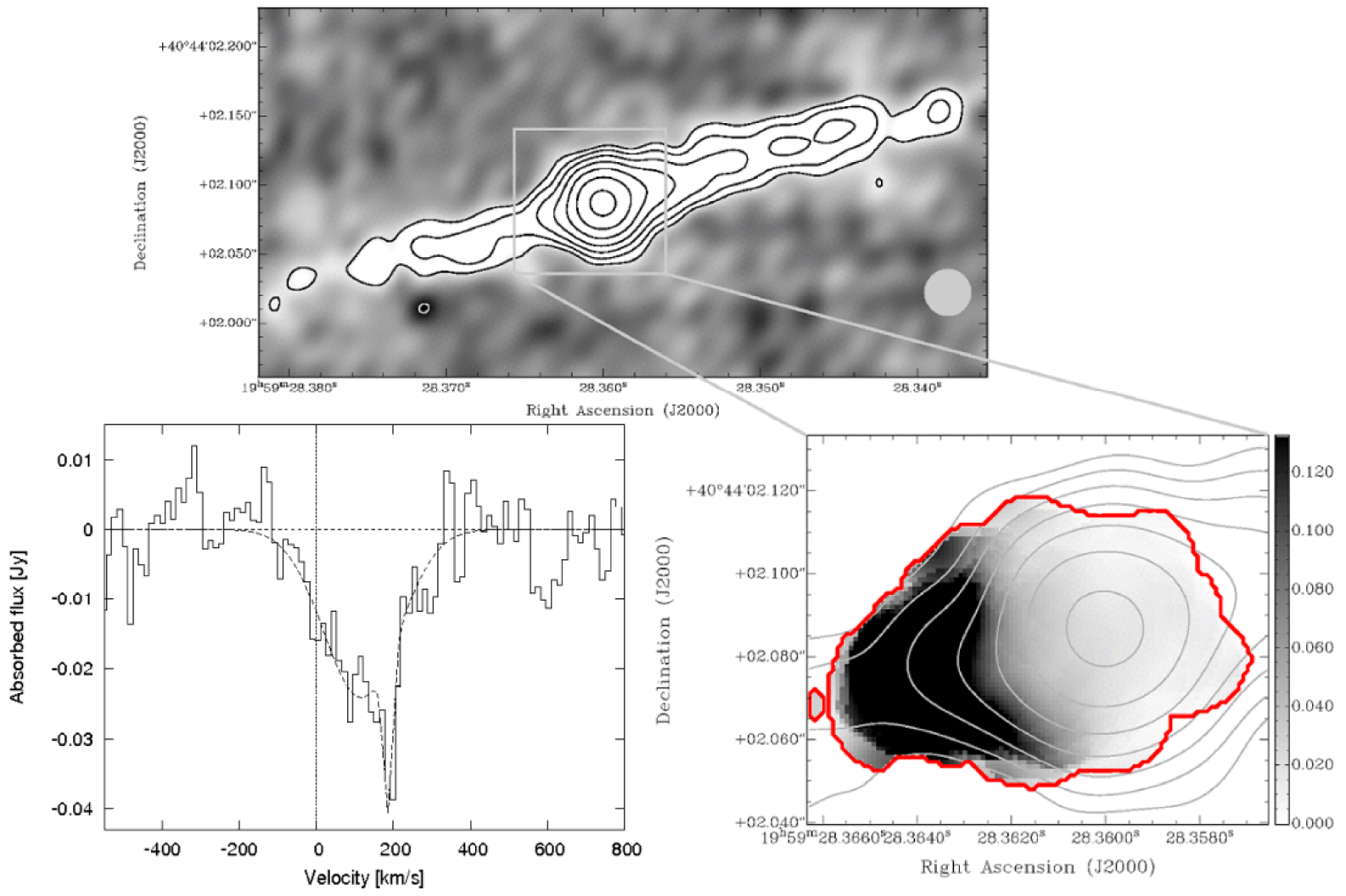

Figure 2.9: The circumnuclear disk of H I gas detected towards the central few parsec in Cygnus A. Credit: Struve & Conway (2010), reproduced with permission ©ESO.

Another important finding is that not all the gas is settled in regularly rotating structures. The search for infalling gas fueling the AGN has been particularly difficult with only a few cases where, thanks to the high resolution of VLBI, this gas could be identified and separated from the settled gas. In PKS 2322 − 123 located in the X-ray cooling cluster Abell 2597, Taylor et al. (1999) detected narrow redshifted absorption indicative of infalling gas in addition to a torus-like structure of H I. A similar situation has been reported for J094221.98+062335.2 which belongs to a galaxy pair that is undergoing a major merger (Srianand et al. 2015). VLBI observations of the central radio emission of the giant radio galaxy NGC 315 spatially resolved H I gas that is likely falling towards the centre of the AGN (Morganti et al. 2009). Other examples are B2352 + 495, 4C31.04, PKS B1718 − 649 and the cluster Abell 2597 (Araya et al. 2010; Struve & Conway 2012; Tremblay et al. 2016; Maccagni et al. 2017). In the latter two cases, the infalling gas has been linked to chaotic cold accretion processes (see also Fig. 2.11).

Finally, the most recent and surprising finding is that H I is also involved in AGN-driven fast outflows. A striking signature of this phenomenon is the presence of broad blueshifted absorption of H I covering up to 1000 km $\mathrm{s}^{-1}$. This has been detected in a number of different AGN such as young or restarted radio galaxies, but also Seyfert galaxies (see Morganti & Oosterloo 2018 for a review). The powerful resolution of VLBI has made it possible to trace the outflows down to parsec-scale.

Morganti et al. (2013) detected a massive $> 10^3 M_\odot$ outflow in 4C 12.50 towards the southern radio lobe about 100 pc away from the nucleus. More recently, Schulz et al. (2018) partially recovered the H I outflow in 3C 236 with VLBI. Rather than a single, slightly extended feature as in 4C 12.50, the outflow seems to be clumpy. The highest velocity clouds are compact and are detected towards the nuclear region while an extended feature is also detected co-spatial to the lobe (see Fig. 2.10). These results indicate that at least in some cases the H I outflows are driven by the jet as it enters a clumpy medium. In other cases the outflowing gas is still likely to be clumpy, but driven by radiation pressure (e.g. Mrk 231, Morganti et al. 2016). The location of the outflow with respect to the radio source provides the best way to distinguish between these driving mechanisms (e.g. Morganti et al. 2013). Interestingly, H I outflows have been predominantly detected in young and restarted radio galaxies.

### 2.2.4 Probing AGN feedback and galaxy evolution with VLBI

Observationally, H I VLBI continues to have a tremendous astronomical impact by utilising the full spectral power of VLBI. A single observation provides not only information on the gas properties but vital complementary data on the continuum source. As such it requires dealing with the particular challenges in calibration, RFI mitigation and data processing. In addition, it is well suited for wide-field observations which require high spectral and time resolution. H I VLBI affects the design of current and future generations of telescopes. As such, it will play an important role in utilising the full potential of VLBI observations with the SKA.

The data itself is vital to constrain theoretical models and numerical simulations of AGN feedback and evolution on galactic scales (Mukherjee et al. 2016, 2017, 2018). Even though these simulations do not fully model the cold gas, in many cases H I VLBI is the only observational technique which can match the resolution of the simulations. This provides at least a qualitative comparison which is particularly important for studying inflows and outflows of gas.

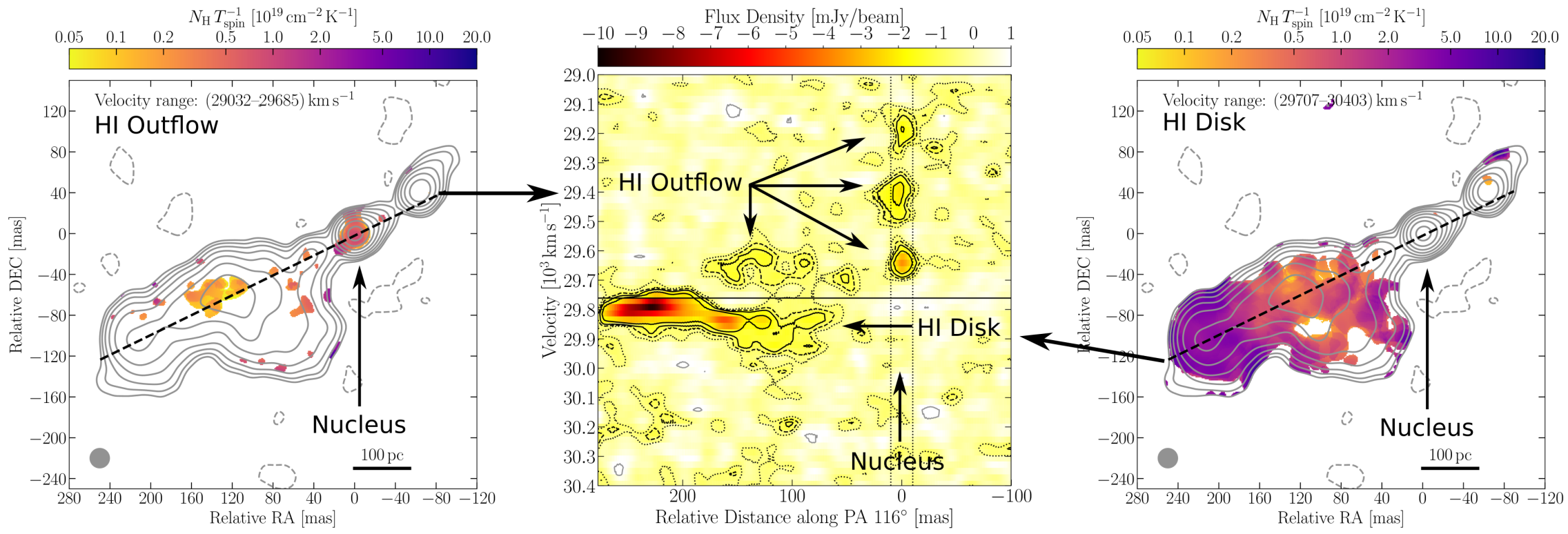

Figure 2.10: The jet-driven outflow of H I gas in 3C 236 has been partially recovered with VLBI revealing a clumpy structure even towards the nuclear region of the AGN. Adapted from: Schulz et al. (2018), reproduced with permission ©ESO, and Morganti et al. 2018b.

#### H I studies with the EVN

So far, H I absorption studies have been limited to a few case studies. This will change in the coming years with the upcoming H I surveys conducted around the world (Table 2.1). They will significantly expand the sample of known H I systems. In the northern hemisphere, the SHARP survey conducted by APERTIF will likely become the reference catalogue for H I VLBI follow-up with the EVN. The detection rate of H I has been estimated for radio galaxies to be 20%–30% (e.g. Maccagni et al. 2017; Curran & Duchesne 2018). Southern hemisphere surveys will also cover areas of the sky accessible to the EVN. VLBI follow-up observation will yield important high-spatial resolution information of associated absorption in a significantly larger number of sources. This will open the possibility of an unprecedented statistical analysis of the properties of H I gas covering a variety of radio AGN. The results will shed light on the gas dynamics for different types of AGN and their evolution.

| Survey | Redshift range | Noise over 5 km/s [mJy/beam] | Sky coverage [deg$^2$] | Number of detections | Detection limit [mJy] |
|---|---|---|---|---|---|
| APERTIF-SHARP | 0–0.26 | 1.3 | 4000 | 400 | 30 |
| ASKAP-WALLABY | 0–0.26 | 1.6 | 30000 | 2300 | 40 |
| MeerKAT-MALS | 0–0.57 | 0.5 | 1300 | 450 | 15 |

Table 2.1: Overview of upcoming H I absorption surveys taken from Morganti & Oosterloo (2018).

The upcoming surveys will also provide a larger sample of intervening absorption. Thus, H I VLBI will be used to gain information on the gas structure of these absorbers (see e.g. Biggs et al. 2016; Gupta et al. 2018 and references therein). As these system are likely non-active galaxies, the data will help understand the evolution of galaxies.

It is less clear whether H I VLBI will become more important for studies of starburst galaxies and supernova remnants, because of the required sensitivity.

There are technical as well as scientific limitations to the study of intervening and associated H I in absorption. The technical limitations are specific for VLBI. The frequency range covered by L-band receivers on VLBI telescopes limits the redshift range to $\lesssim 0.12$. As a result, only a subset of sources detected by upcoming surveys can be followed-up. In addition, since the EVN is a heterogeneous array, not all L-band receivers are able to reach this value while others are technically capable of going to lower frequencies (e.g. *e*-MERLIN). This limits the sensitivity of an observation. As there are no plans to extend to lower frequencies, H I VLBI observations with the EVN will remain focused on the nearby Universe. However, there is an opportunity to significantly improve the scientific output in light of the developments for VLBI with the SKA which may be usable as a VLBI element at frequencies well below 1 GHz.

Another limiting factor is RFI, but its impact can vary. RFI can have a larger affect on short spacing than on longer baselines where it may correlate out. The RFI environment also changes from station to station. On the one hand, this makes RFI mitigation more challenging. On the other hand, it makes it possible to gain information for channels affected by RFI for some stations, by using the data from other stations at a similar spacing.

The scientific limitations are inherent to the nature of H I absorption experiments (e.g. Morganti & Oosterloo 2018 for recent overview). Only gas in front of the radio source can be seen, but this is also a benefit as it allows to differentiate between rotating, infalling and outflowing gas. With a few exceptions, associated absorption is limited to radio-emitting AGN. This is also the case for intervening absorption to some extent, as a background source (even a distant one) is still required. The main uncertainty is the spin temperature which is necessary to calculate the column density[2] and several other properties such as density, mass and other quantities. It is influenced by a range of effects (e.g. Field 1959). In the ISM, collisions of hydrogen atoms with other particles dominate. However, radiative effects will dominate closer to the AGN. Therefore, the spin temperature is considered to vary between a few hundred Kelvin to several thousand Kelvin. This has been confirmed by a few studies (Holt et al. 2006; Struve et al. 2010). Constraints on the spin temperature can be obtained from H I emission measurements, but these are only available for a very limited number of sources and the emitting H I is often on other spatial scales than the absorbing gas.

[2]The column density $N_{\mathrm{H\,I}}$ is defined as $N_{\mathrm{H\,I}} = 1.82 \times 10^{18} T_{\mathrm{spin}} \int \tau(v) dv$, where $T_{\mathrm{spin}}$ is the spin temperature.

While the variety of radio telescopes in the EVN provides some challenges as stated above, it also results in unique capabilities. The EVN includes very large radio telescopes which significantly boost its sensitivity.

In addition, the incorporation of the *e*-MERLIN array in the United Kingdom yields very important short spacings to recover faint and diffuse extended radio continuum emission and extended absorption. Thereby, observation with the EVN in combination with *e*-MERLIN can cover a large range of angular scales from several arcsec to mas. This is particularly important for wide-field VLBI observations. Further strength of the EVN is the strong drive for technological developments and user support from setting up observations to processing the data. This ensures that the H I VLBI capabilities will continue to improve towards the SKA era and that the data are used efficiently.

### 2.2.5 Requirements and synergies

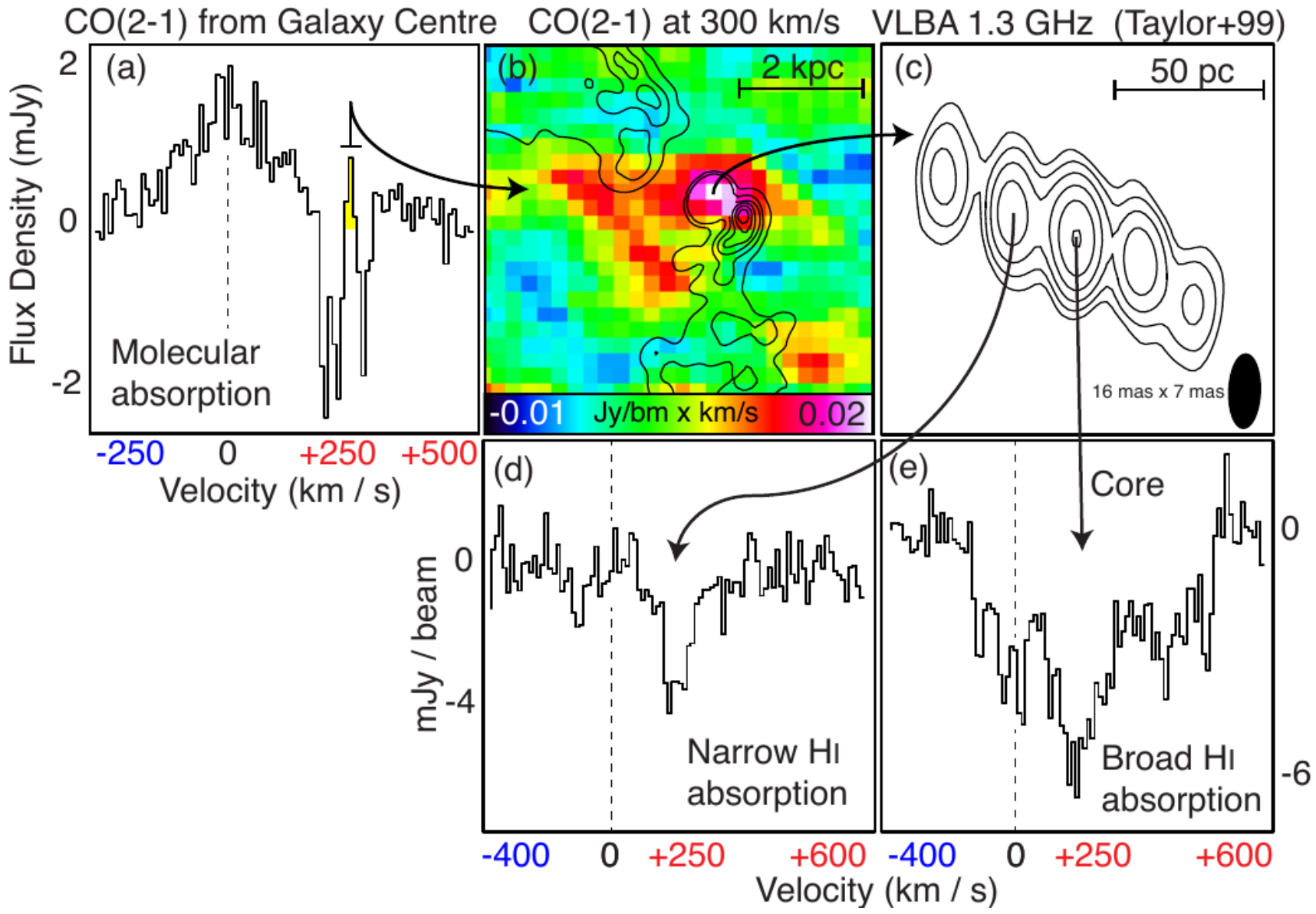

Figure 2.11: ALMA-detected CO(2-1) absorption and emission and VLBI-detected H I absorption show that the inflowing molecular gas clouds in Abell 2597 are likely to be located close to the central black hole (Tremblay et al. 2016).

Utilising the full capabilities of VLBI for H I absorption studies requires primarily an accurate calibration of the bandpass and good RFI mitigation. The latter is especially challenging for the short spacing provided by *e*-MERLIN which is essential to cover a large field of view and recover extended gas as well as continuum emission. In principle the resolution of VLBI is also sufficient to investigate the properties of H I gas over a larger redshift range. This will provide vital insight into the evolution of the gas with cosmic time. However, this can only be achieved by extending the coverage to lower frequencies. Automated processing of the data from the intial calibration, RFI flagging to imaging will become more relevant with the increase of the sample size and data volume.

As mentioned above, upcoming blind surveys such as SHARP with APERTIF in the northern hemisphere are expected to increase the number of H I absorption detections providing the targets for H I follow-up observations. Making use of the wide-field capabilities provides the means to cover multiple sources with a single observation. Since the sensitivity of H I VLBI observations primarily depends on the number of stations and observing time, it can be improved significantly by joint observations with the VLBA, VLA, LBA and Arecibo. This is particularly useful for detailed single source studies. In the SKA era, joint observation with SKA-MID can provide important *uv*-coverage for lower declination sources as well as a significant boost in sensitivity.

Understanding the physics of gaseous matter also requires knowledge of other gas tracers. In terms of the cold molecular gas, they can be obtained almost at a comparable resolution by ALMA and NOEMA. The powerful combination of H I and molecular gas has already been demonstrated in several cases (e.g. Morganti et al. 1998; Tadhunter et al. 2014; Maccagni et al. 2016; Oosterloo et al. 2017). For example, Tremblay et al. (2016) used information from both tracers to show the occurence of chaotic accretion in the Bright Galaxy Cluster Abell 2597 (see Fig. 2.11). The outflowing H I gas in Mrk 231 is likely driven by radiation pressure rather than the jet. Outflows of molecular and atomic gas seem to be more massive than those of ionised gas (e.g. Morganti 2017 and references therein). However, the sample size for outflows of cold gas is still limited and future observations will have to show whether this holds. Molecular absorption has been detected with VLBI, too, for example recently with the KVN in NGC 1052 (Sawada-Satoh et al. 2016).

The radio telescopes comprising the EVN primarily cover the northern hemisphere. As such H I VLBI greatly benefits from the wealth of optical information available from SDSS (Abolfathi et al. 2017) and in the future WEAVE which will perform follow-up of the APERTIF and LOFAR surveys. The gas close to the AGN may also be probed with soft X-ray observations and joint X-ray and H I absorption studies will become more important with the upcoming survey conducted by *eROSITA* (e.g. Moss et al. 2017).

As mentioned earlier, there is a close connection to numerical simulations at high-spatial resolution. H I continues to remain almost the only observational technique to provide constraints at comparable resolution. With the advances in computing power and theoretical models, we are confident that these simulations will be able to better incorporate the cold gas.

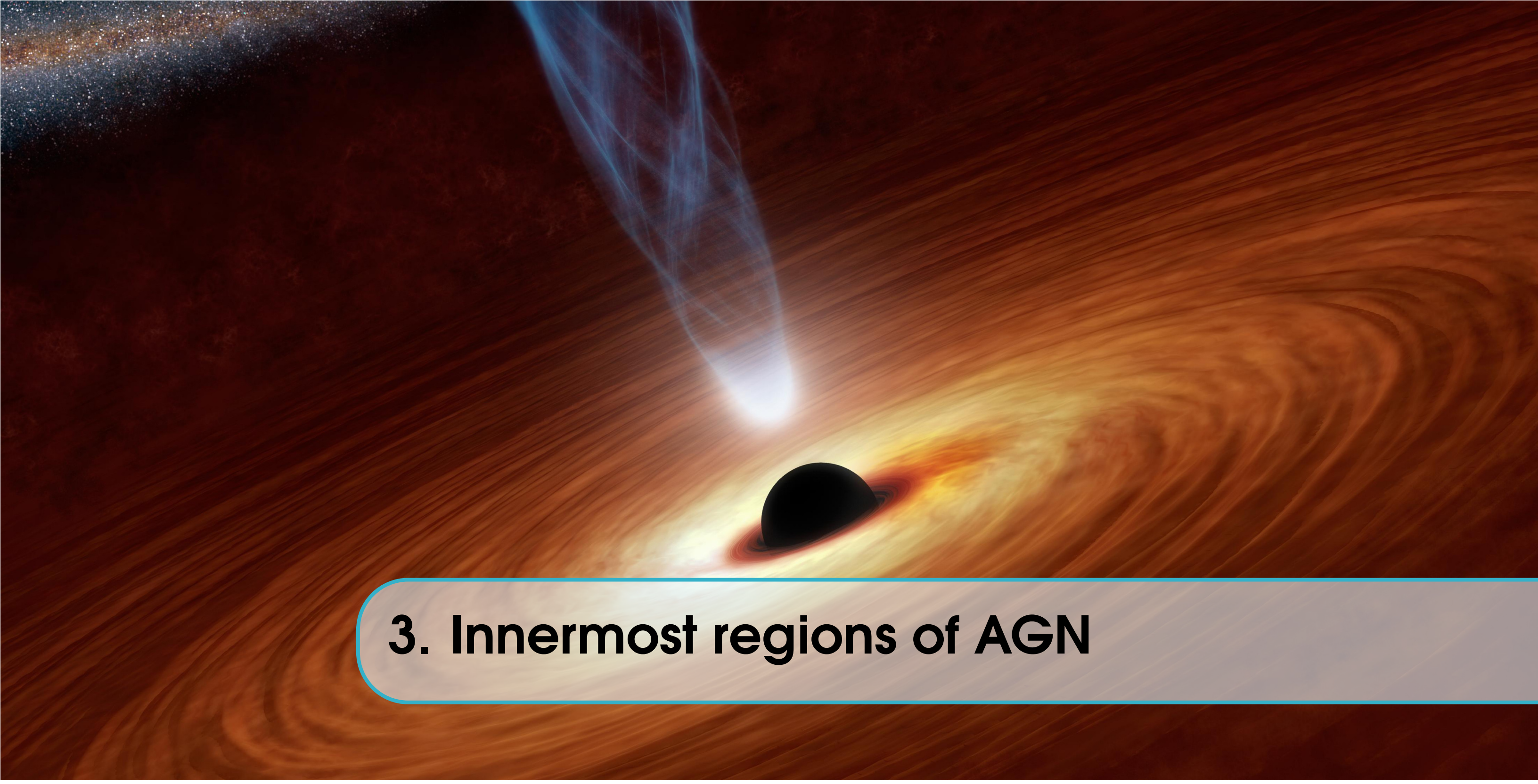

# 3. Innermost regions of AGN

Active galactic nuclei are key targets of very long baseline interferometry. The mas-scale angular resolution provided by VLBI allows not only to unveil the AGN physcis and energy transport in the vicinity of supermassive black holes (SMBH) responsible for their activity, but also to address the question of their evolution with cosmic time and co-evolution with the host galaxy. These two topics are addressed in the following sections.

## 3.1 Active Galactic Nuclei

The current astrophysical paradigm firmly sets galactic activity and active galactic nuclei among the most important factors affecting galaxy evolution and large scale structure in the Universe on cosmological scales. Galactic activity is closely related to the presence of supermassive black holes which are now believed to be residing in the centres of all spiral and elliptical galaxies. Energy generated through matter accretion onto SMBH is released and transported outward in form of broadband continuum radiation and kinetic energy vested into nuclear outflows. In a subset of *radio-loud* AGN, a fraction of the outflowing material is organised in highly collimated, relativistic jets. The jets are responsible for a large fraction of non-thermal continuum emission (particularly during powerful flares), which makes understanding their physics an important aspect of studies of AGN characterised by profound flaring activity arising from extremely compact regions. High-resolution observations of extragalactic jets with very long baseline interferometry (VLBI) offer arguably the best tool to understand their physics down to the smallest spatial scales and to use them as effective probes of the extreme vicinity of central black holes in AGN (see Boccardi et al. 2017 for review).

### 3.1.1 Central regions of radio-loud AGN

The central region of radio-loud AGN manifest a complex interplay between several major constituent parts, including the black hole and its magnetosphere, the accretion disk and the hot, non-thermal

Chapter image credit: NASA / JPL-Caltech.
From: `https://photojournal.jpl.nasa.gov/catalog/PIA16695` .

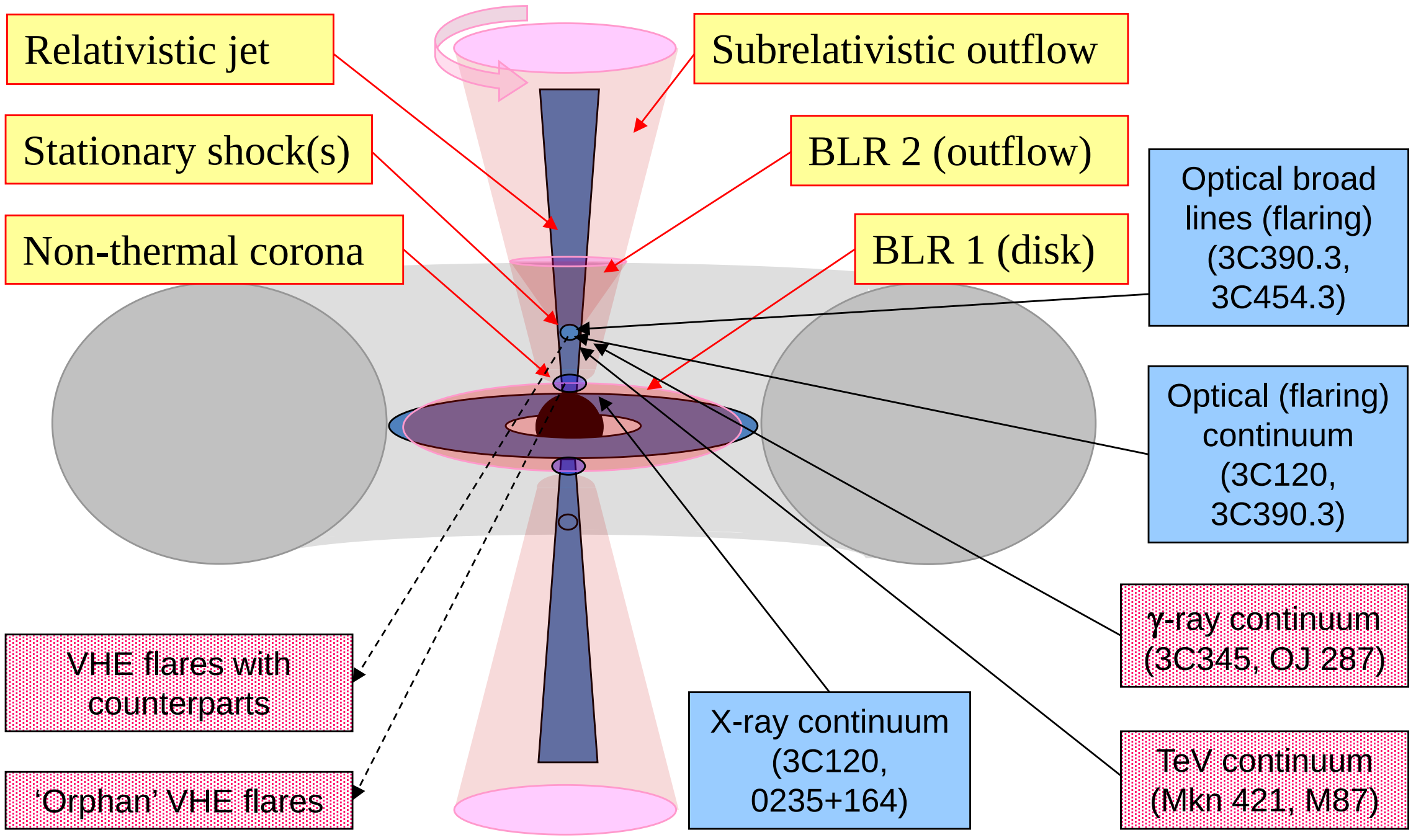

Figure 3.1: Schematic view of the central region of a radio-loud AGN. The overall manifestation of nuclear activity results from a complex interplay between the black hole, accretion disk, non-thermal corona, broad-line emitting region, and relativistic jets. In radio-loud AGN, jets either produce or influence the production of a large fraction of non-thermal continuum, as indicated by the examples shown in the figure and resulting from VLBI-based investigations of individual objects (for details, see Acciari et al. 2010, Arshakian et al. 2010, León-Tavares et al. 2010, 2013, Agudo et al. 2011, Schinzel et al. 2012, Jorstad et al. 2013, Hada et al. 2014).

corona, and the broad-line emitting clouds (Fig. 3.1). Each of the major AGN constituents contributes to a specific domain in the broad-band spectral energy distribution (Ghisellini & Tavecchio 2009), and most of them can be probed through their interaction with relativistic jets produced in radio-loud galaxies.

Relativistic jets form in the vicinity of central black holes, with ample evidence connecting their formation and propagation to physical conditions in the accretion disc and broad-line region (cf., Arshakian et al. 2010, León-Tavares et al. 2010, Hada et al. 2014). Imaging and polarimetry of radio emission on milliarcsecond scales provided by very long baseline interferometry (VLBI) offers a range of possibilities for studying ultra-compact regions in relativistic jets (*e.g.*, Gómez et al. 2016) and relating them to main manifestations of activity in AGN.

Simultaneous monitoring of the optical and high energy variability and the evolution of parsec-scale radio structures yields a detailed picture of the relation between acceleration and propagation of relativistic flows and non-thermal continuum generation in active galaxies (cf., Acciari et al. 2010, Arshakian et al. 2010, León-Tavares et al., 2010, 2013, Jorstad et al. 2013, Hada et al. 2014). Opacity effects provide a measure of magnetic field strength on scales down to $\sim 1000$ gravitational radii, $R_g$, and trace the distribution of broad-line emitting material (Lobanov 1998, Fromm et al. 2013).

Correlations observed between parsec-scale radio emission and optical and gamma-ray continuum indicate that a significant fraction of non-thermal continuum may be produced (particularly during flares) in extended regions of relativistic jet at distances up to 10 parsecs from the central engine (Agudo et al. 2011, Schinzel et al. 2012). Combined with studies of jet component ejections and X-ray variability, these correlations also suggest that time delays, nuclear opacity, and jet acceleration may have a pronounced effect on the observed broad-band variability and instantaneous spectral energy distribution (León-Tavares et al. 2010). A number of recent studies, combining VLBI polarisation and opacity measurements with multi-band information provide an ever growing support for magnetic fields as one of the most important factors governing the generation and outward transport of the energy in AGN (cf., Zamaninasab et al. 2014, Baczko et al. 2016, Mertens et al. 2016, Walker et al. 2018)

These examples demonstrate that VLBI observations of relativistic jets remain a unique tool of choice for localising, classifying, and investigating the broadband continuum and broad spectral line emission produced in AGN. This unique position of VLBI is solidly based on its unmatched capability observation to provide two-dimensional and time-dependent kinematic, spectral, and polarisation information about regions down to the immediate vicinity of the central black holes in AGN. Future directions of the EVN research should fully exploit and enhance this capability.

### 3.1.2 Jet physics from VLBI observations

Jets in active galaxies are formed in the immediate vicinity of the central black hole, at distances of 10–$10^2\,R_g$ (Camenzind 2005, Meier et al. 2009). The jets carry away a fraction of the angular momentum and energy stored in the accretion flow (Blandford & Payne 1982), the non-thermal corona (in low luminosity AGN; Merloni & Fabian 2002), or the rotating magnetosphere of the central black hole (Blandford & Znajek 1977).

Table 3.1: Spatial, temporal, and emission scales in jets

| **Region** | **Characteristic scales** | | | | $\nu_m$ |
|---|---|---|---|---|---|
| | Natural [$R_s$] | Linear [pc] | Angular | Temporal | [GHz] |
| Launching | 10–$10^2$ | $10^{-4}$–$10^{-3}$ | 0.02–0.2 $\mu$as | 1.5–15 h | >100 |
| Collimation | $10^2$–$10^3$ | $10^{-3}$–$10^{-2}$ | 0.2–2.0 $\mu$as | 0.6–6 d | 50–100 |
| Acceleration | $10^3$–$10^6$ | $10^{-2}$–$10^1$ | 0.02–2.0 mas | 0.25–17 y | 10–50 |
| Propagation | $10^6$–$10^{10}$ | $10^1$–$10^5$ | 0.005–20″ | 17–17000 y | 0.1–10 |
| Dissipation | $10^{10}$–$10^{11}$ | $10^5$–$10^6$ | 0.33–3.3′ | 0.17–1.7 My | <0.1 |

**Notes:** Scales are calculated for a $10^8 M_\odot$ AGN at a distance of 1 Gpc (these also correspond to a $10 M_\odot$ XRB at a distance of 100 pc) The temporal scales pertain to the light crossing times of the respective length scales. The peak frequency $\nu_m$ gives an approximate range for the synchrotron turnover frequency.

In all of these mechanisms, electromagnetic processes and magnetic fieldss in particular play a pivotal role in launching, collimating, and accelerating the jets on scales of up to $10^6\,R_g$ (cf., Komissarov et al. 2007, Lyubarsky 2009, Fendt 2011, Marscher 2014). This calls for particular attention to multi-frequency VLBI observations recovering broadband spectrum, polarisation, opacity, and internal structure of the flow on these scales.

A sketch of an extragalactic jet shown in Fig. 3.2 in comparison to the typical resolution of VLBI observations, and basic properties of different jet regions summarised in Table 3.1 indicate that effective studies of jet launching and formation should be done at frequencies above 22 GHz, while the jet propagation is best assessed at lower frequencies.

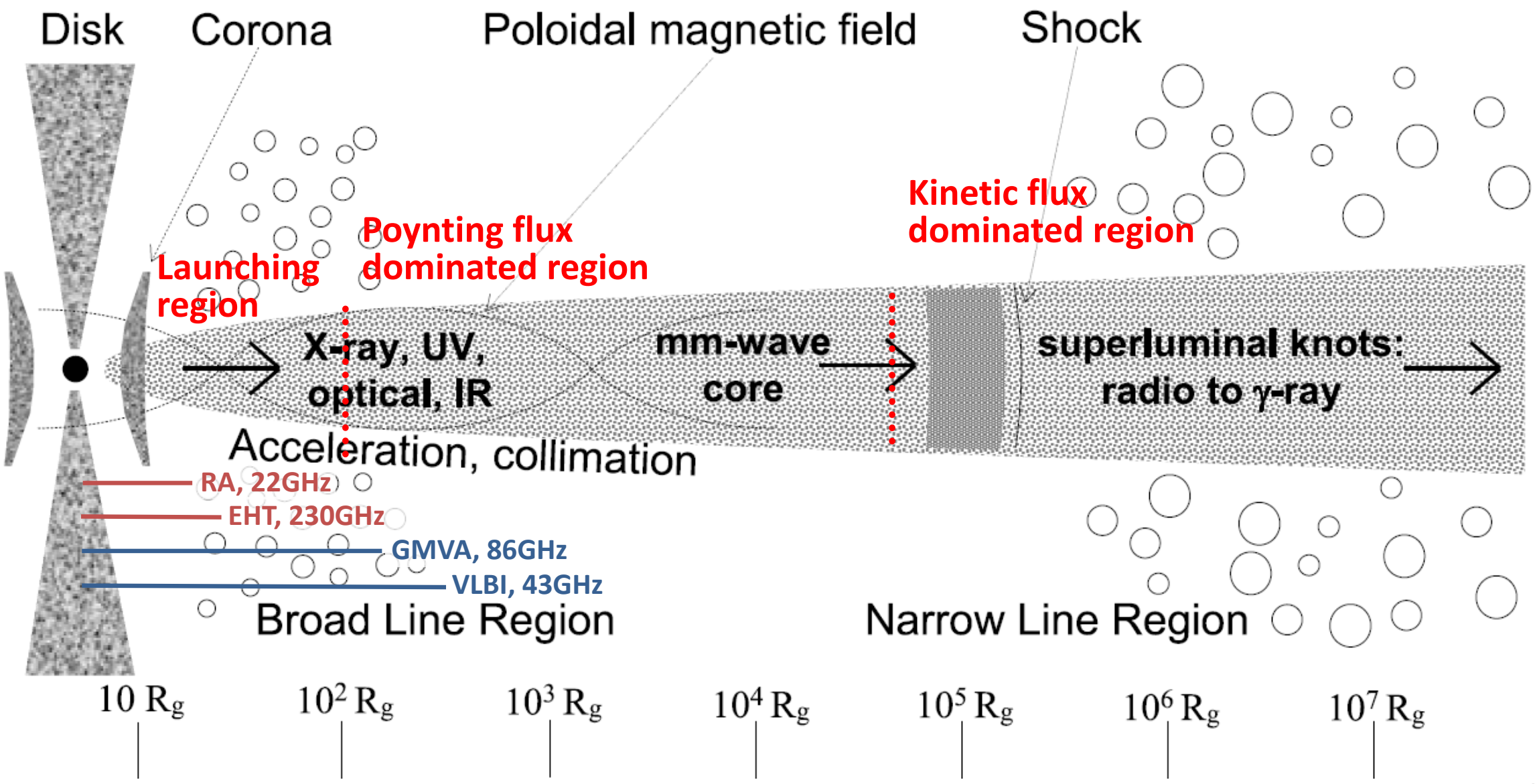

Figure 3.2: Sketch of a relativistic jet in AGN in relation to other constituents of the central region and a typical resolution of different VLBI instruments when targeting a nearby object such as M 87. The global VLBI at 43 GHz is sufficient to probe well into the jet collimation and acceleration region, and space VLBI observations with *RadioAstron* at 22 GHz could resolve even the jet launching region. These observations however may be hindered by the opacity in the jet and in the ambient medium. Observing at higher frequencies with the GMVA and EHT should alleviate this problem.

*RadioAstron* observations of 3C 84 at 22 GHz (Fig. 3.3) and multi-frequency measurements of the jet collimation and velocity field (Fig. 3.4 in M 87 exemplify the effectiveness of high-fidelity VLBI observations for studying these regions of the flow. These observations suggest that the jet may indeed tap their energy from both the accretion disk and the magnetosphere of the black hole (cf., Hardee et al. 2007). Further support for this idea comes from multi-frequency polarisation studies (cf., Gabuzda et al. 2014, Molina et al. 2014) indicating the likely presence of helical magnetic fields in the jets on these linear scales.

To be effective in future studies of this kind, VLBI observations should put the main emphasis on achieving high-fidelity polarisation measurements at multiple bands, warranting a true two-dimensional recovery of the spectral and magnetic field information. With this approach, it would become possible to understand the role and relative contribution of the accretion disk and black hole magnetosphere to the jet formation, and to look for the elusive non-thermal corona in low-luminosity AGN, and to distinguish between different scenarios of jet propagation and energy release.

### 3.1.3 Immediate vicinity of event horizon scales

Direct imaging of the event horizon scales is currently one of the prime goals of VLBI studies of AGN, with the focus of attention firmly laid on the Event Horizon Telescope (Doeleman et al. 2012) observing the central black holes in Sgr A* and M 87 at 230 GHz. While holding the best promise for revealing a compelling evidence for the existence of the event horizon, the EHT observations alone may not necessarily be able to distinguish a true black hole from one of its "mimickers" such as, for instance, the gravastar (cf., Chirenti & Rezzolla 2017).

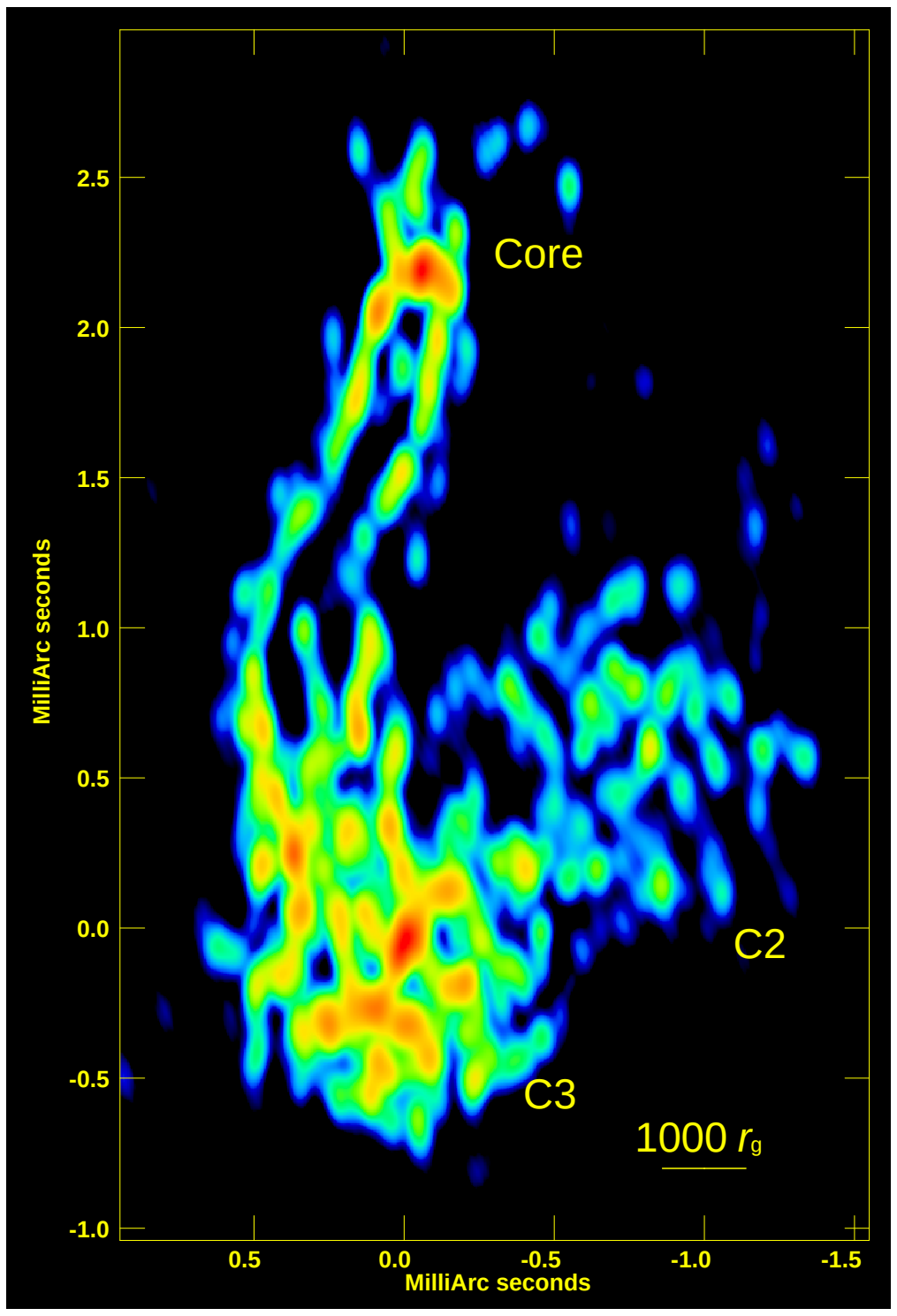

Figure 3.3: Inner jet in 3C 84, observed at 22 GHz with *RadioAstron* (Giovannini et al. 2018). The jet is transversally resolved and shows an extremely rapid collimation into an essentially cylindrical profile with a transverse radius of $\sim 250 R_g$. This suggest either an extremely rapid lateral expansion of a magnetospheric (Blandford-Znajek) jet or the launching of the jet from the accretion disk (via Blandford-Payne mechanism).

High-sensitivity and high-fidelity GMVA observations at 86 GHz already now can help discriminating between different models for the jet formation (Fig. 3.5). These observation also provide a much needed input for physical modelling of the data from the EHT observations at 230 GHz including the information about the jet rotation, and velocity distribution (cf., Mertens et al. 2016, Walker et al. 2018). Further improvements of the dynamic range of GMVA observations will result in increasing the effective image resolution of VLBI images at 86 GHz, providing a much more detailed support for modelling jets and recovering physical information from EHT images.

An even more important contribution to studies of the event horizon scales in AGN would be provided by dedicated measurements of magnetic field strength and structure made on scales below $\sim 1000 R_g$, where the maximum magnetic field of $\sim 10^4$ G expected for the black hole scenario can be effectively contrasted with much stronger dipole fields expected for most of the black hole alternatives (Lobanov 2017). Effective measurements of magnetic fields on scales of $< 1000 R_g$ can be enabled by making detailed VLBI observations of polarisation (cf., Gómez et al. 2016), Faraday rotation (cf., Gabuzda et al. 2014, Molina et al. 2014) and opacity (cf., Kovalev et al. 2008, Pushkarev et al. 2012, Fromm et al. 2013, Plavin et al. 2019).

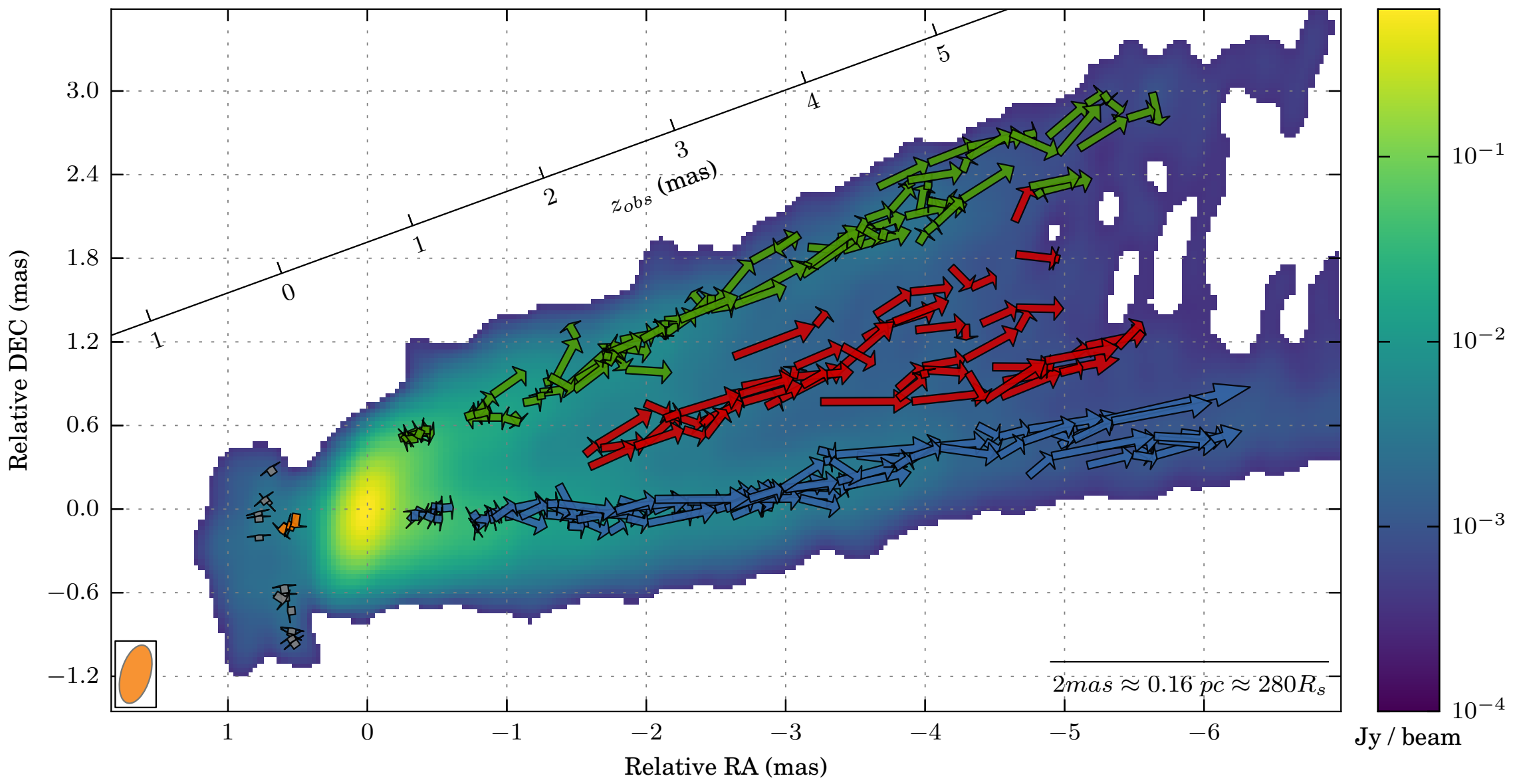

Figure 3.4: Two-dimensional velocity distribution in the jet M 87 obtained from wavelet-based decomposition of a sequence of VLBI images at 43 GHz. The velocity field reveals the presence of two distinct velocity components and, combined with the collimation profile of the flow strongly suggests that the jet launching occurs both at the accretion disk and near the black hole magnetosphere, with the latter flow reproducing the faster spine of the flow. Credit: Mertens et al. (2016), reproduced with permission ©ESO.

### 3.1.4 Main research directions for the coming decade

Enabling robust, frequency agile, polarimetric VLBI operations in the 1.6–86 GHz range of frequencies will provide an excellent foundation for addressing a number of outstanding fundamental questions about physics of relativistic jets and cosmic black holes, as well as the role they are playing in shaping up the galactic activity. Starting from these premises, one can identify three major areas of engagement for VLBI studies of relativistic jets to be made in the coming decade:

1. Multi-band observations at frequencies below 22 GHz would give ultimate answers about the physical mechanisms of jet collimation and acceleration.
2. Systematic studies of internal structure, velocity fields and their evolution would provide the clues necessary to assess the role and relative prominence of the Blandford-Payne and Blandford-Znajek mechanisms of energy extraction from accreting black holes. Extending such VLBI studies to low luminosity jets should also give better understanding of the physical nature of the non-thermal corona believed to be playing a crucial role in shaping the activity in radio-quiet AGN.
3. Dedicated imaging programmes at 86 GHz and systematic measurements of magnetic field strength and structure on linear scales below 1000 $R_g$ would provide crucial information about the jet launching and the physical nature of cosmic black holes.

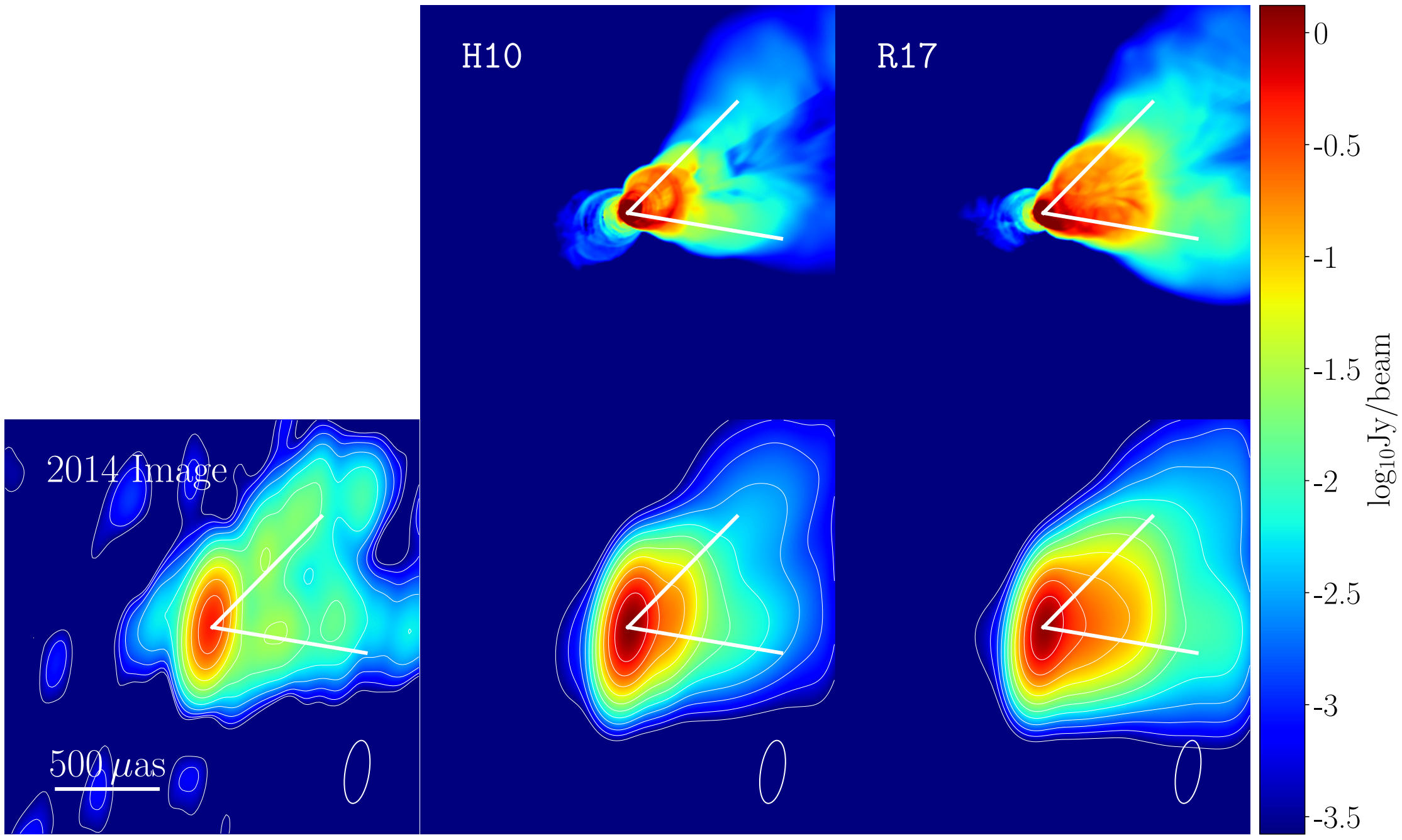

Figure 3.5: GMVA image of M 87 (lower left panel; Kim et al. 2018, reproduced with permission ©ESO) contrasted with results of numerical simulations of a jet produced from a two-temperature magnetically arrested disk (Adapted Fig. 12 from Chael et al. 2019). The full resolution frames of two specific models (top frames) are shown in the bottom row after being convolved with the same restoring beam as in the real VLBI image. Limited conclusions about the preferred scenario can already be made from the analysis of the internal structure of the jet, slightly favoring the H10 model (see Chael et al. 2019 for details of the modelling). Improving the dynamic range and fidelity of GMVA imaging would increase the effective resolution of 86 GHz VLBI images and provide much stricter constraints on the physical models.

### 3.1.5 VLBI at microarcsecond resolution

Successful engagement with the fundamental issues identified above relies on steady improvement of technical and imaging capabilities of VLBI measurements. Effective studies of the event horizon scales require VLBI imaging at a $\sim 10\,\mu$as resolution and sensitive to smooth emission on angular scales up to $\sim 5$ mas. Such requirements impose stringent constraints not only on thermal noise sensitivity but also on the phase stability and *uv*-coverage of VLBI observations.

At frequencies below 15 GHz, the planned development of the BRAND receiver (tuccari et al. 2017) should provide excellent frequency coverage and agility while maintaining the detection sensitivity. At higher frequencies steady increase of the observing bandwidth will bring further reduction of the thermal noise and improve the detection threshold. However, as the image dynamic range equally depends on the phase noise, other factors contributing to it should also be dealt with.

Reducing the amplitude noise, $\sigma_{\rm amp}$, increases the effective resolution $\theta_{\rm res} \propto FWHM_{\rm beam}\sqrt{\sigma_{\rm amp}}$. Conversely, reduction of phase noise, $\sigma_{\rm phas}$ improves positional accuracy, $\Delta_{\rm pos} \propto FWHM_{\rm beam}\sigma_{\rm phas}$. A potentially very powerful way to achieve both these improvements is offered by the multi-frequency (22/43/86/129 GHz) receiver technology (Han et al. 2013) developed at the Korean VLBI Network (KVN). The source frequency phase-referencing technique (SFPR, Rioja et al. 2015) enabled by this receiver has already allowed the KVN to achieve a $\sim 30\,\mu$as astrometric accuracy on a $\sim 500$ km baseline (cf., Dodson et al. 2017). Further improvements are expected on the imaging dynamic range (Zhao et al. 2018), once it is implemented on a full-fledged imaging VLBI array.

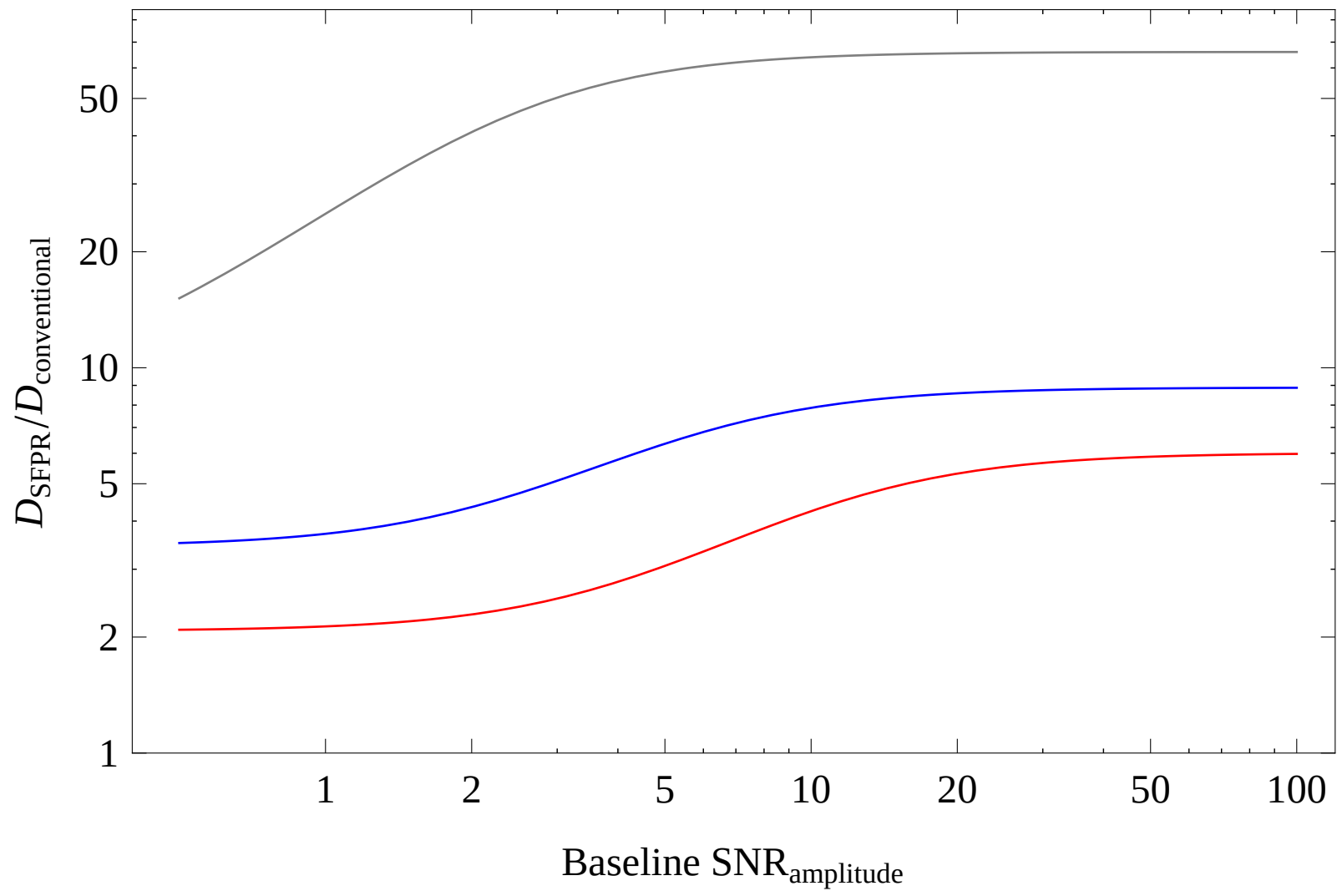

Figure 3.6: Dynamic range of SFPR imaging compared to conventional imaging. The comparison is done as a function of the average SNR of visibility amplitudes in the VLBI data. Quite expectedly, the largest improvement is realised for high amplitude SNR data, and it can reach a factor of 50 for imaging with the EHT.

Improvements of image dynamic range expected from implementation of the SFPR for imaging are shown in Fig. 3.6 for VLBI observations at 43, 86, and 230 GHz. The strongest improvement is achieved for imaging with high SNR of the amplitudes and for EHT it can reach a factor of 50, while an order of magnitude improvement is expected for the GMVA observations.

A minimum option of equipping a few EVN antennas with dual (22/43) or triple (22/43/86) frequency modification of the KVN receiver design would readily improve positional measurements, achieving a $\sim 10\,\mu$as astrometric accuracy. Enabling the SFPR option on an imaging array such as the GMVA, the VLBA, or the EAVN(+EVN) would substantially improve fidelity and effective resolution of imaging. The expected effective resolution of 86 GHz VLBI imaging may even exceed that of the present EHT observations, which would revolutionise both the GMVA and the EHT measurements, as discussed in Jung et al. (2015).

### 3.1.6 Conclusions and recommendations

Dedicated and systematic observations of relativistic jets performed at frequencies between 1.6 and 86 GHz hold an excellent potential to address several outstanding fundamental questions about the jet physics and physical conditions in the immediate vicinity of the supermassive black holes in AGN. These measurements may also hold the best clue about the very nature of the cosmic black holes. The success of these studies relies on the continued technical progress in VLBI observations and imaging. The advances offered by the developments of broadband (BRAND) and multi-band (KVN) receiver technology would provide revolutionise VLBI imaging in the 1.6–86 GHz regime and they should be considered for a broad implementation at the active VLBI instruments.

## 3.2 High-redshift AGN and SMBH growth

Quasars are the most prominent AGN, and in fact they are the most powerful non-transient objects in the Universe. Because of their extreme luminosities, they can nowadays be detected from vast cosmological distances. The current record holder is J1342+0928 at redshift $z = 7.54$, corresponding to the cosmological epoch when the Universe was just at 5% of its present age (Bañados et al. 2018b). Radio-loud (or, by adopting the more physically motivated term, jetted; Padovani 2017) quasars constitute the minority ($\leq$10%) of the population. However, from VLBI point of view, they are important observational targets because their structures can be imaged with the finest details, surpassing the resolution offered by any other astronomical technique. The most distant jetted AGN currently known is J1429+5447 at $z = 6.21$ (Willott et al. 2010; Frey et al. 2011). While high-resolution radio interferometric studies of jetted AGN at any redshift in general provide a wealth of unique information on the physical and geometric properties of jets, the accretion onto SMBHs, the jet launching and the emission mechanisms, these objects at very high redshifts ($z > 3-4$) are of particular interest from a couple of points of view. First of all, they should represent the earliest phases of SMBH activity in AGN. Therefore they are crucial for understanding the cosmological evolution of SMBHs and their co-evolution with the host galaxies. On the other hand, high-redshift AGN are promising probes of cosmological models (see e.g. Melia & Yennapureddy 2018, for a recent review). In the subsequent sections we describe high-redshift quasars, as well as peculiar types of AGN (dual/multiple systems; intermediate-mass black holes) that are important targets for the understanding of SMBH growth and evolution.

### 3.2.1 Blazars as tracers of high-redshift jetted AGN

At low redshifts, the jetted AGN are observed to be $\sim$10% of the total AGN population. Assuming reasonable values for the jet beaming angle ($\sim 3^\circ - 7^\circ$), simple geometrical considerations and an isotropic distribution of jet orientations, we expect that around 1–2 out of 200 jetted sources have their jets aligned close to our line of sight, i.e. can be classified as *blazars*. This source ratio is well proven by comparing high-energy blazar catalogues (such as *Fermi*/LAT catalogues; Atwood et al. 2009, Ackermann et al. 2011) with radio and optical quasar catalogues (such as SDSS+FIRST surveys; York et al. 2000, Shen et al. 2011), that include misaligned blazar counterparts and non-jetted analogous sources. All-sky high-energy catalogues are not able to go beyond $z \sim 3.5$, so to verify this relation, a systematic search of jetted sources was needed. In particular, blazars are extremely efficient tracers for the overall jetted population: the observation of a single blazar under a viewing angle smaller than its jet beaming angle ($\theta_v \leq \theta_b \simeq 1/\Gamma$, where $\Gamma \sim 10-15$ is the bulk Lorentz factor of the emitting region) implies the presence of $2\Gamma^2 \sim 200-450$ analogous jetted sources with their jets directed randomly in the sky. For each blazar with specific mass, accretion rate and jet power, we can infer the existence of hundreds of radio sources with the same mass, accretion rate and jet power, only with their jets directed elsewhere. For a few years, this approach has been applied to high-redshift catalogues, to study the high-redshift jetted AGN population.

From the early 2000s, a large collection of active SMBHs with masses comparable or larger than $10^9\,M_\odot$ has been observed, hosted in bright quasars up to redshift 7. Currently, around 200–300 sources have been identified at $z > 6$, mainly radio-quiet or silent. If the occurrence of jetted sources in the AGN population at such high redshifts is comparable with that of the low redshift population, less than 10% should show a jet, and none of them is expected to point towards our line of sight. At $z > 4$, the number of bright quasars significantly increases up to 1248 objects in the SDSS+FIRST quasar catalogue whose footprint is $\sim 1/4$ of the sky (Shen et al. 2011). For this reason, the search for very high-redshift blazars has been performed starting from this redshift range.

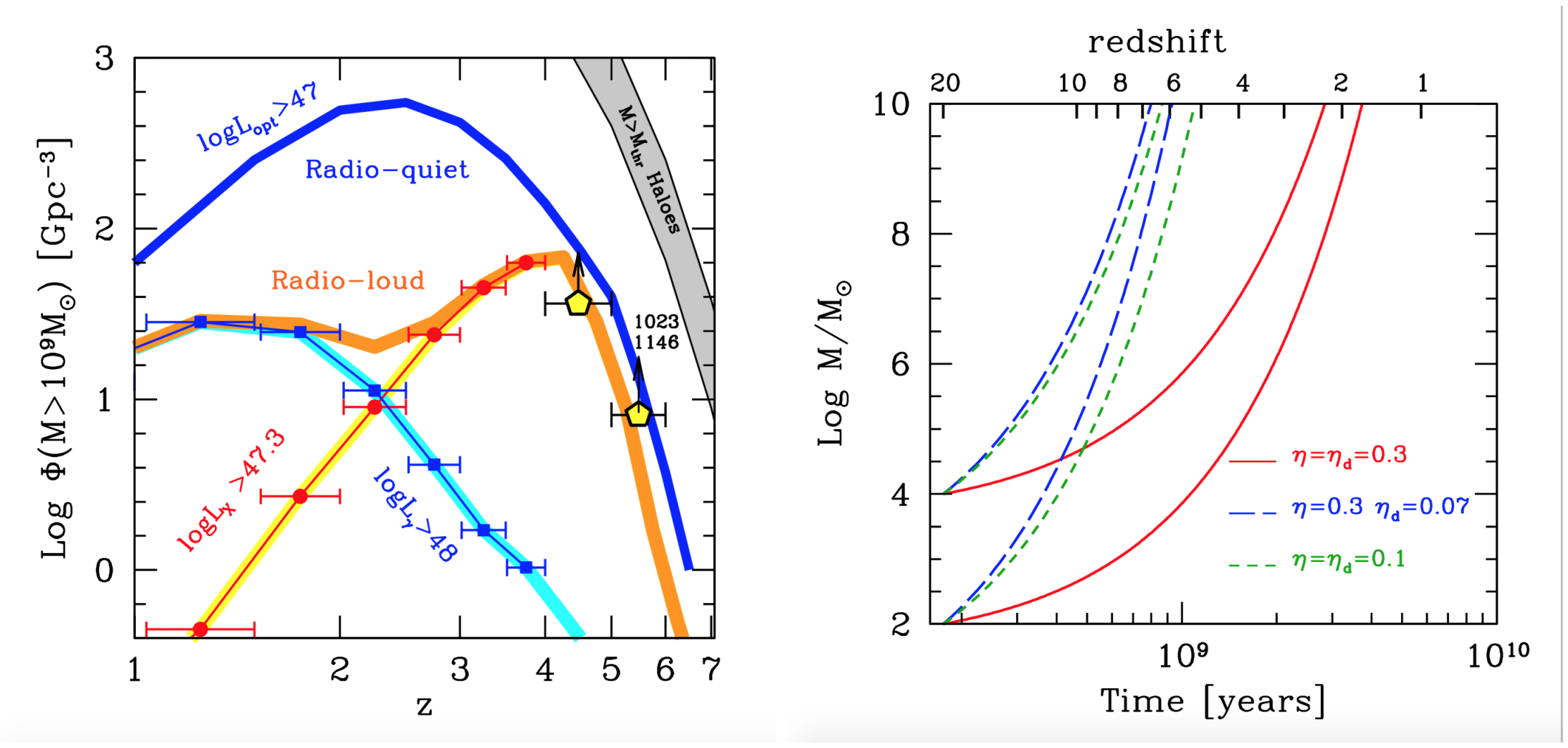

Figure 3.7: *Left panel:* comoving number density of active SMBHs with masses $> 10^9\,M_\odot$ hosted in non-jetted (blue line) and jetted quasars (orange line; Fig. 41 from Sbarrato et al. 2015). Blue points and cyan line are derived from the *Fermi*/LAT all-sky blazar catalogue, while red points and yellow line come from the *Swift*/BAT blazar catalogue. Yellow pentagons are derived from blazars classified through X-ray observations performed on blazar candidates at $z > 4$. *Right panel:* mass of a SMBH accreting at the Eddington limit as a function of time and redshift. Accretion is bound to start at $z = 20$ from a black hole seed of $10^2\,M_\odot$ or $10^4\,M_\odot$ with different efficiencies. Larger $\eta_d$ values correspond to a smaller amount of mass accreted on the black hole itself. If the fraction of gravitational energy released ($\eta$) contributes both to radiation ($\eta_d$) and energy launching a jet, the growth time decreases (Fig. 3, Ghisellini et al. (2013).

Blazar candidates are selected from SDSS+FIRST survey as extremely radio-loud quasars. Dedicated X-ray observations are then performed, in order to confirm their blazar nature. A hard and strong X-ray spectrum is the unmistakable signature of high-energy emission from a jet aligned to our line of sight. A detailed broad-band spectral energy distribution (SED) fitting helps in constraining SMBH accretion and jet emission features, including viewing angle and bulk Lorentz factor. Currently 8 sources have been classified as blazars in the SDSS+FIRST spectroscopic survey (Sbarrato et al. 2012, 2013, 2015; Ghisellini et al. 2014, 2015) that allow to infer the presence of $\sim 2700$ jetted quasars with masses $> 10^9\,M_\odot$, accreting at around 10% of the Eddington limit only in the SDSS+FIRST sky area. The results of these classifications allow us to study the comoving number density of active massive black holes at high redshift, and compare the jetted and non-jetted populations. The left panel of Fig. 3.7 shows that the comoving number density of active $> 10^9\,M_\odot$ SMBHs hosted in jetted AGN peak around $z > 4$, much earlier than non-jetted sources of same SMBH mass, that peak at $z \sim 2.5$. This is extremely interesting and challenging from the point of view of early SMBH formation: is it possible to have more jetted than non-jetted AGN in the early Universe?

The large amount of high-redshift very massive quasars already puzzles the AGN community, since such extreme masses are difficult to assemble in the short amount of time available for these sources to evolve (at $z \sim 6$ the age of the Universe is only 900 Myr). Introducing a predominant presence of jets in this puzzle might complicate the already existing issue. The right panel of Fig. 3.7 shows that a non-spinning SMBH accreting at the Eddington limit from a seed of $10^2\,M_\odot$ in standard conditions (i.e. with a radiatively efficient standard accretion disk that releases a fraction $\eta = 0.1$ of its gravitational energy in radiation) is hardly able to build up $10^9\,M_\odot$ before $z \sim 6$ (green short dashed line). In the case of $z \sim 4$ sources, these assumptions hold. Jets, though, are generally associated to highly spinning black holes, and therefore a release of gravitational energy $\eta = 0.1$ typical of non-spinning black holes is not acceptable. The radiative efficiency associated with a maximally spinning black hole is much larger, of the order of $\eta = 0.3$. This means a much smaller amount of mass accreted on the black hole at fixed luminosity (Eddington-limited) in the same amount of time. In other words, maximally spinning black holes are bound to accrete much slower, and blazars such as the ones classified in the last years cannot be assembled before $z \sim 2.5-3$ (red solid lines). This problem can be solved if a fraction of the gravitational energy released during the accretion goes into launching the relativistic jet, instead of being only radiated away. In other words, the release of gravitational energy follows $\eta = 0.3$, but only a fraction of it is radiated from the accretion disk, mimicking the emission from a non-spinning black hole ($\eta_\mathrm{d} = 0.1$). For an Eddington-limited luminosity, this assumption allows the black hole to accrete faster (long dashed blue line in right panel of Fig. 3.7). This extremely simple toy model gives a new view on the larger comoving number density of massive jetted sources at high redshift: the jet might actually be a way to accrete faster at a fixed accretion luminosity.

The blazars observed up to now in the X-rays have been lately complemented with high-resolution VLBI observations, as summed up in Coppejans et al. (2016). Most sources can be confirmed as blazars with this approach, such as J1026+2542 at $z = 5.266$ (Sbarrato et al. 2012, 2013) that even shows signs of jet component proper motion (Frey et al. 2015). But few of them show different features in radio than at high energies. J1420+1205 ($z = 4.034$) is the most striking example, clearly showing two distinct components in VLBI data (Cao et al. 2017), with clear signatures of being a misaligned source with respect to our line of sight. This inconsistency between high- and low-frequency results might be ascribed to the bending of the jet in a given source, or even to new emitting features for very high redshift sources.

VLBI observations are therefore key in solving these riddles hidden in the early Universe. How many jetted sources are really present? What does the jet emission look like at all scales at very high redshift? High-resolution VLBI observations pushed to lower flux density limits, combined with wider frequency-range X-ray spectroscopy will be instrumental to better understand jet emission and launching mechanisms, hopefully even in connection with SMBH accretion physics.

### 3.2.2 High-redshift AGN observations with VLBI

Since the early 2000s, following their optical spectroscopic discoveries, the majority of the presently known radio quasars at $z \approx 6$ have been observed with the EVN at 1.6 and/or 5 GHz (J0836+0054 at $z = 5.82$, Frey et al. 2003, 2005; J2228+0110 at $z = 5.95$, Cao et al. 2014; J1427+3312 at $z = 6.12$, Frey et al. 2008; J1429+5447 at $z = 6.21$, Frey et al. 2011). Imaging observations of these mJy-level weak radio sources largely benefit from the high sensitivity offered by the large apertures in the EVN. However, the VLBA has also been successfully used for studies of extremely high redshift AGN (Momjian et al. 2008, 2018), including the brightest ($\sim$100 mJy) source J0906+6930 at $z = 5.47$ (Zhang et al. 2017), and even for the radio-quiet source J0100+2802 at $z = 6.33$ (Wang et al. 2017) employing long total integration time. At a somewhat lower redshift limit ($z > 4.5$), a detailed account of VLBI observations is given by Coppejans et al. (2016). At the time of writing this document, there are slightly more than 30 jetted $z > 4.5$ AGN with VLBI imaging observations available, at least at a single observing frequency. Based on sometimes sparse information on their radio spectral indices, flux density variability and brightness temperatures, the ratio of off-axis (unbeamed) jets to Doppler-boosted jets pointing close to the line of sight is about unity (Coppejans et al. 2016).

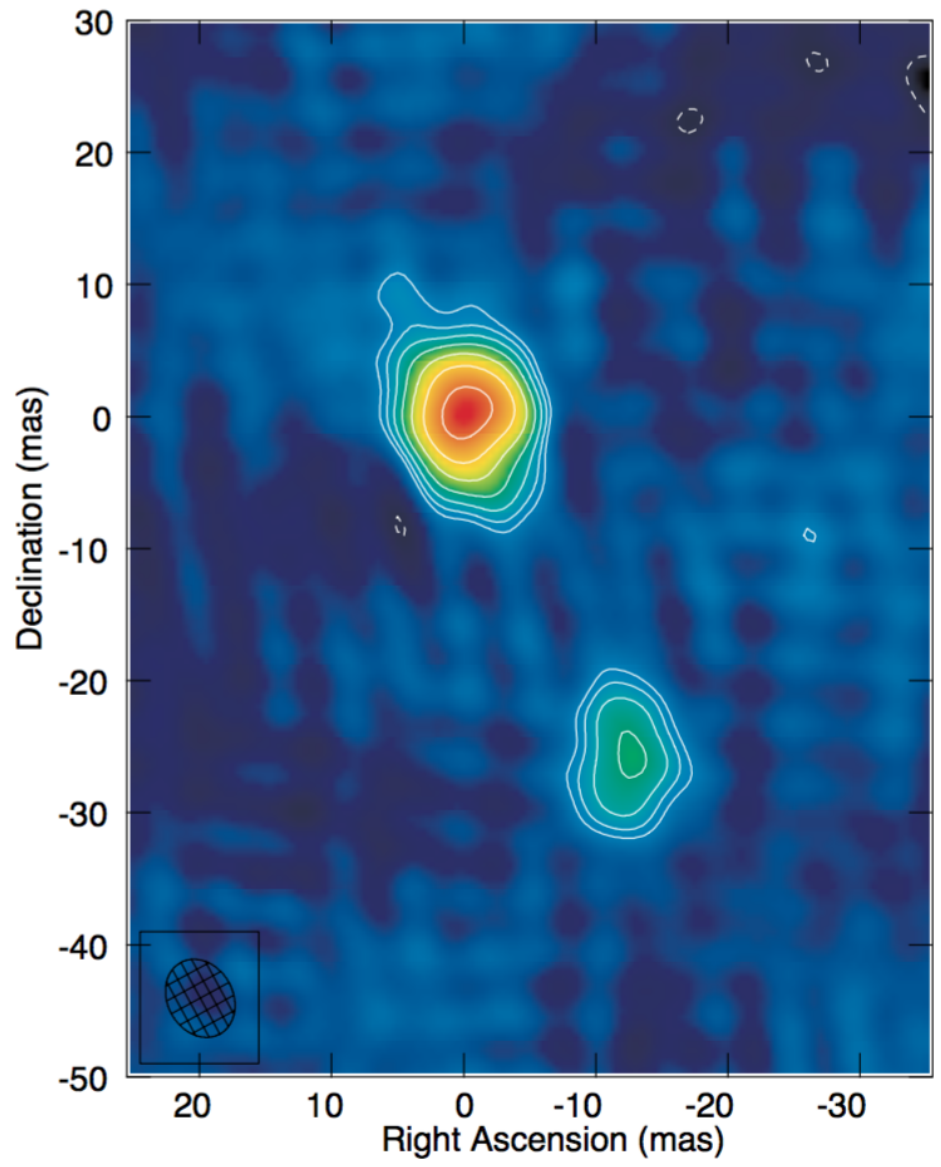

Figure 3.8: An example of a distant radio AGN with off-axis structure: the 1.6-GHz EVN image of the quasar J1427+3312 ($z = 6.12$) shows a double structure typical of young compact symmetric objects (CSOs). The positive contour levels increase by a factor of $\sqrt{2}$. The first contours are at $\pm 50\ \mu$Jy beam$^{-1}$. The peak brightness is 460 $\mu$Jy beam$^{-1}$. Credit: Frey et al. (2008), reproduced with permission ©ESO.

The enormous power of the most distant quasars originates from accretion onto SMBHs with billions of solar masses. The mere existence of such monsters well within 1 Gyr after the Big Bang is a serious challenge for the models that have to explain their initial formation. It is currently not understood in the framework of the ΛCDM cosmology how matter could have assembled so quickly in these black holes. Theoretical models either assume hyper-Eddington accretion rate with $> 100\,M_\odot$ seed black holes (e.g. Inayoshi et al. 2016) or a rapid creation of very massive ($\sim 10^5$ $M_\odot$) black hole seeds (e.g. Alexander & Natarajan 2014). Also unknown is the final redshift frontier of the formation of jetted AGN which could supply information on the accretion process, possibly constraining the models.

Based on growing observing evidence, its seems that blazars, i.e. radio AGN with jets nearly pointing to us, are increasingly over-abundant with respect to unbeamed jetted sources at $z > 3$ (e.g. Volonteri et al. 2011). The reason for this apparent disagreement between measurements and expectations is not found yet. Ghisellini & Sbarrato (2016) proposed a scenario where an obscuring bubble surrounds the AGN at very high redshifts, effectively blocking the lines of sight different from those of the jet direction, preventing their optical detection and spectroscopic redshift measurements. On the other hand, a remarkable fact is that none of the five $z \approx 6$ sources imaged with VLBI to date falls into the blazar category.

Recent EVN observations of $z > 4$ blazar candidates (Cao et al. 2017) revealed two "impostors" which share the properties of blazar broad-band spectral energy distributions but their radio emission is clearly not relativistically beamed. Instead, they show nearly symmetric sub-arcsecond radio structures. This underlines the importance of VLBI measurements in providing the ultimate evidence for the blazar nature of jetted AGN.

For good statistics, we need large samples. Major advances are anticipated in the field of high-redshift AGN research in the next decades. Presently, less than 3000 objects with measured spectroscopic redshifts at $z > 4$ are known, but their number is expected to grow rapidly. For example, the Panoramic Survey Telescope & Rapid Response System 1 (Pan-STARRS1) started to deliver high-redshift discoveries (e.g. Bañados et al. 2018a). It seems that the jetted fraction of AGN does not show an obvious evolutionary trend until at least $z \approx 6$ (Bañados et al. 2015), so we can expect that nearly $\sim$10% of the new discoveries will become suitable VLBI targets. The definition of high-redshift AGN samples for radio studies critically depends on optical spectroscopic measurements and we should understand the associated selection effects.

### 3.2.3 Multiple supermassive black hole systems

#### SMBH growth through mergers: a natural consequence of galaxy evolution

Dual or binary supermassive black hole (SMBH) systems have long been predicted to be common in the Universe (Begelman et al. 1980), the expected combined result of (a) hierarchical galaxy formation (e.g. Springel et al. 2005; Schaye et al. 2015), and (b) all massive galaxies hosting a nuclear black hole (Kormendy & Richstone 1995). Their primary importance, arguably, stems from the expectation that sub-pc (more accurately, $\lesssim 0.01$ pc) binary SMBHs dominate the stochastic gravitational wave background at nHz-$\mu$Hz frequencies (e.g. Wyithe & Loeb 2003; Sesana 2013). However, another major consequence of these systems is on the host galaxies themselves. Hydrodynamical simulations predict significant increases in the star formation rate and AGN accretion rate as a dual/binary SMBH pair spiral in towards one another and disrupt the neutral and ionised gas angular momentum (e.g. Mayer et al. 2007; van Wassenhove et al. 2012).

Despite this forecasted ubiquity and the predicted binary-SMBH impacts, our observations of these systems remain limited. There are still very few compelling binary/dual SMBH candidates with separations $\ll 1$ kpc (e.g. Rodriguez et al. 2006; Valtonen et al. 2008; Deane et al. 2014). The methods used to identify binary SMBH candidates are diverse, spanning the full electromagnetic spectrum, employing direct imaging, temporal variability, and spectroscopic signatures. There has been much activity in the past decade in developing new approaches; however, VLBI has always held a distinct advantage for two primary reasons, (a) its unparalleled angular resolution in astronomy, and (b) its ability to isolate regions of high brightness temperature emission while remaining insensitive to dust obscuration (e.g. An, Mohan & Frey 2018). VLBI can identify binary SMBH candidates through a number of methods, but most prominently through the imaging of two flat-spectrum radio cores, with evidence of nearby ejecta from one or both. An excellent example of this is 0402+379 (Rodriguez et al. 2006). Fainter, higher-redshift examples are expected to increase as more wide-field VLBI surveys are carried out. For example, there were two compelling binary SMBH candidates discovered in a $\sim 10\,\mu$Jy beam$^{-1}$ survey of the COSMOS field (Herrera Ruiz et al. 2017).

#### Synergies with Gravitational Wave Experiments

Given the small number statistics, the properties of low separation binaries themselves are ill determined (e.g. typical binary in-spiral rates and environmental coupling at sub-kpc scales). These are important to constrain as they directly determine the low-frequency gravitational wave spectrum, particularly at nHz frequencies (Ravi et al. 2014). Indeed, without environmental coupling or host galaxy triaxiality, binaries would take of order a Hubble time to merge via gravitational radiation alone, following the ejection of most matter within binary orbital separations of $\sim$1 parsec (Merritt & Milosavljević 2005). Statistics from a large sample of binaries will measure the in-spiral rate and directly address the question of whether binaries ‘stall’ or not (Burke-Spolaor 2011).

In-spiral rates are also determined by orbital eccentricity. As shown in a number of studies (e.g. Sesana 2013; Ravi et al. 2014), stellar scattering driven models predict that if typical binary SMBHs have an initial eccentricity of $e_0 \sim 0.7$ at formation, the expected characteristic strain at 1 nHz is suppressed by a factor of a few. This of course has significant impact on the detectability timescale of the GW signal with current and future pulsar timing array experiments. Therefore, direct constraints from an EM perspective are critically important in attempts to find consistency between black hole growth models, merger rates, and gravitational wave signal detections (or the lack thereof).

In addition to the contribution towards and benefit from pulsar timing array results, VLBI could play an important role in understanding the lower-mass SMBH in-spiral and merger rate. The number of merger events detected by *LISA* in the $10^{-4.5} \lesssim \nu_{\rm GW} \lesssim 10^{-2}$ range will need to be reconciled with the number of spatially-resolved systems in the corresponding mass range of $M_{\rm BH} \sim 10^{5-7}$ M$_\odot$ (Schaye et al. 2015), for lower redshifts where these correspondingly lower luminosity systems might be detectable with VLBI. This will require sensitivity enhancements or increased observing time on VLBI arrays, given the lower mass black holes. Our window of the gravitational wave Universe is set to expand dramatically over the next 1-2 decades and VLBI stands poised to play an important role across the GW spectrum.

#### Leveraging off the radio survey revolution: VLBI follow-up of binary candidates from arcsec-scale imaging

Deep, wide-area surveys at arcsec-scales with new and upgraded radio telescopes are set to revolutionise our view of the radio sky. The log-spiral configurations and large numbers of antennas in new instruments such as MeerKAT and SKA1-MID result in dramatic image fidelity improvements. These traits combined with sheer number of sources expected to be detected will result in new opportunities in the search for exotic radio-jet morphologies, most ably performed using machine learning techniques (Aniyan & Thorat 2017). A subset of these outlier sources may well be caused by binary SMBH-induced precession (orbital or geodetic, e.g. Kaastra & Roos 1992; Krause et al. 2019) as well as mergers (e.g. Merritt & Ekers 2002; Krause et al. 2019). VLBI will be critical in following up this these large samples of exotic radio source morphologies.

Simulations of gas-rich systems suggest that the in-spiral from $\sim$1000 to $<$1 parsec takes of order several Myr (Blecha et al. 2011; van Wassenhove et al. 2012), which is well-matched to a typical radio jet synchrotron lifetime time of order $\sim 10-100$ Myr. Therefore, multiple modulations of the jet axis should be present in a select number of systems. An excellent example of binary-induced precession is possibly seen in Cygnus A (see Fig.3.9), the archetype radio FR-II radio galaxy recently found to host a kpc-scale binary SMBH (Perley et al. 2017). NGC 326 is another low-redshift example thereof, featuring S-shaped jets (or X-shaped, depending on angular resolution and projection effects) that are very likely caused by the presence of a kpc-scale and spatially-resolved black-hole pair (e.g. Merritt & Ekers 2002).

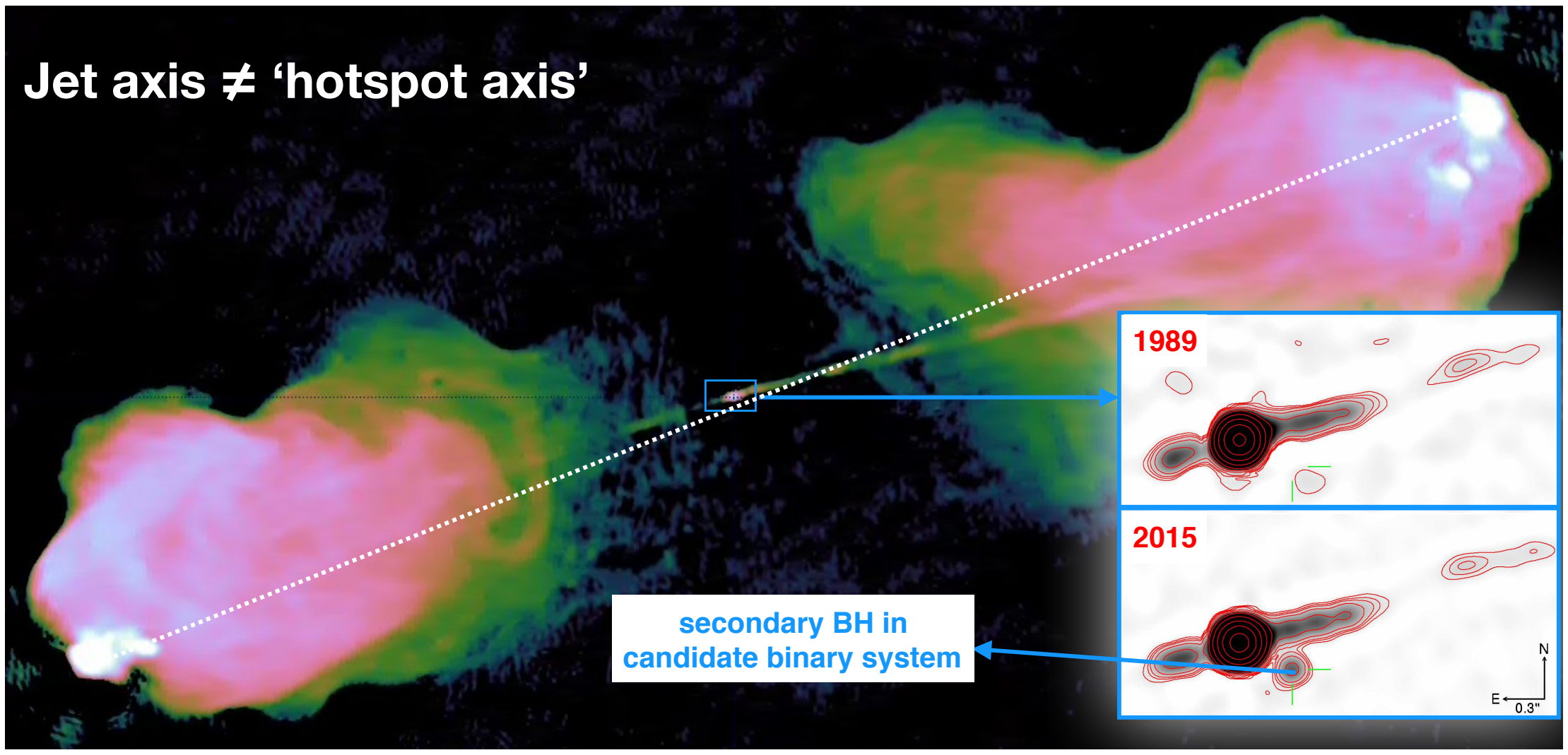

Figure 3.9: Cygnus A (image credit: Smirnov & Perley) jet axis appears misaligned with the vector between the brightest hot spots (white dashed line) – due to precession? (inset) Appearance of a transient radio source indicaties that Cygnus A is in a binary SMBH system (Perley et al. 2017). ©AAS. Reproduced with permission.

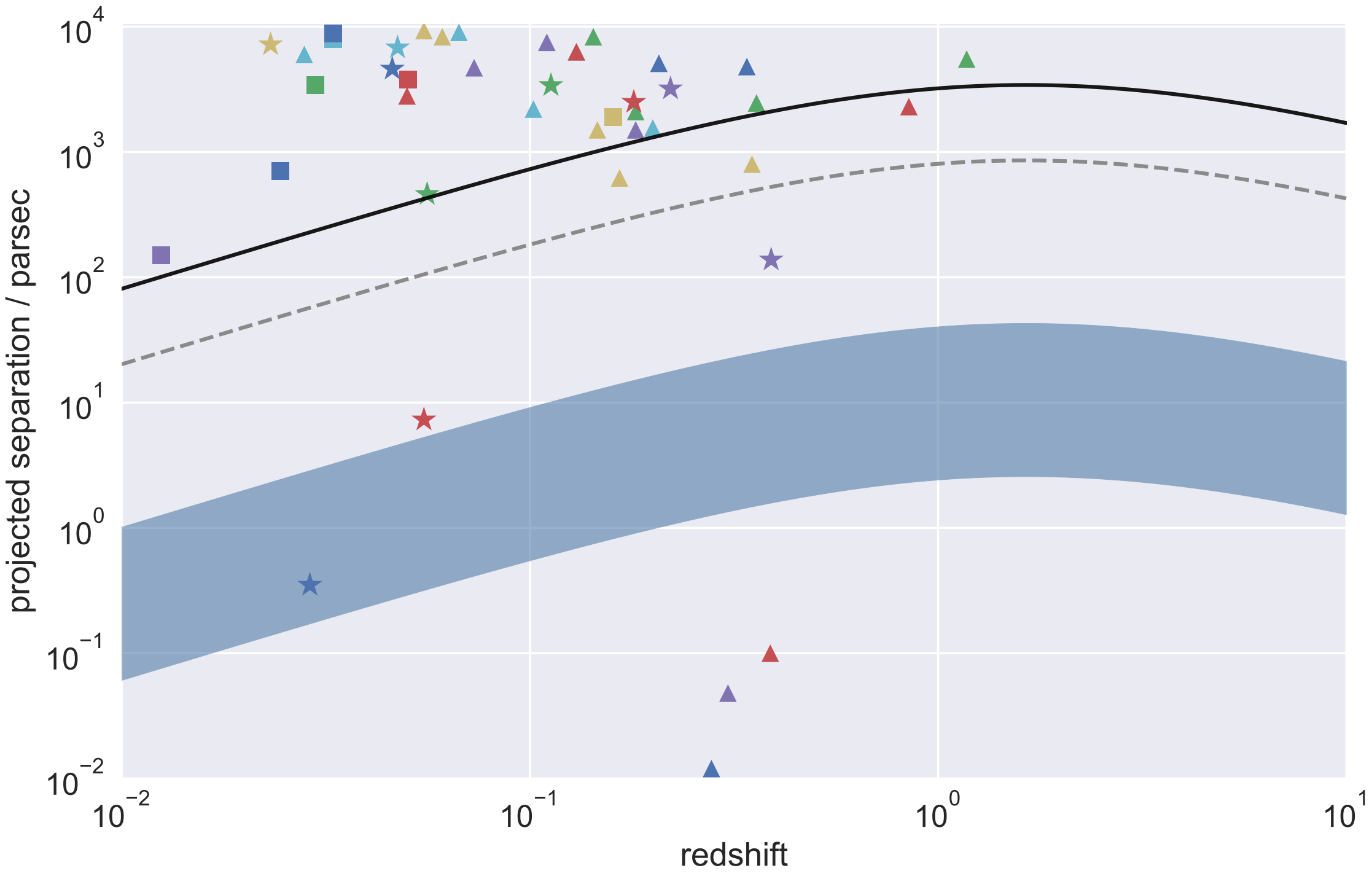

Figure 3.10: Image modified from Deane et al. (2015). Sample of dual/binary AGN candidates as revealed by X-ray (squares), optical/infrared (triangles) and radio (stars) wavelengths. Spatial resolution limits: EVN (for frequency range 1-22 GHz), dark blue region; *HST* – dashed grey line; *Chandra* – solid black line.

All the above examples suggest that a close-pair ($\sim$kpc-scale) dual/binary SMBH could in principle be discovered using the signature modulations imprinted onto the radio jets. Recently, Krause et al. (2019) performed a systematic search for precessing jets in the 3CRR sample (Lang et al. 1983), as well as a few other well-known radio sources. They find such evidence wide-spread, which, if interpreted as a signpost for binary SMBHs, points toward this being a valid tracer thereof. The striking assertion made by Krause et al. (2019) is that the majority of powerful radio sources host binary SMBHs, based on morphological traits consistent with geodetic precession. They make the case for why jet precession should be predominantly driven by binary black-hole systems, rather than jet-cloud interaction, peculiar velocity with respect to an intracluster medium, or tri-axiality of the host halo. The set of signature traits for jet precession will be readily detected in vast numbers, making VLBI critical to test this assertion and potentially expand this window into binary SMBH formation and evolution.

#### The role of VLBI in multiple SMBH research

VLBI has a unique and critical role to play in directly imaging binary SMBH systems towards GW-relevant separations ($\lesssim 0.01$ pc). As is seen in Fig. 3.10, VLBI is able to probe down to $\sim$ pc-scale separation for all redshifts and into the pulsar timing array sensitivity band for low-$z$ targets with high-frequency imaging. The projected separation lower limit in the figure is set to binary SMBH separations with gravitational wave frequencies that fall into the pulsar timing array sensitivity band (for $M_{\odot} \sim 10^8\,M_{\odot}$). The parameter space between 1-100 pc, which is almost exclusively accessed by VLBI, is critical to constrain in-spiral rates, the prevalence of 'stalled binaries', and the impact on the gravitational wave spectrum, as well as SMBH growth.

The advantage of VLBI in binary AGN discovery is the superior angular resolution and its high-brightness temperature filter. However, it is important to consider VLBI's limitations in this particular science case and how these may be alleviated with technological developments and array enhancements. In the past, its disadvantages have primarily been (a) low survey speed (b) low sensitivity, (c) dynamic range limits due to poor *uv*-coverage and imperfect calibration, and, of course, (d) the fact that not all active SMBHs are jetted, let alone have compact, high-brightness temperature components. Many of these limitations will be or to some extent have been significantly mitigated, posing a bright future for multiple SMBH searches with VLBI arrays. In order to identify these seemingly rare systems, wide-field surveys will need to become a more commonly used VLBI observing mode - both for contiguous surveys of legacy multiwavelength fields as well as targeted follow-up of promising samples (e.g. radio-jet or optical host morphology selected).

Just as for the inner-most regions of AGN, VLBI studies of multiple SMBHs require multi-band and polarimetric imaging across about a decade of frequency coverage ($\sim 1-20$ GHz), to ensure the nature of emission components can be decoupled from one another, as well as utilising the angular resolution lever arm of high frequencies to probe smaller separation binaries. This science case will also greatly benefit from wide-band feeds for in-band spectral indices. The higher fractional bandwidths will provide improved *uv*-coverage and hence imaging fidelity and dynamic range, which in turn provide better deconvolution performance. These are crucial aspects for systems with a very faint secondary component.

While VLBI has a number of spectacular examples of binary SMBH candidates, the expectation is that array, technology, and technique enhancements will lead to these outlier examples becoming common-place in the coming 1-2 decades. This will not only serve to improve our understanding of black hole and galaxy evolution models, but more closely tie electromagnetic and gravitational wave observations of the Universe together.

### 3.2.4 Intermediate-mass black holes

The finding of quasars powered by SMBHs of $10^9$ $M_\odot$ when the Universe was only 0.7 Gyr old (e.g. Mortlock et al. 2011; Wu et al. 2015; Bañados et al. 2018b) suggests that these had to grow from seed black holes of 100-$10^5$ $M_\odot$ via accretion and merging (Volonteri et al. 2003). Such seed black holes could form from the death of the first generation of (Population III) stars, from direct collapse of gas in protogalaxies, or from mergers in dense stellar clusters (e.g. Volonteri 2010; Mezcua 2017). Detecting these early Universe seed black holes poses an observational challenge; however, the leftovers of those seeds that did not become supermassive should be found in the local Universe as intermediate-mass black holes (IMBHs) with $100 < M_{\rm BH} < 10^6$ $M_\odot$.

A myriad of studies have hence focused on searching for IMBHs in the local Universe. The first searches aimed at globular clusters, where several IMBH candidates have been proposed based on dynamical mass measurements and pulsar accelerations (e.g. Gebhardt et al. 1997, 2002; Lützgendorf et al. 2013; Kızıltan et al. 2017). However, no conclusive detections of accretion signatures have been achieved (e.g. Wrobel et al. 2015) even when reaching rms sensitivities of $\sim 1\mu$Jy beam$^{-1}$ via stacking of the VLA radio images of 24 globular clusters (Tremou et al. 2018).

Dwarf galaxies with quiet merger and accretion histories are thought to be the best analogues of the first galaxies and therefore an excellent place where to look for the relics of the early Universe seed black holes. A few hundreds of IMBH candidates have been found in dwarf galaxies as low-mass AGN ($M_{\rm BH} \lesssim 10^6$ $M_\odot$) based on optical (e.g. Greene & Ho 2004, 2007; Reines et al. 2013; Baldassare et al. 2015; Chilingarian et al. 2018) and infrared (Satyapal et al. 2008; Sartori et al. 2015; Marleau et al. 2017) emission line diagnostics or the detection of X-ray emission (e.g. Schramm et al. 2013; Lemons et al. 2015; Baldassare et al. 2017; Mezcua et al. 2016, 2018a). In galaxies having recently undergone a minor merger, such IMBHs would be off-nuclear and be detected as ultraluminous X-ray sources (e.g. Farrell et al. 2009; Bellovary et al. 2010; Webb et al. 2012; Mezcua et al. 2013a,b, 2015, 2018b). Very few systematic studies have however been carried out in the radio regime (e.g. Greene et al. 2006).

Detecting jet radio emission from IMBHs/low-mass AGN can tell us about their accretion physics, jet efficiency and power, and whether they are able to impart mechanical feedback on their hosts (i.e. hampering or triggering star formation) as more massive radio galaxies do (e.g. McNamara & Nulsen 2007, 2012; Tadhunter et al. 2014; Morganti et al. 2015; Maiolino et al. 2017). The finding that AGN feedback from IMBHs significantly impact the formation of stars in their host dwarf galaxies and thus the amount of material available for the IMBH to grow would have strong implications for seed BH formation models, as it would imply that the local IMBHs/low-mass AGN so far detected might not be the relics of the early Universe seed BHs (Mezcua 2019). Very few of the IMBHs/low-mass AGN found in dwarf galaxies or ultraluminous X-ray sources do however show jet radio emission (Greene et al. 2006; Wrobel & Ho 2006; Wrobel et al. 2008; Mezcua & Lobanov 2011; Nyland et al. 2012, 2017; Reines & Deller 2012; Webb et al. 2012; Mezcua et al. 2013a,b; Reines et al. 2014; Mezcua et al. 2015, 2018a,b). The radio fluxes of these sources barely reach more than 1 mJy at the sub-arcsecond angular resolutions of the VLA, which makes detecting and resolving the radio emission on VLBI milli-arcsecond scales extremely challenging (e.g. Greene et al. 2006; Wrobel & Ho 2006; Reines et al. 2011, 2014; Mezcua & Lobanov 2011; Mezcua et al. 2013a,b, 2015; Nyland et al. 2012; Reines & Deller 2012; Paragi et al. 2014). In order to disentangle core and jet radio emission and search for outflow morphologies at VLBI scales we need $\mu$Jy sensitivities, which would be achievable with the implementation of wide-band receivers in the EVN facilities. This would allow us to probe whether, similarly to SMBHs, jets are also produced and collimated on sub-parsec scales in IMBHs, supporting the universality of the accretion physics at all mass scales, and thus to apply the fundamental plane of black hole accretion to estimate the black hole mass for those sources with core radio emission (e.g. Gültekin et al. 2009, 2019; Plotkin et al. 2012). In the case of globular clusters, the $\mu$Jy sensitivities would allow us to search for IMBH accretion signatures without the need of multiple observations and posterior stacking techniques (e.g. Tremou et al. 2018).

### 3.2.5 Requirements and synergies – the role of the EVN

As in most fields of astrophysics, observations at multiple wavebands are necessary to paint a full picture of the processes in the early Universe. Radio AGN can also serve as illuminating background sources for HI 21-cm absorption studies of the intergalactic medium around the epoch of reionisation. The unique role of VLBI continues to be the ability of high-resolution imaging of jetted sources. Milliarcsecond-scale compact radio structure is conventionally used as the best evidence for jetted AGN activity, even at very low accretion rates. The improving sensitivity and/or the extension of the available observing time at the EVN would be desirable to allow for a survey of a sizeable sample of traditionally "radio-quiet" sources from these enormous distances. An increase of the length of EVN observing sessions would leave more room for studies of newly-discovered high-redshift sources. Optical spectroscopic discoveries e.g. with the James Webb Space Telescope (*JWST*) Near-Infrared Spectrograph (NIRSpec) will certainly cover the southern sky as well. Because of their geographic locations, most EVN stations will be in a unique position for co-observing with the upcoming SKA and its precursor MeerKAT. The VLBI capabilities with the SKA (Paragi et al. 2015, Agudo et al. 2015) would be essential not only for the sensitivity increase but also for a better *uv*-coverage at low/southern declinations.

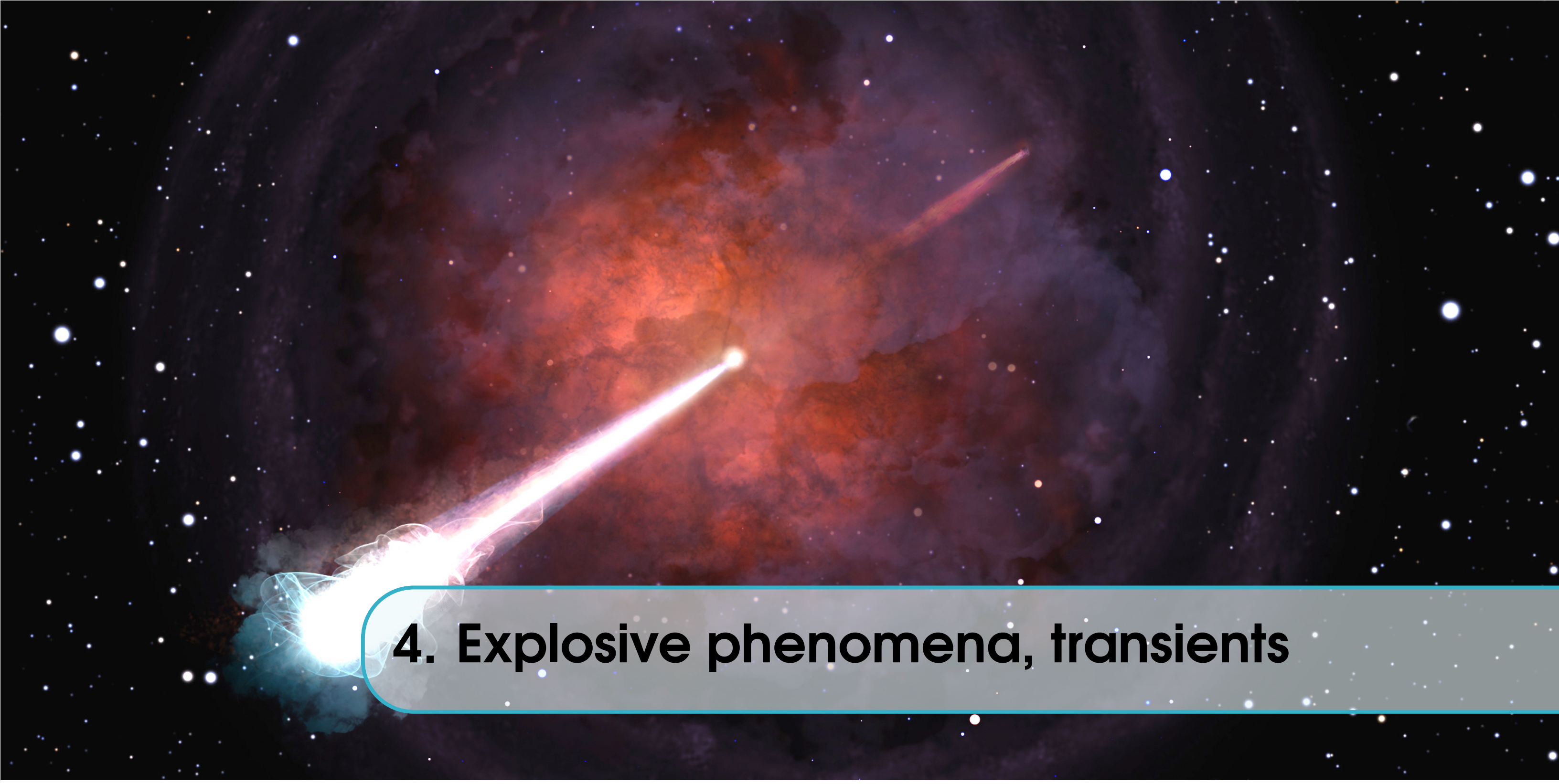

# 4. Explosive phenomena, transients

Transient sources of electromagnetic radiation are ideal sites to probe extreme physics. Some of the transient source classes provide the most luminous objects in the Universe, for the duration of the outburst. In the radio regime we distinguish between two broad types: broad-band, incoherent transient sources of non-thermal (typically synchrotron) radiation with a duration much longer than a few seconds, and coherent fast transients that last shorter than 2s.

Since transients are initially very compact, and can be very bright, they are natural targets for very high resolution observations. Arrays like the EVN, that consist of telescopes that are not fully dedicated to VLBI observations all year round, have had limited capabilities in addressing high impact transient science. This all changed with the development of the electronic VLBI technique, that allowed flexible organisation of regular, as well as Target of Opportunity e-EVN sessions (Szomoru 2008). The EVN transient science, and the first ten years of e-EVN work in particular is summarised by Paragi (2017). Below we describe the transient science the EVN could address in the future, for both incoherent (Sect. 1.) and coherent types (Sect. 2.) of transients.

## 4.1 Synchrotron transients

The first ever published science results using the e-EVN was observations of ‘microquasars’ Cyg X-1 (Rushton et al. 2007) and Cyg X-3 (Tudose et al. 2007). Therefore we start with radio-jet X-ray binaries, and describe other major transient classes in subsequent sections.

### 4.1.1 Black hole and neutron star X-ray binaries

Jets are now believed to be a common feature of accreting neutron star (NS) and black hole (BH) X-ray binaries (XRBs), with properties that depend on the nature of the accretion flow. Although they have only been directly resolved in the most luminous systems (e.g. Tingay et al. 1995, Spencer et al. 2013) their presence is often inferred through broadband spectra (typically including a spectral break between radio and infrared frequencies; e.g. Migliari et al. 2010), polarisation measurements

Chapter image: Jet breaking through the kilonova ejecta following the merger of two neutron stars. Beabudai Design.

indicating synchrotron emission (e.g. Corbel et al. 2000), or through their location and evolution in the radio/X-ray luminosity plane (e.g. Gallo et al. 2006).

While there are a handful of persistent radio-emitting XRBs, the majority of systems are transient. Outbursts of these transient systems show a large dynamic range in accretion rate, leading to significant changes in the geometry of the accretion flow and the morphology of the jets. These changes, typically occurring on timescales of days to weeks, allow us to study the universal coupling between accretion and ejection, which cannot be as easily probed in more slowly varying AGN.

The other key advantage of XRBs in studying jets is the range of accretor properties. Comparative studies of jets from BH and NS systems can elucidate the effect of the stellar surface and the depth of the gravitational potential well on the formation of jets. Furthermore, since magnetic field strengths and spin periods can be directly measured in NSs, we can readily explore their effect on jet formation. However, the relative radio faintness of NS XRBs has meant that to date they have been much less well studied with VLBI than their BH counterparts. Future VLBI sensitivity upgrades would therefore facilitate more detailed comparisons of jets from NS and BH systems.

#### High-resolution imaging

The past few decades of observational effort (primarily on the radio-brighter BH systems) have led to a paradigm in which the compact, steady radio jets that exist at low luminosities are quenched at the peak of an X-ray outburst, giving way to bright, relativistically-moving transient ejecta (e.g. Fender et al. 2004). Understanding how these jets are launched is one of the major open questions in the field. With time-resolved VLBI imaging, we can trace moving jet components back to the time of ejection, enabling us to pinpoint any X-ray spectral or variability signatures associated with the ejection event. This cannot be achieved using radio light curves alone, both due to optical depth effects and the difficulty of distinguishing direct jet emission from interactions with the surrounding medium (e.g. Rushton et al. 2017). While new methods of probing the jet base using high time resolution optical observations can provide complementary probes of the jet launching process, VLBI imaging allows a direct causal connection to be made between jet ejection events and the associated changes in the accretion flow. Future X-ray polarimeters (e.g. *IXPE* and *XIPE*) will also determine the inclination angle of the inner accretion flow. Together with jet proper motions from VLBI and optical constraints on the binary orbit, this will provide a powerful probe of the geometry of the system, including any warping or precession of the disk and their impact on the jets.

The rapid changes in both morphology and brightness on the timescale of a few-hour observation (as shown in Fig. 4.1) pose significant challenges for VLBI imaging of XRB jets. Splitting an observation into short time bins requires a source that is both sufficiently bright and structurally simple that its radio morphology can be reconstructed with minimal *uv*-coverage. Spurred by similar challenges for the Event Horizon Telescope, new algorithms are therefore being developed that will enable much-improved imaging of time-variable sources (e.g. Johnson et al. 2017).

One underexplored area that would benefit greatly from extra sensitivity is VLBI polarimetry of XRB jets. The lower resolution and lower surface brightness of XRB jets (as compared to AGN) has meant that the signal-to-noise ratio is not always sufficient to detect polarised emission, and most of our knowledge of the magnetic field structure in XRB jets stems from integrated radio flux densities from lower-resolution facilities such as ATCA or the VLA.

Another open question is how relativistic XRB jets can be. While a few cases of apparent superluminal motion have been observed (Mirabel et al. 1994; Tingay et al. 1995; Hjellming & Rupen 1995), the Lorentz factors of most XRB jets are very poorly constrained. However, it has been suggested that their Lorentz factors could reach $\Gamma > 10$, rivalling the jets of AGN (Miller-Jones et al. 2006). This question, which significantly affects the inferred energetics of the jets, can be answered by measuring the proper motions of corresponding approaching and receding jet ejecta in sources with accurate distances (Mirabel & Rodríguez 1999; Fender 2003).

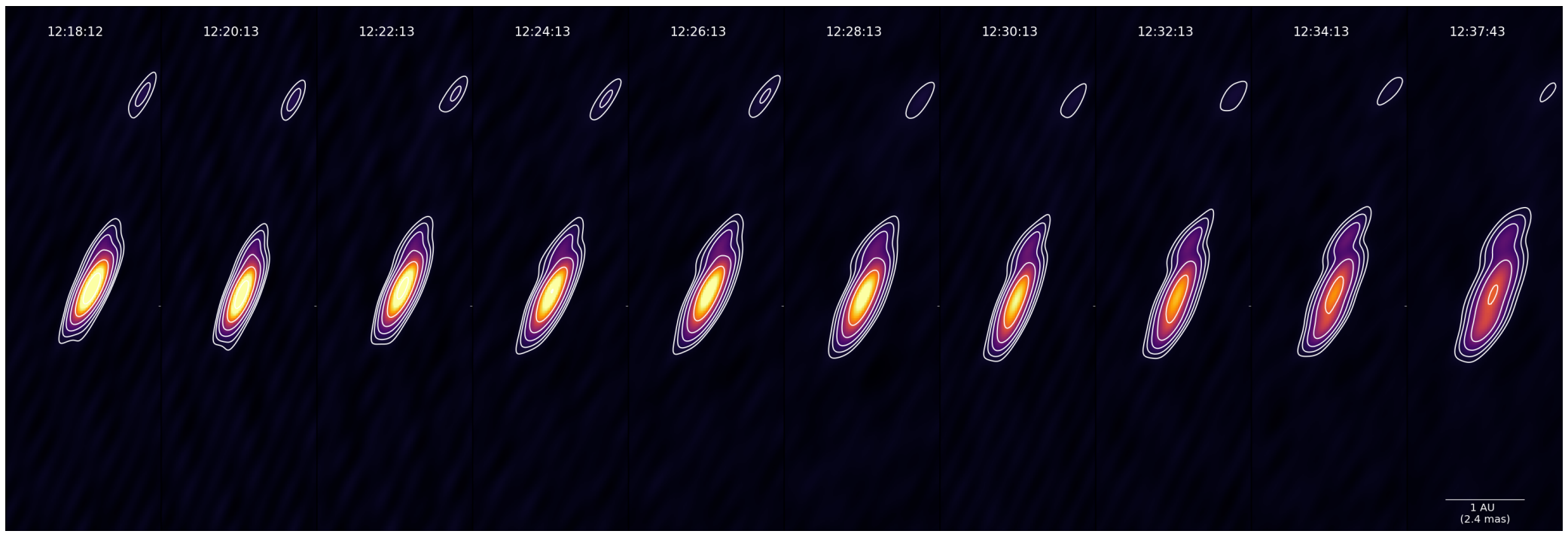

Figure 4.1: Time-resolved VLBA imaging of the BH XRB V404 Cyg on 22 June 2015 (time of each image shown in UT). The source evolves during each 2-minute snapshot (Miller-Jones et al. 2019).

### Astrometry

As core-dominated radio sources, XRBs in their hard or quiescent states provide excellent astrometric probes, allowing parallax and proper motion measurements. Even with the advent of *Gaia* DR2, only eleven BH XRBs have optical parallax measurements, with only three being significant at $\geq 4\sigma$ (Gandhi et al. 2019), such that distance estimates rely heavily on the choice of prior. Since many XRBs lie towards the Galactic bulge, the visual extinction is often too high to detect an optical counterpart, and high-precision VLBI astrometry will continue to play an important role in this area. However, the majority of systems are too radio-faint to be detected in quiescence, so astrometric campaigns must be undertaken during irregular outbursts. The advent of synoptic optical surveys by facilities such as ZTF and LSST will enable earlier detection of new outbursts, allowing us to obtain better astrometric sampling during the rising quiescent and hard state phases.

While distance measurements require extremely high precision and appropriate sampling of the parallax ellipse, XRB proper motions are significantly easier to measure, since the signal is both larger (a few mas $yr^{-1}$) and cumulative. Together with an accurate distance and an optical or infrared measurement of the systemic radial velocity, XRB proper motions can be used to determine the full three-dimensional space velocity of the system. Modelling of the Galactocentric orbit together with the binary evolution of the system (e.g. Willems et al. 2005) can provide excellent constraints on the process in which the compact object formed – whether in a natal supernova, or (in the case of BHs) by direct collapse. In the case of a supernova, both the recoil and any asymmetries in the explosion can lead to significant natal kicks, the distribution of which is still poorly observationally constrained. The first detections of gravitational waves from merging BHs and NSs has given renewed importance to determining the kick distribution, since large kicks will both unbind a primordial binary in the field, or eject a compact object from a globular cluster before it has a chance to dynamically form a binary. Thus, stronger natal kicks should reduce the rate of merger events (Wysocki et al. 2018). While the strongest natal kicks should unbind the progenitor binary system, milder kicks can introduce eccentricity into a binary orbit, thereby increasing the probability of a merger (Giacobbo & Mapelli 2018). Hence the differing biases of gravitational wave observations and X-ray binary astrometry provide different insights into the compact object formation process.

Proper motions can also be used to discriminate Galactic objects from extragalactic background sources, for example in deep radio surveys of dense regions such as the Galactic bulge or the globular cluster population (e.g. Strader et al. 2012). While population synthesis codes (e.g. Yungelson et al. 2006) predict $\gtrsim 10^4$ BH XRBs within the Milky Way, only a few dozen candidates are known. The remainder of the population may be transients that have either been in quiescence since the dawn of X-ray astronomy, or whose outbursts are too faint to trigger all-sky monitors. However, their radio jets could be detected in quiescence, and distinguished from extragalactic background sources via their proper motions and parallaxes, as was done recently for a BH candidate toward M15 (Kirsten et al. 2014; Tetarenko et al. 2016). Sensitive, wide-field VLBI surveys could reveal a larger fraction of this population. Finally, it has been suggested that with sufficient sensitivity, astrometric observations could even detect nearby isolated BHs accreting from the ISM (Fender et al. 2013), thereby revealing some of the closest BHs to Earth.

With sufficiently high precision, astrometry can not only measure the parallax and proper motion of an XRB, but also (for wide binaries) measure the orbit of the compact object. For systems with sufficiently bright optical counterparts, the combination of optical astrometry from *Gaia* and VLBI astrometry in the radio band could measure the orbits of both donor and accretor, thereby enabling us to determine the full set of system parameters, including the masses of both components (e.g. Miller-Jones et al. 2018). Such geometrically-determined masses could provide some of the most robust, model-independent mass estimates available for NSs and BHs.

### 4.1.2 Thermonuclear runaway supernovae

Type Ia SNe are the end-products of white dwarfs (WDs) with a mass approaching, or equal to, the Chandrasekhar limit, which results in a thermonuclear explosion of the star. While it is well acknowledged that the exploding WD dies in close binary systems, it is still unclear whether the progenitor system is composed of a C+O white dwarf and a non-degenerate star (single-degenerate scenario), or both stars are WDs (double-degenerate scenario). In the single-degenerate scenario, a WD accretes mass from a hydrogen-rich companion star before reaching a mass close to the Chandrasekhar mass and going off as supernova, while in the double-degenerate scenario, two WDs merge, with the more-massive WD being thought to tidally disrupt and accrete the lower-mass WD (e.g., Maoz et al. 2013) This lack of knowledge makes it difficult to gain a physical understanding of the explosions, and to model their evolution, thus also compromising their use as distance indicators.

Radio and X-ray observations can potentially discriminate between the progenitor models of SNe Ia. For example, in all scenarios with mass transfer from a companion, a significant amount of circumstellar gas is expected (e.g. Branch et al. 1995), and therefore a shock is bound to form when the supernova ejecta are expelled. The situation would then be very similar to circumstellar interaction in core-collapse SNe (see above), where the interaction of the blast wave from the supernova with its circumstellar medium results in strong radio and X-ray emission (Chevalier 1982). On the other hand, the double-degenerate scenario will not give rise to any circumstellar medium close to the progenitor system, and hence essentially no prompt radio emission is expected. Nonetheless, we note that the radio emission increases with time in the double-degenerate scenario, contrary to the single-degenerate scenario. This also opens the possibility for confirming the double-degenerate channel in Type Ia SNe via ultra-sensitive radio observations of decades-old Type Ia SNe. The best constraints on the mass-loss rate from Type Ia SNe come indeed from radio interferometric observations of SN 2011fe (e.g. Chomiuk et al. 2012), using the VLA, and of SN 2014J (Pérez-Torres et al. 2014; see also Fig. 4.2).

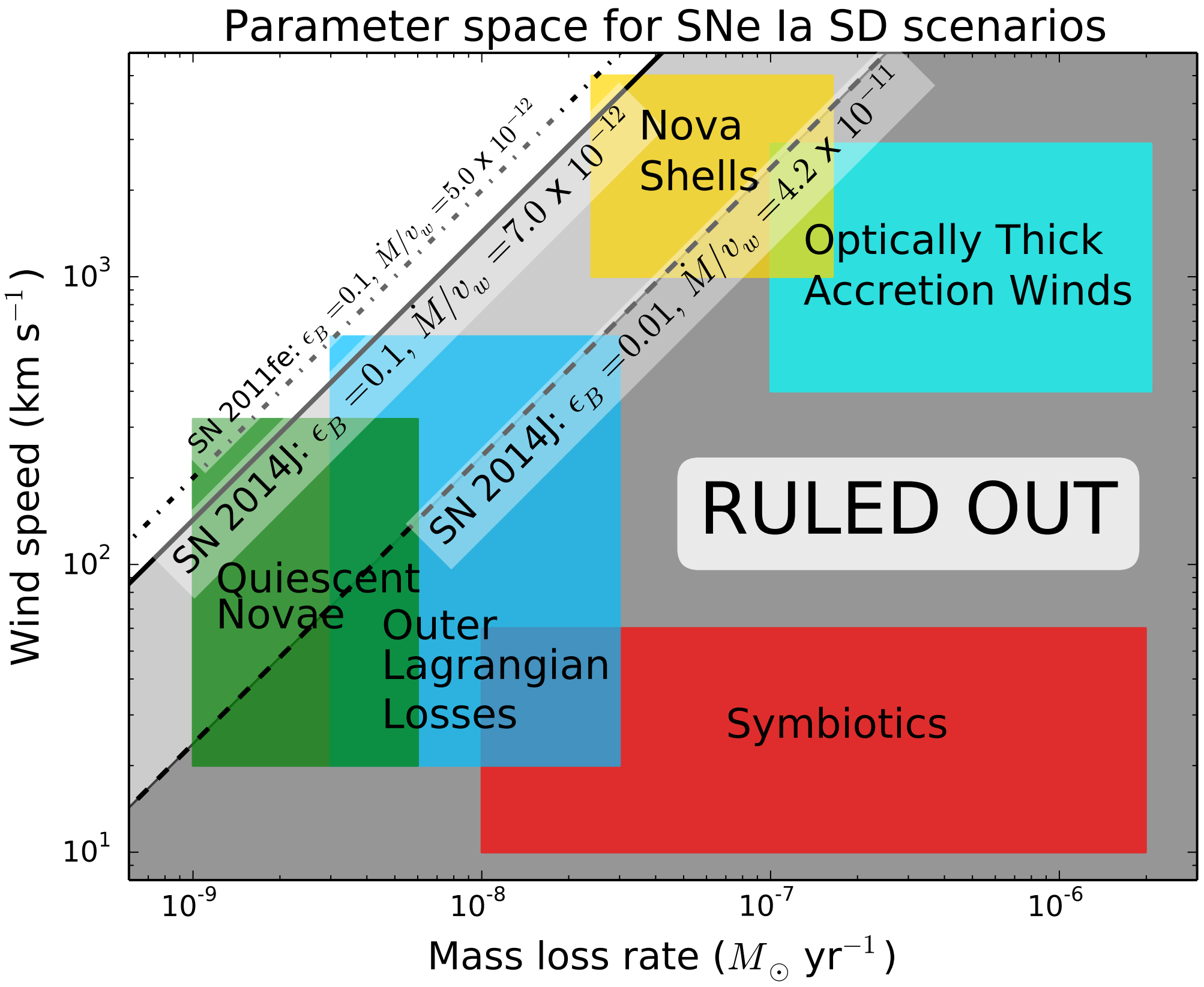

Figure 4.2: Constraints on the parameter space (wind speed vs. mass-loss rate) for single degenerate scenarios for SN 2014J. Progenitor scenarios are plotted as schematic regions. We indicate our $3\sigma$ limits on $\dot{M}/v_w$, assuming $\varepsilon_B = 0.1$ (solid line) and the conservative case of $\varepsilon_B = 0.01$ (dashed line). Mass loss scenarios falling into the gray-shaded areas should have been detected by the deep radio observations, and therefore are ruled out for SN 2014J. For comparison, we have included also the limit on SN 2011fe (dash-dotted line) for the same choice of parameters as the solid line for SN 2014J, which essentially leaves only room for quiescent nova emission as a viable alternative among the single-degenerate scenarios for SN 2011fe (Pérez-Torres et al. 2014). ©AAS. Reproduced with permission.

Unfortunately, the volumetric rate of Type Ia SNe is rather small, $\sim 3 \times 10^{-5}$ SN $\mathrm{yr}^{-1}$ $\mathrm{Mpc}^{-3}$ (Dilday et al. 2010). Therefore we should expect to have on average about 1 SN Ia every year within a radius of 20 Mpc. This illustrates two things: first, the crucial point of following each and every SN Ia that happens to explode nearby, as each one represents a unique opportunity. For example, SN 2011fe in M51 ($D \approx 7$ Mpc) and SN 2014J in M82 ($D \approx 3.8$ Mpc) yielded upper limits to the mass-loss rate of any putative non-degenerate companion of $\lesssim 5 \times 10^{-9}$ $M_\odot\,\mathrm{yr}^{-1}$and $\lesssim 7 \times 10^{-9}$ $M_\odot\,\mathrm{yr}^{-1}$, respectively. Those constraints are about two orders of magnitude deeper than probed by all previous radio interferometer observations together.

While VLA-like radio interferometers are more sensitive than the EVN, the latter will continue to play a very important role in those studies, thanks to its mas-angular resolution, which filters out the diffuse emission.

### 4.1.3 Core-collapse supernovae and long gamma-ray bursts

#### Core-collapse supernovae

The discovery of an expanding radio shell in SN 1993J (Marcaide et al. 1995a) using global VLBI, beautifully confirmed theoretical expectations (Chevalier 1982) for core-collapse supernovae (CCSNe) and opened an avenue for detailed studies of the supernova-circumstellar medium interaction, and the (direct) measurement of the supernova radius and its expansion speed as a function of time has implications on both the CSM and the progenitor's ejecta density profiles (Chevalier 1982).

However, so far only a handful of CCSNe have been imaged with enough detail as to make significant progress, including SN 1993J, a Type IIb SN, still being the best studied case after 25 years (Marcaide et al. 1995a,1995b,1997; Martí-Vidal et al. 2011a,2011b), SN 1979C (Marcaide et al. 2009), SN 1986J (Pérez-Torres et al. 2002), SN 2001gd (Pérez-Torres et al. 2005), SN 2008iz (Brunthaler et al. 2010).

With increased sensitivity and better *uv*-coverage provided by the growing number of antennas, the EVN should be able to probe the SN-CSM interaction for all CCSN types, from the relatively faint Type IIP to the extremely radio bright Type IIn SNe in the local Universe, via directing VLBI imaging of those SNe. Probing the SN-CSM interaction for all CCSN types will allow us to obtain basic, crucial information to characterise their progenitors, including the following: (i) pre-supernova mass-loss rates; (ii) shock radius measurements and radius evolution, allowing to study the different regimes of deceleration in the expansion, and whether self-similar expansion applies, or not; (iii) estimate the magnetic field–directly from VLBI observations–for synchrotron self-absorbed SNe.

Another relevant, yet poorly known aspect in supernova studies is (iv) the transition from (young) supernova to supernova remnant, i.e. the passage from a phase where the emission is driven by the interaction of the SN ejecta with the circumstellar medium to a phase where the radio emission is driven by the interaction with the ISM. To understand better that important phase in the lifes of SNe, we necessarily require long-term monitoring programmes to trace those changes in detail.

Finally, EVN observations of CCSN factories in the central regions of starbursts have yielded a unique opportunity to track the evolution of large amounts of CCSNe (e.g., Parra et al. 2007, Pérez-Torres et al. 2009, Batejat et al. 2011, Bondi et al. 2012, Varenius et al. 2019), compared the more focused, individual studies of SN, at the expense of missing the opportunity of directly imaging their expansion. However, advances in the modelling of the *uv*-data, which carries this fundamental information, is overcoming this limitation (e.g. Marti-Vidal et al. 2012), and will be probably the way to go forward.

In short, even after the advent of SKA1, the use of the EVN, whether as a standalone array, or as part of a global VLBI array, including MeerKAT as a a phased-array, will be crucial to attain the milliarcsecond angular resolution, a mandatory requirement to reach any of the above stated goals. Thanks to the slow evolution of CCSNe (essentially all type IIn, IIL, IIP, IIb SNe), the fact that the EVN does not have multi-frequency simultaneous capabilities is not a dramatic issue, but it must be recognised that this is a drawback, and a serious limitation for studies of the fast evolving CCSNe (Type Ib and Ic SNe). The existence of the e-EVN has improved the serious limitations that the EVN had in those studies, allowing to point to a "good" VLBI CCSN candidate ($D \lesssim 30$ Mpc) within one-two weeks from its explosion, and permitting for a reasonable monitoring of its evolution.

#### Long gamma-ray bursts

It is now widely accepted that long-duration GRBs are ultra-relativistic jets marking the deaths of a certain subset of massive stars. Since the discovery of the GRB–associated supernova (SN) 1998bw (to Galama et al. 1998), about a dozen core-collapse SNe have been identified in connection with GRBs (Modjaz et al. 2016). Intriguingly, all GRB-associated SNe discovered thus far are classified as type Ic with broad lines (BL-Ic), establishing a relationship between GRBs and BL-Ic events (Woosley et al. 2006). According to the fireball model (Piran 1999), for every GRB we see in $\gamma$-rays there should be $\sim 10-100$ more that launch misaligned jets and are missed at high energies, called off-axis GRBs. If an off-axis GRB went off relatively nearby, we would expect light from the associated BL-Ic SN to be visible initially, followed by an off-axis jet emerging as a strong radio source at later times, when the jet decelerates, spreads, and intersects our line of sight (Figure 4.3, panel C; see also Nakar & Piran 2003).

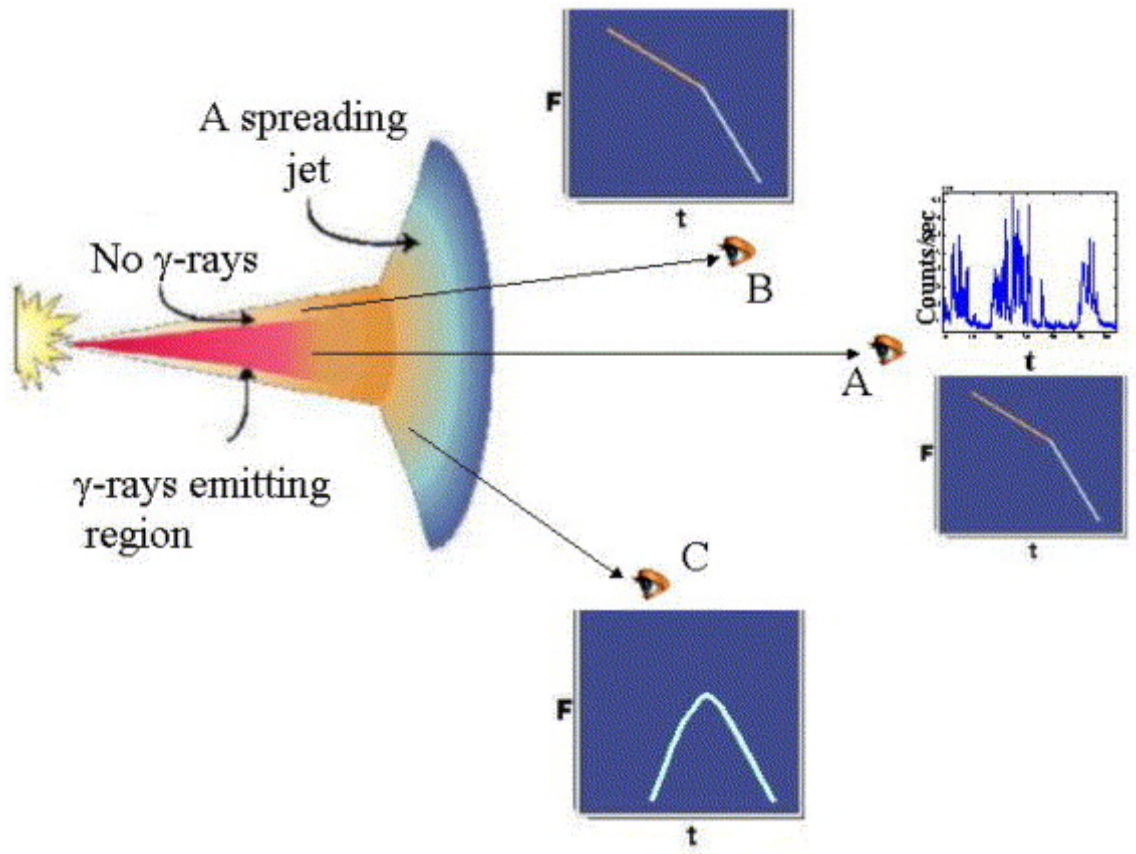

Figure 4.3: Observing long-GRB jets. Observer A is on–axis and detects the GRB and its afterglow. Observer B is on–axis but outside the gamma-ray emitting region, and detects an on–axis orphan afterglow similar to observer A. Observer C is off–axis (i.e. outside the initial jet opening angle) and detects a radio off–axis orphan afterglow that rises and falls and differs from the afterglow detected by observers A and B. Figure from Nakar & Piran (2003).

However, after more than 20 years since the discovery of GRB afterglows, we have, not yet been able to find an off-axis GRB associated with a BL-Ic SN. VLBI observations of long-GRBs are particularly useful, as they provide an extremely accurate and precise location of the GRB afterglow, as well as to monitor how the expansion proceeds. Direct size measurements with VLBI can also provide crucial information on the jet dynamics, in particular the sideways spreading of the jet. While it is well known what happens at early times when the jet is ultra-relativistic, and at late times when the outflow is non-relativistic and quasi-spherical; but what happens in between those two stages is less well established. In early GRB studies, it was assumed that the sideways spreading of the jet was very fast (e.g. Rhoads et al. 1999), but hydrodynamics simulations have shown that the sideways spreading is more gradual (e.g. Zhang & MacFadyen 2009). This is still a point of debate in the GRB community, and can probably only be resolved by accurate VLBI measurements of GRBs (see also Granot and van der Horst 2014).

VLBI observations can also provide crucial information to discern among the GRB scenarios at work, in particular to disentangle whether the GRB ejecta interacts with a steady circumstellar wind ($\rho \propto r^{-2}$), or with a constant density ISM.

Some SN Ic-BL show more than one peak in its radio luminosity curve. This behavior can be due most likely to either extreme CSM density variations, or to an emerging off-axis GRB. In the varying density scenario, a spherical SN shock powers the radio emission, and the double-peaked radio light curve would be due to density variations in the CSM, perhaps related to eruptive progenitor mass loss (e.g., Soderberg et al. 2005, Wellons et al. 2012, Salas et al. 2013, Corsi et al. 2014, Milisavljevic et al. 2015). In the emerging off-axis GRB, the radio emission observed during the first peak is produced by the spherical SN shock, while the second radio peak corresponds to the emerging off-axis jet initially pointed away from our line of sight (Fig. 1, panel-C).

The above scenarios make rather different predictions for the angular diameter size of the GRB at (very) late times. For example, for the case of PTF 11qcj, its angular diameter will reach $\approx 1$ mas around 2500 days post explosion (Palliyaguru et al. 2020). A much larger angular diameter of $\approx 25$ mas is instead expected for an off-axis jet at the same epoch. Therefore, VLBI observations of those GRBs are the only possible way to remove model degeneracies, which are left open by radio light curve monitoring, and ultimately test for the presence of an off-axis jet in this BL-Ic SN. We note that all previous claims of off-axis GRB jet discoveries have indeed been ruled out via VLBI angular diameter measurements (e.g., Granot & Ramirez-Ruiz 2004, Bietenholz et al. 2014). Thus, this method has proven to be extremely effective in the few cases so far observed.

### 4.1.4 Tidal disruption events

Tidal disruption events (TDEs) are transient flares of electromagnetic radiation produced when a star is ripped apart by the gravitational field tides of a supermassive black hole (SMBH; Rees 1988). During the disruption, one fraction of the stellar debris is ejected, whereas the remaining fraction is accreted onto the SMBH, generating a bright flare that is normally detected in the X-ray, UV, and optical part of the spectrum (see, e.g., Komossa 2015 for a review). Early theoretical models of tidal disruptions predicted the accretion rate onto the SMBH and consequently, the tidal disruption flare decay with characteristic time dependence $L \propto t^{-5/3}$ (Rees 1988; Evans & Kochanek 1989). Subsequent studies have found deviations from this power law dependence for different stellar structure of disrupted stars and different initial orbits (Lodato et al. 2009; Guillochon & Ramirez-Ruiz 2013).

In addition to these thermal signatures, some TDEs show evidence for existence of powerful relativistic jets (Burrows et al. 2011; Levan et al. 2011; Zauderer et al. 2011; Bloom et al. 2011; Cenko et al. 2012). Jetted TDEs are expected to produce associated radio transients, lasting from months to years, and form an important population which can be used to study the formation and evolution of relativistic jets in otherwise dormant SMBHs (Giannios and Metzger 2011).

Indirect evidence of such a putative radio jet has been inferred for *Swift* J1644+57 (e.g. Mimica et al. 2015), but the source remains unresolved with VLBI; ultra-precise EVN astrometry constrained the apparent ejection velocity to less than 0.3c, averaged over three years (Yang et al. 2016). ASASSN-14li was another case where indirect evidence pointed to a radio jet (van Velzen et al. 2016), although the nature of the radio emission from this object is controversial, with others interpreting it as being due to a disk wind (Alexander et al. 2016). EVN observations of ASASSN-14li (Romero-Canizales et al. 2016) resolved the radio structure into two components, but the interpretation of the observations was also inconclusive. The direct confirmation of a TDE radio jet has therefore remained elusive until very recently, when Mattila et al. (2018) detected a milliarcsecond-long radio jet in the dust-enshrouded TDE Arp 299-B AT1 in the nearby (D=45 Mpc) galaxy merger Arp 299 (see also Fig.4.4). The role of the EVN in this study was crucially important.

The gravitational field of the SMBH in Arp 299-B, with a mass 20 million times that of the Sun, shredded a star with a mass more than twice that of the Sun. This resulted in a TDE that was not seen in the optical or X-rays because of the very dense medium surrounding the SMBH, but was detected in the near-infrared and radio. The soft X-ray photons produced by the event were efficiently reprocessed into UV and optical photons by the dense gas, and further to infrared wavelengths by dust in the nuclear environment. Efficient reprocessing of the energy might thus resolve the outstanding problem of observed luminosities of optically detected TDEs being generally lower than predicted.

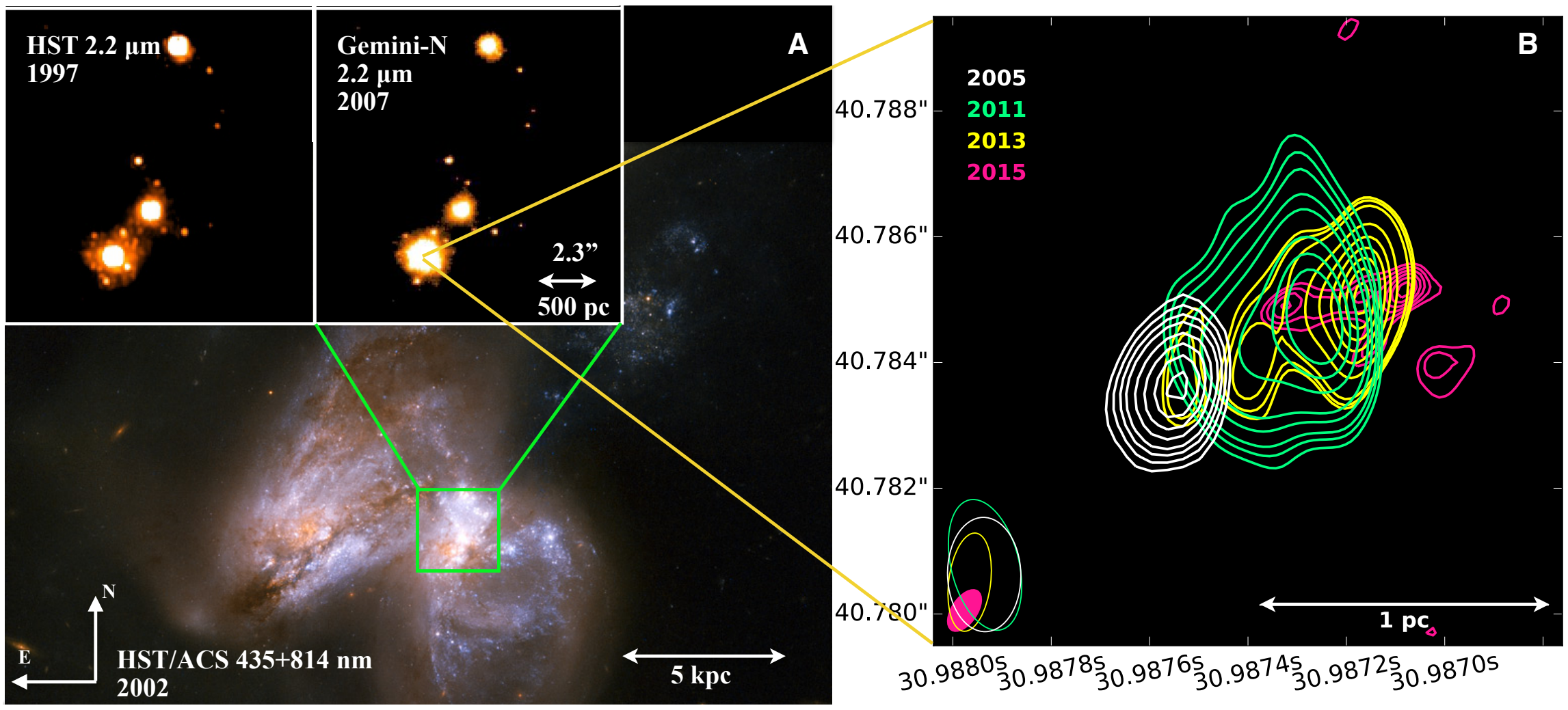

Figure 4.4: The transient Arp 299-B AT1 and its host galaxy Arp 299. **(A)** A colour-composite optical image from the *HST*, with high-resolution, 12.5 by 13 arcsec size near-IR 2.2-mm images [insets **(B)** and **(C)**] showing the brightening of the B1 nucleus (7). **(D)** The evolution of the radio morphology as imaged with VLBI at 8.4 GHz [7 × 7 milli-arcsec (mas) region with the 8.4-GHz peak position in 2005, right ascension (RA) = 11h28m30.9875529s, declination (Dec) = 58°33'40".783601 (J2000.0)]. The VLBI images are aligned with an astrometric precision better than 50 mas. The initially unresolved radio source develops into a resolved jet structure a few years after the explosion, with the centre of the radio emission moving westward with time. The radio beam size for each epoch is indicated in the lower-right corner. From Mattila et al. (2018), Science.

The case of Arp 299-B AT1 suggests that tidal disruptions of young, massive stars may be relatively common. However, events similar to Arp 299-B AT1 would have remained hidden within dusty and dense environments and would not be detectable by optical, UV or soft X-ray observations. Such TDEs from relatively massive, young stars might power strong AGN radiative feedback, especially at higher redshifts where galaxy mergers and luminous infrared galaxies like Arp 299 are more common.

#### The role of the EVN in TDE studies: Lessons learned and caveats.

1. The combination of high-angular resolution near-IR and radio observations will allow for many new discoveries that are hidden by a curtain of dust, so missions like WISE, and radio interferometric arrays like SKA (with baselines of several thousands of kilometres) will result in many more discoveries than in the optical, UV, or X-rays.
2. Direct probing of the formation and evolution of radio jets in TDEs necessarily requires VLBI monitoring of TDE candidates, at high-dynamic range and deep sensitivities.

3. The astrometric capabilities of the EVN are, and will be, crucial for those studies, as the location and motion of a putative TDE jet requires position milliarcsecond accuracies and angular resolutions, or even better, as shown in the cases of *Swift* J1644+57 (Yang et al. 2016) and Arp 299B-AT1 (Mattila et al. 2018).
4. The role of the e-EVN will increase, thanks to an increased cadence of the observations.
5. The bottleneck of the EVN to contribute in an unique way to this field are: (i) its lack of frequency agility, as has been the case since the very beginning of the EVN; (ii) the still rather inhomogenous frequency setup offered by the array, which significantly complicates interpretation.

### 4.1.5 Neutron star and black hole mergers, and gravitational waves

With the discovery of mergers of binary black holes (Abbott et al. 2015), the new era of gravitational wave astronomy started. Soon after the discovery of the first merger of two binary neutron stars GW/GRB 170817 (Abbott et al. 2017b) signed the start of the multi-messenger astronomy era due to the detection of the electromagnetic emission shortly following the gravitational signal (Abbott et al. 2017a).

GW/GRB 170817 showed three types of electromagnetic emission components:

1. two seconds after the gravitational event, detected by LIGO/Virgo (see e.g. Abbott et al. 2017b and references therein), a weak short duration $\gamma$–ray burst triggered *Fermi* (Goldstein et al. 2017) and INTEGRAL (Savchenko et al. 2017). Its temporal and spectral features resembled those of short Gamma Ray Bursts (GRB), expected to arise from a binary compact object merger;
2. less than 11 hours later a bright optical counterpart was discovered (Coulter et al. 2017) in the outskirts of the faint host NGC 4339 at 40 Mpc. Follow up observations of this thermal optical/near-IR emission component over the next 25 days showed a fast decline compatible with the emission from the radioactive decay of the heaviest elements produced by rapid neutron capture in the merger ejecta. About 0.05 $M_\odot$ neutron rich material was ejected in the merger (Pian et al. 2017).
3. starting 10 days post merger and up to more than a year after the merger long–lived non–thermal emission component was detected from the X-ray through the optical to the radio band (e.g. Margutti et al. 2018, Hajela et al. 2019). This was interpreted as the afterglow produced by the (collimated ?) relativistic outflow.

If, similarly to other bright short and long GRBs, the $\gamma$–ray emission of GRB 170817 were produced by a relativistic jet of few degrees aperture ($\sim$5-10°), given the small distance to the source, it should be misaligned ($\sim$20°) with respect to the line of sight. This would account for the observed isotropic equivalent luminosity $L_{\rm iso} \sim 10^{47}$ erg s$^{-1}$ which is four orders of magnitudes smaller than ever measured for cosmological short GRBs. Alternatively, such a low luminosity could be due to a mildly relativistic, nearly isotropic, outflow either produced by the neutron stars' magnetic field (Salafia et al. 2018) or by the interaction of the jet with the merger ejecta. In the latter case, the jet deposited all its energy in the expanding ejecta (choked jet) producing an expanding nearly isotropic mildly relativistic outflow (cocoon).

These two models, different for their geometric and dynamical properties (relativistic narrow jet versus mildly relativistic nearly isotropic outflow), have been competing for almost three months until the intense radio monitoring (Mooley et al. 2018a - confirmed also by X–ray and optical observations - Margutti et al. 2018) showed that the afterglow luminosity increased slowly ($\propto t^{0.8}$) and finally peaked ∼200 days after the event reaching flux densities of $\sim 100\mu$Jy at 6 GHz (Mooley et al. 2018c)[1]. This led to a substantial modification of the above scenarios, which would produce a much steeper luminosity rise, allowing for an angular structure of the jet energy and bulk velocity (structured jet model) or for a radial distribution of the expansion velocity within the cocoon.

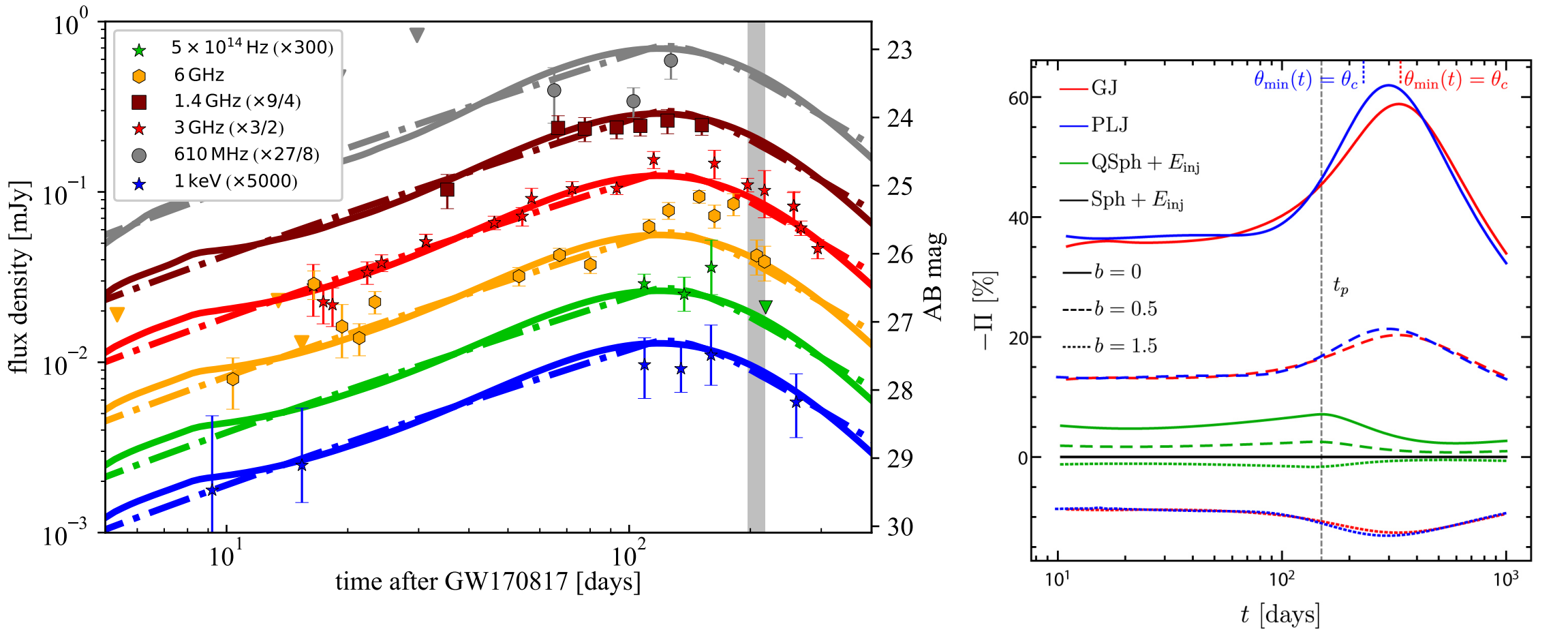

Figure 4.5: (left) Light curve at different frequencies (re-scaled as shown in the legend) with the two competing models (adapted from Ghirlanda et al. 2019): a relativistic jet with an angular distribution of energy and bulk velocity (structured jet - dot–dashed line) and a nearly isotropic mildly relativistic jet with a radial stratified structure in bulk velocity (cocoon - solid line). (right) Linear polarisation as a function of time (Fig. 7, Gill & Granot 2018) for different geometric configuration of the outflow (GJ=Gaussian Jet, PLJ=Power–Law Jet, QSph=Quasi Spherical outflow with energy injection and Sph=Sperical outflow) and different configuration of the magnetic field (b=0 aligned with the shock plane and no component perpendicular to the shock plane.

The angular structured jet (dot–dashed line in Fig.1) and the dynamically structured cocoon (solid line in Fig.1) , however, are compatible (for some sets of parameters) with the observed light curve and cannot be distinguished with multi-wavelength photometry alone.

The different nature of the structure could be distinguished through radio observations providing:

- *continuum polarisation*: a large degree of linear polarisation (≈20%) would indicate a high level of asymmetry, favouring a jet scenario (e.g. Rossi et al. 2004; Nakar et al. 2018). However, polarisation is determined by (i) the geometry of the outflow and (ii) the magnetic field configuration. JVLA observations (at 2.8 GHz) of GRB 170817 240d after the merger (Corsi et al. 2018) rule out linear polarisation larger than 12% (99% confidence level). This result can still be compatible with the structured jet scenario if the magnetic field in the emission region has, also, a component perpendicular to the shock plane (e.g. Gill & Granot 2018 - Fig.4.5, right panel);
- *imaging*: due to its larger expansion velocity and being more collimated with respect to the cocoon, a relativistic narrow jet should produce (e.g. Zrake, Xie & McFadyen 2018 - Fig.4.6) a larger apparent displacement and a smaller size (more compact source). The projected position (with respect to the explosion site) in the plane of the sky of a relativistic jet should be larger than that of the cocoon (eventually with no displacement in the case of a completely symmetric isotropic outflow). VLBA observations (Mooley et al. 2018b) of GRB 170817 showed that its apparent motion is consistent with the predictions of a relativistic structured jet (Fig.4.7 - left panel). A source size (Fig.4.7 - right panel) smaller than 2.5 mas at 207 days obtained through global VLBI observations (Ghirlanda et al. 2019) probes that a relativistic jet emerged from the BNS ejecta.

[1]The spectral energy distribution from the X–ray to the radio is consistent with synchrotron emission from shock accelerated electrons (Margutti et al. 2018).

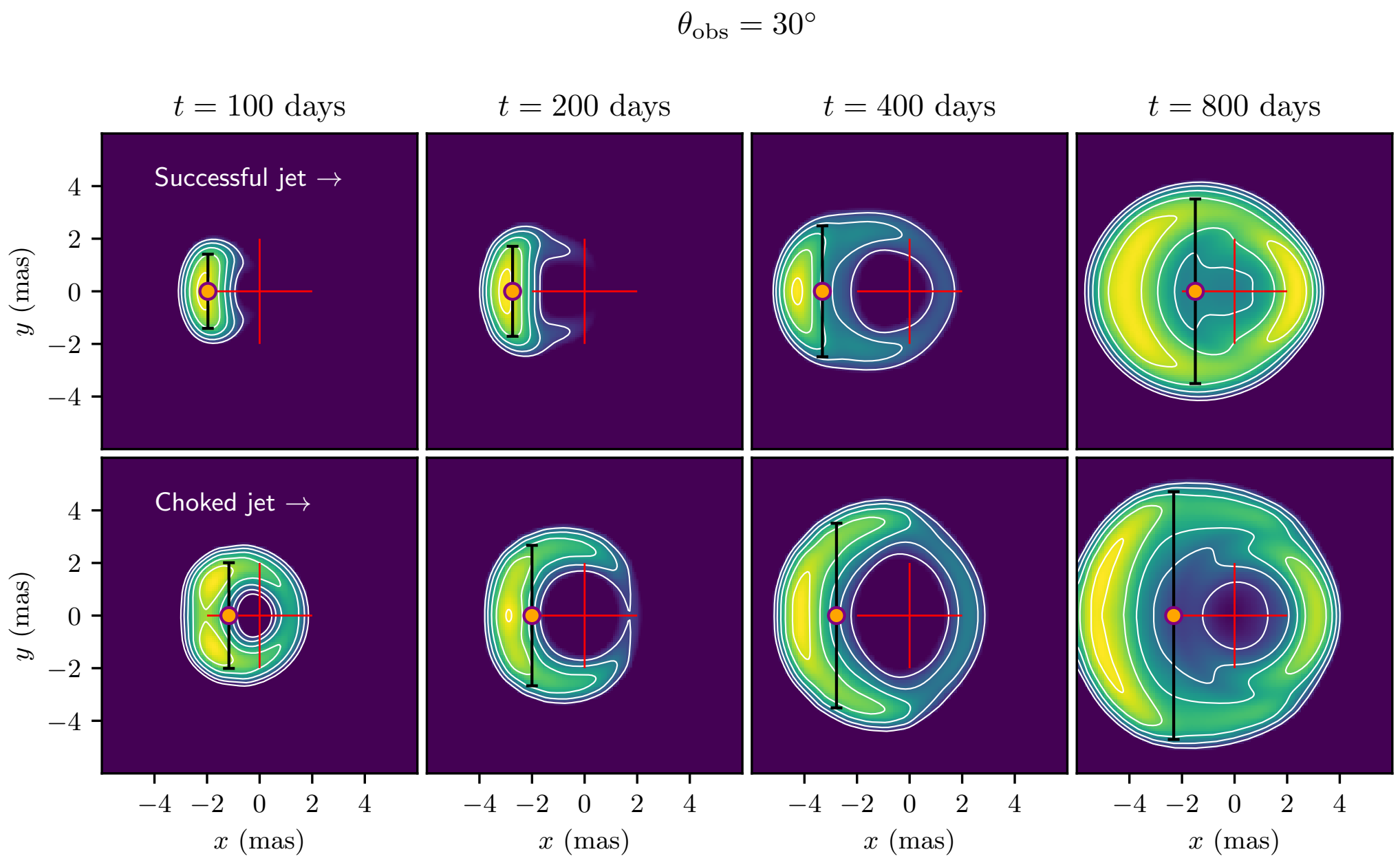

Figure 4.6: Radio images of the structured jet model (top panels) and of the choked (or cocoon) model (bottom panels) at different epochs with respect to the explosion time. The site of the explosion is marked by the red cross (2 mas in size) and the surface brightness (colour coded) spans four orders of magnitudes. The orange dot show the centroid of the surface brightness and the vertical black bar its FWHM (Zrake, Xie & McFadyen 2018). ©AAS. Reproduced with permission.

The reprise of the advanced LIGO and Virgo observing run O3 will offer the opportunity to combine these probes (continuum polarisation measurements and high resolution imaging). The brightness of the new events will be influenced by several factors like the density of the circumburst medium or the jet orientation and the global energy in the relativistic ejecta. Also the polarisation evolution (Fig.4.5 - right) will depend on the jet geometry and on the magnetic field configuration. Polarisation constraints below 10%, attainable with the VLA with events as luminous as twice GRB 170817 at its peak (Corsi et al. 2018), are expected to challenge the jet scenario (Fig.4.5, right panel). However, if a jet is present, such constraints would probe the magnetic field configuration in the emission region. Independent probes of the outflow properties will be provided by high resolution imaging of the source as was the case of GRB 170817 (Mooley et al. 2018b; Ghirlanda et al. 2019). There are great expectations also for the possible first detection, in O3, of BH-NS merger events whose electromagnetic emission is still completely unexplored. The EVN can contribute in exploiting the wealth of information attainable through observations in the GHz range: radio continuum flux measurements can trace the temporal evolution of the flux and, in concert with other wavelengths, e.g. optical and X-ray, to probe the broad band SED and the emission regime of the

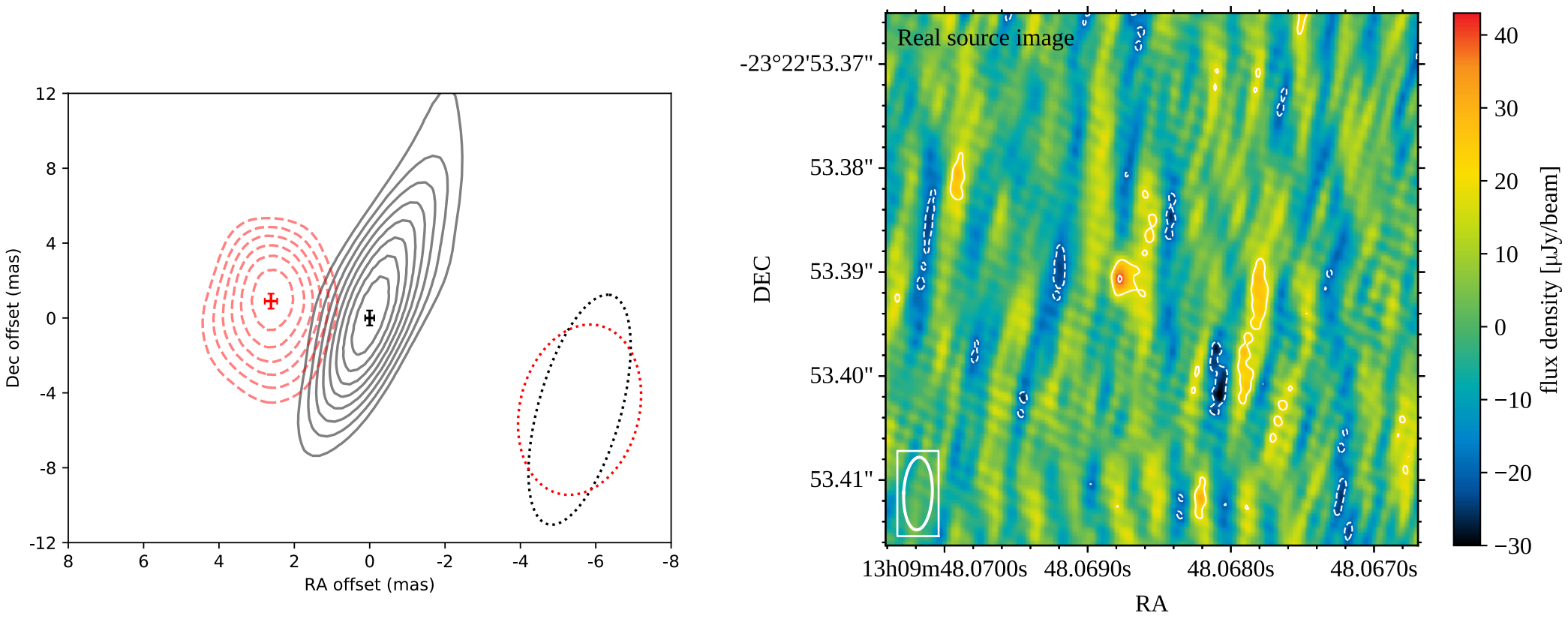

Figure 4.7: (left) Source position at 75d and 230d showing an apparent displacement of 2.7±0.3 mas. The bottom right corner shows the synthesised beam size of the two epochs (Mooley et al. 2018b). (right) Source image at 207 days obtained through global VLBI observations (Ghirlanda et al. 2019).

non–thermal component produced by the deceleration of the relativistic outflow.

## 4.2 Fast transients

A relatively unexplored part of the radio transient parameter space is a phenomena collectively called fast transients. These are very difficult to detect because they have durations much smaller than the typical integration times of a few seconds in most common VLBI applications, and in general they are not trivially distinguished from RFI. They originate in coherent processes and their brightness temperatures may therefore well exceed the inverse-Compton limit for incoherent transients. These short pulses or flares are dispersed by the ionised medium as they travel through it towards the observer. In most cases these can only be detected once the data have been dedispersed to a broad range of dispersion measures (DM); having to search DM-space is one of the reasons why finding fast transients is challenging.

### 4.2.1 Fast radio bursts

#### The FRB phenomenon

The first fast radio burst (FRB), was found in a Parkes pulsar-search data base after an algorithm optimised for finding single pulses using a matched filtering technique was run up to very high DM values. FRB 010824 had a dispersion measure value well in excess of the typical line-of-sight Galactic DM in that direction, and therefore was assumed to be extragalactic (Lorimer et al. 2007). This was followed by a number of other detections (e.g. Thornton et al. 2013), but understanding the nature of FRBs was not possible because single dish detections provide only very poor localisation, down to a few arcminutes at best.

Measuring the DMs for hundreds of FRBs and the redhsifts of their hosts would have a profound impact on cosmology, as it would reveal the baryon distribution in the Universe within redshift of $z \sim 1$ (Macquart et al. 2015). One may constrain the dark energy equation-of-state parameter using FRBs at even higher redshifts, where the DMs are dominated by the IGM (Zhou et al. 2014). Finally, the intergalactic magnetic fields could be constrained from the rotation measure and scattering of FRB signals (Macquart et al. 2015). FRB detection and localisation is one of the highest-ranked high priority science objectives of the SKA, in which therefore VLBI may play a key role, because $< 0.5''$ localisations will be necessary for secure dwarf galaxy host identifications at redshifts $z > 0.1$ (cf. Eftekhari & Berger 2017).

The VLBI data are usually heavily averaged in time and frequency that smears out the signal, and also reduces the field of view drastically. Because of the very short duration, traditional VLBI follow-up observations of single pulses are not possible either. The solution is to apply wide-field techniques and find repeating FRBs that can be followed-up at high resolution. The basic scheme of using the e-EVN for single-pulse localisation and the related LOCATe project work was outlined by Paragi (2017). Around the same time, the first repeating fast radio burst was found at Arecibo (Spitler et al. 2016). FRB 121102 was successfully localised with the JVLA (Chatterjee et al. 2017, $\sim$100 mas precision) and the e-EVN (Marcote et al. 2017, $\sim$10 mas accuracy) within a year of the discovery of its repeating nature. A large number of FRBs have been discovered with ASKAP (Shannon et al. 2018) and CHIME (Amiri et al. 2019) in the following few years. Sub-arcsecond localisation have only been possible for a few cases (Bannister et al. 2019 and Prochaska et al. 2019 with ASKAP; Ravi et al. 2019 with the DSA-10), and the only other mas-precision result was for a second repeater FRB 180916.J0158+65, found by CHIME and localised with the EVN (Marcote et al. 2020).

In the first ten years of FRB research there have been more models proposed to explain the phenomenon than detected FRBs. The first localisation shed light on the extragalactic origin of FRB 121102, which reduced the number of competing models significantly. This is however just the beginning of the story: we still do not know if there are multiple populations of FRBs, and what their true nature is. To be able to reveal their possible progenitors, many more FRB localisations are necessary with interferometers than the above mentioned few cases. In the case of FRB 121102 the host is a metal-poor dwarf galaxy, also typical site for long-GRBs and superluminous supernovae (Tendulkar et al. 2017). The milliarcsecond localisation with VLBI, together with high-resolution optical imaging indicated that it originates in a star forming region within the host galaxy (Bassa et al. 2017). The three localised single-pulse FRBs were related to massive galaxies with little or no star formation, while the second localised repeater was found in a spiral galaxy (Marcote et al. 2020).

Meeting the challenge of detecting short-duration, dispersed astronomical signals will have a major impact on astronomy and possibly beyond, just like in the early days of research into calibrating radio interferometry data and detecting the Hawking-radiation influenced the development of Wi-Fi. FRB searches in VLBI data will be a driver for other developments, because it requires high time/frequency correlation to meet the requirements for finding very short duration, dispersed signals in a large FoV, but at the same time providing very high angular resolution. This science case pushes along all axes of parameter space, which will be beneficial for other applications as well. In addition, it will have an impact on the design of future telescopes, data acquisition systems, search algorithms (machine learning?), RFI detection, and the way we are looking at 'big data' problems. There is considerable synergy between the FRB field and SETI, which has a broad societal impact (e.g. Siemion et al. 2013; see section 4.2.3).

#### Fast transient research with the EVN

The $logN - logS$ distribution of the FRB population is still not known, but the latest research indicates that there may be an FRB in the sky every minute detectable by SKA sensitivities (Fialkov & Loeb 2017). The EVN can make a significant contribution by characterising the known FRB hosts (requiring regular VLBI), getting refined positions of repeating sources (in custom experiments), and detecting/localising unique new single-pulse events (a commensal approach is needed: e.g. Burke-Spolaor et al. 2016; Paragi 2016). The former two should be possible for up to dozens of new FRBs during the next decade, assuming new discoveries will come in great numbers. The chances for unique single pulse detections are less clear.

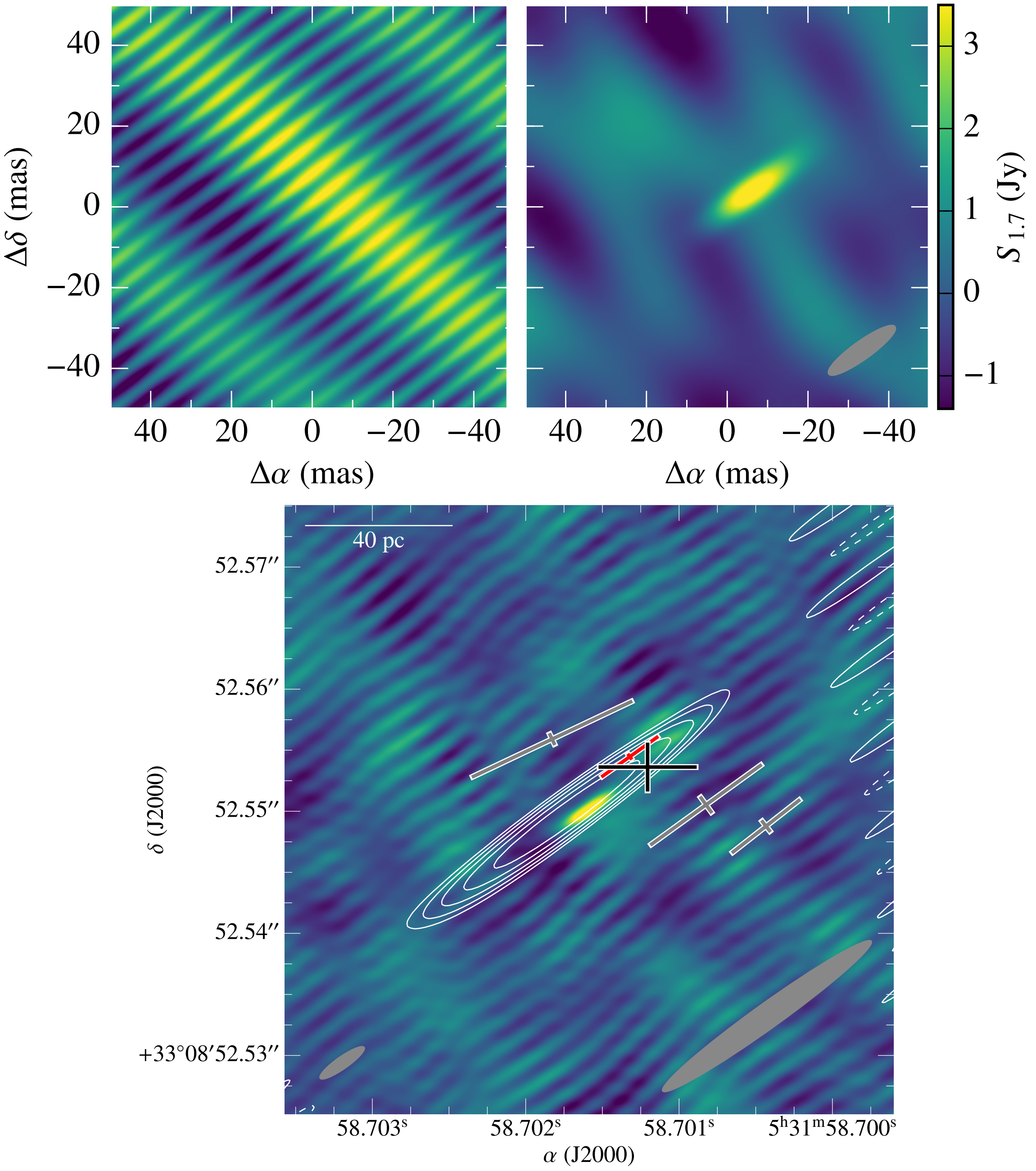

Figure 4.8: (upper panel) The dirty (left) and clean (right) maps of the brightest pulse from FRB 121102 on 20 September 2016. The limiting *uv*-coverage in VLBI observations has an even higher impact on very short snapshot imaging. In this particular case the dominance of Arecibo baselines is clearly visible. (lower panel) The positions of all four pulses detected in the same experiment with their error estimates (red: brightest; black: weighted mean), overlaid on the clean map of the compact, persistent radio source found coincident with FRB 121102 (Marcote et al. 2017). ©AAS. Reproduced with permission.

The major limitation of current instrumentation for FRB searches is the low survey speed for fast transients[2], and it is related to the fact that most of the EVN sensitivity comes from great dishes that limit the field of view. Also, while single-pulse FRBs are not very broad band, pulses may appear at a broad range of frequencies (Gajjar et al. 2018) therefore having narrow band receivers limits the chances of detection. The usual 2-bit sampling employed in VLBI systems limits the dynamic range for single pulses and makes fighting the RFI more difficult, although for the latter there are promising new methods (e.g. Nita et al. 2019). Other limitations include poor instantaneous *uv*-coverage, issues with formatting VLBI data, automatic gain control and $T_{sys}$ measurements (Paragi 2016; Huang et al. in prep.).

The unique capability of the EVN for FRB research is that it combines a collective area that of the planned SKA1-MID with very high angular resolution. To detect short bursts collecting area is very important, since one cannot win extra sensitivity by increasing the observing time. The current fluence detection limit with the greatest telescopes is $\sim$0.1 Jy ms, i.e. a few hundred mJy peak flux density for a millisecond-duration event – for FRB 121102, three out of the four e-EVN-detected pulses could not have been found without Arecibo. So far it has also been an advantage that some of the largest dishes that are member of the EVN have had dedicated (broad-band, high-bit-sampling) pulsar backends to support FRB observations. The flexibility offered by the regular e-EVN sessions was also a plus to be able to carry out monitoring observations of FRB 121102, hoping to catch active periods. The flexibility of the EVN SFXC correlator (Keimpema et al. 2015) for doing wide-FoV style correlation of very short time samples, and the SFXC implementation of coherent dedispersion capability was also essential[3].

### Requirements and synergies

The characterisation of the FRB host galaxies – to better constrain the environments of FRB progenitors – requires a multi-band approach. It mainly implies studying the star formation and AGN activity in the host, and especially near the FRB position. The host of FRB 121102 is a metal-poor dwarf galaxy at a redshift of $z = 0.1927$. It has an off-nuclear persistent, compact radio source of $\sim$ 180 $\mu$Jy coincidental with the FRB position, representing either a magnetar-powered pulsar-wind nebula, or low-luminosity AGN activity. The observed extremely high RM of $\sim 10^5$ radians per square metre (found in the nearly 100% linearly polarised pulses of FRB 121102) have only been seen around (super-)massive black holes – while a scenario of a highly magnetised pulsar-wind nebula or supernova remnant surrounding a young neutron star still cannot be ruled out (Michilli et al. 2018). The persistent radio source may be a representative for a new source population well into the sub-mJy regime. To study this population in detail one just needs regular phase-referencing VLBI, but with $\mu$Jy sensitivity and milliarcsecond resolution. To localise future known sources of repeating FRB pulses at mas resolution, one will need custom experiments in which e.g. Arecibo or Effelsberg piggybacks on the VLBI observations with their pulsar backends. This is essential for optimal sensitivity single-pulse search for bursts. Real-time e-VLBI observations may be required to trigger on active periods of the sources, but the telescope voltage data has to always be recorded. Only this approach allows the FRB timestamp in the VLBI data to be de-dispersed and recorrelated. There is thus a lot of synergy with large single dishes (76m Lovell Telescope, 100m Effelsberg, 500m FAST etc.) or interferometers with PAFs (WSRT-APERTIF, ASKAP), as well as other dedicated FRB search engines (Molonglo Observatory Synthesis Telescope in Australia, and CHIME in Canada) and the SKA1-MID itself.

[2]The survey speed metric for fast transients is the product of the FoV and the limiting survey sensitivity $S^n$, where $n$ is $-1.5$ for Eucledian Universe and no cosmological FRB evolution (Macquart et al. 2015).

[3]The e-EVN upgrade was one of the main requirements in the EVN2015 document. The development of the e-VLBI technique and the SFXC correlator in particular have been made possible by the EC-supported EXPReS and NEXPReS projects.

To summarise, the major requirements are the increase in sensitivity and instantaneous *uv*-coverage by adding new telescopes and significantly inrease the collecting area of the EVN. While increasing the bandwidth by using broad-band single-pixel feeds may have advantages for single pulse detection at higher frequencies (cf. Gajjar et al. 2018), increasing the collecting area will remain very important because both the pulses and the related persistent sources will likely have steep spectra and thus are more likely to be detected at lower frequencies. Flexibility for triggered observations is also very important. The optional switching off of the automatic gain control as well as the continuous Tsys, and good RFI mitigation would greatly improve the data quality for FRB-pulse detection with the EVN. A commensal, real-time detection pipeline (real-time during the correlation) might be considered for the EVN correlator in JIVE, but the majority of new FRB discoveries will be delivered by very large FoV custom FRB-search experiments. A large field of view EVN Archive, storing the raw data for all observations –at least for a prolonged period of time– might be highly useful. This would allow for new searches whenever new algorythms are being developed, with practically unlimited spectral resolution. While chances for detection is low, a wide-field EVN Archive would facilitate mas-localisation for single-pulse FRBs, and this approach seems to be the only way to associate single-pulse events to individual sources within the host, which is crucial for understanding the FRB progenitor population. In addition, it would allow for a direct VLBI measurement of the scatter-broadened size for a high-$z$ FRB to characterise the IGM up to very high redshifts, which would be a major breakthrough. This also requires a robust amplitude calibration (at least to a few percent) and short spacings for the telescopes that provide the longest baselines for the EVN.

### 4.2.2 Neutron stars and pulsars

#### A new renaissance of pulsar research

Half a century after the original discovery, in the last decade pulsar and neutron star science produced more outstanding results than ever. For instance: the discovery of the most massive NSs (Demorest et al. 2010, Antoniadis et al. 2013, Linares et al. 2018), the most extreme relativistic pulsar binaries (Cameron et al. 2018, Stovall et al. 2018), the first multiple system including three compact objects (Ransom et al. 2014), the transitional millisecond pulsars (Papitto et al. 2013, Archibald et al. 2009), a radio magnetar very close to the Galactic centre (Mori et al. 2013, Eatough et al. 2013). Also, the success of many search experiments has more than doubled the number of known millisecond pulsars (MSPs), some of them became regular targets of the regional Pulsar Timing Arrays, as well as of the International Pulsar Timing Array (IPTA), established in 2011 with the primary aim of detecting gravitational waves in the nanoHz frequency regime by using repeated timing observations of a large ensemble of millisecond pulsars.

Building from this very successful decade, the perspectives for pulsar and neutron star research are extremely promising: in fact, it has been selected as key science for many planned large instruments, and most notably for SKA1 (Kramer et al 2015). In this framework, the slogan invoked by the pulsar and NS community since the beginning of the pathway to SKA *"find them, time them, and VLBI them"* is more valid than ever. The use of the VLBI technique, and in particular the capabilities of the EVN, will play a unique complementary role for the full success of the foreseeable future experiments, by providing very precise astrometry, improved timing precision, as well as key information about the NS environment and the material along the line of sight. The study of NS dynamics in the Galaxy (via determination of the proper motion of many targets, including the large number of slow pulsars), detailed imaging of the Pulsar Wind Nebulae in search for the occurrence of possible jets (via application of pulsar gating to subtract the strong pulsar signal and thus unveil either the termination shock or putative jets), and the investigation of the location and properties of the radio emission in NS binaries (via high resolution imaging) are some well known examples of NS science made possible by VLBI.

#### Prime goals for very high resolution pulsar science

1. Some relativistic binaries (e.g. the B1534+12 and the double pulsar systems) have precision tests of general relativity and alternate gravity theories limited by our inability to correct for kinematic effects and/or by poorly constrained distances. This limitation will become more and more important in the next decade, when the quality of the timing observations will be strongly enhanced by the availability of new large telescopes. In fact, although pulsar timing can directly provide a value for the proper motion and the parallax, an increasing corpus of data is manifesting significant discrepancies among the astrometric parameters obtained by pulsar timing and more precise VLBI determinations, e.g. Deller et al. (2016) invokes the solar wind as a likely candidate for the timing model errors when looking at low ecliptic latitude pulsars. Precision timing, coupled with VLBI astrometry will finally enable tests of gravitational radiation emission theories at the $\sim$0.01% level.
2. Precise distances, provided by the EVN, in combination with accurate optical spectrometry (delivered by the new large telescopes which are at the design and/or construction phase, e.g. ELT), will be extremely beneficial in order to constrain the mass, radius and chemical composition of the companion (most often a white dwarf or a swelled main sequence star) to many binary neutron stars. An accurate determination of those parameters will enable strong constraints to theories of gravity and to nuclear physics, as well as providing key information for the study of the evolution of compact systems including a NS.
3. Detection of nanoHz gravitational waves (GWs) could occur within some years using observations of the already existent Pulsar Timing Arrays (PTAs), and all studies concur that is warranted by SKA1. The first detection(s) will result from the observation of the effects of the stochastic GW background at the Earth. On a longer timescale, one will aim to also detect the effects of the GWs produced by single sources (e.g. Supermassive Black Hole Binaries, SMBHBs) at the location of (at least) a subset of the pulsars belonging to the PTAs. That will open up the intriguing possibility to strongly constrain both the sky location and the distance of the individual SMBHB (e.g. Deng & Finn 2011), whence the exciting chance to follow up the SMBHB in other electromagnetic bands. However, in order to achieve that, a very precise knowledge (of order $\sim 0.1 - 1.0$ pc) of the distance of the aforementioned subset of pulsars is required. Joint interferometric EVN-SKA1 (or EVN-MeerKAT) observations and/or interferometry of EVN telescopes combined with other very large instruments (e.g. FAST, ngVLA) will be suitable to deliver the best distance constraints. Besides improving on the achieved rms of the observations, such a high sensitivity system will allow one to also exploit relatively faint ($< 1$ mJy) in-beam calibrators, which should often be close enough on the sky to the targeted pulsar. At the same time, the higher sensitivity will open the possibility to plan EVN pulsar observations at higher frequencies, thus strongly reducing the impact of the scattering, which is a serious issue at L-band for pulsars in the Galactic plane.

4. The quality of the timing data from a given pulsar typically scales linearly with the signal-to-noise of the collected signal profile. Therefore, the higher the sensitivity of the telescope is, the better the data are. A way to create very large (and highly sensitive) telescopes is to simultaneously operate many large dishes in tied-array mode. In pulsar research, this idea has been successfully implemented during past decade in the EU-funded LEAP project (Large European Array for Pulsars, Bassa et al 2016). This involves Effelsberg, Lovell, Nancay, Westerbork and Sardinia radio telescopes, the collecting area of which approaches that of a 200m dish, capable to observe a large portion of the sky. The perspective of adding most of the EVN dishes to the LEAP infrastructure, or of planning EVN+MeerKAT (or EVN+SKA1) tied-array observations, will create a fantastic instrument to perform ultra high precision timing for some tens of the most important targets.
5. Pulsar signals experience a full range of effects due to the intervening partially ionised medium: dispersion, scattering, refractive and diffractive scintillation, as well as Faraday rotation. Thus, single dish pulsar observations are known to be very effective tools for investigating the distribution of the ionised gas in the ISM and the Milky Way magnetic field. Their diagnostic capabilities grew with the discovery of the secondary spectrum *parabolic arcs* (Stinebring et al. 2001). In last decade, a new very powerful tool has been suggested, relying on VLBI observations of those parabolic arcs, via the so-called secondary cross spectra (Brisken et al. 2010). By using this technique, EVN observations (even better if combined with that of other large instruments, e.g. FAST, MeerKAT, SKA1, ngVLA) will allow one to potentially map the scattered brightness of a few pulsars with finer resolution (down to $\lesssim 100\mu$as) than the diffractive limit of the global instrument, shedding light on the sub-mas properties of the ISM along several lines of sight.
6. It has been recently shown that one can exploit the ISM as a giant lens, by performing VLBI imaging of the interstellar scattering speckle pattern associated with a very bright pulsar (a technique dubbed *pulsar VLBI-scintillometry*). It is expected (Pen et al. 2014) that the availability of future larger instruments in the VLBI network will provide the capabilities to measure the motion of the pulsar emission with sub-nanoarcsec accuracy, as well as enabling picoarcsec astrometry. This will improve the constraints on the still strongly debated size and height of the radio emission regions, as well as determine the size of the projected orbit for bright pulsars in binaries. This science case (as well as the one about the study of the ISM) is particularly tailored for VLBI at low frequencies, $\sim 300$ MHz.
7. In next decade there will be the possibility of exploiting the combined radio and optical emission from several binaries composed of a millisecond pulsar and a white dwarf in order to compare the astrometric coordinates derived by independently using the pulsar timing procedure, VLBI observations, and *Gaia* observations. That will lead (Paragi et al. 2015) to finally tie the three reference frames with better than 10 $\mu$as precision.

#### Technical requirements (pulsars)

To achieve these aims will require the EVN to: **(i)** increase the total sensitivity. That can be reached by including more telescopes in the array or by planning dedicated simultaneous observations with other high sensitivity instruments, and by widening the usable bandwidth. The aim should be 2 GHz of bandwidth, with dual pol and 2 bit sampling. That would deliver 16 Gbps, i.e. a factor of $\sim 4$ beyond what has been demonstrated at the moment, but it would technically be possible and would imply many benefits for pulsar-science; **(ii)** set up the correlator for enabling tied-array observations; **(iii)** ensure the availability of the pulsar gating mode, ideally with the capability of observing several pulsars at the same time, as for the many (sometimes tens of) millisecond pulsars included in a globular cluster. In this case, the visibilities should be simultaneously gated in pace with the ephemeris of each of the pulsars in the FoV; **(iv)** providing suitable frequency channelisation, in order to minimise the effects of the dispersion on the signal. In order to enlarge the circle of the users, it will be finally very important to provide and maintain public codes and pipelines for reducing pulsar-VLBI observations.

### 4.2.3 SETI

Over the last few years, the field of SETI (Search for Extraterrestrial Intelligence) has undergone a major rejuvination. The discovery by the *Kepler* mission that most stars host planetary systems, and that around 20% of these planets will be located within the habitable zone, plus the continually growing evidence that the basic pre-biotic constituents and conditions we believe necessary for life are common and perhaps ubiquitous in the Galaxy, has brought new focus to one of the most important questions that human-kind can ask itself - Are We Alone?

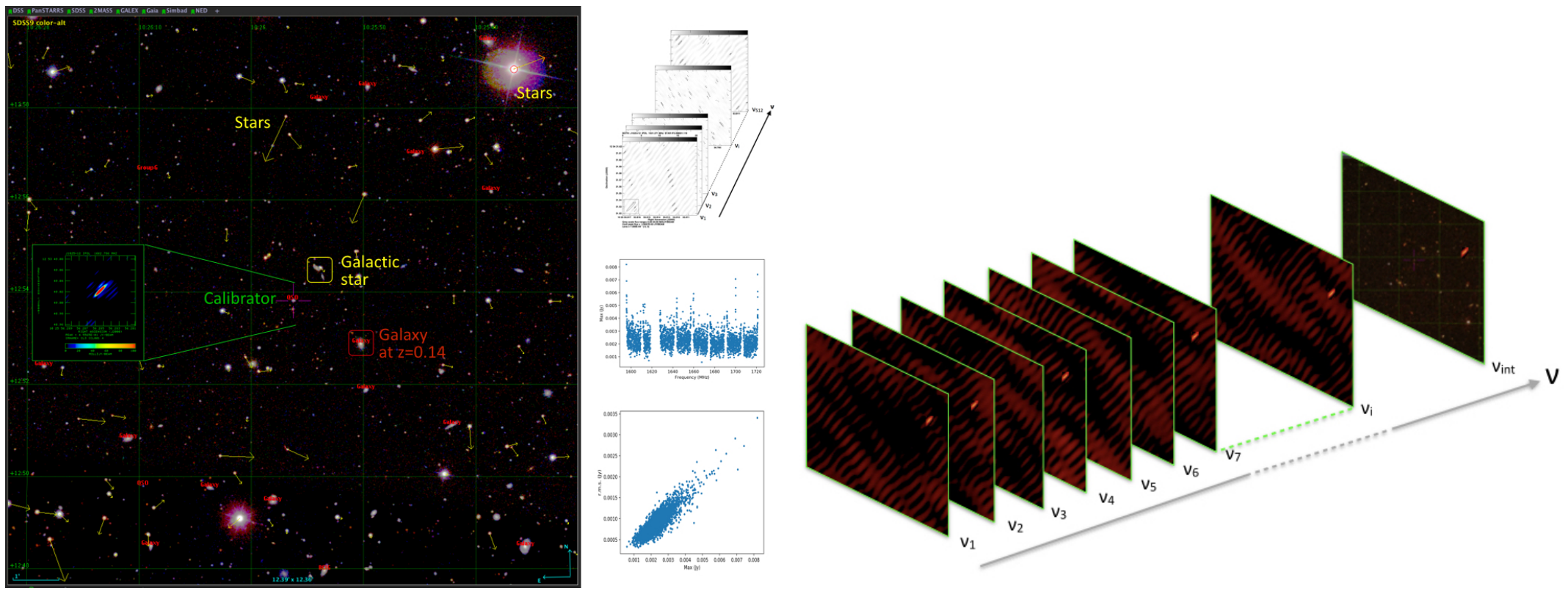

Figure 4.9: (left) The SDSS field centred on the calibrator J1025+1253, extracted from high time/frequency resolution EVN archive data. To the extreme right, the main results of searching for outliers in the $8192 \times 32$ kHz multi-channel images associated with a galactic star in the field. (right) The position of a SETI signal is likely to be invariant in its location on the sky. While everything else might be changing (e.g. central frequency and/or intensity with time) the signal location is expected to remain fixed, at least within the duration of a short observation. Credit: Garrett (2018).

One significant problem with all SETI searches to date (incl. the state-of-the-art Breakthrough Listen programme (Enriquez et al. 2017) is the number of false positives generated by both terrestrial radio frequency interference (RFI) and space-borne satellite communication systems. This is a situation that continues to grow steadily worse with time. Single dish telescopes are particularly vulnerable, compared to distributed arrays that can form an interferometer network. Interferometer arrays, such as the EVN/*e*-MERLIN, distributed on scales of 100-1000s of km are particularly good at suppressing RFI signals because the response to any signals arising outside of the relatively narrow field of view (e.g. any region outwith the primary beam response) decorrelate rapidly as one moves away from the correlated phase-centre of the observations (e.g. Rampadarath et al. 2012). Although the SETI community has been reluctant to adopt interferometers as SETI search instruments, they also offer several other advantages:

- *Detection significance, confidence and redundancy:* a major issue with all SETI surveys historically is the significance of any claimed detection and the overall confidence in the result. This is particularly true for non-repeating signals. A radio interferometer generates multiple independent baselines, and since SETI signals are very likely to be unresolved, a signal detected in one baseline must also be detected in another. The redundancy and greater confidence made available via interferometry is an important advantage over single-dish and beam-formed instruments.
- *The invariance of position on the sky:* a general characteristic of a narrow-band SETI signal is that it is likely to drift in frequency due to the relative Doppler accelerations between the transmitter and receiver. Temporal variations in intensity might also arise due to scattering effects in the ISM or intrinsic periodicty or modulations of the transmitter. In short, for an artificial narrow-band signal, everything could be changing, with the only guaranteed invariant being the location of the transmitter on the sky (at least on short timescales- minutes to hours). Since interferometers can make images, and also locate the position of a source in the *uv* or delay-rate plane, the invariance of a SETI signal's position on the sky can be an important constraint on confirming a bona fide candidate signal. The concept is shown in Fig. 4.9.
- *Field of view:* the time and frequency resolution required by SETI observations, naturally leads to an interferometer array in which the entire field of view (only limited by the extent of the primary beam response) is available for analysis. The result is that one can descriminate between thousands of potential SETI targets in the field of view, with the prospect of studying all the targets simultaneously.
- *Sensitivity:* last but not least, it is well known that the most sensitive radio telescope arrays in the world are delivered (for unresolved sources) by VLBI. Bringing together the largest telescopes in the world into a single SETI array e.g. FAST, GBT, Arecibo, Lovell, Effelsberg, SRT, etc) would greatly improve the sensitivity provided by single dishes or small arrays.

The EVN (including *e*-MERLIN) is a facility that can have a major impact on SETI research. Although an interferometer based approach is probably not suitable for conducting large, systematic surveys of the sky (at least not yet), it can be extremely useful in (i) following-up interesting candidate signals discovered by large scale surveys (e.g. the Breakthrough Listen million star survey, Isaacson et al. 2017) and (ii) making independent observations of interesting targets (exotica) - recent examples would include the interstellar object Oumuamua, KIC 8462852 (also known as "Tabby's Star" or "Boyajian's Star") etc e.g. Enriquez et al. (2018). Observations of such exotica engage the general public in astronomical research in general, and in SETI in particular.

Some initial progress in using the EVN as a SETI search instrument has been made from the analysis of EVN archive data (ca. 2012) correlated with very fine time & frequency resolution (Garrett 2018). With only coarse data editing employed (established via obvious apriori station-based problems), RFI is essentially absent from the data at the $4\sigma$ level (see Fig. 4.9). This permitted simple limits to be placed on the presence of any artificial extraterrestrial signals in the data, with the sensitivity to a SETI signal limited by the frequency resolution of the data (32 kHz). This demonstration shows the huge potential that the EVN, *e*-MERLIN and other long-baseline arrays possess in terms of extending SETI research. With the development of the JIVE software correlator (SFXC) routinely permitting the production of correlated data with very fine frequency/time resolution, the EVN is sensitive to both narrow-band and transient signals - the characteristics one might expect from an artificial radio source. It should also be noted that in the event that an artificial signal is detected, VLBI observations could precisely locate the position of the source (at 1 kpc, 1 AU subtends 1 mas on the sky) and reveal important details of the dynamics of the transmitter (rotation, proper motion etc.).

The detection of intelligent life outside of the Earth would be one of the most profound discoveries in the history of humankind. Such a discovery is likely to leave no aspect of human life untouched. The discovery of an independent biogenesis would provide strong evidence that intelligent life is common, and we can only guess at the ways society would be transformed if it was also possible to decode such signals. As Breakthrough Listen, and more recently NASA begin to fund the search for 'techno-signatures' at significant levels, it's important that Europe also contributes to the field. The EVN is well placed to make a real contribution to this important field.

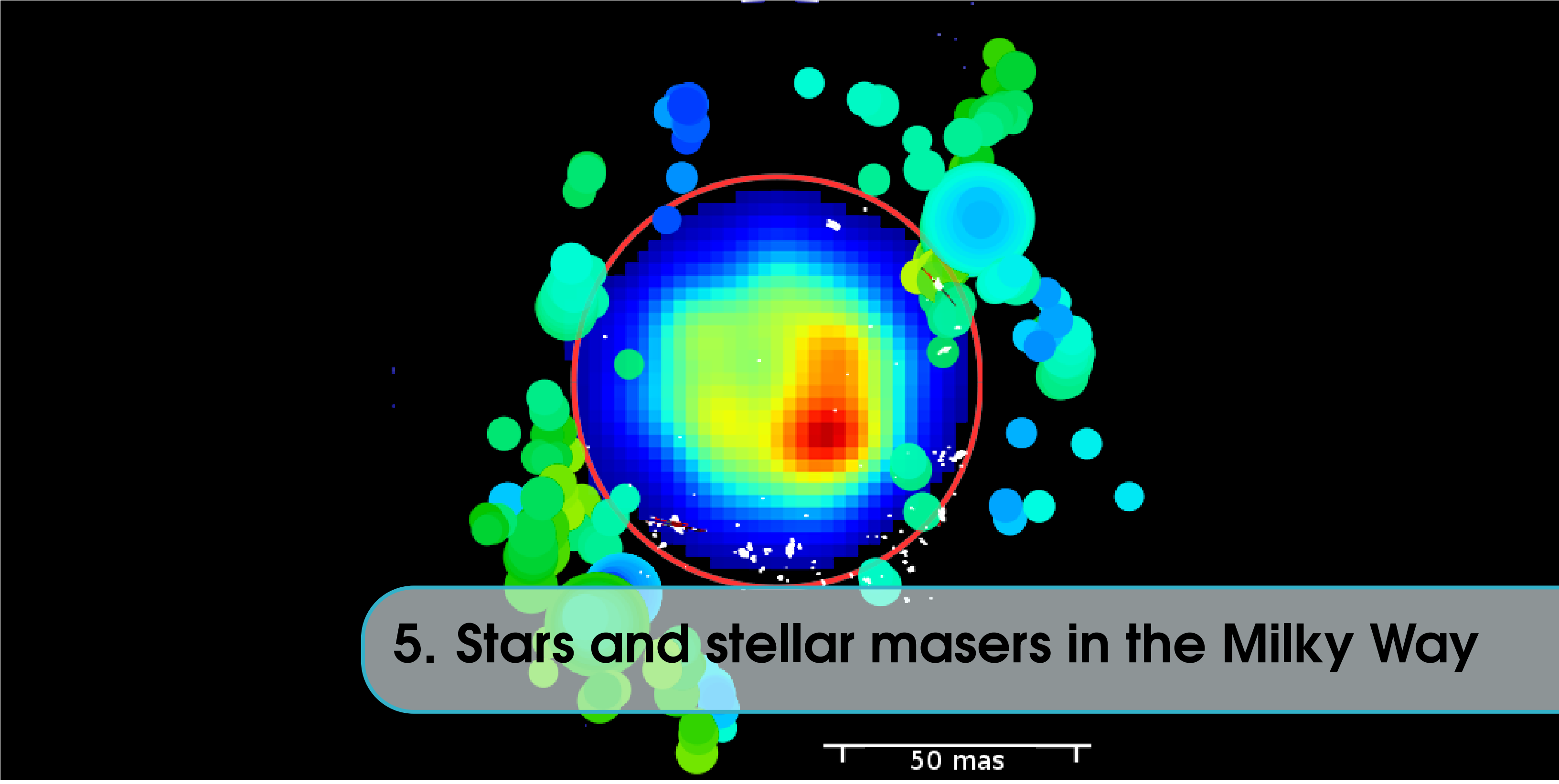

# 5. Stars and stellar masers in the Milky Way

Over the last decades, our understanding of the stellar astrophysics has been greatly enhanced by observations of radio interferometers. Although stars are not profuse radio emitters, the increasing resolution and higher sensitivity of new instruments have provided valuable insights on every stage of the stellar evolution. Accordingly, the contribution of VLBI in this field materialises in a variety of discoveries including stellar flares, structures of the magnetic coronae, colliding-wind binaries, low-mass objects near the substellar transition, as well as the measurements of positions, proper motions, parallaxes, and orbital motions with accuracies of a fraction of a milliarcsecond. The first part of this chapter is therefore devoted to the wealth of astrophysical phenomena, and other spin-offs, related to stellar continuum studies. The second section of this chapter will describe spectral line observations of various masing molecules that constitute the VLBI Galactic maser science. Masers appear mostly in star-forming regions and in envelopes of evolved stars. High sensitivity studies of maser emission at milliarcsecond resolution, particularly from OH, $H_2O$, $CH_3OH$ and SiO molecules, in the radio and millimetre domain provide one of the best existing tools for uncovering the kinematics and the physical conditions of regions that are hidden in dense, high extinction, environments. Maser structures, being compact and non-thermal, are also ideal astrometry targets with which one can study the furthest regions in the Galaxy and Galactic structure itself. Finally, the technical requirements and synergies for star and stellar maser observations are presented.

---

Chapter image credit: Diverse radio emission of W Hya, the Mira-type variable star in the constellation Hydra. The 338 GHz continuum emission detected by ALMA (Vlemmings et al. 2017), the 22 GHz continuum marked by the red circle (Reid & Menten 1990. 43 GHz SiO masers marked by white contours with polarisation “E” vectors (Cotton et al. 2008) and 22 GHz water masers marked by colourful spots (Richards et al. 2012) – reproduced with permission ©ESO.

## 5.1 Stellar evolution and planetary systems

### 5.1.1 Pre-main sequence stars

#### Protostellar radio jets

Accretion disks and collimated outflows (jets) are intrinsically associated with the star-formation process. In particular, the youngest, deeply embedded protostars are usually radio sources of free-free emission that trace the (partially) ionised region, very close to the exciting star (10-100 AU), where the outflow phenomenon originates, and are referred to as "thermal radio jets" (see Anglada et al. 2015, 2018 for recent reviews). In more evolved low-mass stars the radio emission is dominated by a non-thermal (gyrosynchrotron) process (Feigelson & Montmerle 1985) produced in their active magnetospheres; these compact non-thermal sources are excellent tools for the determination of accurate parallaxes using VLBI observations (e.g., Kounkel et al. 2017; see below).

Thermal radio jets are present in young stars across the stellar spectrum, from O-type protostars (Garay et al. 2003) and possibly to proto-brown dwarfs (Palau et al. 2014). This radio emission is relatively weak, and correlated with the bolometric luminosity. The observed centimetre radio luminosities (taken to be $S_\nu d^2$, with $d$ being the distance) go from $\sim$100 mJy kpc$^2$ for massive young stars to $\sim 3 \times 10^{-3}$ mJy kpc$^2$ for young brown dwarfs (Anglada et al. 2018). Nevertheless, given the large obscuration present towards the very young stars, the detection of the radio jet provides so far the best way to obtain their accurate positions. Dynamical timescales of these radio jets are short, and these observations provide information on the physical properties, direction and collimation of the gas ejected by the young system in the last few years.

VLBI observations could in principle trace details of the jets at very small scales. An angular resolution of 1 mas in a nearby radio jet ($d$= 200 pc) would resolve details of the jet structure with a physical size of 0.2 AU, which is of the order of the size of the launch region. However, the thermal nature of the emission hinders the study with the highest angular resolution. A maximum intensity of only $\sim$7 $\mu$Jy/beam (assuming optically thick emission at 10,000 K) is expected for a beam of 1 mas, which would require very sensitive observations.

Interestingly, non-thermal synchrotron emission is also present in several protostellar jets. This non-thermal emission is usually found in relatively strong radio knots, away from the jet core, showing negative spectral indices at centimetre wavelengths (Rodríguez-Kamenetzky et al. 2017). The synchrotron nature of the emission was confirmed with the detection and mapping of linearly polarised emission in the HH 80-81 jet (Carrasco-González et al. 2010), establishing a link with the AGN and microquasar relativistic jets. Although velocities of protostellar jets are non relativistic, several theoretical studies show that the observed synchrotron emission can be produced by a small population of relativistic particles that have been accelerated in the strong shocks in protostellar jets (e.g., Araudo et al. 2007). Possible compact knots in these non-thermal knots are potential targets for future VLBI studies.

#### Protoplanetary / debris disks

The radio regime is unique for studying proto-planetary discs, the material out of which protostars and their planets are formed. Data from the Atacama Large Millimeter/submillimeter Array (ALMA) has proven that discs are not smooth flat plates of dust and gas, but highly complex in structure, and stirred by planet formation. The inner few AU of these disks, however, are opaque at millimetre wavelengths, and despite ALMA's high angular resolution, mm/submm interferometers cannot detect emission from the mid-plane of the disk, where planets are expected to form. In contrast, radio continuum arises from optically thin blackbody radiation, directly tracing the solid mass held in large grains. Grains are poor emitters at wavelengths above their circumference, so only radio emission can tell if ‘pebbles’ are present, crucial to further aggregation into planets. The current state-of-the-art is to map the inner disc, and spatially separate the radio emission of the jets of young stars (a complexity that has plagued low-resolution studies). This arena with JVLA and especially *e*-MERLIN will progress into SKA (and precursors) studies. The next phase is to study the very fine details of material flowing into planetary cores, hinted at by ALMA, and ideally matched to milliarcsecond EVN scales. The science goals are to understand how planets build up mass and migrate within the disc, via resolving the Hill accretion sphere, spiral arms around the planet within disc gaps, etc. As the dust signal is very faint because of the frequency-squared or steeper spectrum, such observations are very challenging for present-day EVN. For example, the HL Tau b proto-planet candidate (Greaves et al. 2008) is expected to have $\sim 5\mu$Jy total flux at C-band. Zhu et al. (2018a, b) discuss the detectability of disks and jets associated with proto-planets. Their results show that sensitivities well below 1 $\mu$Jy are required to attempt this kind of studies at centimetre wavelengths. The combination of *e*-MERLIN with the shortest EVN baselines may help to resolve the detailed processes. Increases to collecting area and bandwidth are therefore highly important.

In a few massive young stellar objects there is evidence that the associated centimetre source actually traces a photoionised disk (e.g., Hoare 2006, Reid et al. 2007). These objects show a similar centimetre spectral index to that of jets and are identified in terms of other criteria, such as their orientation with respect to the outflow axis or their radio luminosity relative to the bolometric luminosity. Ionised disks can also be present in the case of young low mass stars. The high resolution images of GM Aur presented by Macías et al. (2016) show that, after subtracting the expected dust emission from the disk, the centimetre emission from this source is composed of an ionised radio jet and a photoevaporative wind arising from the disk perpendicular to the jet. It is believed that extreme-UV (EUV) radiation from the star is the main ionising mechanism of the disk surface.

### Clusters and star forming regions

Although star-formation has been observed and studied throughout the Milky Way and in many external galaxies, our current understanding of star-formation is largely based on the careful study of less than a dozen benchmark regions located within a few hundred parsecs of the Sun –Taurus and Orion are, perhaps, the best known of these regions. Indeed numerous detailed surveys have been performed at X-ray, infrared, microwave and radio wavelengths to characterise the stellar populations and the ISM content of these regions (e.g. Bally 2008, Muench et al. 2008, O’Dell et al. 2008 and Kenyon et al 2008 for reviews of such surveys of Orion and Taurus). As a result of the recent detailed imaging of protoplanetary disks in the same regions by ALMA (e.g. HL Tau in Taurus or Elias 2-27 in Ophiuchus; ALMA partnership et al. 2015, Perez et al. 2016), they are also quickly becoming the benchmark regions for the study of planet formation.

VLBI astrometric observations can readily provide position measurements accurate to better than 0.1 mas even for modest signal-to-noise detections (Reid & Honma 2014). As a consequence, multi-epoch VLBI observations appropriately scheduled over a time period of a few years can provide trigonometric parallax measurements accurate to 10 $\mu$as or better, and proper motions accurate to a few tens of $\mu$as yr$^{-1}$. This is comparable with the expected performance of the *Gaia* satellite. For objects within a few hundred parsecs, this yields distances and tangential velocities accurate to better than a few percent. In addition, VLBI observations at radio wavelengths are immune to dust along the line of sight, so accurate distance and velocity measurements can be obtained even for objects that are deeply enshrouded in dust, such as young stars in star-forming regions.

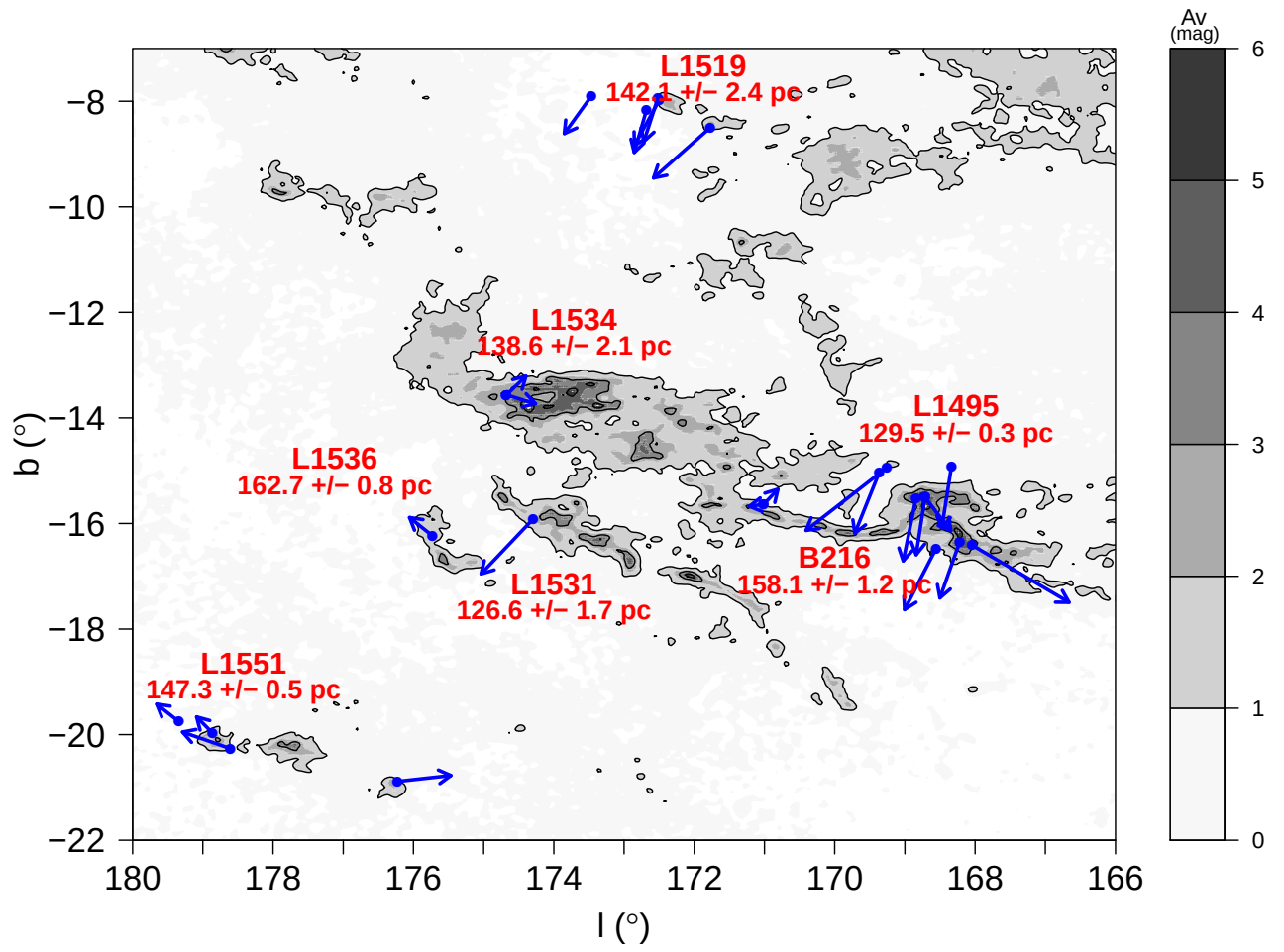

Figure 5.1: Three-dimensional structure of the Taurus star-forming regions (adapted from Galli et al. 2018). The underlying grey-scale image shows the extinction map derived by Dobashi et al. (2005) in Galactic coordinates. The different dust clouds in the region are indicated, together with their distance measured though VLBI astrometric observations. The blue arrows correspond to the proper motions (also determined from VLBI observations) measured in the Local Standard of Rest (LSR). Adapted from Galli et al. (2018). ©AAS. Reproduced with permission.

Taking advantage of this situation, the Gould's Belt Distances Survey (GOBELINS; Loinard 2013) was carried out to measure the distance and proper motion of a representative sample of young stars in five of the most prominent star-forming regions within 500 pc of the Sun (Taurus, Perseus, Ophiuchus, Orion and Serpens). From these observations, the mean distance, the three-dimensional structure, and the internal kinematics of each region could be obtained (Ortiz-León et al. 2017a, 2017b, 2018; Galli et al. 2018; Kounkel et al. 2017). Additional regions, such as Monoceros or LkH$\alpha$101, were considered in complementary observations (e.g. Dzib et al. 2016, 2018). Together, these observations have resulted in great improvements in our knowledge of the distribution of nearby star formation (see Fig. 5.1 for the example of the three-dimensional structure of the Taurus region), and have sometimes revealed large errors in previous estimates. For instance, the Serpens core was usually accepted to be at a distance of 260 pc (Straižys et al. 1996, 2003) but the GOBELINS survey revealed that it is located nearly 70% farther at 436 $\pm$ 9 pc (Ortiz-León et al. 2017b). This implies, in particular, that previously estimated luminosities for objects in that region are underestimated nearly by a factor of 3.

One of the interesting by-products of the GOBELINS survey has been the resolution of nearly two dozen very tight young binary stellar systems (with separations between a few and a few tens of mas – Ortiz-León et al. 2017a, 2017b; Galli et al. 2018; Kounkel et al. 2017).[1] In such cases, the displacement of the stars on the celestial sphere can be modelled as a combination of their trigonometric parallax and their systemic and orbital motions. The characterisation of the orbital motion enables the determination of masses through Kepler's law. Using orbital motions to estimate masses of stars in binary systems has, of course, been used for many decades. For young stars, near-infrared observations have typically been preferred (e.g. Ghez et al. 1993), and increasingly sophisticated techniques (speckle interferometry, adaptive optics or aperture masking) have been developed to resolve ever tighter systems (currently down to about 10 mas). VLBI observations at centimetre wavelengths have a typical resolution of about 1 mas, so they can resolve systems that near-IR observations cannot. VLBI observations have two additional fundamental advantages over near-IR imaging. The first one is that they enable a simultaneous determination of the orbital motion **and** the trigonometric parallax. As a consequence, the binary systems resolved by VLBI observations automatically also have very accurately known distances. Near-IR imaging, on the other hand, measures the angular semi-major axis of the orbits, but needs to rely on independently measured (and often quite uncertain) distances to transform this angular measurement to a physical distance. Since the mass depends on the cube of the semi-major axis, even a small error on distance can result in a very significant mass error. The second important advantage of VLBI over near-IR imaging is that it delivers masses for the individual stars in the system and not only the total system mass. This is because VLBI data provide positions of the individual stars measured relative to a background calibrator (normally a quasar). This implies that the relative motion between the individual stars and the centre of mass of the system can be measured, and therefore also the mass ratio. The mass measurements of young stars provided by VLBI observations are fundamental to constrain pre-main sequence evolutionary models (see below).

[1] See also Torres et al. (2012) for an earlier case.

The GOBELINS results described above are based on only about 100 individual VLBI parallax and proper motion measurements. An increase in sensitivity of VLBI arrays would enable a major step forward. At the current level of sensitivity, VLBI observations can only detect 10 to 20% of the young stars known to exist in star-forming regions at a few hundred parsecs. Given the known luminosity function of these sources (Ortiz-León et al. 2017a), an increase in sensitivity by a factor of 10 would roughly multiply by three the number of detectable sources, and bring the detection fraction to about 50%. This would have fundamental implications for the determination of the three-dimensional structure and internal kinematics of nearby star-forming regions, and would enable similar measurements to be carried out for regions that are significantly farther.

#### Calibration of PMS evolutionary models

Stellar evolution models allow us to understand the different phases that the stars cross throughout their existence. As a general rule, the predictions provided by these models fit the observations well and, therefore, can be considered a reliable source of scientific information. This is particularly useful to estimate fundamental parameters of the stars, such as the mass and radius, from theoretical luminosity-based relationships (e.g., Baraffe et al. 1998; Chabrier et al. 2000). The calibration of these stellar evolution models is important, but it is crucial in the case of young, low-mass objects, since these models are frequently used to determine the masses of planets and brown dwarfs. The study of binary stars belonging to young, moving groups (whose main feature is the common age of their members; Zuckerman & Song 2004; Torres et al. 2008) is a reasonable approach to increase the number of PMS stars with dynamically determined masses, as these have shown to be suitable benchmarks to calibrate models. In recent years, several of these moving groups have been discovered (Pictoris, Tucana-Horologium, TW Hydrae, Columba, Carina, Argus, and AB Doradus moving groups); both LBA and EVN astrometric observations have determined or refined the distance and orbital motion of some multiple systems belonging to the previous groups (AB Dor A/C, AB Dor Ba/Bb, or HD 160934 A/c; Guirado et al. 1997; Azulay et al. 2015; 2017), helping to alleviate the deficient observational status of dynamical masses of PMS stars, and therefore imposing strong constraints to stellar evolutionary models (see Fig. 5.2).

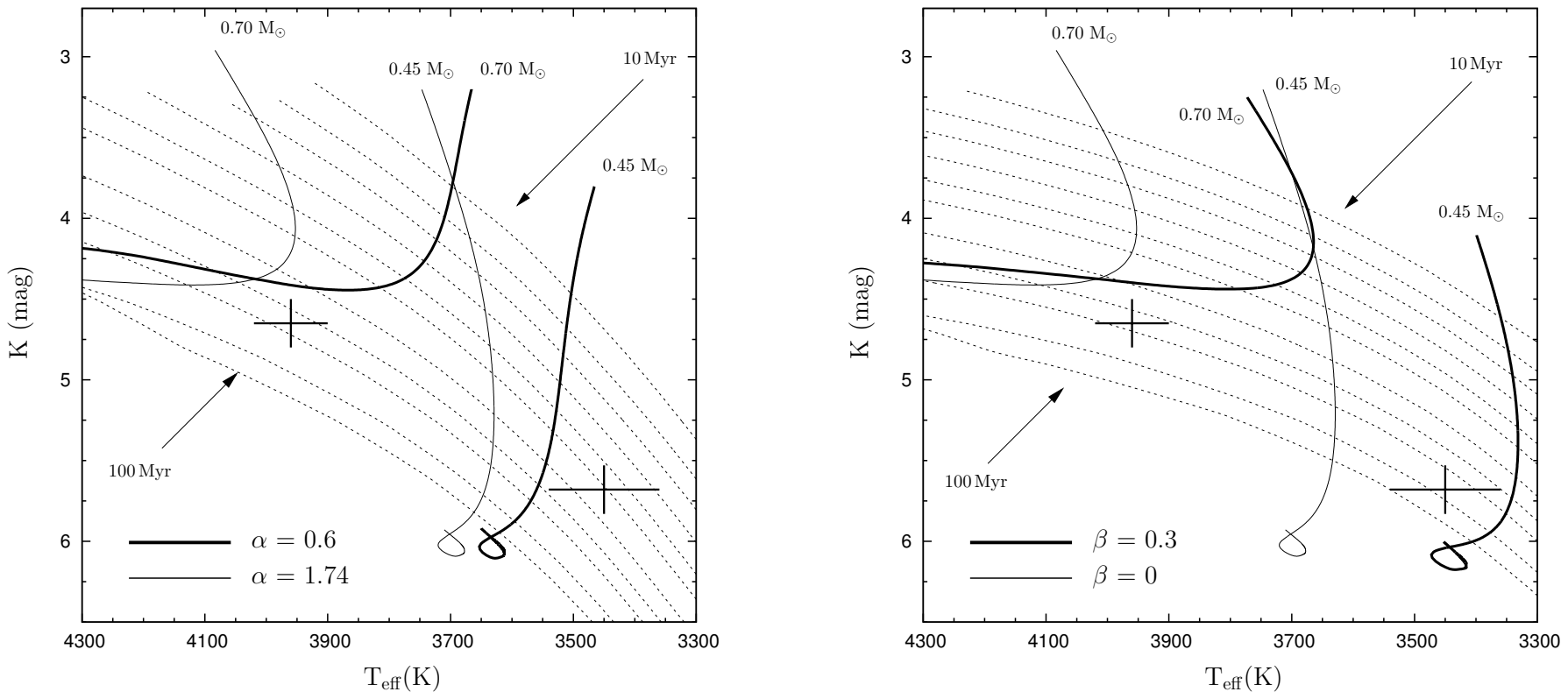

Figure 5.2: Example of comparison of the dynamical masses of the HD 160934 A/c system with the predictions of the models of Tognelli et al. (2011). Both isomasses (solid lines) and isochrones (dashed lines) are plotted. The highlighted tracks are those nearest to the dynamical mass values ($0.70\pm0.07$ and $0.45\pm0.04\,M_\odot$ for components A and c, respectively). (left) Effect of an internal magnetic field, simulated by using a value of the mixing length $\alpha = 0.6$. The isomasses of a standard solar value $\alpha = 1.74$ are also plotted for comparison. (right) Effect of stellar spots, computed for an effective spot coverage of $\beta = 0.3$ (Somers & Pinsonneault 2015). The isomasses for $\beta = 0$ (standard models with no spots) are also plotted for comparison. These and other model comparisons can be seen in Azulay et al. (2017). Reproduced with permission ©ESO.

Although the previous projects are limited to a handful of stars, it constitutes a foundation for future studies of the stellar and substellar evolution using the new, more sensitive radio interferometers. At present, the EVN is an excellent tool for these studies, although higher resolution, better sensitivity, and larger monitoring speed will allow the access to a much larger number of members belonging to the moving groups discovered. Extension of the EVN to the southern hemisphere, through the development of the African VLBI Network (AVN) is particularly welcome, as the majority of the moving groups are located in the southern sky.

### 5.1.2 Main sequence stars

Stellar rotation coupled with convective motions generate strong magnetic fields responsible for a multitude of phenomena including coronal loops, radio flares, or stellar spots.

#### Flares / coronal mass ejections

There is substantial bibliography showing that many late type (F, G, K, M) stars display incoherent radio flares (e.g. Güdel et al. 1998), which evidences the presence of midly relativistic electrons in magnetic fields. On the other hand, flares produced by a coherent radiation mechanism, abundant on magnetically active M dwarfs, carry profound information on the conditions of the corona during magnetic reconnections or coronal mass ejections (CMEs). The studies of coherent radio bursts have been bandwidth limited, at least until recent upgrades of instruments as the JVLA, which have allowed detailed stellar dynamic spectroscopy (studies of the spectral evolution over time) spanned over a wide spectral range (Güdel et al. 1989; Bastian et al. 1998). These studies have provided evidence of both plasma density and magnetic field gradients which translate to coronal plasma motion on active M dwarfs (Osten & Bastian 2006, 2008). Hence, the radio emission can be monitored across multiple coronal scale heights, which helps to identify the mechanism producing these bursts. The role of a sensitive VLBI array would be to image the non-thermal corona, seeking for resolved gyrosynchrotron emission, and tracing the possible motions of coronal plasma. In fact, JVLA and VLBA combined observations have already been carried out on popular flare stars such as AD Leo, UV Ceti and YZ CMi (Villadsen et al. 2017), enabling the simultaneous measurement of the dynamic spectra and the spatial offset between flaring and quiescent emission. This high-resolution search for spatial signatures of CMEs is therefore a perfect complement to the dynamic studies of the stellar corona. Villadsen et al. (2017) also emphasise the relevance of CME's for future astrobiology projects, as they are predicted to cause significant mass loss from planetary atmospheres (Khodachenko et al. 2007, Lammer et al. 2007) of possible companions to M dwarfs. The emission of energetic particles may produce drastic changes in molecular chemistry of planetary atmospheres, affecting in particular biomarkers like ozone (Segura et al. 2010).

#### Ultracool Dwarfs

Radio observations play an important role in understanding the processes involved in the formation and evolution of stellar and substellar objects. In particular, radio emission studies of ultracool objects (late M, L, and T objects; e.g. Matthews 2013) are relevant to probe the magnetic activity of these objects and its influence on the formation of disks or planets. Moreover, the study of ultracool dwarfs may open a suitable route to the detection of radio emission of exoplanets: meanwhile no exoplanet has yet been detected at radio wavelengths, an increasing number of ultracool objects (Hallinan et al. 2008; Pineda et al. 2017; Guirado et al. 2018) show substantial evidence of radio emission at GHz frequencies in objects with spectral types as cool as T6.5 (Kao et al. 2016; 2018). This radio emission is consistent with incoherent gyrosynchrotron with at least another pulsed, auroral emission, produced by electron cyclotron maser (ECM) instabilities, whose periodicity matches the stellar rotation rate. This evidences the persistence of magnetic activity ($\sim$kG) in very low mass objects. Moreover, radio observations become a unique tool to probe the magnetic properties of these objects, as other indicators such as H$\alpha$ or X-rays decline for such late spectral types.

However, although rapid rotation seems to be an essential ingredient to generate kG magnetic fields, the precise mechanism still remains unresolved. The fully convective nature of ultracool dwarfs (UCDs) does not allow the solar-type $\alpha\Omega$ dynamo, operating at the shearing interface between the radiative and convective zones (Parker 1955), to generate and amplify the observed magnetic fields. Since UCDs simulations have not been investigated so far, much emphasis is needed on finding observational constraints for the magnetic field properties. Zeeman spectroscopy requires very high signal-to-noise ratios and low/moderate rotation velocities (Reiners & Basri 2007). These conditions are hardly ever met in UCDs, therefore, observational constraints on the scale, geometry, and origin of the magnetic fields, such as those provided by VLBI, are essential. In this context, VLBI observations of ultracool dwarfs have contributed to establish direct limits on the radio emission brightness temperature of ultracool objects such as TVLM 513–46546 (Forbrich and Berger 2009; 2013). Actually, this object has been the target of an astrometric campaign using the EVN to find the strongest constraints on possible companion masses and orbital periods (Gawroński et al. 2017; see Fig. 5.3). Other systems studied at high-resolution are the multiple substellar system VHS 1256-1257, where EVN observations helped to determine the spectral index of the radio emission (Guirado et al. 2018).

Clearly, more objects will be detected as the sensitivity of the arrays improve. The EVN (C-band) appears as an excellent tool to find compact radio emission in these unique systems which will be essential to observationally constrain the brightness temperature and, therefore, the nature of the radio emission. Of particular interest are double or multiple systems. In addition to enable astrometric studies aimed at determining the dynamical masses, high-resolution observations will determine whether the radio emission originates either in one or in both components. If both objects are radio emitters, they will probably have a similar rotation history (velocity and/or axis orientation), which has evolved to produce a magnetic field with similar intensity. On the contrary, if the radio emission comes essentially from one of the components, it will indicate the presence of underlying differences in the stellar rotation and/or different magnetic field configurations. An interesting avenue is the possibility stressed by Pineda et al. (2017), who suggested that the different radio emission properties in physically similar dwarf binary components (having same mass, luminosity, and temperature) might be due to the existence of a rocky planet around (only) one of them, which may induce a different magnetic configuration; therefore, the presence of the planet is the triggering factor of the radio emission. This is the case in at least two ultracool dwarf binaries: 2MASSI J0746425+200032 (Konopacky et al. 2010), and LSPM J1314+1320AB (Forbrich et al. 2016; see Fig. 5.4), which present radio emission in only one of the components, being primary targets in searches of Earth-mass planets.

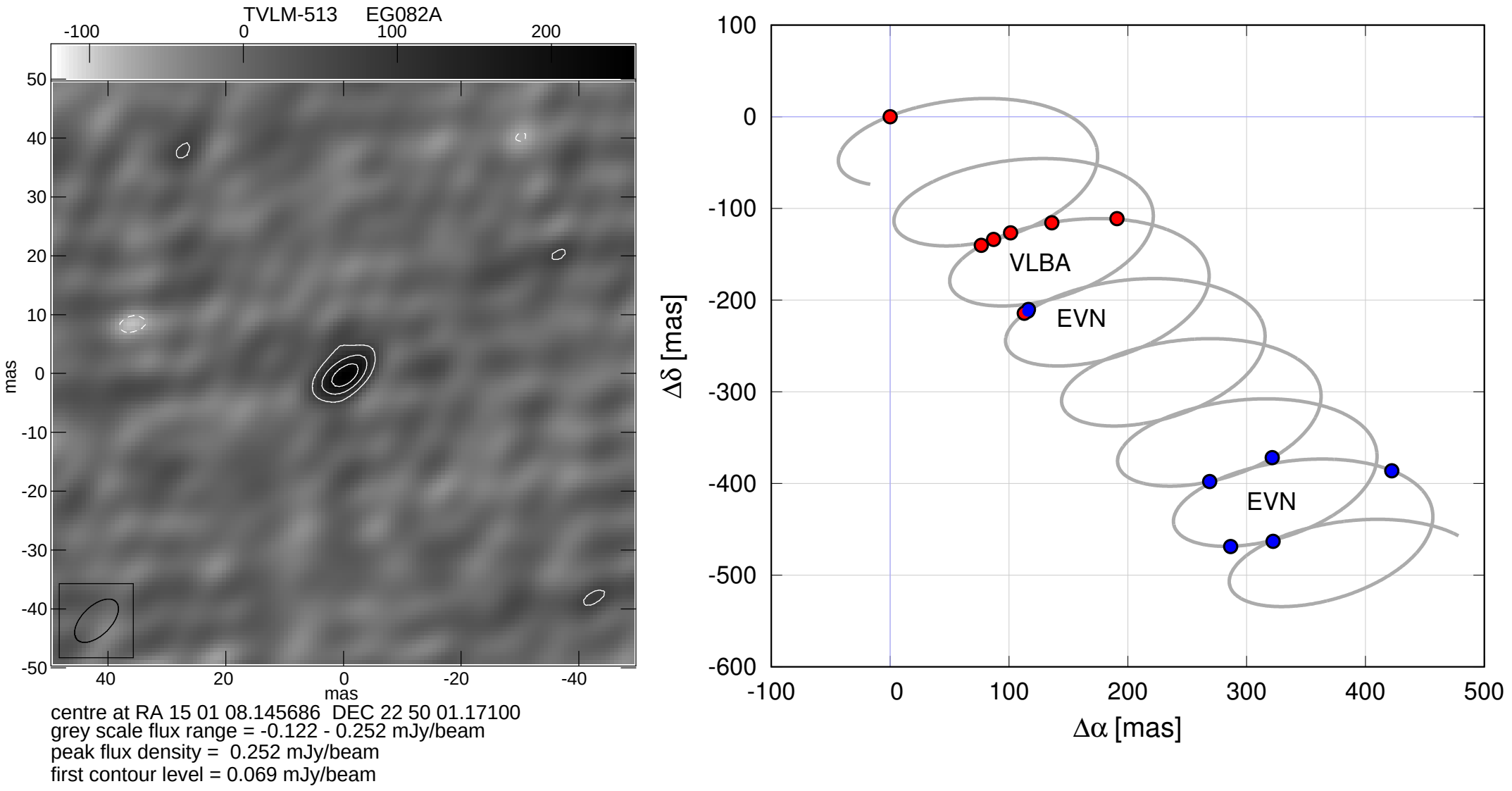

Figure 5.3: (left) 5 GHz EVN map of TVLM 513-46546. The first contour corresponds to the 3$\sigma$ detection limit. (right) Sky-projected (proper motion + parallax) trajectory of TVLM 513–46546 (grey curve) overplotted with VLBA (red filled rectangles) and EVN detections (blue filled circles; adapted Fig. 1 and 5 from Gawroński et al. 2017).

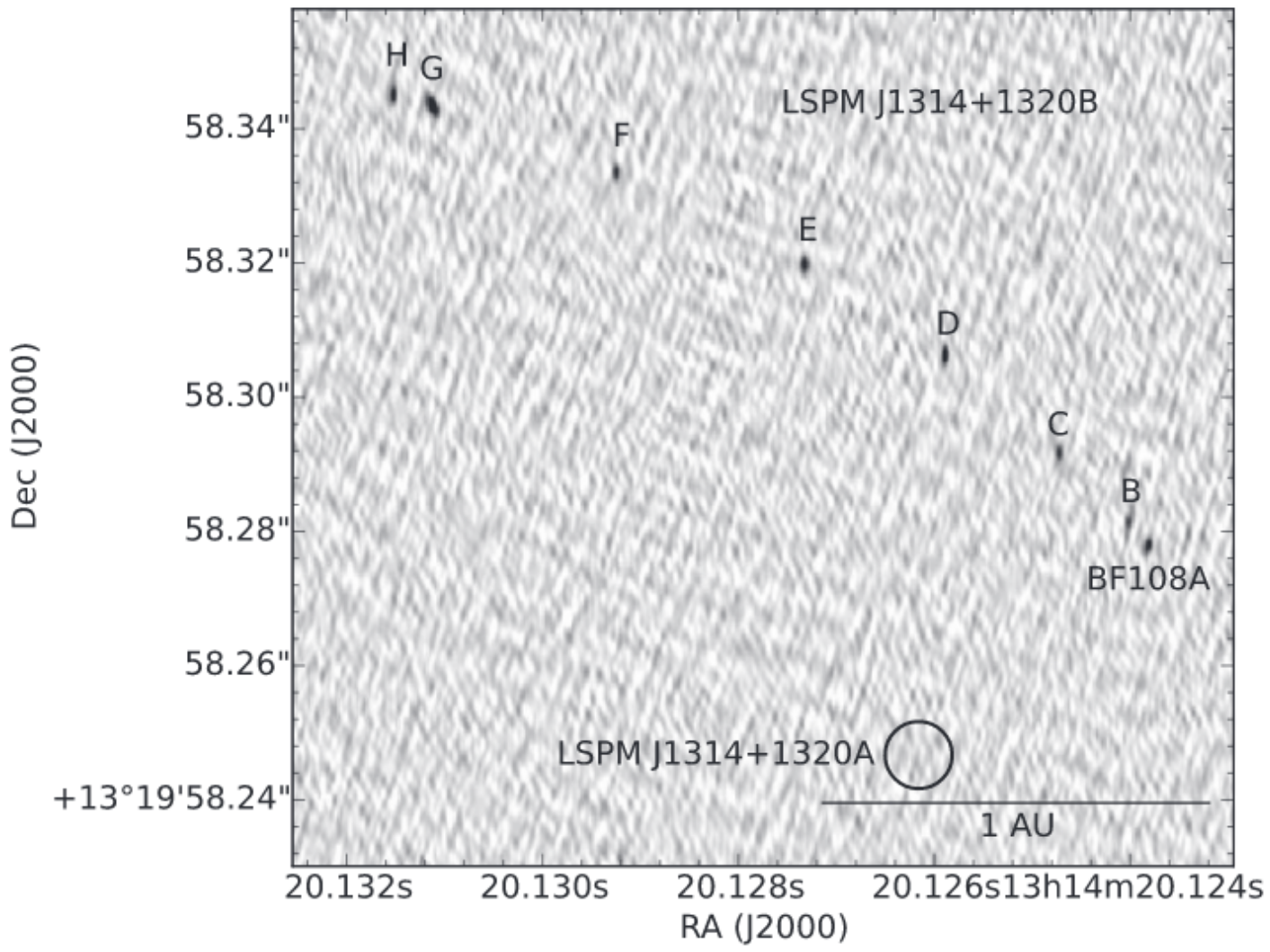

Figure 5.4: Co-added images of VLBA epochs of the ultracool binary LSPM J1314+1320A . The primary remains undetected at the circle of 5-mas radius. All eight detections of the secondary are clearly visible, and the shape of the orbit is recognizable (Forbrich et al. 2016). ©AAS. Reproduced with permission.

#### Exoplanets

Radio wavelengths may provide direct detection of extrasolar giant planets (EGPs). While the ratio of the luminosities between an EGP and its parent star is very low in the optical, the radio luminosity ratio is not necessarily low. In principle, and mostly based on our knowledge of the Sun and Jupiter, this contrast is most favourable at decametre wavelengths, with the emission dominated by the cyclotron maser emission. This route to direct detection of exoplanets looks the most suitable, specially given the new, dedicated instruments at very low frequencies such as LOFAR or ASKAP. Complementary, EGP radio emission at GHz frequencies appears as a promising approach, as proved by the increasing number of radio detections of ultracool dwarf objects (Hallinan et al. 2008; McLean et al. 2011) with spectral types as cool as L3.5 (Router & Wolszczan 2012).

The angular resolution of VLBI systems can reach a few milliarseconds depending on the observing frequency. It provides an opportunity for direct observations of extrasolar planets in nearby planetary systems, which are similar to our Solar System. Using the direct detection/image of an extrasolar gas giant it could be possible to estimate its magnetic field strength and/or dynamical mass. This knowledge is crucial in our understanding of evolution and creation of Jupiter-like gas giants in planetary systems similar to ours. However, the main problem of detecting sub-stellar companions is the expected frequency of radio emission, if the emission processes are similar to what we observe in the Solar System. Radio observations of gas and ice giants in the Solar System show that dominant radio emission extends from kHz to 40 MHz. If the emission range is typical for all massive planets, only low frequency systems like LOFAR or SKA could likely detect radio emission from exoplanets. The frequency range of radio emission strongly depends on the exoplanet magnetic field strength and MHz waves are the result of G magnetic fields found in gas/ice giants in our Solar System.

Among a few possible processes that can generate radio emission in the planetary systems, three of them seem to be the most promising in terms of direct observations. The first mechanism assumes an interaction between the stellar wind and the planetary magnetosphere. The power of the radio emission in this process depends on the kinetic energy flux and spatial density of particles, that are impacting on the planet magnetopause (e.g Farrell, Desch & Zarka 1999). The second emission process is an effect of the interaction between the magnetic energy flux of the interplanetary magnetic field with the planetary magnetosphere (e.g. Farrell et al. 2004). The third scenario assumes the existence of a moon around a planet. The volcanic activity of the moon fills the magnetosphere with matter that is ionised and accelerated by electric currents. The currents are inducted because of the difference between the gas giant rotation velocity and the moon orbital velocity (e.g Nichols 2012). In the Solar System it is impossible to distinguish which emission process dominates (the first or the second).

The theoretical models of the MHz radio emission from massive exoplanets were already investigated by several authors (see for a review Griessmeier, Zarka & Girrard 2011). It was shown, that it should be possible to detect low frequency radio emission from a few known massive exoplanets (e.g. $\varepsilon$ Eri b or $\tau$ Boo b). However, despite many observational projects there is no confirmed detection of radio emission from exoplanets (e.g. George & Stevens 2007, Lecavelier des Etangs et al. 2013). Recent results presented by Kao et. al (2018) may shed some light on this mystery, as it seems the rapid rotation of sub-stellar objects is important in producing strong kG magnetic dipole systems. These authors present detections of radio emission from selected brown dwarfs which are strong and direct constraints on dynamo theory in gas giants located at the substellar-planetary boundary.

Recently it was also postulated that massive and young gas giants (<1 Gyr) could produce kG magnetic fields, which are able to produce radio emission at cm radio bands. It offers the possibility of the direct detection/discovery of exoplanets by existing VLBI systems (Katarzyński et al. 2016). The authors concentrated on the sample of young massive stars from the Solar neighborhood and showed that under reasonable assumptions it is possible to detect radio emission from putative massive gas giants located in those systems (Fig. 5.5). Actually, a recent result from Bastian et al. (2018) show a possible detection of radio emission from the planetary system $\varepsilon$ Eri at cm-wavelengths. The results presented by Kao et. al (2018), Bastian et al. (2018), and Katarzyński et al. (2016) strongly indicate that VLBI systems could have a great importance and impact on current and future exoplanets studies.

### 5.1.3 Evolved stars

#### Mass loss / stellar winds

Radio observations have been a powerful tool to study, among other things, the atmosphere, winds and mass-loss of evolved stars, including Red Supergiants (RSGs), asymptotic giant branch stars and their evolution into planetary nebulae (PNe) as well as massive OB and WR stars. The evolution of AGB, RSG and more massive evolved stars is dictated by their mass-loss through strong stellar winds and is responsible for determining their final end-point. If ionised, these stellar winds produce detectable thermal radio emission. Mass-loss rates measured using a number of indicators (e.g. H$\alpha$/UV/radio) appear to disagree by significant amounts (e.g. Fullerton, Massa & Prinja, 2006) casting doubt on the expected evolution of the star. It is now believed that small-scale structure in the form of clumping is the reason for these discrepancies (e.g. Puls et al. 2006). Radio observations however, provide a distinct advantage in determining accurate mass-loss rates as they offer by-far the least model-dependent method and these can be interpreted directly from the radio flux density. Though massive stars are relatively faint, upgrades to existing interferometers such as increasing the bandwidth have improved the situation significantly. Radio surveys of Galactic clusters are now beginning to perform a census of radio emission from massive stellar winds through both focused surveys such as the *e*-MERLIN study of Cygnus OB2 (COBRaS: Morford et al. submitted; Fig. 5.6) and the ALMA investigation of Westerlund 1 (Fenech et al. 2018) as well as larger area and all-sky surveys such as the ASKAP SCORPIO (Umana et al. 2015) and EMU projects (Norris et al. 2011). Though observations of thermal stellar winds are brightness sensitivity limited, increased bandwidths along-side the correct combination of baselines (e.g. *e*-MERLIN with the inner EVN antennas) can be used to do more targeted observations of brighter objects to study potential variability and structure in the stellar winds.

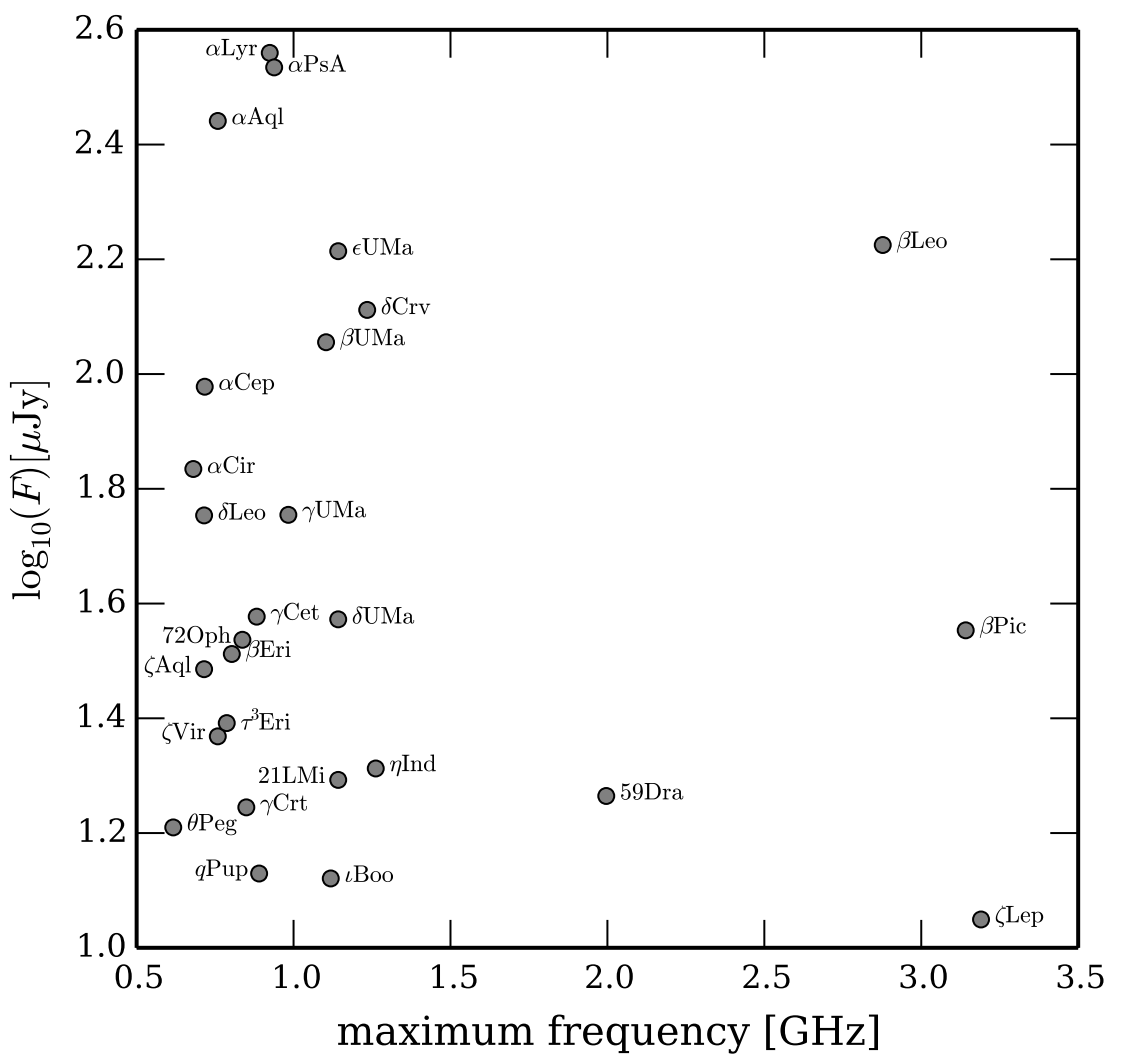

Figure 5.5: Expected radio emission from hypothetical object with masses M = 15 $M_{\rm Jup}$ located at the distance d = 1 AU around selected main-sequence A-type stars (Fig. 5, Katarzyński et al. 2016).

### Star spots

Evolution of starspots on various time scales allows us to investigate stellar differential rotation, activity cycles, and global magnetic fields, as well as to constitute the basis for our understanding of the stellar dynamo mechanism.

Starspots have been best-studied on Betelgeuse, starting ~20 years ago with optical and IR interferometry. The radius of Betelgeuse has been resolved out to ~ 125 mas (25 AU, or over 5× the optical radius) at $\lambda$6 cm using the VLA and *e*-MERLIN. The latter shows several surface brightness fluctuations of order 10% against the average temperature ~ 2300 K at 5 cm. At wavelengths around ~ 1 mm, ALMA has resolved one or two spots on Betelgeuse, Mira and a few other objects. In all these cases, the instrumental resolution is pushed to the limit and the spot sizes, and hence brightness temperature fluctuations, are not well defined; there could be smaller, brighter spots or even localised clusters of larger numbers of spots. It is intriguing to postulate a link with the low-filling-factor chromosphere measured for Betelgeuse (Harper et al. 2006) as well as with convection cell models (Chiavassa et al. 2010), but this has not been possible thus far. Adequate sensitivity at higher resolution can tackle these problems: 1) Tighter constraints on the numbers and nature of star spots. The optimal combination of resolution and sensitivity (considering atmospheric transmission as well as source properties) is around C-band, where Betelgeuse has a flux density of ~ 2 mJy at 6 GHz. The *e*-MERLIN resolution places an upper limit of 50 mas on the spot sizes. If 10% fluctuations were smooth, this corresponds to 18 $\mu$ Jy per 15-mas resolution element, or 6$\mu$ Jy rms for a 3$\sigma$ detection. This would be practical by combining *e*-MERLIN with EVN baselines out to ~ 1000 km

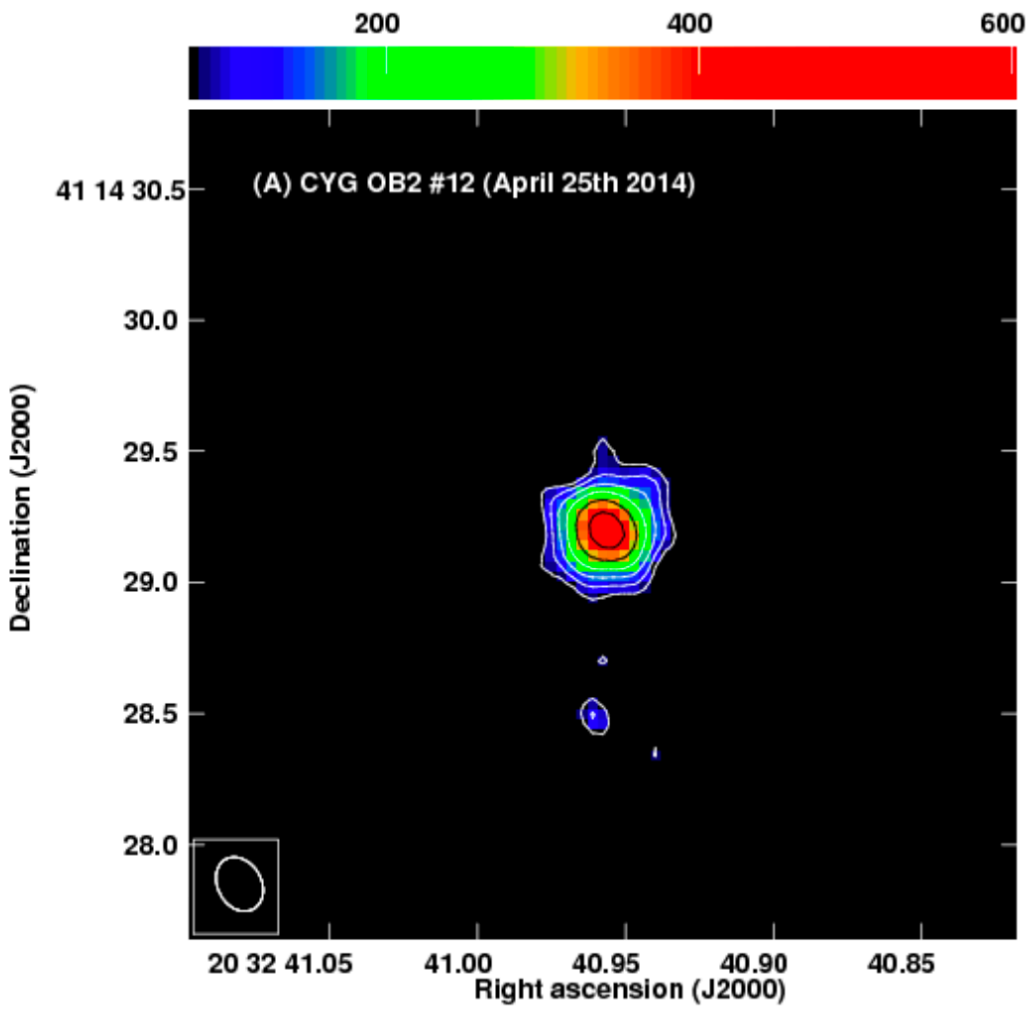

Figure 5.6: Image of one of the target sample stars from the COBRaS 21 cm *e*-MERLIN Legacy observations, Cyg OB2 #12, showing the first ever resolved image of its thermal emission at 21 cm (rms = 24 $\mu$Jy/beam; Fig. 1 in Morford et al. 2016).

–of course, longer baselines would be valuable to investigate the existence of more compact, more extreme hotspots. 2) The hot-spots and stellar flux are variable and may have distinctive spectral indices. This is complicated by the wavelength dependence of the imaged surface, but this does offer the possibility to track disturbances by observing later epochs at longer wavelengths. Hence, for a single image, a fractional bandwidth of not more than 10% is best, from observations taken over a few days or weeks at most. On the other hand, if it was possible to observe at several frequencies using suitable combinations of EVN, *e*-MERLIN and the VLA, at suitable time intervals, this could trace disturbances propagating from 2–6 $R_{\star}$, on timescales from a week to many months, depending on whether radiative/recombination, shocks or bulk transport are involved (Harper & Linsky 1999).

There is evidence for a very low filling factor chromosphere in Betelgeuse (Harper et al. 2006) and the current estimates of radio hot spot brightness temperatures are based on > 50 mas resolution images. A very hot spot (~50 000 K, size of 2–3 mas) was measured on the AGB star W Hya by Vlemmings et al. (2017) using the long baselines of ALMA. It is possible that these mas-size features may be collections of much smaller, possibly even hotter spots with high enough non-thermal brightness to allow detection at cm-wavelengths with *e*-MERLIN combined with the shortest EVN baselines.

### Colliding winds

Understanding binary interaction is also key for understanding stellar evolution and mass-loss and this is especially important considering the significant fraction of stars expected to undergo some binary interaction during their lifetime (50–70%). Interaction of stellar winds of massive binaries (OB type or WR) can produce shocks, defining a wind collision region (WCR) where particles are accelerated up to relativistic velocities. Within the WCR thermal radio emission is typically produced, but also non-thermal radio emission (from synchrotron radiation) which constitutes a suitable scenario for EVN/VLBI observations to spatially resolve such emission. Hence, precise distance and orbit, critical for modelling the WCR, have been determined for archetypical objects as WR 140, a WR+O binary system which shows a bow-shaped arc emission that rotates as the highly eccentric orbit progresses (Dougherty et al. 2011). Recent results on the hierarchical triple system HD 167971 (an O-spectroscopic binary with a third O-star in a wider orbit; Sanchez et al. in press) demonstrate the ability of high-resolution VLBI maps to determine the orbit of the components relative to the bow-shock position, allowing to find an absolute astrometric solution for the complete system. The quadruple system in Cyg OB2 #5 is also a good example where the radio observations have been key to understanding the system as not all of the stars have been optically/spectrally identified and yet the non-thermal emission from the WCR clearly indicates their presence (e.g. Dzuib et al. 2013). In some cases, such as in the massive binary HD 93129A, VLBI observations were key to unveil the binary nature of the system due to the discovery of a powerful WCR (Benaglia et al. 2015).

Other systems include a variety of evolutionary stages e.g. O+O binaries, WR systems and probably shorter-lived stages such as Be systems. This will of course also include neutron-star and black-hole binary systems as well. As shown in some studies (De Becker et al. 2017), it is considered that WCR should not be a rare phenomenon in massive binary stars. Actually, these authors estimate that colliding-wind binaries may constitute a significant contributor of the Galactic cosmic rays; this finding increases the interest to widen the present census of massive binaries. Obviously, sensitive interferometers would be ideal, as VLBI should reveal the kinematics of the systems. Sampling at different baseline lengths, from tens to thousands kilometres to avoid resolving the emission, would be essential, as well as flexibility in frequency and time for multi-frequency and multi-epoch observations to observe at various orbital phases.

## 5.2 Stellar masers

The importance of spectral line observations concerning maser emission will be presented in this section. Owing to the increasing spectral resolution and sensitivity of VLBI arrays, many important results have been obtained in the last decades based on the VLBI observations of masers, naturally occurring sources of stimulated spectral line emission in the ISM. In the Milky Way, masers are typically found in star-forming regions and evolved stars. Being compact, bright, non-thermal emission structures they are ideal VLBI targets to be used for understanding physical parameters such as density and temperature, but also for accurate position measurements making them ideal astrometry targets. Maser astrometry is used for measuring the spiral structure of the Milky Way and the Galactic rotation curve. In the following subsections Galactic VLBI maser science of star-forming regions, evolved stars and astrometry are discussed. A summary of the astrophysical masers can be found in Gray (2012).

### 5.2.1 Masers in Star Forming Regions

High-mass stars have a large influence on the energetics of the gas and dust in the Galaxy via their bright UV emission and their strong winds which increase in their evolved stages, and finally by supernova explosions, through which they also dominate a galaxies' metal abundance. Understanding high-mass star formation is thus important for galaxy evolution on the large scales, and for a detailed view of stellar formation, including planetary system formation, on the small scales. In high-mass ($\geq$8$M_\odot$) star formation the accretion of matter onto the star has to overcome the strong radiation pressure released when the star commences hydrogen fusion. Among the various existing theories that explain the continuation of accretion or agglomeration of mass, high-resolution observations of masers in rotating disk-like structures are favouring disks and accretion bursts implying an up-scaled version of low-mass star formation.

Molecular maser emission at cm wavelengths has been most useful to trace the early stages of the evolution of the protostar-disk-jet systems at milliarcsecond scales using VLBI. Sensitive VLBI observations show, mainly in association with intermediate- and high-mass protostars, thousands of maser spots forming microstructures that reveal the 3D kinematics of outflows and disks at small scale. This kind of observations have given a number of interesting results: the identification of new protoplanetary disks in YSOs at scales of tens of AU (Torrelles et al. 1998; see Fig. 5.7); the discovery of short-lived, episodic non-collimated outflow events (e.g., Torrelles et al. 2001, 2003; Surcis et al. 2014); detection of infall motions in accretion disks around massive protostars (Sanna et al. 2017); the imaging of young ($<$ 100 yr) small scale (a few 100 AU) bipolar jets of masers (Sanna et al. 2012, Torrelles et al. 2014); and even allowed us to analyze the small-scale (1-20 AU) structure of the micro bow-shocks (Uscanga et al. 2005; Trinidad et al. 2013).

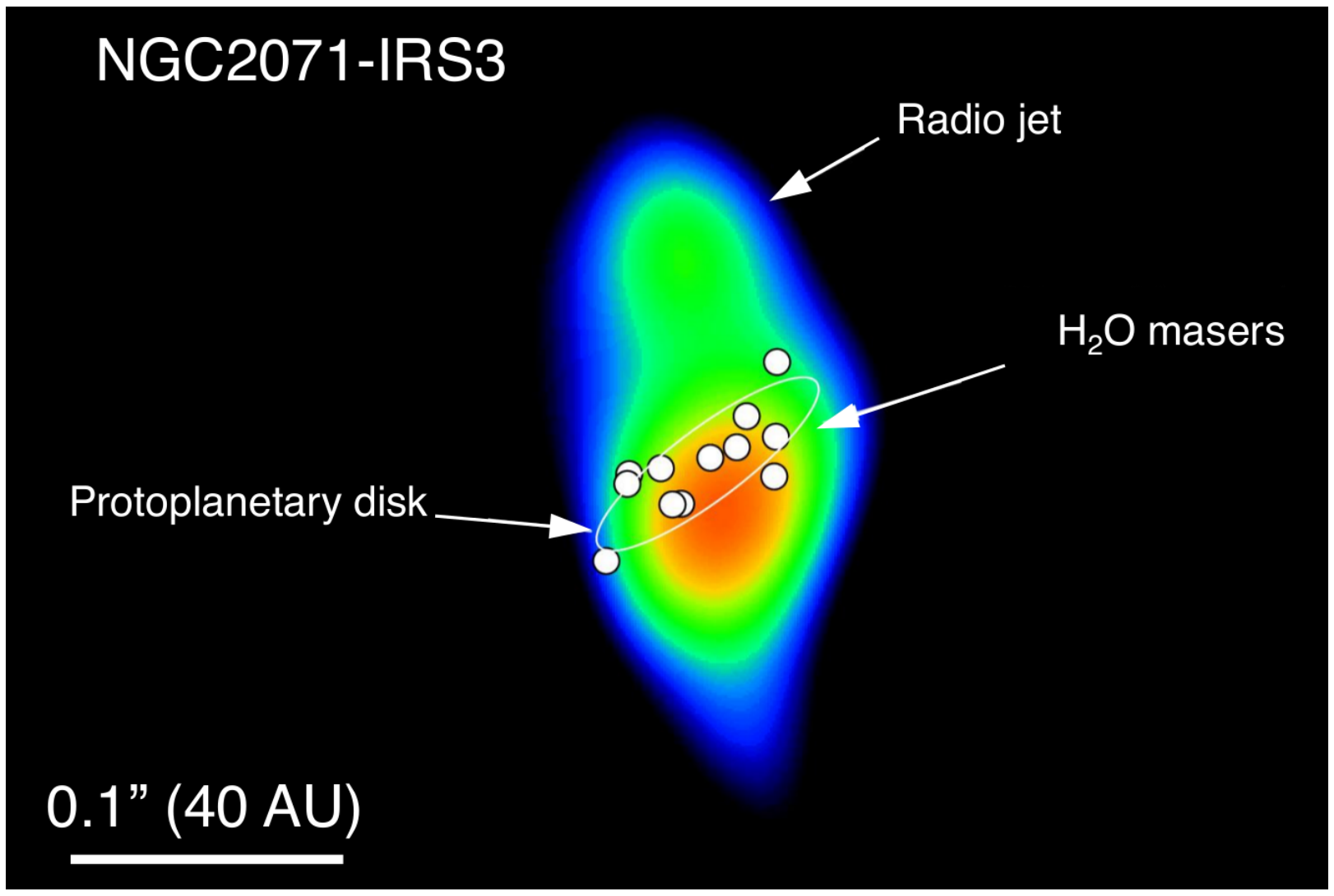

Figure 5.7: $H_2O$ masers (white circles) tracing a protoplanetary disk of radius 20 AU, oriented perpendicular to the thermal radio jet associated with the protostar NGC 2071-IRS3 (Torrelles et al. 1998). ©AAS. Reproduced with permission.

With the rise of polarimetric measurements of star-forming cores, filaments, and, on small angular scales, disk-like structures and outflows, the importance of the magnetic field in star formation has been firmly established, though *how* and *how much* the magnetic field influences the star formation is still elusive to us. These open questions can be largely ascribed to the large difficulty of observing high-mass star-forming regions (HMSFRs). High-mass stars are rare, thus their birth-sites are generally found at large distances and therefore at small angular scales. In addition, they are enshrouded in large amount of natal dust and gas thus causing a high level of extinction and optical depth problems. Maser observations at the radio wavelengths ranges are not hindered by dust. The masers' compactness and high brightness temperatures allow one to use VLBI to accurately trace motions very close to the protostar, providing an unique way to search for rotating structures, such as disks, or outflows. VLBI polarimetry enables via morphology mapping and Zeeman splitting measurements to determine 3D magnetic fields to investigate the connection between matter and magnetic field at 10 – 100 AU scales.

Recently, the interest in the variability of maser emission associated with HMSFRs has skyrocketed by the discovery that many class II methanol masers vary periodically (Goedhart et al. 2014). Multi-transition maser observations (Fig. 5.8) yield precious information on the physical conditions where these masers operate, provided that good pumping models exist. The variability of astrophysical masers introduces a time-domain aspect which provides additional, and possibly unique, information of the star formation environment. Maser super bursts have now been observed in a few HMSFRs, such as the 6.7 GHz methanol maser emission flare in S255 NIRS3 (Moscadelli et al. 2017, Szymczak et al. 2018b), and appear to be connected to accretion bursts which could provide a solution to the radiative pressure limiting the accretion of matter onto the young stellar object (YSO) allowing it to grow in mass beyond $\sim$10 $M_\odot$. Water masers in jet-driven bow shocks provide additional information to episodic accretion (Burns et al. 2016). Only one case has been found so far where the 6.7 GHz methanol maser alternates with the 22 GHz water maser, that is the intermediate-mass young stellar object G107.298+5.639 (Szymczak et al. 2016). However, the origin of the anticorrelated variability of both lines is still under debate.

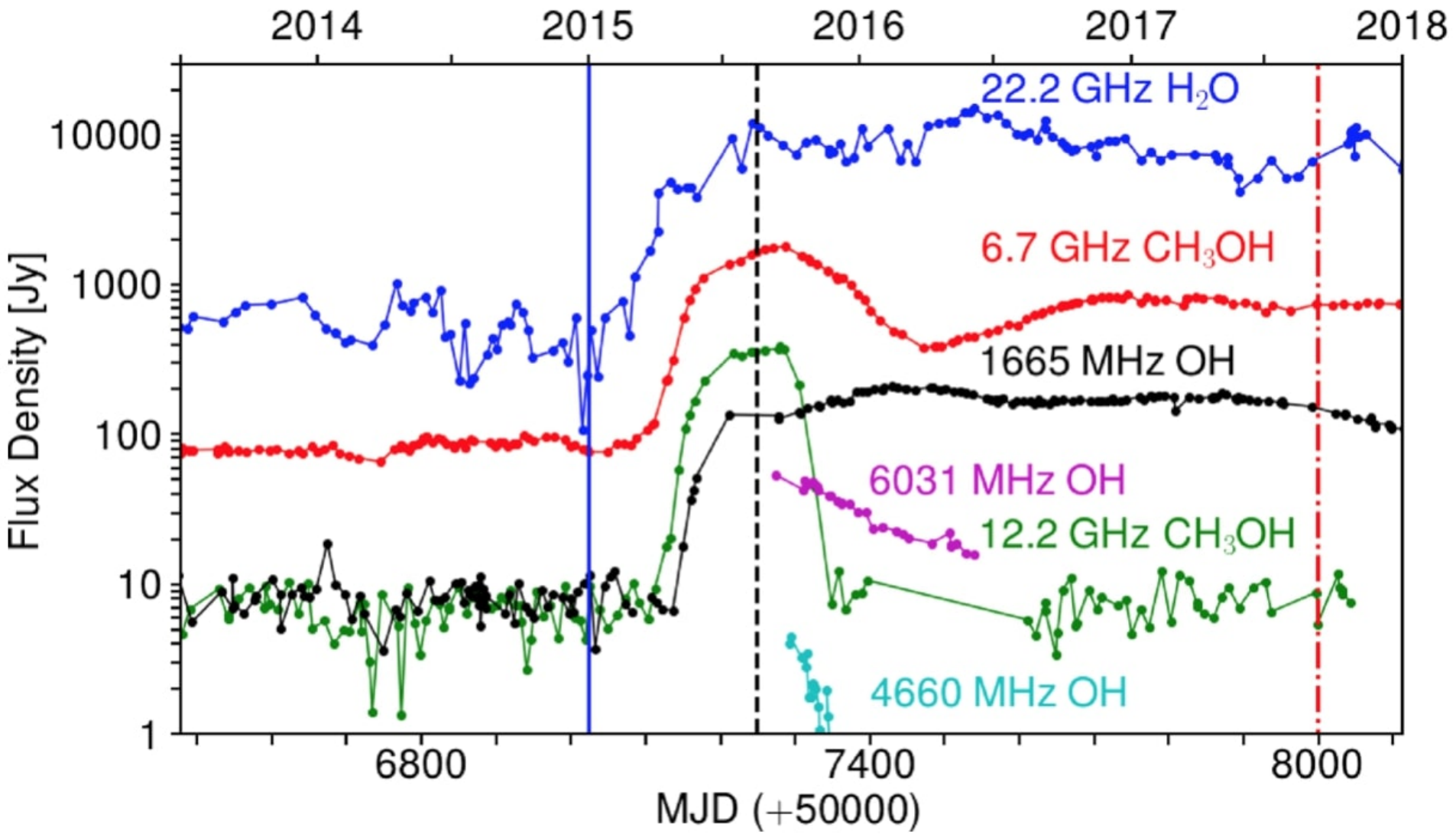

Figure 5.8: Time series of selected velocity channels of various maser species towards NGC 6334I measured with the 26 m telescope of the Hartebeesthoek Radio Astronomy Observatory (HartRAO) (Fig. 9, MacLeod et al. 2018). The blue vertical solid line indicates the onset of the outburst and the black dashed line marks the initial peak.

### EVN research of masers in high-mass star-forming regions

High angular resolution observations of intense molecular masers, in particular SiO, $H_2O$, and $CH_3OH$, are crucial to resolve the gas kinematics at 10 – 100 AU (requiring angular resolutions < $0\farcs1$ at the typical distances > 1 kpc of HMSFRs) around high-mass YSOs, where the accretion and ejection processes operate in close proximity. Combining maser VLBI data with (subarcsecond) interferometric observations of thermal (continuum and line) emission provides a very detailed view of the gas kinematics and physical conditions near the forming star. In general, it is found that water masers trace mass ejection in outflows or jets, while the methanol masers are more often located closer to the YSO, possibly in the disk-plane (perpendicular to the outflow axis) with kinematic patterns suggesting rotation. Recent EVN kinematics studies revealed that methanol masers might actually be related to an interface of the jet and disk regions rather than the disk-plane itself (Bartkiewicz et al. 2018). Unfortunately, detailed VLBI studies, which combine also other interferometric observations, have been performed only towards a small number of objects, so far, but have yielded very important observational clues. For example, submillimetre array observations reveal that the high-mass YSO G23.01−0.41, expected to be a ZAMS O9.5 star of $\approx$20 $M_\odot$, has a collimated bipolar molecular outflow at 0.1 pc scales, and, at the centre of the outflow, on a few $10^3$ AU, the gas and dust distribution is flattened and oriented perpendicular to the outflow axis (Fig. 5.9). This is fully consistent with multi-epoch VLBA 22 GHz water and EVN 6.7 GHz methanol maser observations. On the one hand, the elongated distribution and collimated proper motions of the water masers suggest that they are tracing the YSO's jet driving the motion of the large scale molecular outflow. On the other hand, the 6.7 GHz methanol masers are distributed at radii from a few 100 AU to a few $10^3$ AU around the YSO, and trace the envelope-disk kinematics (Fig. 5.9). Their proper motions are quite complex and can be explained in terms of a composition of slow (ca. 4 km $s^{-1}$ in amplitude) motions of radial expansion and rotation about an axis approximately parallel to the water maser jet.

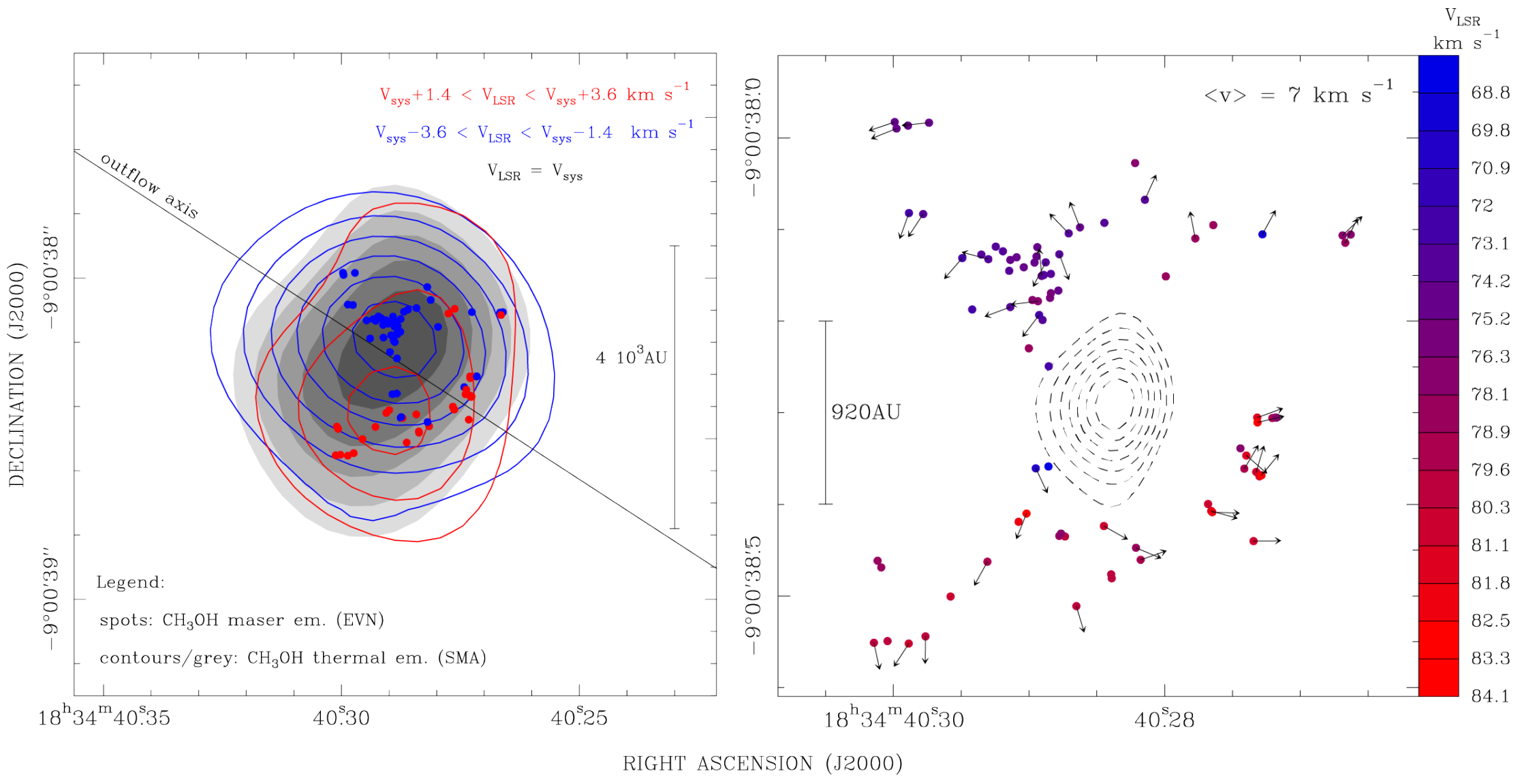

Figure 5.9: (left) Map of the $CH_3OH$ ($15_4$-$16_3$) E line emission observed with the SMA towards G23.01-0.41. Adapted from Sanna et al. (2014), reproduced with permission ©ESO. (right) Positions and proper motions of the 6.7 GHz methanol masers (EVN) and VLA 1.3 cm continuum (dashed contours). Adapted from Sanna et al. (2010), reproduced with permission ©ESO.

The EVN has given a particularly strong contribution to the magnetic field studies in HMSFRs. Observations of the polarised emission of different maser species have provided magnetic field measures in different ambient conditions within HMSFRs, at a spatial resolution unachievable with any other kind of observations. Perhaps, the most groundbreaking and unique results have been the EVN observations of polarised emission of 6.7 GHz methanol masers. Surcis et al. (2013, 2015, 2019) and Dall'Olio et al. (2017) showed that these masers, located close to the accretion disks and in the ambient medium between disks and outflows, have preferred magnetic field orientations along the outflow axis. By combining the magnetic field morphology with the gas motions, measured simultaneously from the same data set, Sanna et al. (2015b) have proved, for the first time, that gas with low turbulence moves along the magnetic field lines (Fig. 5.10). In the last 10 years Zeeman splitting measurements of 6.7 GHz methanol masers (in particularly with the EVN) have become more common. However, only since 2018 these measurements can be further used to estimate the magnetic field strength (measured to be of the order of 1-10 mG), thanks to the newly calculated Landé g-factors for all the hyperfine transitions of the methanol molecule (Lankhaar et al. 2018). This provides important input for theoretical models of star formation.

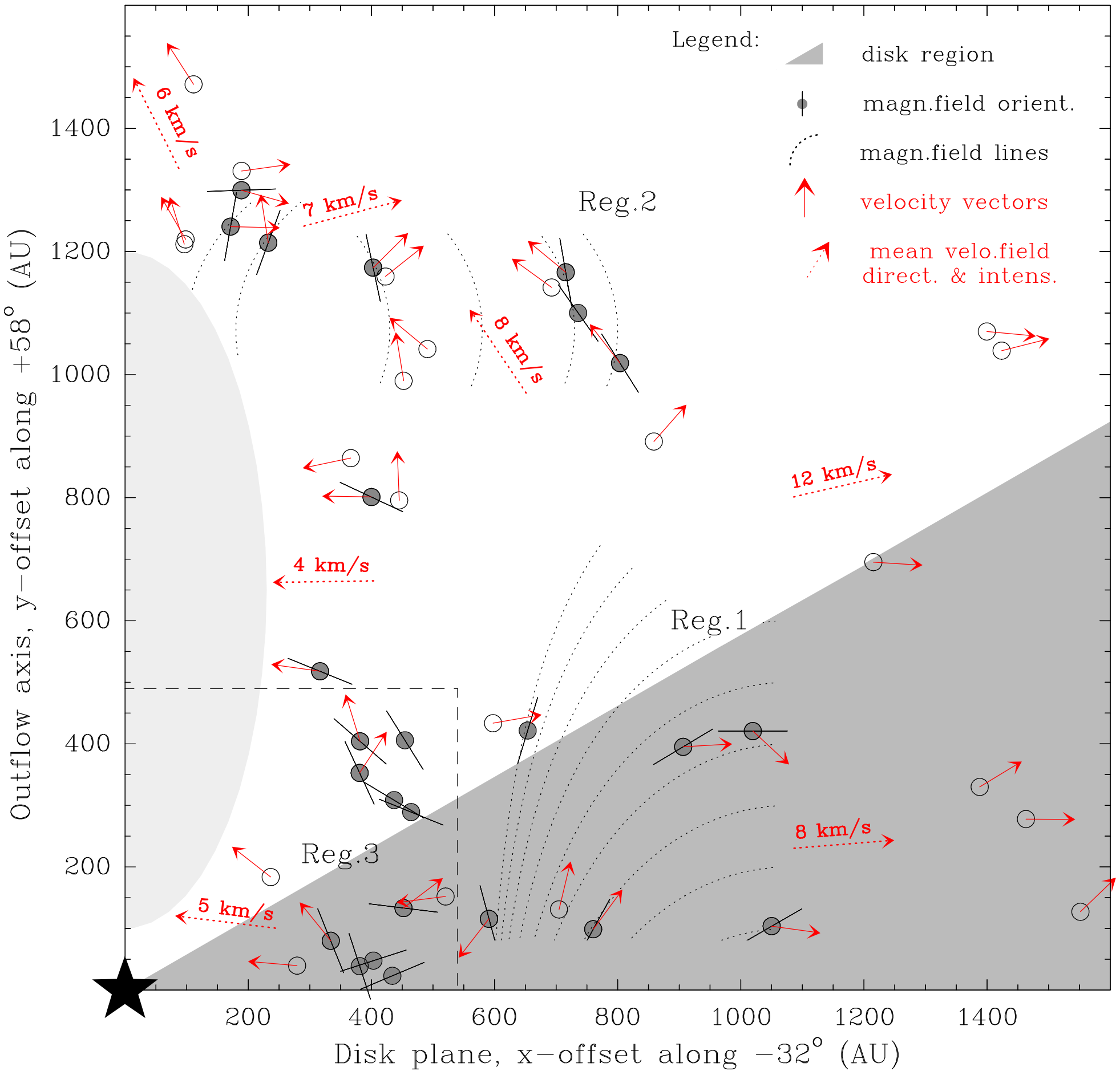

Figure 5.10: Gas dynamics and magnetic field configuration revealed within the inner 2000 AU of a high-mass YSO in G023.01−00.41 (Sanna et al. 2015b). Empty and solid dots in the vicinity of the hot molecular core centre (star) mark the 6.7 GHz $CH_3OH$ masers emitting unpolarised and polarised light, respectively. Arrows correspond to the local direction of the velocity of the masers and bars represent magnetic field vectors. Magnetic field lines (dotted lines) are extrapolated from the average behaviour of the magnetic field vectors. Adapted from Sanna et al. (2015b), reproduced with permission ©ESO.

Small scale magnetic fields can be also studied with water maser VLBI polarimetry. Goddi et al. (2017) revealed a complex picture of water maser polarisation using full polarimetric VLBA observations: on scales of thousands AU, the magnetic field traced by polarised water maser is aligned with the synchrotron jet in W3($H_2O$), though, on 10s to 100s of AU, a misalignment between the magnetic field and the velocity vectors was revealed, and ascribed to the compression of the field component along the shock front. Similar studies are also done using the EVN, of particular interest is the case of W75N(B): a 6-year EVN monitoring project has been investigating the correlation between outflow collimation and magnetic field, both measured using water maser emission, around a number of massive young stellar objects in the region (e.g. Surcis et al. 2014).

The EVN is the proper instrument to take up the challenge of understanding the cause of maser variability. The focus of high resolution maser studies should be directed first to those star forming regions where the maser variability is predictable and repeatable, e.g. the periodic class II methanol masers. The first of such observations, a single epoch VLBA observation, was done towards G9.62+0.20E, where not only class II methanol masers show periodic flaring but also the 1665 and 1667 MHz OH masers (Sanna et al. 2015a; Goedhart et al. 2018). Although this source is not observable with the EVN, there are a few periodic methanol maser sources that have similar methanol maser flare profiles which can be studied with the EVN (Szymczak et al. 2016, 2018a; Olech et al. 2018).

In all 6.7 GHz observations the software XF-kind correlator (SFXC) at JIVE allowed easily to achieve a sufficient spectral resolution of 0.9 km $s^{-1}$ (e.g. the 2 MHz bandwidth divided onto 1024 spectral channels). However, future studies with a higher spectral resolution may bring surprises like discovering amplification-bounded masers (Bartkiewicz et al. 2016). Recently, Takefuji et al. (2016) measured temporal coherent lengths of water maser emission in bright HMSFRs using the 34-m dish, indicating that coherent maser emission might exist. This could be confirmed by high spatial resolution (to isolate individual masers) and ultra-high spectral resolution to reveal the underlying (not Gaussian/broadened) line shape. Measuring coherent maser emission would reveal fundamental properties of the maser phenomenon itself.

### 5.2.2 Masers around Evolved Stars

Most circumstellar masers are associated with the Asymptotic Giant Branch (AGB) evolutionary phase, when low and intermediate mass stars climb the AGB, or with more massive stars in the Red Supergiant stage. In the process of transforming into PNe or supernovae (SN) such stars evolve within a few years and lose their final outer layers very rapidly. In the molecular material that is flowing away from variable, O-rich stars, OH, $H_2O$ and SiO masers occur. The various maser transitions sample different parts of the circumstellar envelope (CSE) tracing different physical processes (Gray et al. 2016). Aside from understanding this evolutionary stellar phase per se, studying the expulsion of matter from evolved stars is important since it is responsible for the bulk of dust and many CHNOPS-containing molecules in the ISM. Currently, it is suspected that stellar pulsations, large convective cells and magnetic fields are important, but no conclusive observational tests have yet been performed on a wide enough sample. Once enough dust forms, radiation pressure drives the wind away from the star – which continues accelerating for 100s $R_\star$ (the optical photospheric radius), beyond the ability of current models to explain. High-resolution imaging shows that much of the wind is concentrated in dense clumps. Distinguishing their properties (density, temperature etc.) from the surroundings is vital to deduce the chemistry, composition and total mass loss, and the prospects for survival of dust and molecules into the ISM. Also in this topic, masers are an ideal tool for these studies as the bright, compact emission allows positions and line widths to be measured with an order of magnitude more precision than thermal lines.

### EVN research of masers in evolved stars

SiO masers (43, 86 and 129 GHz) arise at 2–5 $R_\star$, where the stellar atmosphere becomes optically thin in the IR and internal pulsations give rise to shocks, and are thus used to study the impact of stellar pulsations. Proper motions of SiO masers reveal complex motions including outflow and infall. Large samples of SiO proper motions are required to disentangle various velocity fields from the net radial expansion, ballistic ejection and correlate these with polarisation measures and dust formation. Furthermore, the comparison of the absolute positions of the emission distribution of different SiO maser lines, may reveal whether such masers are radiatively or collisionally pumped and constrain the IR radiation field (Fig. 5.11, left panel).

Water masers at 22 GHz (collisionally pumped) trace rapid changing conditions in evolved star CSEs from about 5–50 $R_\star$. They occur in clouds of typical radius $R_\star$, which, if traced back to the star assuming radial expansion, implies a birth radius of 0.05–0.1 $R_\star$. If individual (sub-mas) maser components can be resolved without losing more extended flux, the nature of maser beaming can be investigated and clouds can be distinguished as either quiescent or a shocked slab. The latter are rarer and variability often affects the entire maser shell, on a timescale too fast for shocks to propagate, implying radiative heating effects. Changes between tangential and radial beaming are due to fluctuations in the velocity field, in turn a product of the efficiency of conversion of radiative energy into acceleration, also demonstrated by velocity drifts of maser features. More detailed water maser beaming and proper motion studies are needed to investigate these mechanisms.

OH 1612-MHz masers form shells, undergoing uniform expansion, at hundreds $R_\star$. In many cases their intensity has a straightforward relationship with the stellar period allowing the phase-lag method to be used for distance determination (Etoka et al. 2018). The precise determination of distances with this method relies on an accurate estimate of the size and geometry of the OH 1612 MHz shell, which is often aspherical. Appropriate resolution and high-sensitivity imaging are required in order to constrain these two parameters. While intermediate/short baselines are important to image, in particular, the faint external emission, proved to be extended towards OH/IR stars, the zooming power of the EVN is crucial to image and study the intrinsically compact regions associated with e.g. flaring events. Surprisingly, the OH mainline (1665/7 MHz) masers often overlap the outer 22 GHz $H_2O$ maser shell (e.g. Etoka et al. 2017), although the former require much cooler, less dense conditions, probably emanating from inter-clump gas. The magnetic fields in the CSE are found to be important, as can be seen from the highly ordered linear polarisation vectors observed in the OH maser components of $o$ Ceti (Fig. 5.11, right panel).

The expanding frequency coverage of VLBI, alongside ALMA, provides observations of multiple maser transitions in CSEs. Computational advances have led to recent maser models which can infer physical conditions such as temperature and number density, based on which lines co-propagate or are segregated. Maser polarisation and Zeeman splitting show that there is a stellar-centred magnetic field. This could be poloidal, Solar-type or toroidal (field strength $\propto r^\alpha$, where $\alpha$ is 3, 2 or 1 respectively), and could be well-constrained by measuring the field strength using maser species sampling a wide range of distances from the star. Thus far, only a few measurements are available, mostly from different objects or without accurate distances, favouring a lower value of $\alpha$ (Vlemmings et al. 2005, Leal-Ferreira et al. 2013).

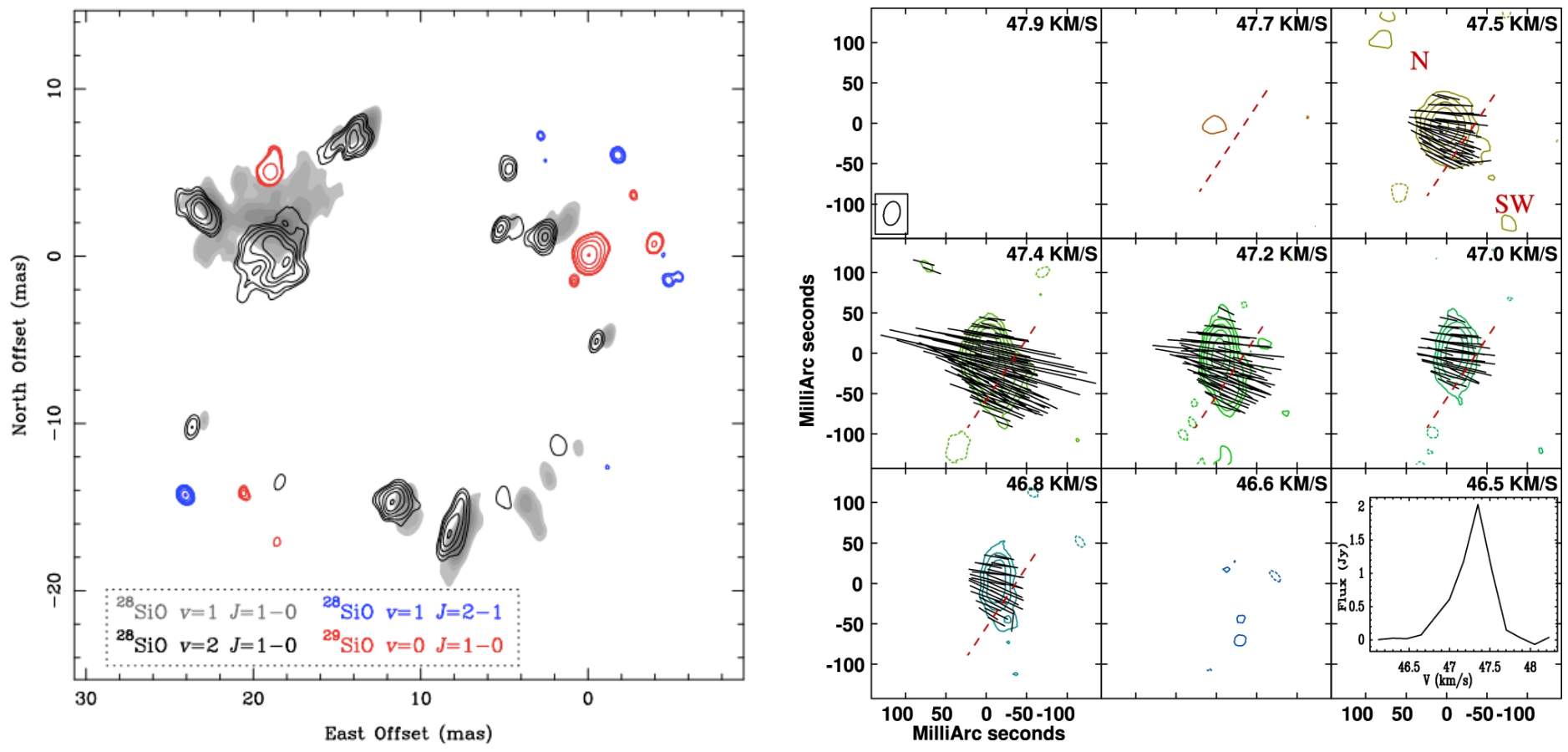

Figure 5.11: (left) Comparison of several maser lines of SiO in the AGB star IRC +10011. Adapted from Soria-Ruiz et al. (2004), reproduced with permission ©ESO. (right) EVN-*e*-MERLIN Stokes I contour maps of the 1665 MHz flaring emission towards o Ceti with the relative polarimetric information overlaid (Fig. 11, Etoka et al. 2017).

### 5.2.3 Maser Astrometry

The Milky Way offers unique possibilities to study structure formation in the Universe. Despite the location of our Solar System in the Galactic plane, we can infer how the Galaxy evolved and possibly was subject to minor and major mergers. In principle, this can be done by measuring the dynamics of stellar populations and characterising their age and metallicity. Astrometric campaigns in the radio, using VLBI, can address these topics, and are complementary to the impressive output data coming from ESA's optical astrometry satellite *Gaia*. Besides the structure of the Galaxy, direct distance measurements are also of fundamental importance to derive fundamental properties of astrophysical objects. Mass, luminosity, accretion and mass-loss rates, inferred pressures and energy balances all depend critically on accurate distances.

In the plane of the Galaxy, precise distances and proper motion measurements are very difficult to obtain, and *Gaia* distances are limited to a few kpc due to interstellar extinction. Thus, optical astrometry will not have the capability to measure the spiral structure (or infer the Hubble type) of the Milky Way, and neither will it resolve the kinematics of the inner Galaxy, including the Galactic bar and bulge. These structures are of particular interest to understand the nature and history of the Milky Way. VLBI astrometry is therefore fundamental for measuring the large-scale parameters of the Galaxy, such as the distance to the Galactic centre and rotation curve (Reid & Honma 2014). For the latter, measuring stellar motions and distances in the outer Galaxy is also important.

The distance and proper motions of HMSFRs, which are deeply embedded in dense molecular clouds, can be measured through water, methanol and SiO maser astrometry. Such campaigns map out the spiral structure of the Galaxy (Fig. 5.13, left panel) and can also provide a census of the current high-mass star formation across the Galaxy. The tremendous reach of VLBI astrometry includes masers in the far side of the Galaxy, with distances up to 20 kpc (Sanna et al. 2017). Water and in particular 6.7 GHz methanol masers are closely associated with those sites where high-mass stars are born (e.g. Menten et al. 1991, Breen et al. 2013), and these sites are located along the Galactic spiral arms, maser observations also yield the dynamic scale (size and rotation) of the Galaxy. It should be noted that low mass star formation does not produce bright enough (stable) masers for astrometry, though VLBI astrometry can be performed by observing the non-thermal emission from pre-main sequence stars (see more in Section 5.1.1).

Many AGB stars also contain maser emission and could be used for VLBI astrometry (see Fig. 5.12). Unlike HMSFRs that are still confined to the natal molecular clouds, evolved stars move through the Galaxy on dynamically relaxed, collision free orbits that may still carry information on their birth events, possibly Giga years ago. The least obscured AGB stars, the Mira variables and carbon stars, are observable by *Gaia* too, but *Gaia*'s astrometry cannot be expected to provide much information about the enigmatic dynamics of the inner Galaxy.

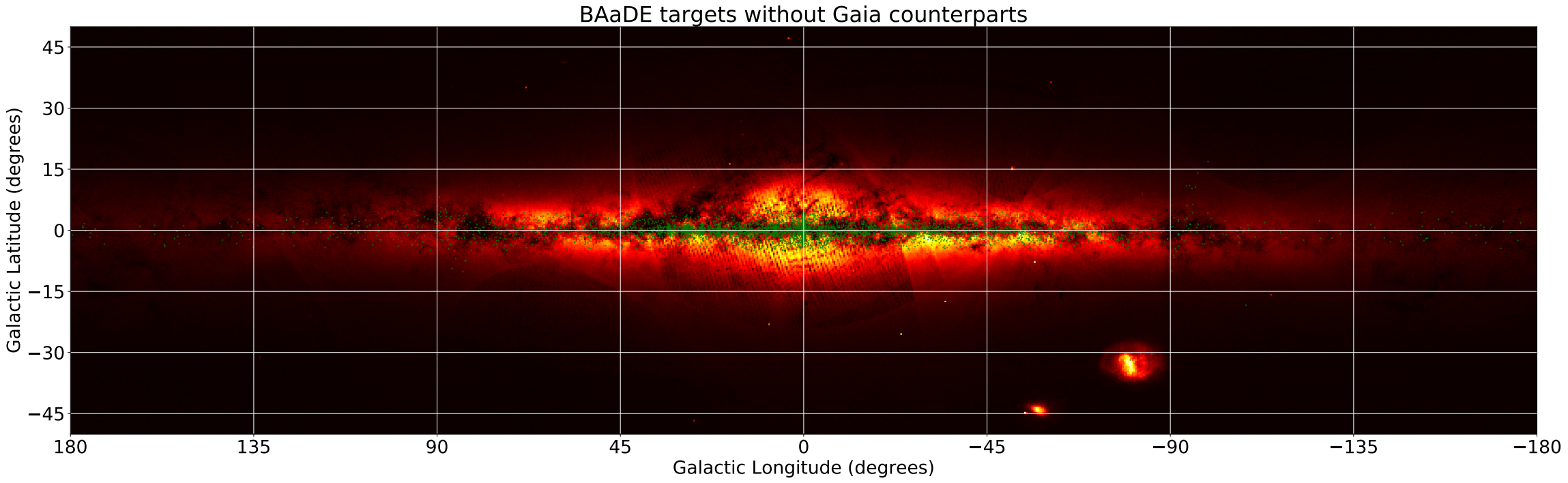

Figure 5.12: Plot of *Gaia* detections towards the inner Galaxy. The green symbols are BAaDE, SiO maser star targets for which no *Gaia* counterpart can be identified. Adapted from Quiroga-Nuñez et al. (2018), reproduced with permission ©ESO.

### EVN research of maser astrometry

The scientific impact of maser astrometry has been very substantial; papers that summarise the Galactic structure are widely recognised across many astronomical fields (Reid & Honma 2014). In addition, results that give robust distances for otherwise obscured targets, often have very high citations. For example, many recent ALMA papers quote distances for evolved stars and star-forming regions based on VLBI. For such embedded objects, it can be expected that *Gaia* will only improve the distances for a very limited number of them, and recent comparisons suggest that *Gaia* and VLBI results are consistent (*Gaia* Collaboration, 2016) and complementary .

The highest impact programme on maser astrometry so far have been the BeSSel and VERA projects (Brunthaler et al. 2011, Reid et al. 2014, 2019, Kobayashi et al. 2003). They started by targeting the brightest methanol (12 GHz) and water masers (22 GHz) with the VLBA (BeSSel) and water (22 GHz) and SiO (43 GHz) masers with the Japanese VLBI array (VERA). It was complemented by some 6.7 GHz masers observed with the EVN (Rygl et al. 2010, 2012). With the new C-band receivers on the VLBA the BeSSel sample has been doubled by including observations of ubiquitous 6.7 GHz masers. We must note, some of this work could in principle have been done equally well with the EVN, which is maybe harder to calibrate for astrometry but possibly more sensitive. But, it requires a large commitment in observing time outside the normal EVN sessions.

The astrometry of evolved stars started with results on OH masers (Vlemmings & van Langevelde 2007). But, at 1.6 GHz (18 cm) the distance range accessible is restricted, because of the limited (intrinsic) brightness of the masers. However, higher accuracy measurements will become possible when multiple close calibrators can be observed simultaneously with high signal-to-noise ratio (Rioja et al. 2017). The most relevant results are however still obtained through water masers, most of these observed with VERA and the VLBA. SiO masers may be more relevant in the future. Recently, the VLA and ALMA were used to establish a very large sample ($> 10,000$) of SiO masers (the BAaDE project, Sjouwerman et al. 2017). In principle, VLBI could uniquely measure the stellar motions in the inner Galaxy through these sources, although the requirements for calibration and monitoring are very severe; switching times would be under a minute and source variability is expected to occur on time scales of a month. Absolute positions of SiO and water masers in AGB stars and proto PNe can be used to study various structures, such as bipolar jets or the stellar envelope (see e.g. Desmurs et al 2007), to understand the evolutionary scenario of these stars. Also, a lack of water in the equatorial regions of the inner CSE may be explained by the presence of a companion star.

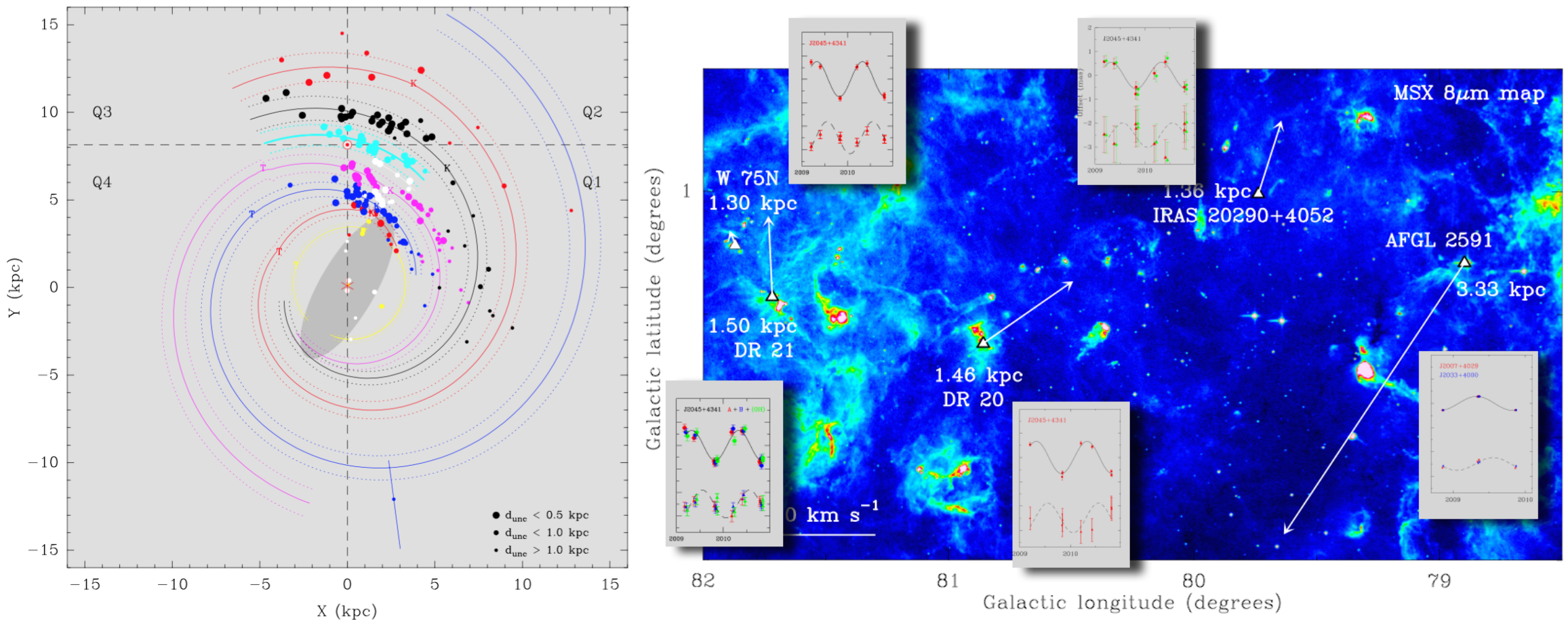

Figure 5.13: (left) The location of high-mass star forming regions in the Galaxy outlines a spiral arm pattern (Reid et al. 2019). ©AAS. Reproduced with permission. (right) Distances to star-forming regions in Cygnus X complex: the insets show the fitted parallax sinoids in R.A. and declination of the star-forming regions, while the white arrows (on the *MSX* map) indicate the 3D space motions of the star-forming regions (indicated by white triangles). Adapted from Rygl et al. (2012), reproduced with permission ©ESO.

The EVN made a considerable impact by providing astrometry of methanol masers at 6.7 GHz (Rygl et al. 2010, 2012, see Fig. 5.13, right panel), when it was still the only VLBI array with many receivers covering the 5 cm transition. Various other projects measured the kinematics of methanol masers, by monitoring experiments not optimised for parallax as described in Section 5.2.1. Here the EVN excels by making many weak features visible and tracing them over a few years.

### 5.2.4 Technical requirements and synergies

The three most-pressing limitations of the EVN for stellar studies nowadays are:

(i) The array's sensitivities, especially for continuum and polarisation studies. Quiescent, disc-averaged stellar radio emission is very weak. Most of the projects described above require continuum sensitivities better than a few $\mu$Jy/beam, and they are natural targets for connected interferometers as the JVLA or *e*-MERLIN. Present EVN sensitivity may reach 5-10 $\mu$Jy ($1\,\sigma$) for reasonable integration times at cm-wavelengths and assuming the object is not overresolved. This leaves little room for the EVN in terms of sensitivity to thermal stellar emission, restricted to the brightest objects. Nevertheless, VLBI appears as the only technique able to provide the submilliarcsecond resolution necessary to map the fine details of coronal or collinding-wind structures combined with the submilliarcsecond precise astrometry needed for measuring the distance to star forming regions or the reflex motion induced by substellar objects. Sensitive observations with good spectral resolution are necessary for maser observations especially for polarisation studies. To detect at $5\sigma$ typical low circular polarisation fraction (>0.3%) for maser emission with flux $>1$ Jy beam$^{-1}$ it is necessary to reach an rms level of 0.6 mJy beam$^{-1}$ per channel that currently requires 167 hr and 417 hr of observing time at C- and K-band, respectively (see Surcis et al. 2014, 2019).

(ii) The poor *uv*-coverage in the north–south direction; this is the same orientation as the Galactic plane where many HMSFRs and AGB stars are. Extending the EVN baselines to AVN telescopes will be a giant step in terms of building the *uv*-coverage, providing access to a complete census of star forming regions and stellar moving groups, and opening the inner Galaxy up for unparalleled VLBI exploration.

(iii) The lack of short baselines. EVN technical improvements should be addressed to provide $\mu$Jy/beam sensitivity in short-spaced baselines not to overresolve the stellar objects. Compatibility with *e*-MERLIN is a sensible recommendation which would add baselines of $<100$ km, allowing significant contributions in every single stellar project; e.g. the EVN plus shorter spacings will be invaluable in resolving non-thermal flares or possibly chromospheric hot spots with sufficient detail to monitor their origins and role in binary evolution and mass loss.

#### Further technical considerations

The EVN's limitations for some projects like astrometry stem mostly from the sparse telescope availability outside of the EVN sessions that is required to sample the parallax sinusoid properly to distinguish the parallax imprint from the proper motion, the high slewing times that make phase-reference observations very slow, and the currently scarce high frequency coverage. For these reasons most astrometric results today are coming from the US VLBA or the Japanese VERA, which have antennas able to perform rapid phase referencing and operate all year. The EVN, providing an unparalleled sensitivity at the lower frequencies thanks to 70–105 m dishes, is frequently used for astrometric projects that do not require special scheduling constraints. Improvement on the astrometric precision (down to a few $\mu$as) may come more from the use of techniques as the Multiview phase-referencing (Rioja et al. 2017), which achieves a more effective treatment of (atmospheric) systematic errors, relaxing the constraints of the angular separation of the reference source(s). On the longer term, EVN antennas equipped with phase array feeds will allow multiview-like observations for virtually any target-calibrator separation.

The BRAND broad band receiver should cover the range from 1 to 15 GHz to permit simultaneous observations of methanol (6.7 and 12 GHz) in full polarisation mode allowing for the derivation of physical conditions such as temperatures and densities in dense regions of high-mass star-formation sites. However, one must remember, the line sensitivity will be worse. Ideally, the receiver coverage could include also the 22 GHz water masers and 1.6665/7 GHz OH masers as these masers are often coexistent in HMSFRs and AGB stars and provide information on various structures with different kinematic properties. Simultaneous imaging of multi-maser transitions for a number epochs during a flare will permit to accurately track the flare evolution at various spatial positions without any alignment complications (in time and space) and derive physical conditions. Also, for AGB stars, where water and OH are spatially in close proximity, here proper motions might reveal how velocity gradients, number densities and temperature contrasts constrain pumping. The broad band receiver will improve significantly continuum measurements, which perhaps might not be enough for detecting the stellar continuum in AGB stars, but it would enable the obtention of rigorous spectral information both for coherent flaring and incoherent gyrosynchrotron or synchrotron radiation. Additionally, the astrometric precision of the EVN would be improved via more precise atmospheric calibrations and the possibility of using weaker position reference sources with a smaller separation to the target. It should be underlined that still much is to be gained by increasing the spectral resolution (contrary to the wide band observations mentioned above ), in particular for maser polarimetry, which requires spectral resolution of 100 kHz to resolve the methanol hyperfine structure and improve Zeeman splitting measurements, and for detecting coherent maser emission, for which the spectral resolution has to be as fine as 1 kHz. Observations of coherent maser emission require a spectral resolution range of $<1$ Hz to many kHz to resolve the homogeneous width at all levels of saturation.

Most SiO maser VLBI studies have been carried out with the VLBA, the KVN and the VERA (e.g. Gonidakis et al. 2010, Oyama et al. 2016, Yun et al. 2016). But, the future installation of new K/Q/W receivers on several EVN antennas, especially the large dishes, will boost the SiO science. Simultaneous maser observations of water and several lines of SiO, will help understanding the mass loss mechanism at work in AGB shells. Playing a role in possible future SiO maser astrometry may require additional receiver efforts for those EVN telescopes that are able to reach millimetre wavelengths with capability of fast-switching. This proposed stellar astrometry is a different application from simultaneously observing $H_2O$ and SiO masers, it would aim to use more sensitive K band calibrator observations to measure the relative position of SiO masers at Q band. Such a technique for transferring phase corrections between bands was developed by Rioja & Dodson (2011). SiO astrometry is interesting for the CSE per se, but will be an almost unique method to measure accurate distances and proper motion of AGB stars in the Galactic bar and bulge (see section Astrometry). Also higher Class I methanol masers transitions (e.g. 25 GHz, 36 GHz) are of great scientific relevance for HMSFR studies, though they might be resolved on the longest (intercontinental) EVN baselines.

For a complete picture of the circumstellar envelopes, covering a wide range in size scales from tens of mas down to hundreds of mas, VLBI has to be combined with shorter baselines, and complemented with single dish and connected interferometers. *e*-MERLIN can provide the short baselines at 22 GHz, but for the higher SiO frequencies intermediate baselines, between 36 km (VLA) and a few hundred kilometres are a world-wide problem. Connected mm interferometers (ALMA/NOEMA) can observe several SiO and water maser lines together with the stellar continuum, and provide a unique method to spatially correlate the maser emission with the stellar photosphere. In general, there is the importance of combining interferometric continuum and molecular emission with maser VLBI measurements to obtain a complete picture for the (usually) dense regions where optical observations are impossible. In addition, much is expected from the future synergy with the SKA, if it will have VLBI capabilities, improving the sensitivity and *uv*-coverage towards the inner Galaxy. In particular, the SKA-MID which would include the 6.7 GHz methanol (band 5a) and the 12 GHz methanol transition (band 5b, possibly extending to the 22 GHz water maser line) could allow for new methanol astrometry of southern targets (see Loinard et al. 2015).

Classical synergies between single dish and VLBI keep being vital for the variability of masers, with different classes of variability like periodically variable and bursts. Therefore, the rapid VLBI response organised through trigger proposals is important.

Finally, intensive and complex data processing schemes to derive parallaxes from VLBI monitor campaigns as well as the bulk of the all EVN data reduction, rely mostly on classic AIPS and ParselTongue. It is hoped that such processing schemes are more easily implemented and executed in the future with the growing VLBI functionality of CASA for which VLBI reduction pipelines are already existing (Janssen et al. 2019).

### 5.2.5 Summary

The use of the VLBI technique is already extended throughout the complete H−R diagram, enabling studies of the radio emission of all sorts of stellar objects from protoplanetary disks to evolved stars, including main-sequence stars, brown dwarfs, and exoplanets. Maser science has provided fundamental insights in the workings of high-mass star formation by measuring the magnetic fields and gas dynamics down to size scales of 100 AU, the mass expulsion in evolved stars from a few $R_\star$ out to hundreds $R_\star$ together with magnetic field measurements, and has mapped out the spiral structure of our Galaxy and improved our knowledge of fundamental parameters such as the distance to the Galactic centre and the Galactic rotation curve. The EVN has played a role in many stellar projects, however, still it can be used more. These studies are rather time-consuming, often requiring more than one epoch with rather long on-source time, resulting in a limited number of stellar targets that are studied by VLBI. Often, in astrometry studies there is a need to observe outside the standard EVN sessions. Future EVN, and VLBI in general, observations with improved sensitivity through the inclusion of more (and larger) dishes, improved *uv*-coverage (in particular in the north-south axis via AVN and SKA), increased frequency coverage, increased receiver bandwidth allowing for simultaneous observations of several maser transitions will allow VLBI science to extend to fainter targets, improved astrometry, and a better Galactic plane coverage which is crucial for understanding Galactic dynamics. Thus, there is still much to be explored in the Milky Way using the EVN.

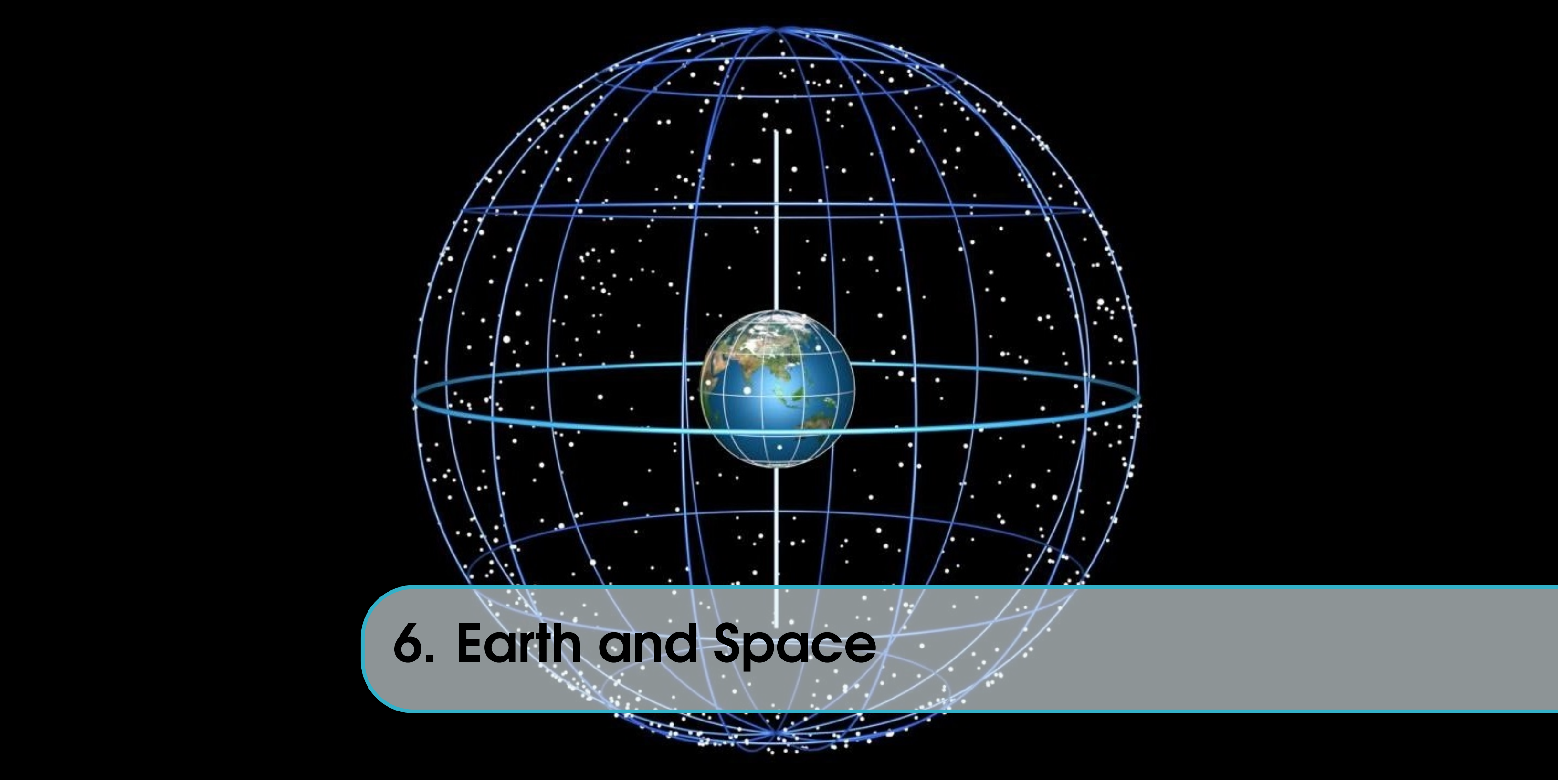

# 6. Earth and Space

## 6.1 Celestial reference frames

### 6.1.1 A unique capability for positioning

The determination of the motion of celestial bodies through repeated measurements of their position using astrometric techniques constitutes the foundation of our present understanding of the Universe. For consistency in time and space, and hence comparisons for different celestial bodies, such determinations must be established in well-defined coordinate (or reference) systems. In practice, coordinate axes for celestial systems are not defined directly but only implicitly through the adoption of a set of fiducial directions, precisely identified and highly stable over long timescales. Specific celestial bodies possessing the required properties are used to materialise such directions.

VLBI is a unique tool that allows astronomers to study the compact radio emission of celestial bodies in extreme details and to pinpoint their direction in the sky with unprecedented accuracy. Thanks to this capability, the technique has been used for the past twenty years to establish fundamental celestial reference frames. The objects targeted for this purpose are black-hole powered AGN. These possess highly compact central emission (of non-thermal synchrotron origin), ensuring well-defined fiducial directions. Additionally, they have no detected proper motions due to their location at cosmological distances. The number of objects measured in this way grew from a few hundreds in the 1990s to several thousands today. On the celestial sphere, they form a grid of points whose two-dimensional coordinates define a reference frame. Such a grid is also the basis for ultra-precise relative VLBI astrometry, e.g. to determine distances and transverse velocities of Galactic objects out to tens of kpc through measurements of proper motions and parallaxes.

The most comprehensive VLBI reference frame to date is the third realisation of the International Celestial Reference Frame, ICRF3 (Charlot et al. 2020), adopted by the IAU in August 2018 to replace the previous realisation, ICRF2 (Fey et al. 2015), built ten years earlier, as the new fundamental celestial reference frame. ICRF3 comprises positions for 4536 extragalactic sources, as measured at 8 GHz (Fig. 6.1), 303 of which, uniformly distributed on the sky, are identified as *defining sources* and as such serve to define the axes of the frame. Positions at 8 GHz are

Chapter image credit: E. Siegel, ScienceBlogs.

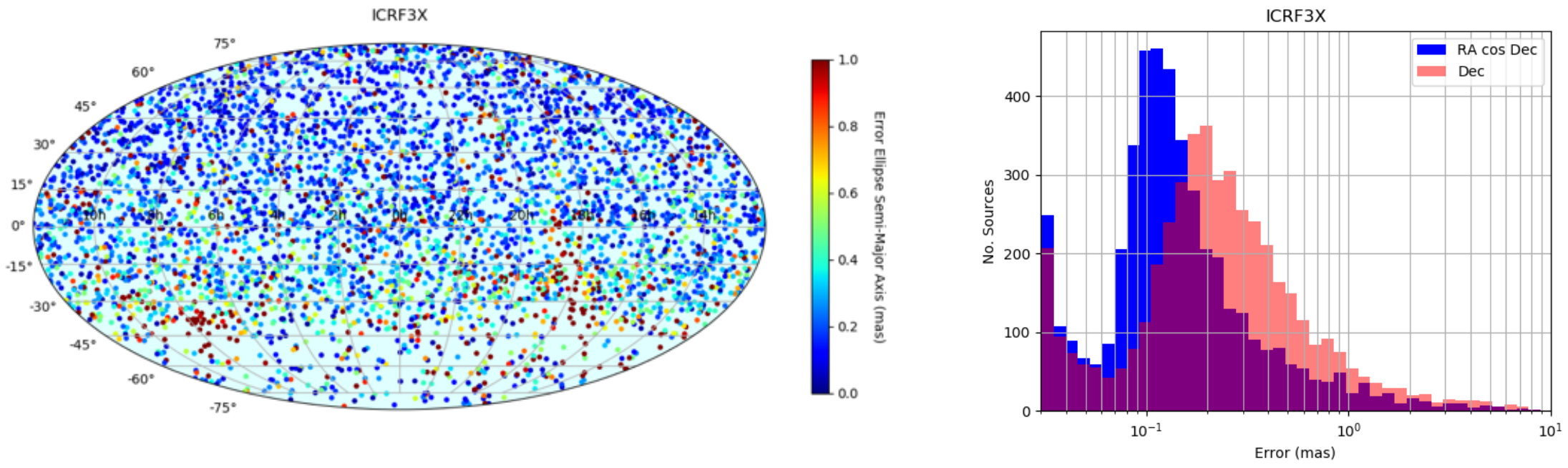

Figure 6.1: (left) Sky distribution of the 4536 extragalactic sources comprised in ICRF3 (8 GHz frequency) with colour coding according to position accuracy. (right) Histogram of errors in right ascension and declination for those sources plotted with a log-scale. Charlot et al. (2020).

supplemented with positions at 24 GHz for 824 sources and at 32 GHz for 678 sources. In all, ICRF3 includes a total of 4588 sources, 600 of which sources have three-frequency positions available. The positions have been estimated independently at each of the frequencies in order to preserve the underlying astrophysical content behind such positions. The frame shows median positional errors of the order of 100 $\mu$as in right ascension and 200 $\mu$as in declination with a noise floor of 30 $\mu$as in the individual coordinates. Among these, a subset of 500 sources is found to have extremely accurate positions, in the range of 30–60 $\mu$as (Fig. 6.1).

### 6.1.2 Fundamental physics and astronomy

The newly-released ICRF3, with its increased positional accuracy and its observing span approaching 40 years (1979–2018), will allow the scientific community to tackle new questions in astronomy and fundamental physics. The long time base now makes mandatory the modelling of Galactocentric acceleration, a secular effect introduced by the rotation of the Solar System barycentre around the Galactic centre. This effect, first detected by Titov et al. (2011), has emerged from the ICRF3 data set with a magnitude of $5.8 \pm 0.3$ $\mu$as/yr and manifests itself through apparent long-term proper motions of the radio sources if not accounted for in the modelling. Continued observing and accumulation of data in the future will further improve that value and determine whether the motion of the Solar System is purely towards the Galactic centre or whether it has an off-plane component. In a similar way, the low-frequency ($< 10^{-9}$ Hz) gravitational wave background, although not detected at present, may be revealed in the future through such quasar proper motions (Gwinn et al. 1997, Titov et al. 2011, Darling et al. 2018). VLBI is also essential for testing General Relativity, e.g. through the determination of the relativistic parameter $\gamma$ (Lambert and Le Poncin-Lafitte 2011) or for trying alternate theories (Le Poncin-Lafitte et al. 2016). Here also, accumulation of data and extending the time base will further improve the level of such tests.

### 6.1.3 Astrophysics of active galactic nuclei

The multi-frequency positional information in ICRF3 together with the optical positions recently derived with the *Gaia* space mission which show similar accuracies (Gaia collaboration et al. 2018) provide new insights into the physics of AGN. At radio frequencies, these objects generally feature a bright compact core and a single-sided relativistic jet with blobs of emission moving away from the core on time scales of months to years (Fig. 6.2). Future reference frames will have to account for such time-varying extended internal structures for the highest accuracy, a perspective that motivates systematic VLBI imaging programmes to monitor the structure of the ICRF sources. In the light of such extended emission, comparison of the ICRF3 and *Gaia* positions becomes essential to understand whether the radio emission and optical emission are superimposed in these objects. Initial estimates of such radio-optical "core shifts" indicated that they amount to 100 $\mu$as on average (Kovalev et al. 2008), which is significant considering the VLBI and *Gaia* position accuracies. While potentially affecting the alignment between the two frames, the radio-optical positional differences also offer a unique opportunity to directly determine those core-shifts and probe the geometry of quasars in the framework of unified AGN theories. In particular, such measurements may help to locate the optical region relative to the relativistic radio jet and determine whether the dominant optical emission originates from the accretion disk or the inner portion of the jet. Taking advantage of the first and second *Gaia* data releases, it was found that significant VLBI-*Gaia* offsets do exist for 10–20% of the sources (Mignard et al. 2016, Petrov & Kovalev, 2017a, Gaia collaboration et al. 2018, Petrov et al. 2019, Charlot et al. 2020) and that these occur preferably along the jet direction, which was interpreted as a manifestation of the presence of bright optical jets (Kovalev et al. 2017, Petrov & Kovalev 2017b, Plavin et al. 2019). Future *Gaia* data releases may reveal offsets for an increased fraction of objects thanks to further improved positional accuracy, in synergy with VLBI measurements.

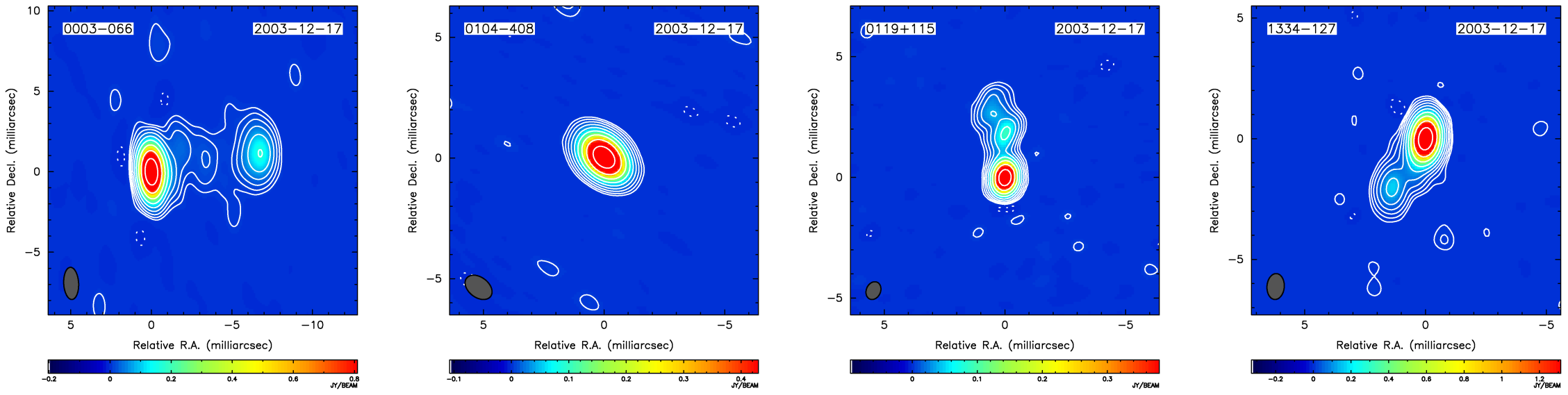

Figure 6.2: A sample of VLBI maps from the Bordeaux VLBI Image Database (Collioud and Charlot 2019) showing the predominantly core-jet structure of the ICRF sources on milliarcsecond scales.

### 6.1.4 Rotational motion and dynamics of the Earth

Another unique capability of VLBI is its ability to track the rotational motion of the Earth in the quasi-inertial frame defined by the distant quasars. This motion includes a secular drift and periodic oscillations of the Earth's rotation axis (i.e. precession and nutation, see Fig. 6.3) along with a daily rotation around it. The latter, which may be expressed as the length of day, is irregular and unpredictable at some level since it is closely tied to the atmospheric conditions, thus requiring continuous VLBI observations to be followed. The nutational motion depends on the geophysical properties of the Earth and allows one to learn about the Earth's interior (Mathews et al. 2002, Rosat et al. 2017). Key challenges in this area are the detection of the Earth's solid inner core, independently of seismic data, and the understanding of the origin and variability of the free core nutation. The latter also requires progress in global circulation models and the theory of the Earth's rotation (Ziegler et al. 2019). These challenges necessitate VLBI monitoring over long time scales and with high accuracy. In the future, this should be accomplished with the VLBI Global Observing System, a new array consisting of fast-moving 12 m antennas that is currently set up by the International VLBI Service for Geodesy and Astrometry and designed to observe 24 hours a day all year long. On the geodesy side, the goal of such an array, permanently observing, is to reach millimeter positional accuracy. Achieving a terrestrial reference frame at that level is essential to understand deformations of the Earth crust (e.g. seasonal signals, seismic and post-seismic effects) and more generally to monitor global changes that affect our planet, among which sea level rise.

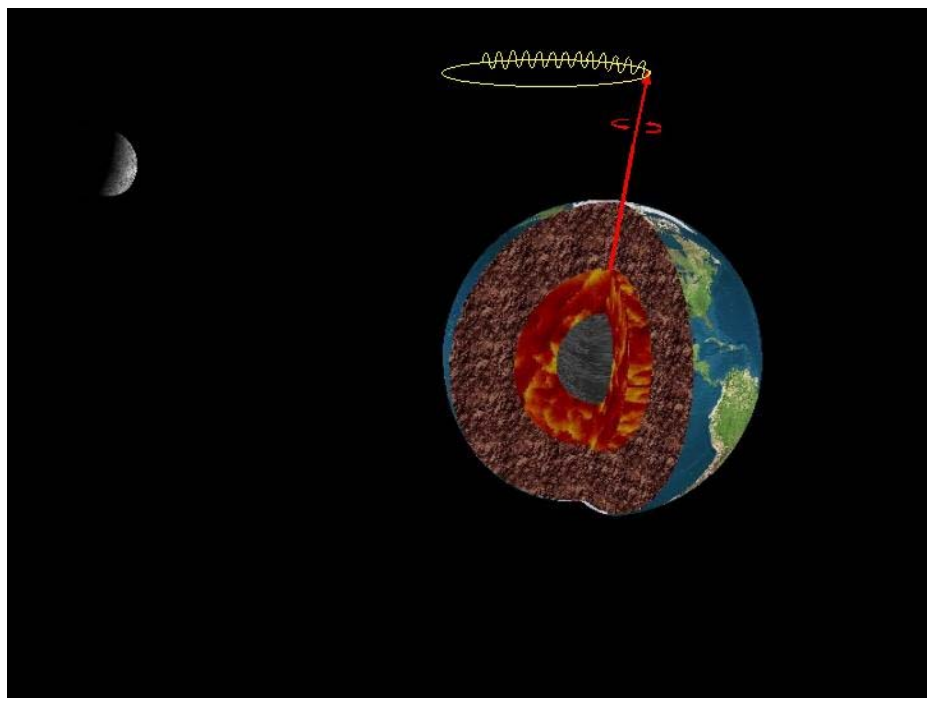

Figure 6.3: Schematic representation of the Earth's precession and nutation motion with the different components of its interior structure depicted (mantle, liquid outer core, solid inner core). Credit: IVS.

### 6.1.5 Contribution of EVN

All scientific topics addressed above should benefit from future improvements of the VLBI celestial reference frame. As shown from Fig. 6.1, there are two obvious areas of improvements: (i) position accuracy in the southern sky, which is typically degraded by a factor of 2 compared to that in the northern sky since the VLBI networks used to build ICRF3 were predominantly East-West, and (ii) lower sky density in the far South (i.e. below $-45^\circ$ declination) due to the sparseness of VLBI radio telescopes in the southern hemisphere. In respect of these improvements, the EVN has a significant role to play. By providing long North-South baselines, from Europe to the Hartebeesthoek antenna in South Africa and in the future to antennas of the developing African VLBI Network (Gaylard et al. 2011), it can help to break the North-South asymmetry in position accuracy. Incorporating SKA1-MID as an element of the array will also largely improve its sensitivity, hence permitting to expand considerably the celestial frame, with the goal of obtaining a more complete radio counterpart of the *Gaia* celestial frame. With its location in the southern hemisphere, SKA1-MID will further help to correct the currently uneven sky distribution, as noted above. Also to be mentioned is the capability of the EVN to observe at higher frequencies, especially at 22 GHz, which is one of the three ICRF3 frequencies, and hence its potential to contribute to the development of the celestial frame at this frequency. Incorporation of observations on long North-South baselines to Hartebeesthoek, here again, would be especially useful to enhance the geometry of the frame.

## 6.2 Near-field VLBI for space and planetary science

### 6.2.1 Spacecraft as a VLBI target: science applications

As outlined above (and also earlier in this document), astrometric and geodetic applications of the VLBI technique are well-established scientific disciplines. They are based on the fundamental assumption of the 'traditional' radio interferometry: the source of the emission is located at an infinitely large distance from the observer, thus the light paths from the source to the elements of the interferometer can be assumed parallel, and the wave front – planar. Such the assumption can be accepted if the distance to the source is (much) greater than the so-called Fraunhofer distance, $B^2/\lambda$, where $B$ is the projected baseline and $\lambda$ the observing wavelength. In astrometry, an ideal source of emission is point-like. Spacecraft, as radio astronomy targets, are indeed point-like sources. In principle, they might be treated as targets of VLBI astrometry and therefore their celestial coordinates could be estimated with the VLBI precision, as is the case for galactic and extragalactic radio sources. However, for typical deep space communication frequencies (e.g. 2.3, 8.4 or 32 GHz) practically the entire Solar system is within the Fraunhofer distance for global VLBI baselines. Thus, implementation of VLBI astrometry for spacecraft requires development of a special technique, the so-called near-field VLBI.

VLBI observations (sometimes called 'VLBI tracking') of spacecraft have been suggested in the early 1970s (e.g. Ondrasik & Rourke 1971) and demonstrated on a number of deep space missions, such as *Apollo 16* and *Apollo 17* (Salzberg 1973), *Pioneer Venus probes* (Councelmann III et al. 1979), *Voyager* (Border et al. 1982), *VEGA* (Preston et al. 1986), *Huygens Titan Probe* (Pogrebenko et al. 2004, Lebreton et al. 2005), *Cassini* (Jones et al. 2011), *SELENE* (Goossens et al. 2011), *IKAROS* (Takeuchi et al. 2011), *Chang'E* (Li 2008 and references therein), *Venus Express* (Duev et al. 2012), *Mars Express* (Duev et al. 2016) and others. In combination with other trajectory measurements techniques, e.g. range and range-rate (Doppler) tracking, VLBI tracking enables estimates of the spacecraft state vector as a function of time – the main task of the orbit determination (OD).

Nominal OD assets of major space agencies, such as the Deep Space Network (DSN) of NASA and European Space Tracking (ESTRACK) of ESA, exploit the Delta-Differential One-Way Ranging, or DeltaDOR technique – an offspring of VLBI (Maddè et al. 2006, Curkendall & Border 2013, and references therein). The latter involves a single baseline configuration of two widely separated (order of thousand kilometres) tracking stations and enables a nanoradian (0.2 mas) precision of lateral positioning of spacecraft. While being the most accurate operational technique for lateral measurements of the spacecraft celestial position, DeltaDOR does not provide as much versatility and robustness as multi-element near-field VLBI tracking. The advantages of the latter technique are in a better instrumental calibratability, redundancy of baseline solutions, ability to exploit larger (more sensitive) radio astronomy antennas, ability to choose the best baselines in terms of local ionosphere turbulence and best target visibility conditions (e.g. a larger elevation at interferometer elements). Besides, flexibility of observing setup, data handling and processing algorithms in VLBI tracking provide for a broad variety of science applications.

Over the period 2005–2018 the methodology of addressing various scientific interests with near-field VLBI observations has grown into the concept of Planetary Radio Interferometry and Doppler Experiment (PRIDE, Gurvits et al. 2013; see Fig. 6.4). The technique of PRIDE has been described by Duev et al. (2012) and Bocanegra-Bahamon et al. (2018); its value for contributing into orbit determination of interplanetary spacecraft and planetary probes is addressed in Dirkx et al. (2016, 2017, 2018).

In addition to the immediate needs of OD, which is required for mission operations (e.g. navigation), near-field VLBI offers a wide range of science applications demonstrated recently in many experiments conducted by EVN and global VLBI networks. These include:

- Ultra-precise tracking of spacecraft during fly-by of a gravitating body (in application to the Mars Express Phobos flyby in December 2013), Duev et al. 2016. Fly-by observations contribute into estimating parameters of the gravitational field and interior of the gravitating body (e.g. internal structure of planetary bodies).

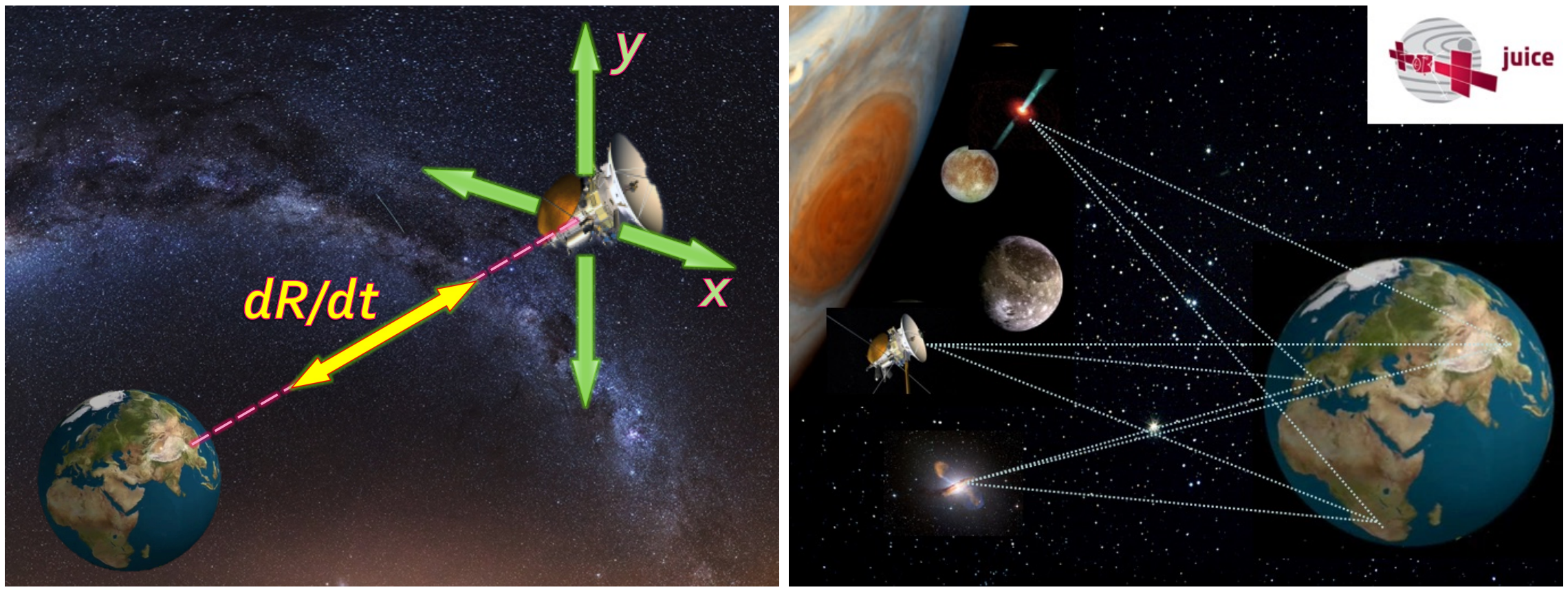

Figure 6.4: (left) Generic configuration of the Planetary Radio Interferometry and Doppler Experiment (PRIDE). Its main measurables are the radial Doppler shift of the spacecraft radio signal (dR/dt) and lateral celestial coordinates x and y. (right) PRIDE configuration for the ESA's Jupiter Icy Satellites Explorer (*JUICE*) mission. From Gurvits et al. (2013).

- Contribution into ultra-precise 'non-standard' OD for scientific missions, e.g. *RadioAstron*, in support of its space-ground VLBI observations (Duev et al. 2015).
- Improvement of ephemerides of planetary satellites (Dirkx et al. 2016, 2017).
- Interplanetary plasma diagnostics (Molera Calvés G. et al. 2014), including serendipitous detection of coronal mass ejections (Molera Calvés G. et al. 2016).
- Deep sounding of (dense) planetary atmospheres by radio occultation measurements (e.g. *Venus Express* occultations, Bocanegra-Bahamon T.M. et al.).
- Contribution into fundamental physics experiments by supplementary Doppler measurements and PRIDE-based data processing (Litvinov et al. 2018).

### 6.2.2 Near-field VLBI in the EVN context

The EVN as a network began operating in the near-field VLBI mode in 2003 in preparation for the Huygens VLBI tracking experiment (Pogrebenko et al. 2004, Witasse et al. 2006). As part of this preparation, the first incarnation of the SFXC correlator has been developed at JIVE, which later has become and remains to date the main operational EVN data processor (Keimpema et al. 2015).

Based on the know-how developed at JIVE with its EVN partners, an implementation of PRIDE for a Jovian mission has been selected by ESA in 2012 as one of eleven experiments of the Jupiter Icy Satellites Explorer, the *JUICE* mission (Hussmann et al. 2014; Fig. 6.4). The mission is scheduled for launch in 2022 and arrival to the Jovian system in the end of 2029 followed by three-year-long science operations in the vicinity of Jupiter and its satellites. PRIDE-*JUICE* will address a number of science topics listed in the previous section.

Several other planetary missions, such as *ExoMars-2020* (Mars), *Chang'E-4* (Moon), *EnVision* (Venus), *Europa Clipper* (Jupiter and Jovian system) are slated to exploit near-field VLBI and PRIDE techniques in order to maximise their science return. PRIDE involvement in planetary science missions addresses most topical issues in the area of origins and evolution of Solar and other planetary systems. As such, the outcome of near-field VLBI tracking addresses the topics of the highest scientific priority in the strategic outlooks of all major world space agencies.

It is important to underline that VLBI tracking experiments with these and other planetary and space science missions require close cooperation with the respective project teams of scientists and engineers for longer than usual in "traditional" VLBI timespans. Typically, a lifetime of a modern planetary science mission is $\sim 20$ years, sometimes longer. This requires a well-balanced foresight of relevant developments of VLBI instrumentation and involved person-power.

EVN is well positioned to address the needs of PRIDE observations in the next one or two decades. Its geographical distribution, frequency coverage and superb sensitivity plus the already demonstrated flexibility of the EVN data processing facilities at JIVE provide a solid basis for multi-disciplinary research in the area of space and planetary science applications.

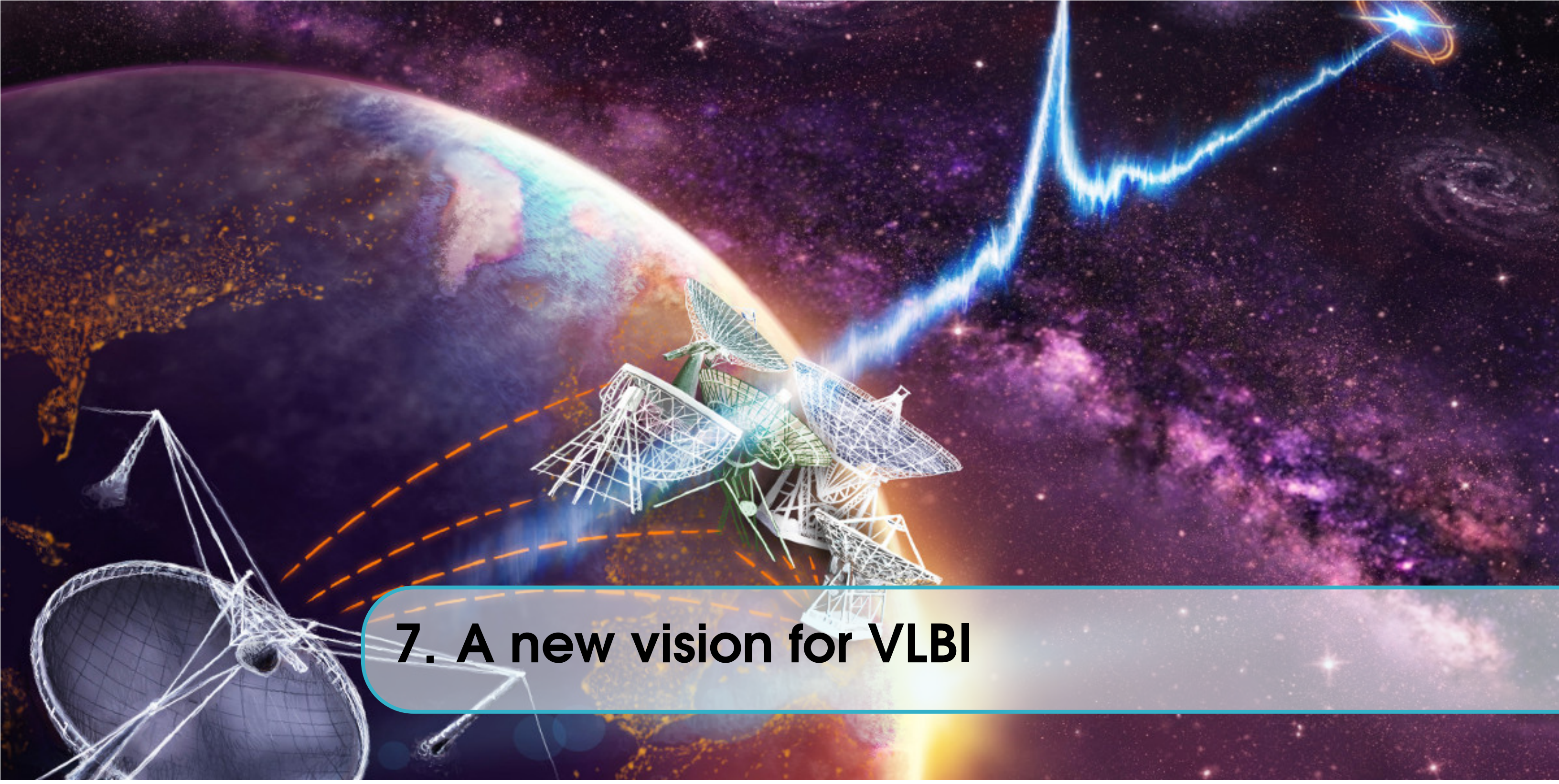

# 7. A new vision for VLBI

The science cases described in previous chapters put forward a range of prerequisites for the EVN to continue its role as the most sensitive cm-wave VLBI array in the world, in all different astrophysical areas. This chapter makes recommendations in terms of needed technical developments to telescopes and correlators, array operations and data archiving, for the EVN to best realise this scientific potential. Since much of the impact of the EVN will be made in tandem with other existing or planned astronomical instruments, we first in Section 7.1 survey the potential synergies between the EVN and those instruments and how those drive future technical, operational and data requirements of the EVN. In Section 7.2 we combine these requirements with the science cases described, to make specific recommendations for EVN development. These recommendations are divided in each case as on-going efforts, to enhance the EVN performance step by step, and long term development goals to be completed by the end of this decade and beyond. Section 7.3 makes some concluding remarks. An overview of the current world radio facilities most relevant to this document, as well as the current technological framework for VLBI in Europe and realistic prospects for its development are reported in Sect. A.1 and A.2 respectively.

## 7.1 The European VLBI Network within the future Astronomical Landscape

### 7.1.1 Synergy with global cm/mm VLBI – Global VLBI Alliance

VLBI is one of the most successful examples of international scientific and technological cooperation. Several independent VLBI networks are operational nowadays (see Sect. A.1): the EVN, VLBA, EAVN, LBA and GMVA offer observing time in the 1-100 GHz frequency range; the International LOFAR Telescope operates at a few hundred MHz; EHT has successfully combined high-frequency telescopes to observe at 230 GHz; the International VLBI Service (IVS) for Geodesy and Astrometry operates in the frequency range 2–14 GHz. All in all, VLBI allows milliarcsecond to $\mu$arcsecond observations in a frequency range which spans 3 orders of magnitude in frequency, from a few

---

Chapter image credit: The globally distributed dishes of the European VLBI Network are linked with each other and the 305-m William E. Gordon Telescope at the Arecibo Observatory in Puerto Rico. Together they have localised FRB 121102's exact position within it's host galaxy. Artwork: Danielle Futselaar.

hundred MHz to a few hundred GHz; finally SKA will include VLBI modes with outer antennas in phase 1, while long (thousands of km) baselines will be part of the phase 2 development.

Such networks usually operate independently of each other, partly because of the different observing frequencies, and partly because of the different simultaneous sky coverage. At the same time, if we look at the VLBI facilities throughout Earth as a whole, the contribution of VLBI at any given time is amazing. The efficiency and impact of VLBI would further increase with a global coordination effort to facilitate the flow of information among the several VLBI networks.

The concept of a Global VLBI Alliance is under development, formally as a working group within IAU, with the scope to provide such overarching activity. Its main purposes would be sharing strategies, technical developments for compatibility, logistics, operations, and user support. It would further promote, propose and coordinate common observational campaigns with these existing networks, and ensure that adequate information is provided to and from the users.

The VLBI technique is still considered complicated ('black belt') by many, and appropriate user support is essential. This is currently offered by each network independently, with the exception of some training events which involve several networks (e.g. the European Radio Interferometry School, ERIS). A unique common portal would explain the characteristics of the different networks, and the options for users to access them or in combination. Moreover, since the VLBI networks are 'open sky' facilities, the community of VLBI users is global. It would then be beneficial that each network has sufficient knowledge of the characteristics of the other networks in house, so that users can receive support locally.

In practice the Global VLBI Alliance would serve as contact point and framework of collaboration of the VLBI networks, and would encourage and support new VLBI activities, such as the AVN, Iniciativa VLBI IberoAmericana (IVIA), and developments in southeast Asian countries.

With the advent of the SKA and its precursors, and in the light of the next generation Very Large Array (ngVLA), such global coordination of the various networks and their participating telescopes will be a strong requirement.

### 7.1.2 Synergies between EVN and LOFAR

At present LOFAR (see A.1.9) mostly works with very wide field imaging and arcsecond angular resolution while the EVN/JIVE works with small field of view imaging and millarcsecond resolution. The gradual extension of ILT to longer and longer baselines meets the trend of densifying the EVN with short spacings, increasing its brightness temperature sensitivity. Given that LOFAR observes down to few tens of MHz, this offers the possibility to study phenomena over a very broad frequency range at similar angular resolution, which puts strong constraints on emission processes and physical conditions. The ILT imaging capabilities at sub-arcsecond resolutions will be further enchanced by the dramatic advances in software and pipelines that exploit the international baselines, and the future inclusion of Nenufar in France, providing a new large station for the ILT which will significantly increase the signal to noise ratio and image fidelity. In short, the ILT will evolve to have angular resolutions more comparable to EVN+*e*-MERLIN, enabling more joint science projects. As for EVN+*e*-MERLIN, ICT advances leading to increase in correlation capacity, data storage and imaging capacity are expected to dramatically increase the field of view of VLBI observations in the cm-bands. In the future, wide-field EVN images are expected to become the norm as opposed to being special projects as at present. The prospects for wide field VLBI imaging are such that projects to image the whole Northern sky at VLBI resolution are presently being planned. Such wide field imaging will be directly comparable to the LOFAR survey data.

In terms of particular common science areas one exciting prospect is to use international LOFAR wide field images to find large populations of compact objects in galaxies consisting of Radio Supernova/Supernova Remnants and weak AGN - which can then be followed up at higher resolution with the EVN. In the field of AGN, the physics of radio lobes, and the detection of HI absorption at high redshifts on sub-kpc scales have already been considered as example areas where both instruments can make a major impact. Other obvious targets are the local LLAGN population and radio galaxies at very high redshifts. These are two populations of particular interest that LOFAR surveys are sensitive to (for different reasons), and which would be ideal targets for deep EVN follow-up observations as well. In the area of transients, there is synergy in being able to observe and potentially image evolving transient structures (including in polarisations) and their environment over a wide range of radio wavelengths. Identifying short transients at LOFAR frequencies would be particularly interesting, but joint FRB studies would be extremely valuable also for low-frequency characterisation of persistent counterparts, that will help reveal the nature of FRB progenitors. A phenomenon that would hugely benefit from observations in a broad range of frequencies is scintillation and scatter broadening (pulsars, AGN). Simultaneous observations of temporal scattering of signals (LOFAR) and scatter broadening (EVN) would place strong constraints on the location and physical conditions of the scattering medium. Studies of interstellar scattering of AGN would be possible in which the EVN can measure intrinsic sizes of sources that are scatter-broadened at LOFAR frequencies. Likewise EVN+*e*-MERLIN 18 cm observations at matched resolution to international LOFAR will allow the study within galaxies of the distribution of free-free absorption and the spatial distribution spectral index of synchrotron emission. There is also significant overlap between the two instruments in observations of radio emission from radio stars and potentially from extrasolar planets of star-exoplanet interactions. Finally, LOFAR with its large field of view and survey data acting as a baseline can identify and locate new radio sources detected at very low resolution in other wavebands (such as gamma-ray or X-ray) which can then be located even more precisely and potentially have their internal structures imaged by the EVN.

Finally there is potential synergy between LOFAR and EVN in terms of data handling and archiving. Especially given that both instruments have their operational headquarter for user support at ASTRON/JIVE, in the same building, there are obvious advantages in coordinating archive data experience for users of both instruments, so that while remaining separate archives, solutions found for one instrument can be used for the other.

### 7.1.3 Synergy with SKA

The first-phase mid-frequency telescope of the SKA, SKA1-MID will be based in South Africa, with maximum baseline length of about 120 km. Since the science cases have been originally developed for the SKA with 2000-3000 km baselines in mind, there is a strong interest in doing VLBI with SKA1-MID that would cover a broad range of angular scales (Paragi, Chrysostomou & Garcia-Miró 2019; see Fig. 7.1). This will be done by phasing up the SKA1-MID core, just like it was done with the phased-array Westerbork Synthesis Radio Telescope within the EVN. The difference is that SKA1-MID will provide multiple phased-array beams, and the use of local interferometer data (e.g. SKA1-MID data products) will be better streamlined for SKA-VLBI (Paragi et al. 2016).

While SKA-VLBI requires a global collaboration, the EVN (and especially JIVE, as potential SKA-VLBI data centre) would play a major role in this. The Horizon 2020 JUMPING JIVE project work package “VLBI with SKA” (WP10) was established in order to help define the key SKA-VLBI science goals and use cases, the operational model, as well as the VLBI interfaces and specific requirements for the SKA telescopes (Garcia-Miró et al. 2019). The SKA will provide calibrated VLBI beams, but it is up to a future (SKA-)VLBI consortium to deliver VLBI terminals, handle the proposals, schedule and carry out the observations, do the correlation in an external data centre and provide user support.

The science with the SKA-VLBI is naturally aligned with that of the current EVN research. There are a few areas though that receive particular attention, especially those that require ultra-precise astrometric precision: key science projects on pulsar VLBI astrometry and 6D tomography of the Galaxy will support some of the highest ranked science objectives of the SKA. Other areas are very high resolution observations of specific deep fields in support of SKA continuum, spectral line or transient surveys. The latter requires both near real-time e-VLBI correlation and data buffering solutions for SKA-VLBI. The current e-EVN is thus a natural pathfinder not only in a technological sense but also for SKA-VLBI operations.

Besides dedicated VLBI projects, another way of realising SKA-VLBI is commensal VLBI observations, piggybacking on some of the SKA1-MID surveys (for constraints see Garcia-Miró et al. 2019). To fully exploit this option a Southern expansion of the EVN and support for setting up an African VLBI Network would be necessary. It is also essential to make the telescope IF systems compatible with the large bandwidth of SKA1-MID (2.5 GHz IFs), to maximise the sensitivity. The correlator will have to be able to deal with the extra load, also in terms of independent VLBI beams from SKA1-MID[1]. In addition, an SKA-VLBI requirement on the EVN archive will be science ready products derived directly from visibilities rather than just images (e.g. full Bayesian characterisation of source position, size and shape from sparse *uv*-coverage data).

The first phase of SKA will thus bring a lot of new opportunities as well as challenges for doing science with the EVN, primarily at frequencies below 15 GHz. How the landscape will develop on longer timescales is less clear at the moment. But the focus of VLBI technology and research might turn towards employing the very wide bandwidth (>>5 GHz) receivers and going to higher frequencies.

### 7.1.4 Synergy with multi-wavelength/multi-messenger instruments

An ultimate goal for the EVN (and VLBI arrays in general) is to exploit their unique capabilities in a way that it can serve a very broad user community. The proposed improvements in technology (to expand the EVN capability), in the operational model, and in the ways we will approach data processing and archive developments all point in this direction. In particular, it is essential to position VLBI in the multi-messenger landscape. The science chapters in this document have already pointed out important synergies with existing or future facilities. Most of these are related to facilities/missions operating in various parts of the electromagnetic spectrum: radio (ALMA, SKA, CHIME etc.), optical/infrared (*Gaia*, *HST*, LSST, *JWST*, ELT), and the high-energy bands (CTA, *Fermi*, *Swift*, *eROSITA*). To maximise the science output of the EVN it is essential to further strengthen these ties, and look for new opportunities as the multi-messenger landscape evolves. Below we mention four areas of particular interest, where VLBI could make a great impact together with emerging new facilities.

#### Ground- and space-based GW observatories

The role of VLBI observations in Mooley et al. (2018) and Ghirlanda et al. (2019) has already been quite prominent by showing that relativistic jets were produced in the binary neutron star merger related to the GW 170817 gravitational wave event (see also Sect. 1.2.3, Sect. 4.1.5)[2]. The fact that VLBI imaging may resolve degeneracies between jet viewing angles and GW standard siren distances (Hotokezaka et al. 2019; Sect. 1.2.3) opens new avenues for high angular resolution radio observations; however, these observations are currently challenging, partly due to the (so far) very faint nature of GW counterparts. With SKA-VLBI and/or a broad-band EVN it will be possible in the future to detect a number of GW events, sufficient for constraining $H_0$ below $\sim 1\%$ precision.

[1] The Science Use Cases deliverable of JJ WP10 lists a number of use cases, with current maximum number of required beams of N=14. This will result in a factor of N times more load on the correlator.

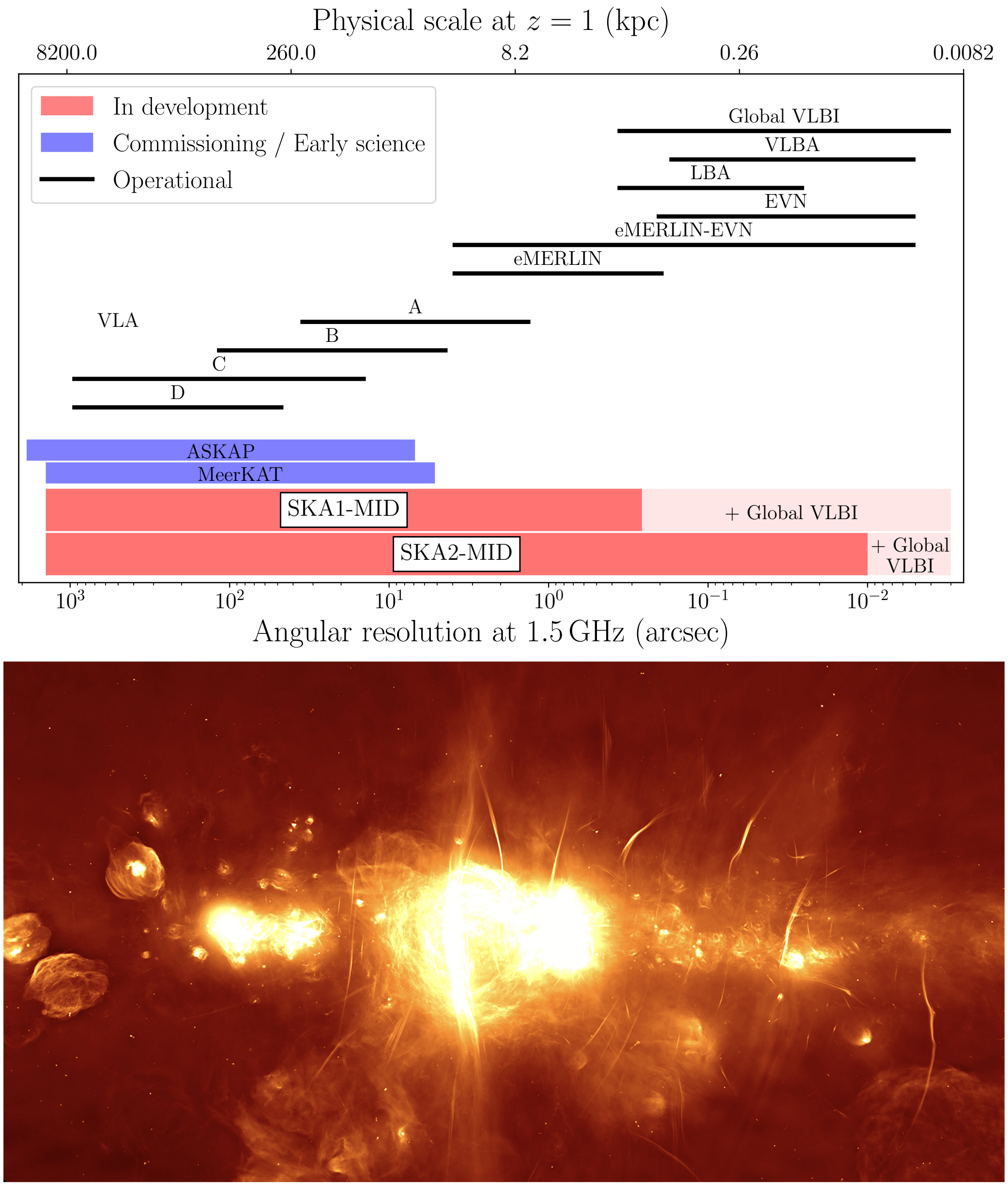

Figure 7.1: (upper panel) A broad range of spatial scales will be probed by a global VLBI array that includes SKA1-MID (courtesy of Jack Radcliffe and Cristina Garcia-Miró). (lower panel) The first public image from the 64-element SKA1-MID precursor MeerKAT, showing the Galactic Centre. (Courtesy of SARAO).

The second generation ground based GW observatories (LIGO, Virgo, INDIGO and others) will continue to play an important role in detection of GWs from compact objects of stellar origin. The third generation ground based GW observatories (e.g., the Einstein Telescope) are currently under study. Pulsar Timing Arrays will continue to increase in importance after 2020, as the sensitivity of the arrays increase. The detection of GWs from individual massive BH binaries is expected to happen within the next 10 years. *LISA* should be a game changer in 2030s, when it is expected to detect GWs from galactic stellar binaries, massive BH binaries and extreme mass ratio in spirals with unprecedented precision. Coalescing massive BH binaries will be detected by *LISA* well in advance of the merger, providing a unique opportunity to multi-messenger follow-up (Kocsis, Haiman & Menou 2008; Lang & Hughes 2008). Another area to explore with *LISA*+EVN is IMBH (Sect. 3.2.4). While these are very difficult to find in their quiet state, they may produce spectacular shows of outbursts during a tidal disruption event (TDE, Sect. 4.1.4). Gravitational wave radiation from an IMBH+WD encounter might be detectable in the Local Group (Rosswog et al., 2009; Anninos et al., 2018), which may produce a jet similar to what was observed in Arp 299-B by Mattila et al. (2018) with the EVN.

Space-based GW astronomy will revolutionise our understanding of these fields. There have been a number of new missions proposed for the ESA call for white papers "Voyage 2050"[3]. There are also two suggested EM missions in particular that are relevant to GW science and highly synergistic with the EVN. One is a space-based X-ray interferometer (Uttley et al. 2019), the other is a space mm-VLBI mission concept (THEZA, Gurvits et al. 2019; see Fig. 7.3). In very different wavelength regimes, both of these would achieve hitherto unprecedented angular resolutions down to the (sub-)$\mu$as regime. Several EVN observatories have been involved with developing the THEZA concept.

### Synergies with ELT

The European Extremely Large Telescope, ELT, is at the frontiers of the investigation of the Universe in the optical and infrared bands, from 0.8 to 13.3 $\mu$m. The telescope, with a 39 m mirror and implemented adaptive optics, is under construction in Chile, on Cerro Amazones, and is expected to provide the first light in 2025. One of the most exciting and novel features of ELT is the diffraction limit mas-scale FWHM which the telescope can provide. Going from 0.88 to 2.20 $\mu$m, the imaging angular resolution will be in the range $\sim$5 to 11 mas, perfectly matching that of the European VLBI Network in the two most sensitive bands, e.g. 1.4 and 5 GHz (as an example, see Fig. 1 in Dravins et al. 2012, and upper panel of Fig. 7.3 in this document). This implies that any radio-optical investigation resulting from ELT observations will hardly be possible with the SKA only, since the high angular resolution of VLBI will be a mandatory requirement.

[2]It is interesting to note that the paper that describes the initial multi-messenger follow-up of GW 170817 by Abott et al. (2017) has become, during a very short time, the most cited paper in which EVN observations are reported. At the time of writing, it has more than 1200 citations.

[3]`https://www.cosmos.esa.int/web/voyage-2050/white-papers`

It is noteworthy that the science case of ELT[4], includes most of the topics in the present document, from exoplanets to transients, from stellar to galaxy evolution, all the way to the very first stars and seeds of galaxies. The scientific potential of the future synergy between ELT and VLBI is thus bound to be transformational. Just to mention some of the several cases, the possibility to resolve binary black holes on $\sim$10–100 parsec scales in the optical and radio bands will revolutionise our understanding of the formation of massive and supermassive black holes, and their observational footprints (de Rosa et al. 2019). The high spectral resolution ELT spectral line observations will map the kinematics and magnetic fields of accretion disk atmospheres around young stars; this combined with VLBI maser astrometry it will strongly constrain models of high-mass star accretion at 10 to 100s of AU scales. VLBI and ELT are both able to directly image nearby exo-planets, the combination of magnetic field measure (from radio) and atmospheric spectral line characterisation (from ELT) will be a great leap forward in the understanding of planetary systems and their formation.

In order to fully exploit such potential, it is crucial to ensure that the portion of sky simultaneously accessible by the ELT, the EVN and the other most sensitive VLBI arrays, such as the VLBA, is as broad as possible. At present it is limited to a stripe of declination $\pm 30^{\circ}$. This will considerably improve in the SKA era, with the SKA-VLBI mode implemented and more AVN antennas in operation.

### Synergies with CTA

The Cherenkov Telescope Array (CTA) is the next generation ground-based observatory for gamma-ray astronomy at very high energies (VHE, $E > 100$ GeV). Currently in pre-construction phase, CTA operations are expected to begin in 2022, with the array construction ending in 2025. With more than 100 telescopes located in two sites in the northern and southern hemispheres, CTA will be the world's largest and most sensitive high-energy gamma-ray observatory. It will provide dramatically improved access to the most extreme accelerators in the Universe, whose understanding will necessarily require coordinated complementary efforts in other observing wavelengths.

The vast majority of the VHE sources revealed so far are also detected in radio: at high galactic latitude, blazars, in particular of the BL Lac type, dominate the census of the currently known population; in the Galactic plane, besides supernova remnants and pulsar wind nebulae (PWN), several compact objects (binaries) have been detected, and more could be hidden among the many sources that have remained unidentified. Based on our current understanding of the physical processes responsible for the emission of VHE gamma rays, it seems inevitable that the majority of the sources detected with CTA will also be ideal targets for VLBI observations.

Whereas CTA will be able to discover and characterise the emission from a large number of sources, its angular resolution will remain limited to a few arcminutes. Therefore, the localisation of the gamma-ray emitting region will only be possible through coordinated VLBI campaigns. At present, the only source for which such a coordinated study has been performed is the bright radio galaxy at the centre of the Virgo Cluster, M87 (MAGIC Collaboration, 2020; Hada et al. 2014; Giroletti et al. 2012). Increasing the number of VHE sources and flares, combined with dense monitoring campaigns with VLBI arrays is considered the only way to constrain the location, and thus the physical conditions, required for the production of VHE flares.

Even in the quiescence state, VLBI observations are crucial to characterise the physical parameters of the sources detected in VHE gamma rays, allowing a proper classification, as was the case of HESS J1943+213 (BL Lac object vs. planetary wind nebula: Akiyama et al. 2016; Straal et al.

[4] https://www.eso.org/sci/facilities/eelt/science/doc/eelt_sciencecase.pdf https://www.eso.org/sci/facilities/eelt/science/doc/

2016) or IC 310 (blazar vs. head-tail radio galaxy: Aleksić et al. 2014). It is to be expected that the number of VHE sources discovered by CTA with an uncertain nature, if not totally unidentified, will see a huge increase, and thus it will be critical to start a systematic process of high angular resolution imaging of these sources with VLBI.

This new forthcoming synergy is an upgraded extension of projects already undertaken with VLBI, including the EVN, aimed at revealing the parsec scale properties of blazars detected in the highest part of the *Fermi* energy range, such as the 2FHL based on $E > 50$ GeV data (Ackermann et al. 2016). Lico et al. (2017) have shown that the highly significant correlation between radio and MeV/GeV gamma-ray emission breaks down when higher energy gamma rays are considered, possibly as a consequence of the correlation between blazar luminosity and spectral properties. The deep survey carried out with CTA will greatly improve our chances to extend such studies, both going to the actual VHE domain and improving the statistics for the spectral types currently more challenging to reveal.

Finally, at lower $\gamma$-ray energies, there remain opportunities for continued synergies with wide-field facilities such as Fermi-LAT and AGILE. In the longer term, future missions will open a new window at lower gamma-ray energies where many radio sources are expected to emit the majority of their output.

#### Detection of ultra-high energy cosmic rays and neutrinos

One of the unsolved mysteries of today's astrophysics is the origin of ultra-high energy cosmic rays UHECR. Although these are charged particles, UHECR are little affected by the magnetic fields in the Milky Way galaxy and thus their source of origin could be traced back. The Pierre Auger Observatory (Pierre Auger Observatory) found statistical evidence that the sky position of UHECR with energies of $\sim 10^{20}$ eV correlate with nearby ($d < 75$ Mpc) AGN (Pierre Auger Coll. 2007; Abraham et al. 2008).

The same processes that create cosmic rays produce high energy neutrinos as well, and these travel undisturbed through the cosmos. There is evidence that the sources of the highest energy neutrinos are blazars (Kadler et al. 2016, Ros et al. 2020). The IceCube detector recorded a 290 TeV event, IC 170922A, which was followed up by a number of instruments from the gamma rays to the radio regime; the neutrino's position was coincident with the BL Lac object TXS 0506+056, showing gamma-ray flaring activity at the time (Icecube Collaboration 2018). The broad-band electromagnetic observations are explained by a novel one-zone lepto-hadronic model, as co-accelerated electrons and protons interact in a relativistic jet, surrounded by a slow-moving plasma sheath as a source of external photons (Ansoldi et al. 2018). The multi-messenger approach to reveal some of the most extreme environments in the Universe is important because TeV gamma rays are an excellent proxy for photo-hadronic processes in blazar jets, which also produce neutrino counterparts (Ojha et al. 2020). While high energy observations provide important constraints on emission models and can also probe sub-horizon scales by means of variability studies, the EVN will also have a unique role in resolving the blazar inner core-jet region where high energy cosmic rays and neutrinos originate from (Aleksic et al. 2014). Recently, analysis of a complete VLBI-flux-density limited sample revealed AGN that are positionally associated with the highest energy neutrinos (energies above 200 TeV) are typically more core-dominated than the rest of the sample (Plavin et al. 2020). This, and the correlation of $> 10$ GHz radio activity with neutrino detections (Kun et al. 2019) provide additional evidence for AGN with Doppler-boosted jets being an important population of neutrino sources.

Blazars are not the only possible source of extragalactic high energy neutrinos. Kun et al. (2018)

describes a scenario in which a supermassive binary black hole merger could produce neutrinos while a reoriented relativistic jet is generated by the spin-flipping of one of the SMBH. As mentioned above, early EM localisation from *LISA* triggers, in which EVN observations will have a role, will provide a unique opportunity to start broad-band observations of these transient events before they actually happen! Hadronic processes may also occur in cataclysmic events that also produce a fast radio burst (non-recurrent FRBs; e.g. following BH coalescence or during the collapse of a supramassive NS), which raises the intriguing possibility that FRBs may contribute to the high-energy cosmic-ray and neutrino fluxes (The ANTARES Collaboration 2019). This is an opportunity to explore further, in a field where the EVN has already made a breakthrough contribution.

The next generation neutrino telescopes (IceCube-Gen2, Super-Kamiokande-IV and successors) will address cosmic ray physics, the nature and properties of dark matter particles, solar neutrinos, atmospheric neutrinos, supernova explosions, and will search for products of proton decay in order to test the Grand Unified Theory (GUT), which unifies strong, weak and EM interactions.

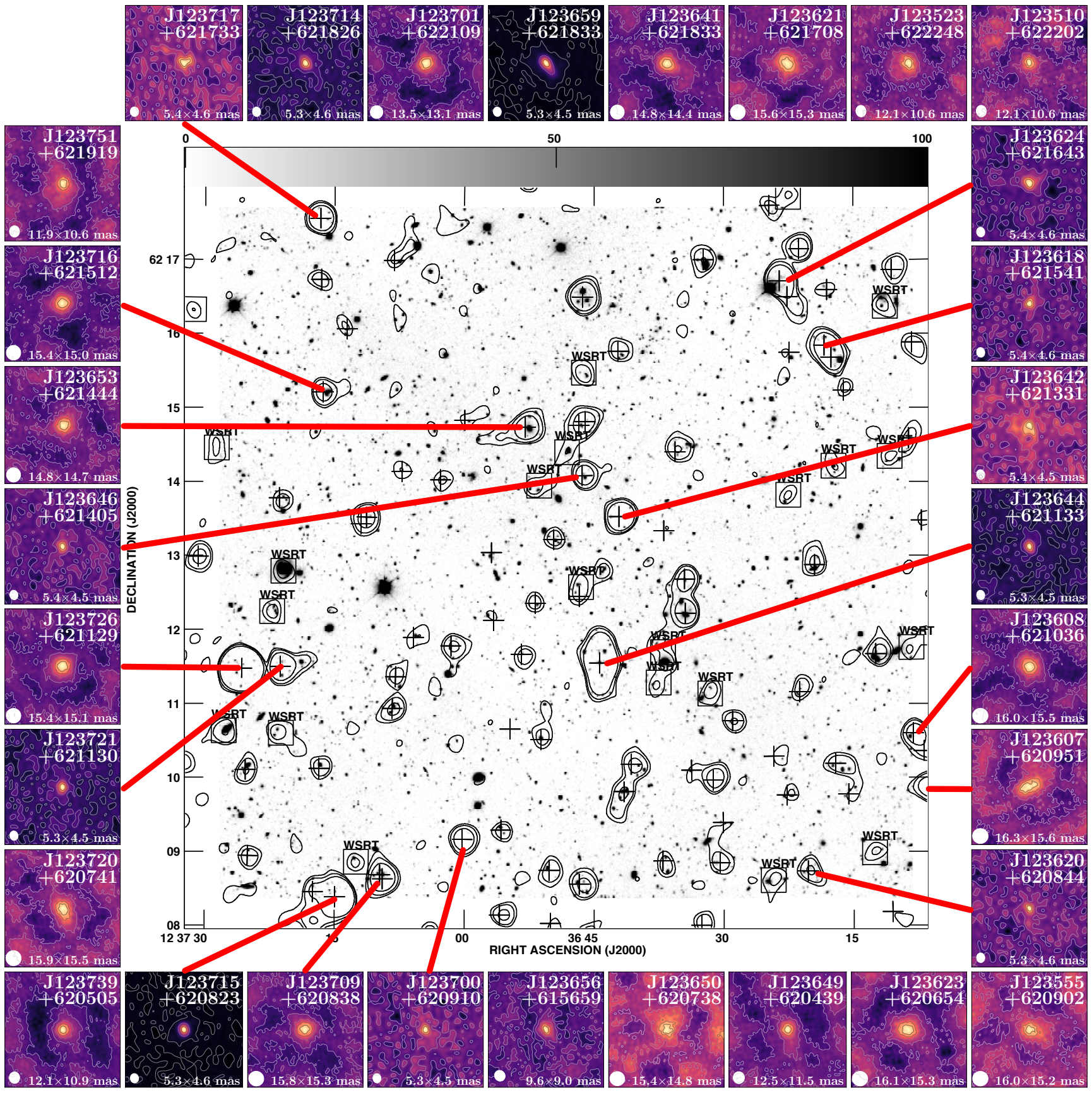

Figure 7.2: Composite image of 1.4 GHz WSRT radio-KPNO optical overlay of the GOODS-N field, centred on the HDF-N (Garrett et al. 2000), surrounded by postage stamp images of the 1.6 GHz 31 VLBI-detected sources. Credit: Radcliffe et al. (2018), reproduced with permission ©ESO.

## 7.2 Recommendations for future EVN Developments

In order to maximise the science synergy with the instruments mentioned in the previous sections, several enhancements in EVN technology, operations, data products and archive need to be explored.

### 7.2.1 Technical enhancements

*On-going efforts*

All the science goals proposed in this document call for improved image sensitivity — at least by one order of magnitude — and image fidelity. These are the first and most urgent needs to to ensure the current high astrophysical impact of the EVN is maintained and expanded. The SFXC correlator at JIVE can already do amazingly flexible operations (e.g. e-VLBI), new solutions might be needed (see A.2) to be able to keep up with the increasing number of telescopes, the increased bandwidth, and the inclusion of telescopes with multiple beams (e.g. SKA-VLBI). The specific science requirements include special fast transient and pulsar processing modes (like a number of simultaneous independent pulsar gates, see Chapter 4.), and the desire to image more steradians on the sky (important for cosmology, galaxy evolution and spectral line VLBI, see Chapters 1., 2. and 5.). The key actions to achieve these requirements are listed below.

- The improvement of image sensitivity requires a continued effort to increase the recording bit-rate. At present the EVN delivers 2 Gbps recording, and this should increase at least by a factor of 4 in the near future. Such a step has obvious implications in terms of data storage at each station and at the correlator. In addition, the connectivity between each station and JIVE needs to be addressed, both for e-VLBI and e-transfer.
- The full integration of *e*-MERLIN in the array is considered very important as it provides users with a much improved *uv*-coverage, from few tens to many hundreds of kilometres. The need for short baselines is particularly relevant for cosmology, galaxy evolution and studies of AGN in continuum and spectral line mode.
- Increasing the image sensitivity and fidelity by adding new and also larger antennas to the array is a continuous effort to increase the EVN capabilities and grow the global VLBI alliance. In particular maser polarimetry requires an increase in antenna surface as higher bit-rates are not the bottleneck for spectral line data. The amazing 500 m FAST and MeerKAT are already operational, and several telescopes are under construction or being refurbished in different countries (some examples are those in the Azores Islands and Thailand, see Sect. A.2.2). It is crucial that the VLBI operations at new observatories are supported adequately to ensure success; this requires a continuous effort.
- The EVN should curate its very sensitive observations in such ways that a larger area of the sky can be imaged (see Fig. 7.2). Not only should it be possible to propose for larger sky coverage, but the EVN should also try to align with 'survey-modes' that are being introduced for many astronomical observatories. Therefore it is necessary to build up an archive that can be used for sensitive, high resolution statistical studies.

Although these are all 'on-going' activities, it should not be underestimated that they require considerable resources, long-term planning, project management and in some cases technical development.

*Goals for new development*

The goals listed below are the core of this document: they are the technological innovations the EVN should focus on. These projects require considerable efforts to design and test, and most of all substantial investments to get the EVN ready for the next decade.

- Broad-band receivers offer extremely important scientific and operational advantages. This development will allow the EVN to make use of digital techniques that cover more bandwidth, providing improved sensitivity, but also instantaneous coverage for wide frequency ranges. It will increase the efficiency of observations as well. Development for a C-X band (covering at least ∼4-8 GHz as a first goal) receiver has started at many stations and it will be important that a system is defined that can be compatible across the EVN and with other VLBI arrays. Caution should be given that polarisation properties and sensitivity are not compromised.
- Extending the frequency coverage is also very important. There is a clearly formulated need to have a better 22 and 43 GHz coverage, both in terms of increased bandwidth and number of telescopes in the array. Improving the capabilities of the EVN at 22 GHz is strongly required both for water maser and cosmological studies. Moreover, AGN and maser (SiO) science would broaden their scope opening up to millimetre wavelength observations. It seems attractive to accommodate in particular the triple band receivers to make progress in high-frequency imaging fidelity. At the same time, follow up of cosmological surveys will require the capability to probe HI at higher redshifts, hence observing well below 1 GHz. A strong requirement in a number of science cases, especially those in need of spectral line observations, is that of homogeneous receivers over as many antennas as possible in the network.
- Improved *uv*-coverage, especially in the Southern direction, is highly desirable. This would increase the prospects of galactic science, as well as improve the accuracy of the terrestrial and celestial reference frames. The EVN should therefore give special attention to initiatives to make antennas available through the AVN (Gaylard et al. 2011, Gurvits et al. 2020) which will include antennas in Namibia, Zambia, Botswana, Mauritius, Kenya, Ghana, Madagascar, Mozambique, but also in the Middle East and in the Canary Islands. Such expansion is a key step forward for the optimisation of the portion of sky common to VLBI arrays and to the SKA.
- Wide-field imaging and survey-mode observing are well established facilities already available to the radio astronomical community, thanks to very large field of view of aperture arrays, such as LOFAR (see Sect. A.1.9), to the large primary beam of small antennas operating at relatively low frequencies, as is the case of MeerKAT, and the use of focal plane array feeds (PAF) as is the case of ASKAP. Increasing the field of view of VLBI imaging with the installation of PAFs on a number of antennas would be an enormous step forward. It would allow fast follow-up of classes of objects over large areas of the sky (a request which is becoming more and more pressing these days), and it would make large VLBI projects (such as gravitational lens surveys for cosmological studies) feasible over limited time scales.

### 7.2.2 Changes in the operations

*On-going efforts*

Since the advent of e-VLBI, the EVN has been adopting new observing modes and proposal types to serve specific scientific goals, such as observing transient events and other time-critical observations. A continued effort to offer observing modes that better match user requirements is needed, which can often happen without requiring upgrades at the telescopes. The most urgent actions are listed below.

- Some science cases – transient and multi-messenger science in particular – call for innovative operational modes, in particular, more observing time and more flexible scheduling (EVN-light). The availability of sub-arrays would allow to accommodate a broader variety of science cases. Short- and long-term monitoring projects would benefit from this change, too. Maser astrometry requires an array that is available in specific seasons for optimal parallax sensitivity.
- A continuing effort is needed to improve the quality of amplitude calibration and to ensure homogeneous set-up within the array.
- Users' support is essential not only to ensure that VLBI is accessible to a broad fraction of the astronomical community, but also to experienced users whose projects require new and more refined observing modes and data processing. An enduring effort in this area is vital to ensure that VLBI is used at the best of its potentials.

*Goals for new development*

Future observatories and observing facilities will most likely broaden the science and require new approaches to observations. For this reason further changes in the operational modes should be considered, to ensure that the EVN keeps and strengthens its role in the future multi-wavelength and multi-resolution framework. In particular:

- In the next decade, the EVN will need to accommodate search for transients in commensal modes. Collaborations with other observatories will be needed, for projects which require fast response to follow-up observations. New proposal methods and operational modes will be needed to allow observations on (a sub-set of) the array.
- The EVN will need to cater for key programmes, presumably delivering public data products. This will maximise the science return for large field of view observing modes.
- In the SKA era, we can foresee that astronomers accessing the SKA will expect to have access to VLBI data through similar procedures and services.

### 7.2.3 Data products and Archive

*On-going efforts*

The overall accessibility of the array strongly depends on the data analysis support. In particular:

- It is essential that the EVN continues to deliver pipeline products, allowing fast access to the data and initial calibration, as well as support to continue the development of VLBI processing in CASA.
- New observing modes may require other data products (e.g. time series) and it is mandatory that the range of observing modes that are covered by the pipelines continues to increase.
- The EVN archive at JIVE is a key interface between the EVN and its users for all types of data access. It is where the users who proposed observations initially find their data. After a propriety period, which is usually 1 year, the data are publicly available to all users. An essential goal for this archive is to deliver well-calibrated data, and even science-ready data, since the EVN wants to attract users that are experienced in other disciplines. The EVN archive is also an invaluable source to access historical observations for new investigations. Specific examples are variability studies, completion of samples, or even reprocessing of published data. The current interface to the archive is however ten years old, and requires new developments.

*Goals for new development*

Following the developments at other facilities, future data activities will be centred on an archive that eventually implements observatory-side data processing. In the long run, the goal is to reach a homogeneous access model for all astronomy. To note:

- Archiving solutions are under study in the framework of the European Open Science Cloud,

whose main directions for science archives can be summarised as "findable, accessible, interoperable, re-usable" (FAIR). A better interface would make VLBI data more re-usable, saving the effort that EVN users make regularly to derive publishable data from the archive. Open science aims at making it easier and more transparent to verify and reproduce the data that led to a scientific conclusion. Thus, it could be considered beneficial if the EVN could offer a method to associate results from datasets (including intermediate steps, storing scripts and calibration) with publications. For these reasons, there is a pressing need to make the archive more integrated with VO methods, and also to provide technology and incentives that allows users to document their data processing efforts.

- The EVN should welcome the development of a data processing platform in common with other major facilities (e.g. VLBI capabilities in CASA), focusing on ways to define and maintain high level scripts and recipes that can also be used to drive observatory side computing. This recommendation is independent of the need of an inventory of the algorithmic processing functionality for all of the EVN science cases. Such efforts would align quite well with the big data challenges that the community is facing with the advent of the SKA. Presumably, SKA users will have to rely on observatory-side computing for anything but their image analysis. Similar data processing recipes would then have to be manipulated on-line for controlling the SKA or SKA data centre compute clusters; note that using the same recipes locally and interactively is very important for prototyping and optimising. In this sense, it must also be ensured that the existing archive is interoperable with future CASA processing.

## 7.3 Concluding remarks

The previous chapters have demonstrated the very broad range of science the EVN addresses. The distribution of the top-100 most cited publications with EVN data (including global VLBI and multi-messenger papers) per research area, organised based on which chapter they would fit the most, and the citation history to these papers have been shown earlier in this document in Fig. 1. This demonstrates how the EVN evolved to become a multi-disciplinary facility during its 40 years of existence. Almost 2/3 of the most cited papers appeared after 2000, and they show a much more balanced distribution between science areas. Past technological advances made this possible: at the end of 1990s and early 2000s the EVN introduced the Mark4 and then the Mark5 systems, allowing high data rates of 512-1024 Mbps, which resulted in a great increase in sensitivity. The EVN MkIV Data Processor and later the EVN Software Correlator at JIVE (SFXC) became operational, and the EVN Archive came online with pipeline data products.

The unique capabilities of the network resulted in high impact discoveries. The very sensitive short spacings compared to other VLBI networks were crucial to show that low-power radio galaxies can also produce relativistic jets (Giovannini et al. 2001), and that CSOs represent a class of very young radio sources (Owsianik & Conway 1998). The very sensitive baselines, especially when combined with sensitive telescopes in other networks have been fundamental to e.g. detecting outflows of neutral atomic hydrogen (H I) from active galactic nuclei, demonstrating jet-driven AGN feedback (Morganti et al. 2013). The advance in galactic science is related to the start of maser astrometry programmes. Observing the 6.7 GHz transition of the $CH_3OH$ line (for a long time unique to the EVN), allowed studies of star formation and Galactic structure (e.g. Rygl et al. 2010, 2012). The introduction of real-time e-VLBI during the mid-2000s brought forward another revolution. The scope for transient science has increased and a number of high impact results followed, like revealing the likely origin of $\gamma$-rays in classical novae (Chomiuk et al. 2014), and eventually the first 10-mas scale localisation of FRB 121102, providing the ultimate evidence that fast radio bursts are indeed extragalactic and may reside in extreme astrophysical environments (Chatterjee et al. 2017; Marcote et al. 2017). The final peak marks the start of the gravitational-wave astronomy era as well (see Fig 1).

As we have seen earlier in this document, the scientific potential of very high angular resolution radio astronomy is still enormous. Emerging new facilities bring forward new synergies and actually strengthen the impact the EVN will have. The science case of the SKA as well as forthcoming facilities in other bands (e.g. CTA and ELT) overlaps with that of the EVN, but some of the highest priority science cases actually do depend on complementary VLBI observations. Furthermore, SKA-VLBI will bring forward new possibilities in the field of VLBI astrometry and in the study of the faint radio source populations[5]. *Gaia* is providing stellar astrometry comparable to what can be achieved with VLBI, but aligning the *Gaia* frame with ICRF and by studying individual sources brings forward interesting new science cases (e.g. Paragi et al. 2016; Kovalev, Petrov and Plavin 2017; van Langevelde et al. 2019).

In summary, the next revolution will be a very sensitive, broad-band, flexible EVN with a wide range of *uv*-spacings and archival/user services compatible with other instruments in the era of VO archives. This will also bring forward exciting synergies for the science addressed by future multi-messenger instruments. A technology roadmap for the EVN will detail the steps necessary to achieve the exciting science described in this document.

[5]The science cases and operations of SKA-VLBI is subject to WP10 "VLBI with the SKA" of the JUMPING JIVE project, parallel to the WP7 "VLBI Future" efforts, which resulted in this EVN Vision Document.

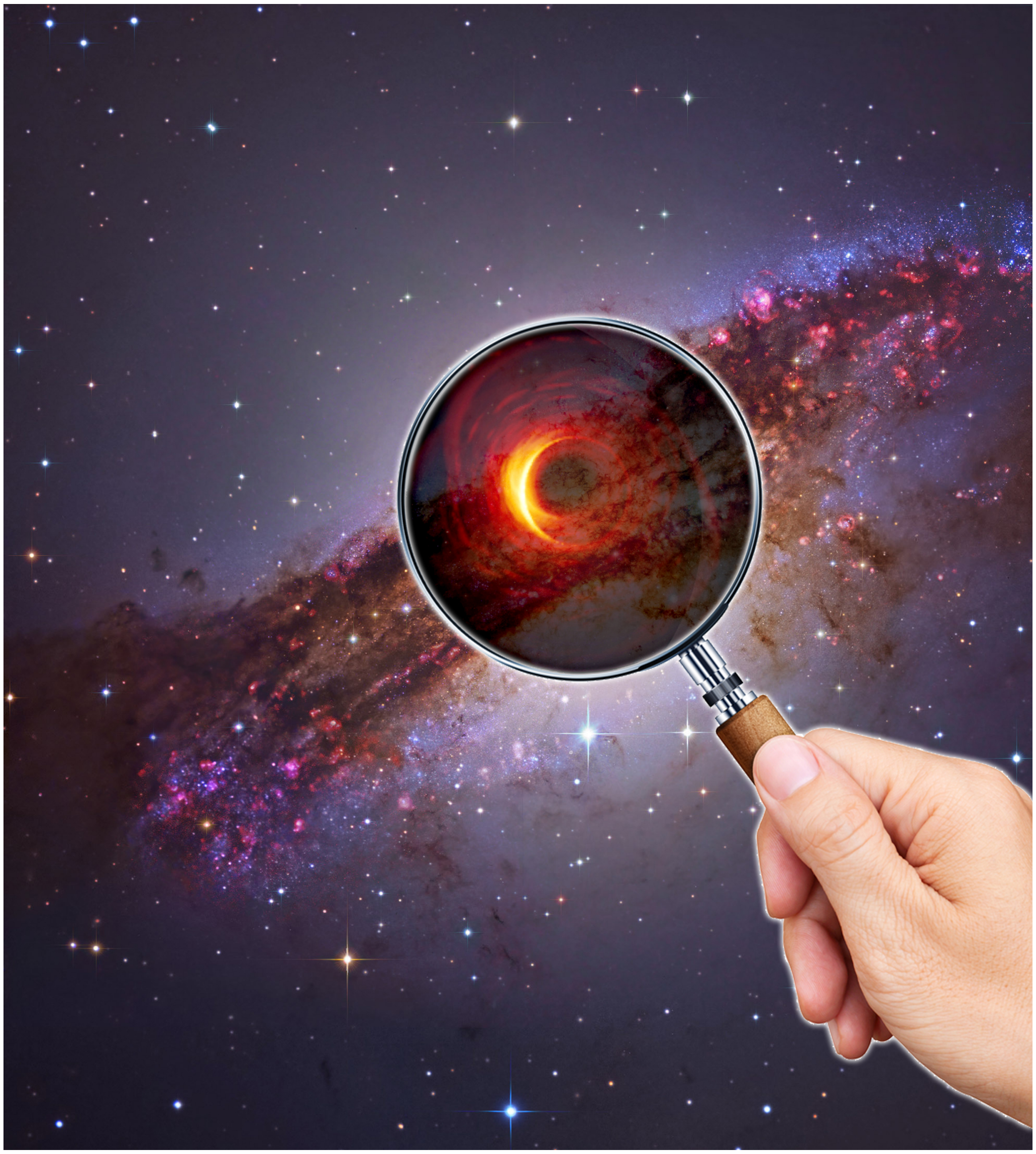

Figure 7.3: THEZA is a space mm-VLBI concept to directly image supermassive black holes in the nearby Universe, and massive black hole binaries up to cosmological redshifts (Gurvits et al. 2019). Figure credits: BH simulations – Monika Mościbrodzka et al. (2014) & Freek Roelofs. Beabudai Design.

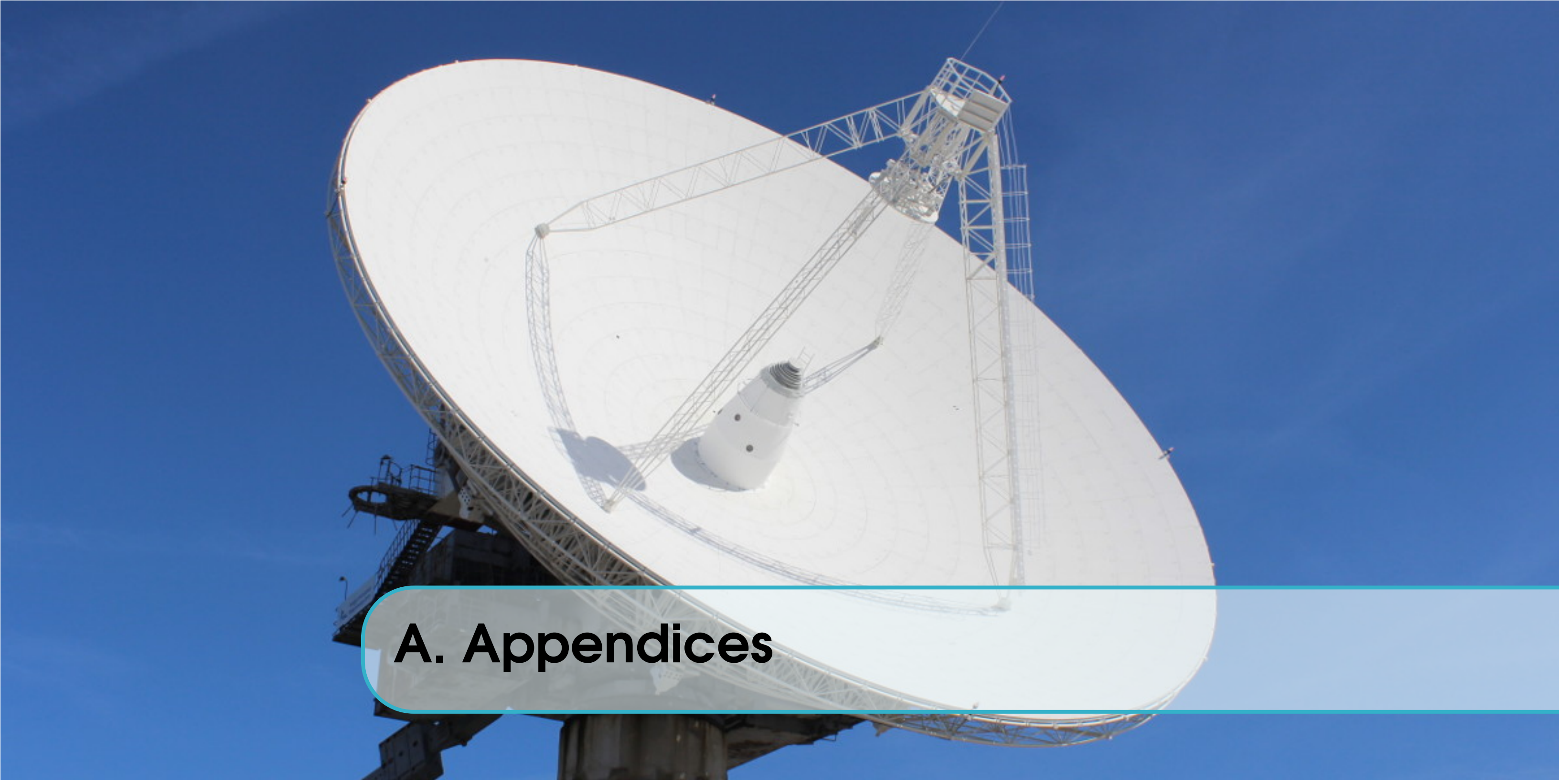

# A. Appendices

## A.1 Present and future VLBI arrays and other radio facilities

In the following we aim to capture the trajectory of technical capabilities of major radio astronomy facilities around the world and other instruments in other wavebands that are expected to play a major role in the evolution of VLBI science during the first half of the 2020s.

### A.1.1 EVN and JIVE

The European VLBI Network (EVN) was formed in 1980 by leading radio astronomy institutes in Europe (see review by Schilizzi 1995). Today it is a joint facility of independent European, African, Asian, and North American radio astronomy institutes with 32 telescopes that cover a broad range of wavelengths from 92 cm to 0.7 cm. The EVN offers an angular resolution down to (sub-)milliarcsecond in the main observing bands 21/18 cm, 6/5 cm and 1.3 cm. It is the most sensitive regular VLBI array that employs a fully open skies policy. The total aggregate bit rate per telescope has increased from the initial 4 Mbit $s^{-1}$ to 2 Gbit $s^{-1}$ in the past four decades. Its collecting area is comparable to that of SKA1-MID.

It operates in three major observing sessions through the year, but a limited number of out of session observations are carried out to support Target of Opportunity (ToO) or other multi-band observing campaigns, and until quite recently, space-VLBI experiments carried out jointly with the *RadioAstron* mission (concluded in 2019). In addition to that, there are 10 days a year dedicated to real-time electronic-VLBI (e-VLBI) observations.

The EVN has a governing body formed by the Consortium Board of Directors (CBD), a Technical and Operations Group (TOG) and a time allocation committee referred to as Programme Committee (PC). The chairs of the two latter groups report to the CBD every 6 months. Beyond the chairs of the three groups mentioned before, there is a fourth officer, the scheduler, who is in charge of scheduling all EVN observations and coordinates with other arrays, like the global VLBI array, the GMVA and the IVS. The CBD is formed by the directors of the EVN institutes who set the technological and

---

Chapter image credit: The freshly refurbished and modernised 32m telescope in Irbene, Latvia during the inauguration ceremony in 2015. Photo by Zsolt Paragi.

scientific strategy of the network for the future. The TOG is in charge of the operations and technical developments of the network. It is composed by VLBI friends at the stations and personnel at the correlators. The composition of the PC is determined by the CBD who invites experts in different fields both from EVN and non-EVN institutes.

The EVN science is very broad, with the highest impact in the fields of radio galaxies, AGN, and studies of Galactic sources of methanol masers. The latter was a unique EVN capability at 5 cm until the mid-2010s (Zensus & Ros 2015). Since the development of e-VLBI, transient phenomena have become a very competitive EVN research field. The latest highlights are the milliarcsecond localisation of the repeating fast radio burst FRB 121102 (Chatterjee et al. 2017, Marcote et al. 2017), the EVN and the VLBA (see below) monitoring of a tidal disruption event in a nearby galaxy (Mattila et al. 2018), and the detection of the first electromagnetic counterpart to a gravitational wave source GW 170817, a merger of binary neutron stars (Abbott et al. 2017). Global VLBI data on the radio afterglow provided key insights to the nature of the explosion (Ghirlanda et al. 2019).

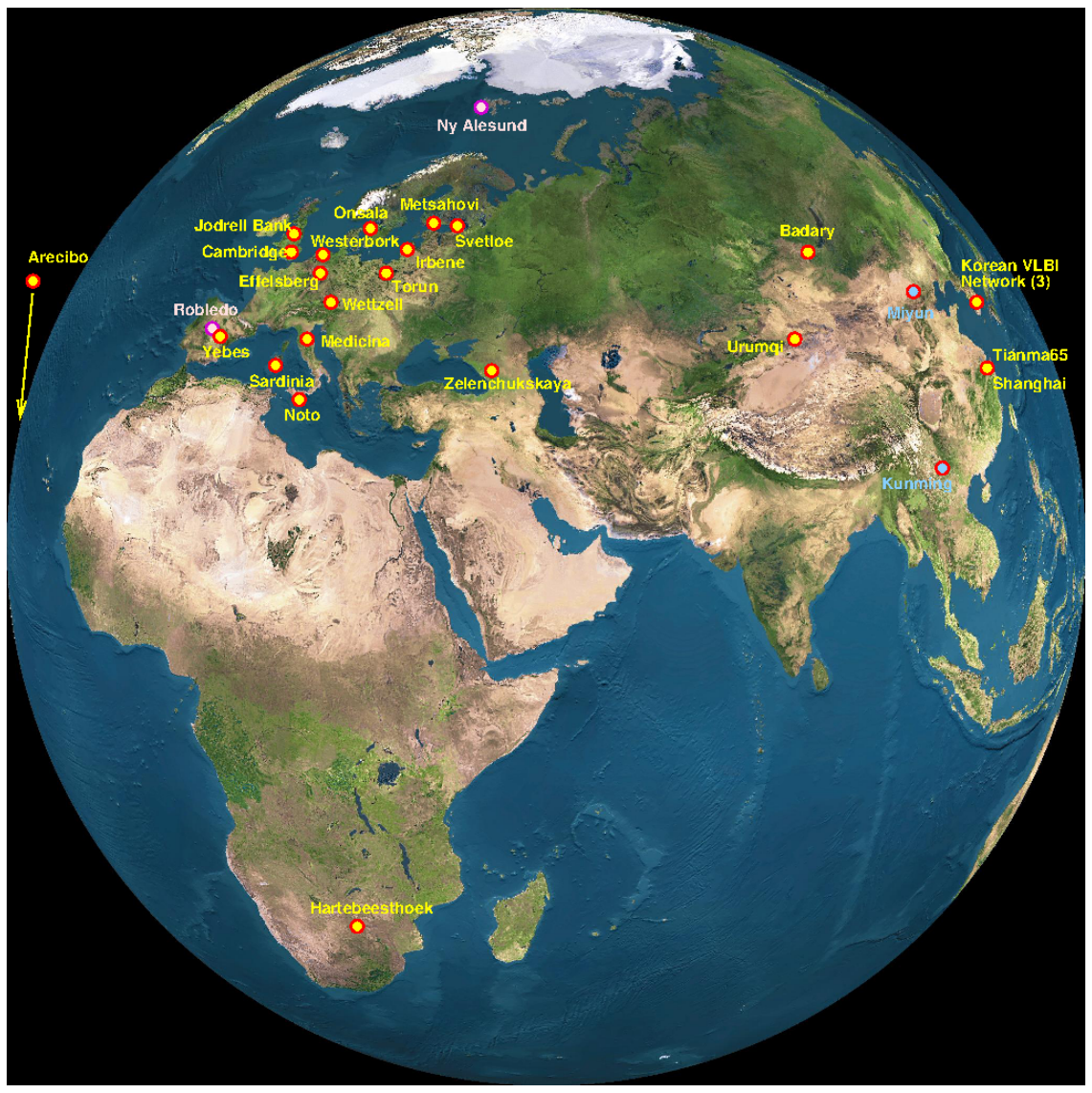

The current members of the EVN (various levels of membership). Image: `http://evlbi.org`.

From the 1980s until the end of 1990s the EVN data were correlated at the Max Planck Institute for Radio Astronomy (MPIfR) in Bonn. In 1993 the EVN established the Joint Institute for VLBI in Europe (JIVE) in Dwingeloo, the Netherlands (formally as a Dutch foundation). A new, broad-band, 16-station MarkIV correlator began operations in 1999. In 2015 JIVE became a new legal entity, a European Research Infrastructure Consortium (ERIC); first (and only so far) of its kind in the field of astronomy. The main mission of JIVE is to support EVN operations, support the EVN users, as well as develop next generation VLBI correlators and the VLBI technique in general. JIVE has led the e-VLBI developments in the early 2000s, and real-time e-EVN observations have been offered routinely since 2006 (Szomoru 2008). The MarkIV correlator has been replaced with the EVN Software Correlator (SFXC; Keimpema et al. 2015), which is very flexible for special VLBI applications (wide-fields of view, pulsar binning observations etc.). The network support includes regular network monitoring experiments, rapid (ftp-)fringe tests during observing sessions, and additional tests for new hardware/firmware and aspiring new EVN members. The user support is offered for all stages of the research from developing a proposal idea through scheduling observations to help with data processing. JIVE maintains the EVN Archive, where all data become available following a proprietary period of one year (half a year for ToOs). All the data are pipeline processed (Reynolds, Paragi & Garrett 2002); the latest version of the pipeline script makes use of ParselTongue, a scripting environment developed at JIVE (Kettenis et al. 2006).

The EVN has grown considerably during the past decade. The possible evolution of the EVN with regards to technologies and expansion by including new antennas, as well as the future possible relations with SKA1-MID and as a standalone very wide-band VLBI array on the long run are described later in the document.

### A.1.2 *e*-MERLIN

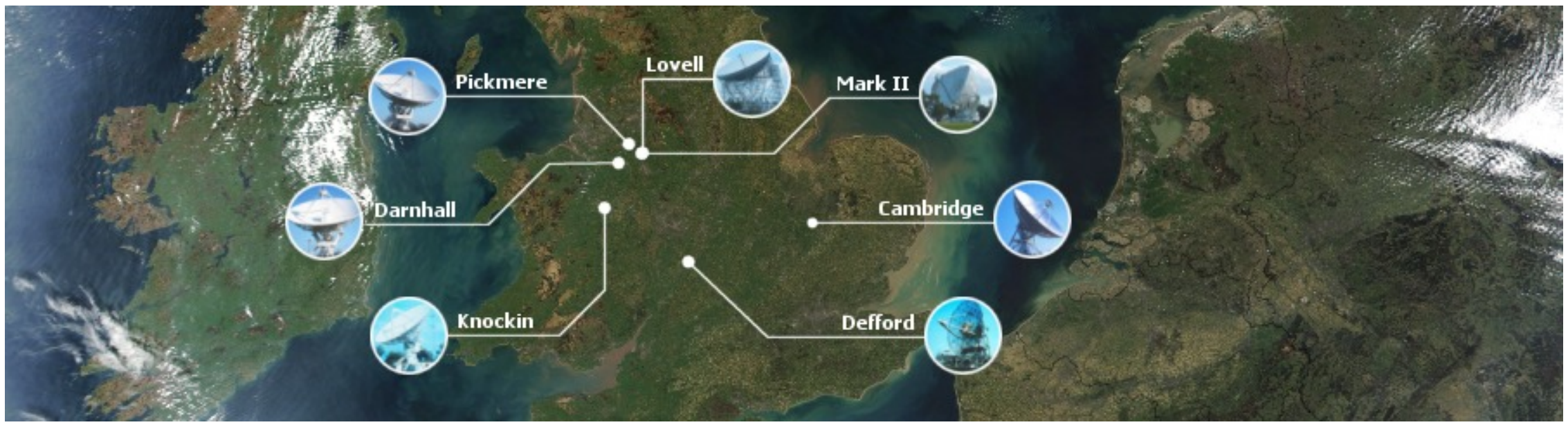

The distribution of the *e*-MERLIN telescopes in the UK. Image: `http://www.e-merlin.ac.uk/`.

In the UK, *e*-MERLIN with baselines ranging from 10 km to 220 km in length provides imaging capability covering a unique range of spatial scales, overlapping with the JVLA (see later) at lower angular resolution and extending up to the EVN with ultimate angular resolution $< 1$ mas. In addition to being a dedicated compact VLBI imaging array with sub-arcsecond angular resolution and $\mu$Jy sensitivity, *e*-MERLIN is used in combination with the JVLA for increased angular resolution whilst retaining superb surface brightness sensitivity—and in combination with the EVN to provide short-spacing coverage to the EVN to both place the mas-scale VLBI images in context with regard to any extended radio structure present. Also, *e*-MERLIN has the capability to directly image such structures at intermediate angular resolutions through integrated high fidelity combination imaging over a range of angular resolutions and spatial scales between 5 mas and 150 mas, providing a unique region of radio imaging capabilities at centimetre wavelengths which have not been available before. This new capability will enable transformational scientific investigations of planetary formation around nearby young Galactic stars at the shorter wavelengths, and probe detailed AGN jet feedback and its interplay with surrounding host galaxies at higher redshifts, including the role of feedback in those systems with intense nuclear starbursts at longer wavelengths where the superb surface brightness sensitivity of EVN+*e*-MERLIN combined can detect and image this steep-spectrum emission to high redshifts.

The additional short baselines for high fidelity imaging at angular resolutions between *e*-MERLIN and the EVN has long been promised and is now finally achieved. Recent tests have confirmed that data for the *e*-MERLIN antennas can now be output in VLBI Data Interchange Format (VDIF) and stored locally at JBO enabling $2 \times 128$ MHz of data (e.g. both circular polarisation states) for 6 *e*-MERLIN antennas to be recorded for correlation with other EVN telescopes at JIVE. In March, 2018, a successful test demonstrated interferometric fringes between Cambridge and Effelsberg (Germany), later followed by participation of other *e*-MERLIN outstations in e-VLBI mode with the EVN in September 2018.

The addition of telescopes at Goonhilly Earth Station in Cornwall to *e*-MERLIN at both L- and C-Band is planned. The benefit of this additional antenna site to *e*-MERLIN and to the integrated EVN+*e*-MERLIN has been well described by Heywood et al. (2011) and Kloeckner et al. (2011) respectively. For *e*-MERLIN in standalone mode Goonhilly nearly doubles the angular resolution in fields at northern declinations, and for equatorial fields significantly circularises the beam and improves image fidelity. In combination with the EVN and *e*-MERLIN, Goonhilly adds additional intermediate spatial frequency coverage to ensure seamless integration of *e*-MERLIN with the EVN over a wide declination range, which will be enhanced still further in the southern skies by the addition of telescopes in the AVN in the next few years. Initial data tests to Goonhilly are expected in the early 2020s.

### A.1.3 CVN

The proposal to build the Chinese VLBI network (CVN) was first raised by Prof. Shuhua Ye of the Shanghai Astronomical Observatory in the 1970s (Ye, Wan & Qian 1991). The first VLBI station comprising of a 25 m telescope was completed in 1986 in Sheshan, Shanghai. The current CVN includes 5 antennas (Seshan 25 m, Urumqi 26 m, Kunming 40 m, Miyun 50 m and Tianma 65 m) and one data processing centre in Shanghai. The longest baseline is 3249 km between Shanghai and Urumqi telescopes. It is one of the few VLBI networks having very short baselines ($\sim$ 6 km between Sheshan 25 m and Tianma 65 m), which is essential for imaging extended emission structures.

The CVN was initially designed for geodetic and astrophysical observations. A notable application of the CVN is the tracking of the Chinese lunar satellites, offering accurate position measurements beyond the reach of other techniques. In the latest *Chang′E-3* mission, the time delay between data acquisition at the telescopes and orbit measurement at the data analysis centre is less than one minute. A relative position accuracy as good as 1 m between the *Chang′E-3* rover and the lander was achieved in post-correlation data analysis (Tong, Zheng & Shu 2014). The CVN is also expanding its applications in astrophysics and astrometry (An et al. 2012; Shu et al. 2017).

The Seshan 25 m and Urumqi 26 m joined the EVN since 1993 and 1994, respectively. The two Chinese telescopes importantly contributed to the increase of the EVN baseline from $\sim$ 3000 km to $\sim$ 9000 km in the east-west direction, improving the angular resolution of the EVN by a factor of three. The Kunming 40 m joins the EVN observations when requested. The Tianma 65 m has participated in EVN observations since 2014. It significantly increases the longest-baseline sensitivity, creates the opportunity of detecting weak sources, and also allows for super-resolution imaging (An, Sohn & Imai 2018).

### A.1.4 EAVN

The EAVN plays an important role in promoting regional VLBI cooperation, including the major facilities of the CVN, the Korean VLBI Network (KVN), the Japanese VLBI Network (JVN) and VLBI Exploration of Radio Astrometry (VERA). The EAVN consists of 21 radio telescopes distributed over a maximum baseline of $\sim$6000 km, offering a highest resolution of 0.5 mas at 22 GHz. The first EAVN observing session started in the second half of 2018. The operational frequency bands are 22 and 43 GHz, and will be expanded to cover lower frequencies in the near future.

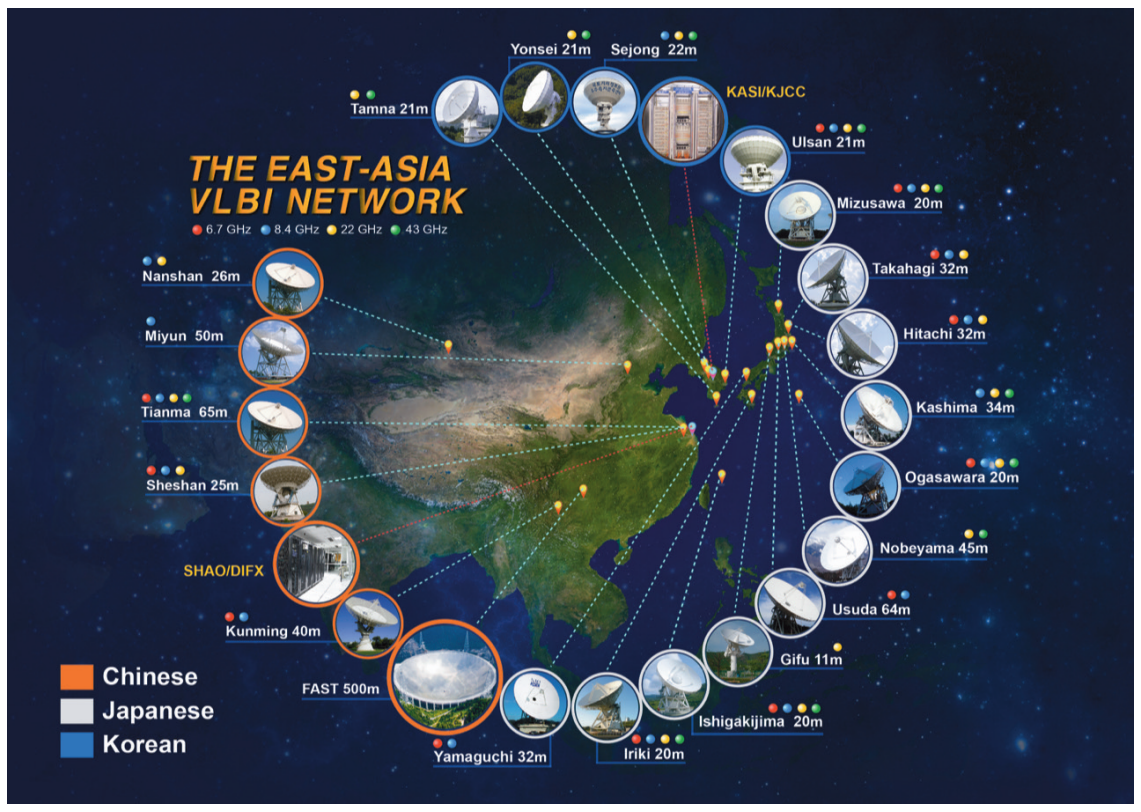

The East Asia VLBI Network (EAVN). Image from An, Sohn & Imai (2018).

The EAVN includes 21 geographically distributed telescopes with sizes ranging between 11 m and 500 m, with baselines ranging between 6 to 5000 km, typically operational in the 2.3 to 43 GHz radio frequency range. More telescopes (e.g., under-construction 110 m telescope in Xinjiang, China, and the planned Thailand VLBI network) in the near future will further broaden the science capability of EAVN (An, Sohn & Imai 2018).

New facilities and development for the EAVN are currently ongoing. The Five-hundred-meter Aperture Spherical radio Telescope (FAST 500 m; Nan & Zhang 2017) in southwest China witnessed the discovery of a dozen new pulsars since its inauguration in 2016 September. A 110 m radio telescope (QTT) has been funded for construction in Xinjiang, China. It will be the largest fully steerable single-dish radio telescope operational in the 0.15-115 GHz bands. The FAST and QTT can enormously increase the sensitivity of the global VLBI. The planned Thailand VLBI Network will join the EAVN on completion. The connection of the EAVN and the Long Baseline Array in Australia forms a transcontinental Asia-Oceania VLBI network at centimetre and long millimetre wavelengths (Li et al. 2018). The Korean and Japanese telescopes have been upgrading their millimetre-wavelength equipment to enhance the observational capability of the global mm-wavelength VLBI network, and especially the EHT, that has recently imaged the event horizon of the central supermassive black hole in our Galaxy.

### A.1.5 VLBA

The Very Long Baseline Array (VLBA; Napier et al. 1994) was inaugurated on Aug 20, 1993 as a stand-alone dedicated VLBI array. In the 25 years since, many aspects of the VLBA have been upgraded, including installation of a $\lambda 3$ mm receiver, widebanding of the $\lambda 6$ cm receiver to span the entire 4 to 8 GHz octave, and bandwidth increase from 16 MHz to 256 MHz per polarisation, with a further doubling to be complete by early 2019. The VLBA's dedicated hardware correlator was upgraded to a software correlator based on the DiFX software correlator (Deller et al. 2007), and shortly thereafter DiFX2 (Deller et al. 2011). DiFX2 supports geodetic processing, pulsar binning, massive-multi-phase-centre correlation (enabling survey observations), and frequency matching capability (allowing flexible correlation of antennas observing with mismatched frequency bands).

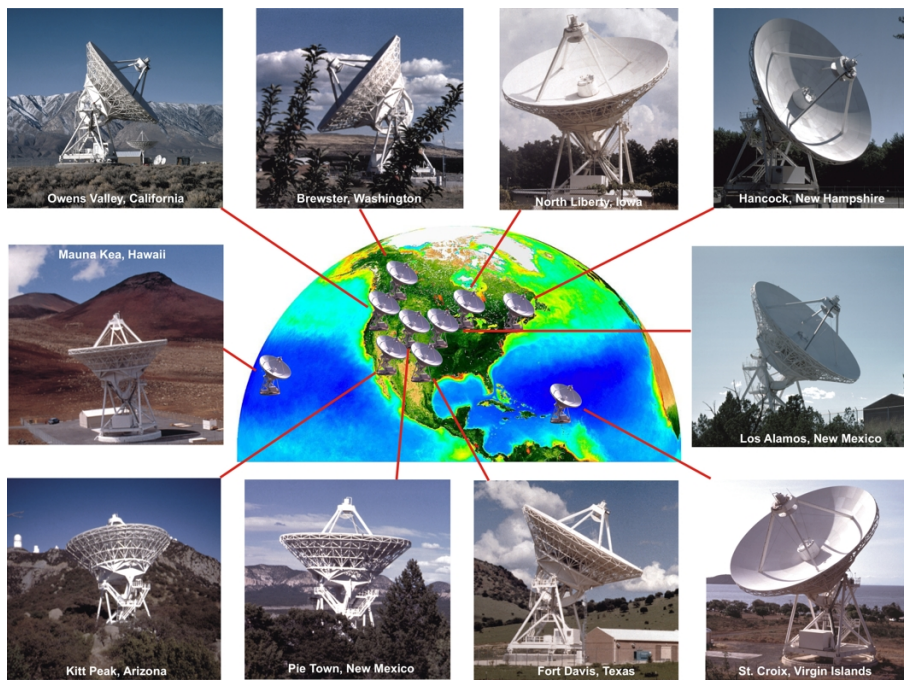

The VLBA consists of 10 identical 25 m dishes spread across the continental US, Hawaii, and US Virgin Islands and is operated by NRAO from Socorro, NM. Image: AUI/NRAO

The VLBA routinely observes as part of a larger network. Four such networks are, the IVS), the EVN, the GMVA, which includes ALMA, and the HSA. The HSA is coordinated by VLBA staff and includes the VLBA, Effelsberg, Greenbank, JVLA, and Arecibo.

To illustrate some of the key capabilities of the VLBA, three recent results are highlighted here. Excellent astrometric precision over 12 years demonstrated relative motion between two black holes (Bansal et al. 2017). The results confirm the existence of supermassive binary black holes with a separation small enough to imply coalescence; objects such as this are likely targets for low frequency gravitational wave detectors such as *LISA*. VLBA relative astrometry of a water maser has reached a precision suitable for direct parallax measurements directly across the Milky Way (Sanna et al. 2017). This particular result was preceded by numerous other VLBA measurements of methanol maser positions and velocities, allowing the creation of a map of the galaxy, demonstrating a barred-spiral plan and refining the Oort constants in the process (Reid et al. 2014). The scheduling and frequency agility of the HSA, combined with imaging and astrometric performance, demonstrated that the radio emission from neutron star merger event GW 170817 arose from a relativistic jet (Mooley et al. 2018). Jet parameters made from these measurements provide substantial evidence that short gamma-ray bursts are indeed neutron star mergers.

New VLBA capabilities will continue to be developed in the 2020s. A VLBA techical roadmap is under development which outlines a series of upgrades that will transform the instrument. The roadmap includes:

- Wider bandwidths: The goal is to increase bandwidth by another factor of 16, bringing it to 4 GHz per polarisation, or a factor of four in raw sensitivity.
- New and upgraded receivers: A top priority is a Ka-band receiver, operating in the 27 GHz to 40 GHz range. If deployed, this receiver would include a dual-frequency capability, allowing simultaneous observation at X-band (8 GHz to 9 GHz) and Ka-band.
- Real-time correlation: Options for deployment of high-speed fiber-optic networks to each VLBA station are being explored. In addition to offering new user capability, such an option would simplify VLBA operations and would increase its interoperability with other VLBI arrays around the world.

### A.1.6 LBA

The Long Baseline Array is the only astronomical VLBI network in the Southern Hemisphere and is operated as a National Facility by CSIRO Astronomy and Space Science in collaboration with the University of Tasmania, Auckland University of Technology and Hartebeesthoek Observatory. The LBA observes about 30 days per year with time being awarded by the Australia Telescope National Facility Time Allocation Committee (ATNF TAC) based purely on scientific merit and applications are open to all.

Comprising a core of 7 stations (ATCA, Parkes, Mopra, Hobart, Ceduna, Warkworth and Hartebeesthoek) supplemented by up to 3 more at certain frequencies and with some restrictions on availability (Katherine, Yarragadee, Tidbinbilla), the LBA also frequently operates in conjunction with telescopes of the EAVN and EVN (in particular the co-longitudinal Kunming and Tianma telescopes in China). A joint application procedure with the EVN already exists, with plans to unify this further so that only a single time request and TAC review will be required in order to secure resources on both arrays, in a manner analogous to the system currently employed for the Global VLBI Network.

The LBA is heterogeneous in nature with telescopes ranging from 12 m to 64 m and including one phased array. The inclusion of the Parkes 64 m and ATCA phased array ($5 \times 22$ m), means that the array has good sensitivity, especially at frequencies below 10 GHz. Data rates up to 1 Gbps are currently supported with plans to increase this to 4 Gbps in the near future. Standard observing bands are 1.4, 1.8, 2.3, 5.4, 6.6, 8.4 and 22 GHz, but observations up to 34 GHz are possible with a subset of the array. The ATCA is operational up to a maximum of 115 GHz and is occasionally used in conjunction with the KVN or other high-frequency telescopes at frequencies above 35 GHz.

The Southern Hemisphere location of the LBA provides it with a special niche for observing southern sources in general, with particular advantages for high precision astrometry of objects in the region of the Galactic centre and Magellanic clouds (e.g. Krishnan et al. 2017; Miller-Jones et al. 2018). LBA telescope positions are known to sub-centimetre accuracy (many participate in the IVS Geodetic network) and the high quality timing systems available allow for very accurate astrometry. Although only a part-time array, observing sessions are planned with consideration for the requirements of ongoing parallax campaigns. Target of opportunity and rapid response observations are also supported on a best-efforts basis.

LBA data are correlated on the Magnus supercomputer at the Pawsey Centre for SKA Supercomputing using DiFX2 (providing essentially all the capabilities described in the VLBA section above). The 1500 nodes available on this shared resource provide an essentially unlimited correlator resource allowing extremely computationally demanding correlation modes to be supported.

Efforts are currently under way to enhance the sensitivity of the array. The newly deployed

Ultra Wideband Low (UWB-L) receiver at Parkes will have a native VLBI mode enabled in its GPU post-processing backend, capable of providing up to 2 GHz of bandwidth in VDIF format at frequencies between 700 MHz and 4 GHz. A planned Ultra Wideband High (UWB-H) receiver will make up to 4 GHz of bandwidth availabale between 4 and 25 GHz (expected on a timescale of a few years). The ATCA is also planned to receive an upgraded GPU correlator which will include a tied array capable of delivering up to 2 GHz of bandwidth in any of the ATCA's available receiver bands (this is almost continuous from 1 to 115 GHz). Telescopes operated by University of Tasmania and Auckland University of Technology are increasing their potential recording bandwidth with the deployment of DBBC3 and Flexbuff recorders, with 4+ Gbps likely to be available in the near future. A currently unfunded, but eagerly anticipated, enhancement is the development of a tied-array system at ASKAP which will create a 36×12 m VLBI element in Western Australia (some 5000 km from Parkes) at frequencies between 700 MHz and 1.6 GHz.

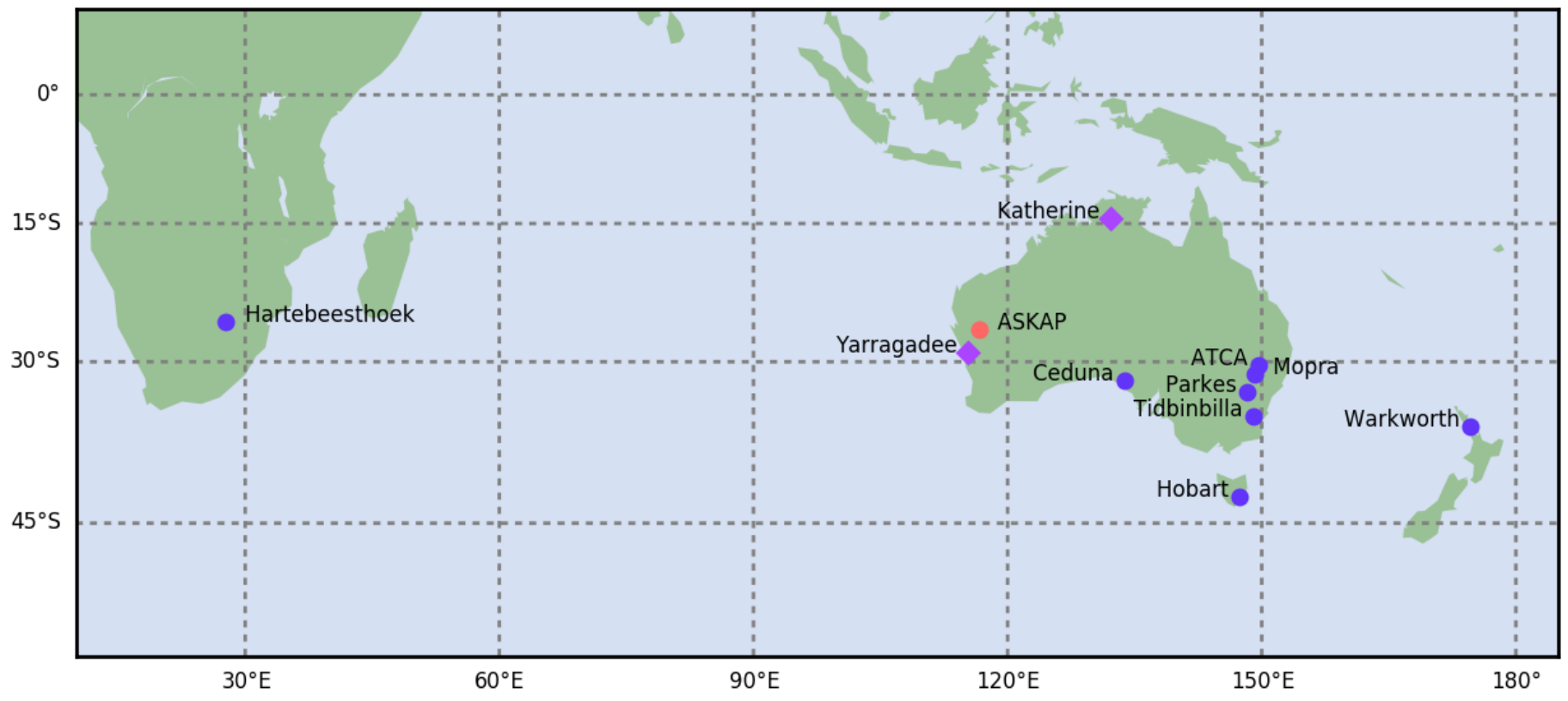

The locations of the LBA stations, distributed across the Southen Hemisphere from South Africa, through Australia to New Zealand. Image: `https://www.atnf.csiro.au/`.

## A.1.7 The Global mm-VLBI Array and the Event Horizon Telescope

The Global Millimeter VLBI Array (GMVA) combines the European telescopes capable of 3mm-VLBI, the VLBA, the GBT, the KVN, the GLT, and beam formed ALMA; additionally 7-mm observations are performed for the equipped telescopes, with the addition of the Noto telescope. The GMVA operates under a Memorandum of Understanding between the partner institutions (see https://www3.mpifr-bonn.mpg.de/div/VLBI/globalmm/mou.html), and has also developed an ALMA concept (completed in 2016) to observe jointly after the ALMA Phasing Project provided green light for joint observations, which are now possible since 2017. Data are post-processed at the DiFX correlator cluster at the Max-Planck-Institut für Radioastronomie. The GMVA offers 3-4 times more sensitivity and a factor of 2 higher angular resolution than the stand-alone VLBA or the HSA. For logistical reasons GMVA observations cannot be "dynamically" scheduled and are scheduled in time blocks, which combine the individual proposals and optimise the use of the available observing time and recording media. The GMVA block observations are scheduled in special observing sessions, performed twice per year, typically in spring (April, May) and autumn (September, October). The actual duration of each session depends on proposal pressure and ranges between 2 and 5 days. Proposal submission for the GMVA is synchronised with the VLBA. Proposals are reviewed by the NRAO and the programme/time allocation committees of the participating observatories. Recent scientific highlights of the GMVA are the studies of the jet in M 87 paving the way for the Event

Horizon Telescope (EHT) (Kim et al. 2018) and the best imaging results to date of the Galactic Centre (Issaoun et al. 2019).

The EHT is an experimental VLBI array that comprises millimetre- and submillimetre-wavelength telescopes spread across the globe. The nominal EHT angular resolution is 25 $\mu$as at 1.3 mm observing wavelength. This is sufficient to resolve the two most nearby supermassive black hole candidates on spatial and temporal scales that correspond to their event horizons. The EHT scientific goals therefore are to probe general relativistic effects in the strong-field regime, and to study accretion and relativistic jet formation on event horizon scales. The key developments that have facilitated the robust extension of the VLBI technique to EHT observing wavelengths were high-bandwidth digital systems that process data at rates of 64 gigabit/s, exceeding those of currently operating cm-wavelength VLBI arrays by more than an order of magnitude, development of phasing systems at array facilities, new receiver installation at several sites, and the deployment of hydrogen maser frequency standards to ensure coherent data capture across the array. The recent publication of the first image of the shadow of a black hole in the heart of Messier 87 is one of the scientific highlights of VLBI science in the recent years (EHT Collaboration et al., 2019a, 2019b, 2019c, 2019d, 2019e, and 2019f).

### A.1.8 JVLA and ngVLA

The Jansky Very Large Array (JVLA) is a connected-element radio interferometer array of twenty seven 25 m antennas operating at frequencies between 74 MHz and 50 GHz. While first available for science in the early 1980s, it continues to deliver spectacular scientific results, with users showing increased interest in triggered observations and astronomical transients, such as the localisation of the repeating FRB 121102 (Chatterjee et al. 2017; Marcote et al. 2017) and the detection of the radio afterglow of the neutron star merger, GW 170817 (Alexander et al. 2017). A phased-array sum can be formed and recorded as a VLBI element with an equivalent diameter of 130 m.

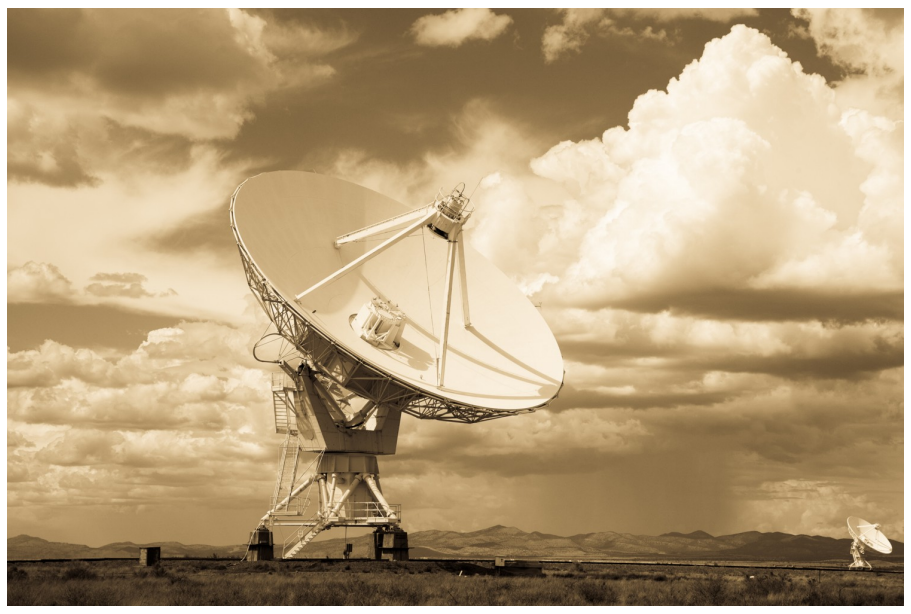

Two VLA antennas on the Plains of San Agustin, New Mexico. Image: AUI/NRAO.

Three major initiatives are underway at the VLA. The first is the VLA Sky Survey (VLASS), which is the highest resolution survey ever undertaken of the radio sky. The survey is being conducted at a frequency of 3 GHz in the B-configuration, giving an angular resolution of 2.5 arcseconds. The survey began in September 2017 and will be carried out in three epochs over seven years. It will use $\sim$ 5500 hours of VLA observing time. The survey will identify numerous sources worthy of follow-up with VLBI observations. Additional details on VLASS can be found at https://science.nrao.edu/science/surveys/vlass (Lacy et al. 2019). The second initiative is a programme to improve the 40+ year old infrastructure at the VLA site. The programme entails replacing heavy vehicles, track maintenance equipment, rail track, machine shop mills and lathes, building roofs, and the overhead crane in the Antenna Assembly Building. The programme's main objective in 2018 is to replace the VLA's electrical switchgear and backup generator. The third initiative is the development of a scientific and technical concept for a next generation VLA (ngVLA) for the US Astro2020 Decadal Survey. The ngVLA (ngvla.nrao.edu ) is envisioned to have 10 times the sensitivity and 10 times the angular resolution of the VLA. It will be located in the southwest US centred on the present location of the VLA, and operate over a frequency range of 1.2-116 GHz. The ngVLA concept includes a long baseline component ($\sim$ 1000 km), and incorporates transformative technology relevant to VLBI, such as LO and time distribution, wideband feeds, and economic cryogenic systems. If the concept is endorsed at the U.S. Decadal Survey, we envision a detailed design and development phase in 2020-2024 followed by a construction phase in 2025-2034.

Community enhancements of the VLA also continue to be made. In collaboration with NRAO, the US Naval Research Laboratory (NRL) expanded its 360 MHz VLITE commensal observing system (Clark et al. 2016) from 10 to 16 VLA antennas. The University of California Berkeley is leading the installation of realfast (Law et al. 2018), a commensal, real time, transient detection system on the VLA. The University of New Mexico is leading the eLWA project, which connects its 74 MHz Long Wavelength Array (LWA; Taylor et al. 2012) with the VLA.

### A.1.9 LOFAR

The LOw Frequency ARray (LOFAR; van Haarlem et al. 2013) is an interferometric array of dipole antenna stations centred in the Netherlands and scattered throughout Europe. LOFAR operates from the ionospheric cutoff of the "radio window" near 10 MHz up to 240 MHz. In total, 51 stations are currently operational. 38 of them are located in the Netherlands and spread out from a core near the village of Exloo, in the northeast of the country. Additionally, the system includes 13 international stations, which are located in the United Kingdom (1), France (1), Sweden (1), Germany (6), Poland (3), and Ireland (1). To date, funding has been secured to build additional stations in Latvia (1) and Italy (1) in the next few years.

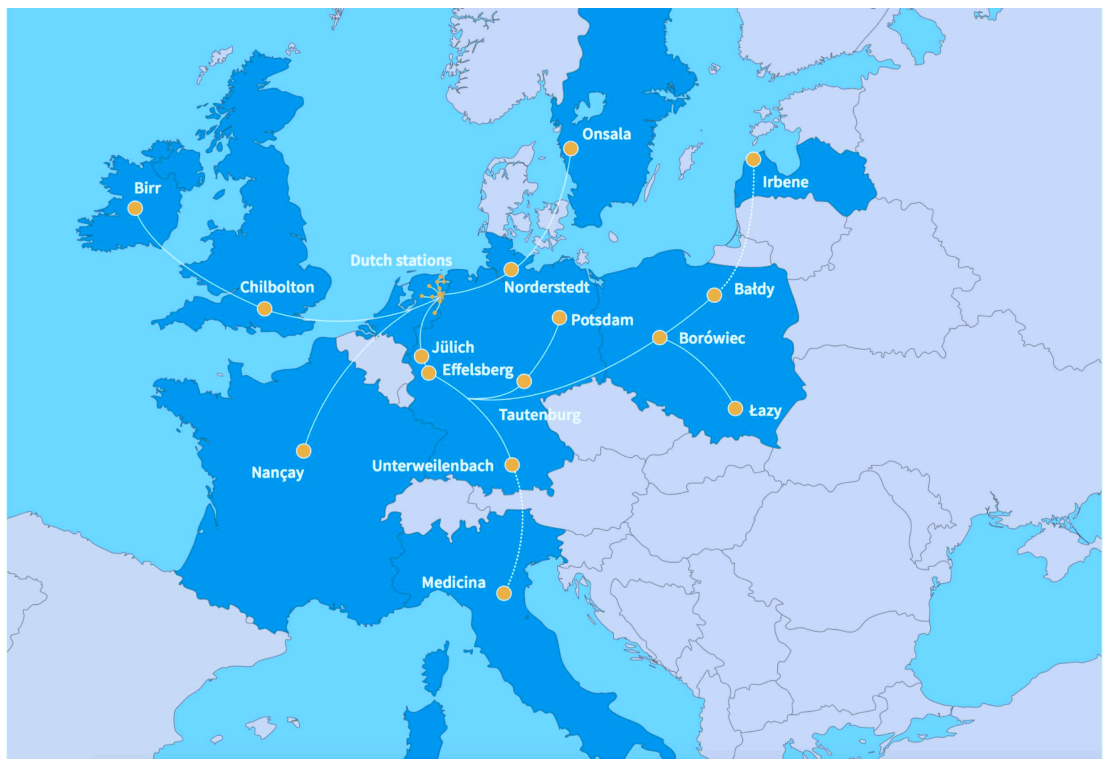

The layout of LOFAR, comprising 51 operational stations and the soon-to-be-built stations in Latvia (near Irbene) and Italy (near Medicina).

To cover the very broad frequency window, the system adopts different and relatively low-cost antenna designs. Specifically, the Low Band Antennas (or LBAs) cover the range between 10 to 90 MHz, while the High Band Antennas (or HBAs) span 110 to 240 MHz. HBAs and LBAs are grouped together to form a LOFAR station.

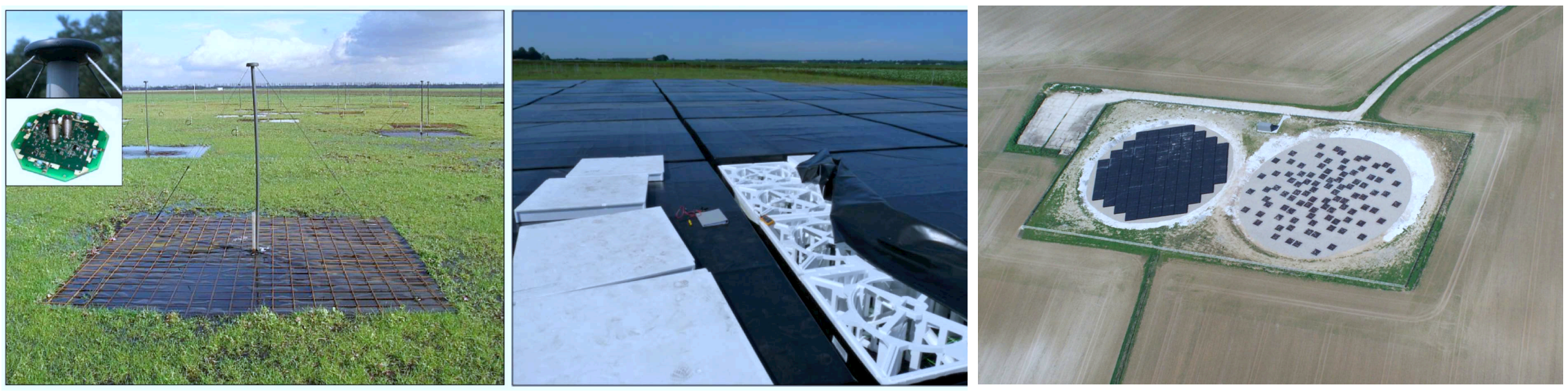

*Left*: a LOFAR LBA dipole. The inset images show the molded cap containing the Low Noise Amplifier electronics as well as the wire attachment points. *Middle*: a LOFAR HBA tile, clustering 16 antenna elements together. *Right*: layout of a LOFAR station (credit: G. B. Gratton/STFC).

LOFAR stations have no moving parts and, due to the effectively all-sky coverage of the component dipoles, the system has a large field of view. At station level, the signals from individual dipoles are combined digitally into a phased array. Electronic beamforming techniques make the system agile and allow for rapid repointing of the telescope as well as the simultaneous observation of multiple, independent areas of the sky using the available 96 MHz instantaneous bandwidth.

Before the construction of LOFAR, subarcsecond imaging was possible only down to 325 MHz. With LOFAR, this has been enabled for the first time at frequencies below 300 MHz. With a maximum baseline of approximately 1980 km, angular resolutions of $\sim 250$ mas are possible at frequencies around 150 MHz, enabling a variety of astronomical applications. These include measuring the angular broadening of galactic objects due to interstellar scattering, spatial localisation of low-frequency emission identified in low-resolution observations, or studying the evolution of black holes throughout the Universe through high-resolution low-frequency surveys.

LOFAR is running production observing since 2012 and is using transformational technologies and novel software approaches to deliver unique data to the community with increasing observing efficiency approaching 70%. Efforts are already ongoing to upgrade the current system to a new version: LOFAR 2.0. This will enhance the imaging capabilities of the system at LBA frequencies and provide a crucial data set for astronomy, which will remain the state-of-the-art for at least the next 20 years. Many of the scientific deliverables of LOFAR 2.0 will come from a 10 to 90 MHz all-northern-sky survey, which will be over 100 times more sensitive (reaching $1\sigma$ sensitivities of 1 mJy beam$^{-1}$ at 60 MHz and 5 mJy beam$^{-1}$ at 30 MHz) and will have a more than $5\times$ higher resolution (15 arcsecond at 60 MHz) compared to any previous or planned survey at these exceptionally low frequencies. The new system will address a broad range of scientific topics, such as (i) the formation and evolution of the earliest massive galaxies, black holes, and protoclusters, (ii) the nature of galaxy clusters and the steep-spectrum sources therein, including the influence of magnetic fields, shocks and turbulence, (iii) the Milky Way galaxy, including the topology of its magnetic field, (iv) exoplanets and their magnetospheric properties, (v) the composition of high-energy cosmic rays, and (vi) the structure and properties of the ionosphere. With an angular resolution over $10\times$ higher than the proposed low-frequency array for the SKA, SKA1-LOW, and also accessing the largely unexplored spectral window below 50 MHz, LOFAR 2.0 will continue to be a unique instrument through the next two decades.

### A.1.10 uGMRT

The GMRT (Swarup et al. 1991) is one of the largest and most sensitive fully operational low frequency radio telescopes in the world today that employs a full open skies policy. The array was commissioned in 2001. It consists of 30 antennas (each of 45 m diameter) spanning over 25 km, provides a total collecting area of $\sim$30,000 m$^2$ at metre wavelengths with arcsecs-scale angular resolution. It nicely bridges the VLA and LOFAR in frequency at comparable angular resolution. In addition to the regular Earth rotation aperture synthesis mode, the GMRT provides incoherent and phased-array beamformer options, which allow for high quality observations of compact objects like pulsars. This phased-array mode can be formed and recorded as a global-VLBI station with an equivalent diameter of $\sim$250 m.

Shortly after comissioning of GMRT, new facilities such as LOFAR (e.g. van Haarlem et al. 2013), MWA (e.g. Tingay et al. 2013), LWA (e.g. Ellingson et al 2013) were conceived and started becoming operational. Keeping in mind the growth of low frequency radio astronomy in the world, and learning from our own efforts and experiences of building and using the GMRT, a plan to upgrade the GMRT was proposed during the period 2007–2012 and the first serious work in this direction was initiated around 2010. The main goal was to add extra capability to the existing GMRT array in terms of frequency coverage and sensitivity, which would allow to open new windows of research in astrophysics and the study of the Universe.

Therefore, following are the key aspects of the upgraded GMRT (uGMRT):

1. Nearly seamless frequency coverage from 50 MHz to 1,500 MHz;
2. maximum instantaneous bandwidth of 400 MHz;
3. improved receiver systems with higher $G/T_{sys}$ and better dynamic range;
4. versatile digital backend correlator and pulsar receiver catering to the 400 MHz bandwidth;
5. revamped, modern servo system;
6. sophisticated next generation monitor and control system; and
7. matching improvements in mechanical systems, electrical and civil infrastructure and computing resources.

In addition to several new and interesting results in a wide range of topics in astrophysics, one area where the GMRT has made a difference is that of low frequency all sky surveys. The 150 MHz survey, the TGSS carried out with the GMRT (Intema et al. 2016), providing a sensitivity complementary to that of the NVSS at 1,400 MHz. To illustrate some of the new science capabilities of the uGMRT, a few science cases, including new recent results are highlighted below.
*Spectral line science:* The increased frequency coverage of the uGMRT allows searches for redshifted HI 21 cm absorption from damped Lyman-$\alpha$ absorbers and neutral hydrogen associated with active galactic nuclei out to the highest redshifts at which these systems have been discovered (see also Kanekar 2014, for new detections of redshifted HI 21 cm absorption in two high-$z$ galaxies).
*Continuum imaging science:* The improved sensitivity for continuum imaging with large bandwidths has great potential for new science with the uGMRT. A good sampling of the *uv*-plane over a large range of angular scales is necessary to image diffuse, extended, low-surface brightness emission. The large fractional bandwidths of the uGMRT bands provide excellent *uv*-coverage to map such diffuse structures. This together with the improved sensitivity and arcsecond-scale resolution will open up several areas of study, going from galaxy clusters to high-$z$ steep spectrum sources, from Galactic Plane studies to the Transient Universe and pulsar studies.

One of the major threat faced at low frequency is of RFI. At the observatory, this has required a multi-pronged approach, in particular,

- to reduce the RFI – coordinating with mobile phone operators, power utilities agencies, etc.,
- to block the strongest RFI – accomplished via feed designs having appropriate cut-offs in frequency with notch filters in the receiver,
- to filter RFI in the digital domain – schemes are implemented after the digitisation stage in the digital back-end for the signal from each antenna at very high time resolution (Buch et al. 2016), and
- to avoid the RFI in space and time – locations and trajectories of satellites having transmissions in the observing band of uGMRT are determined along with the angular separation of the satellite from the beam of the uGMRT antenna, a.k.a. ‘zone of avoidance’ thereby avoid non-linear effects due to RFI in the data.

In summary, the upgraded GMRT, which promises to open exciting new windows on the low frequency radio Universe, is now a reality available to the global astronomy community as a competitive facility.

### A.1.11 ALMA

The Atacama Large Millimeter/submillimeter Array (ALMA) is a connected element interferometer consisting of the 12-m Array, made up of fifty 12 m diameter antennas, plus the Atacama Compact Array (ACA) made up of twelve 7-m antennas packed closely together (the 7-m Array) and four 12-m antennas (the Total Power or TP Array). ALMA is a complete imaging and spectroscopic instrument operating at wavelengths of 3 to 0.3 millimeter, capable of doing polarimetry, mosaics, and combining the data from the various arrays (12-m Array, 7-m Array, TP Array) to examine the observed source structure on various spatial scales. The 12m-array antennas are movable and can be arranged into a number of configuration permitting baselines lengths from 0.16 km to 16 km.

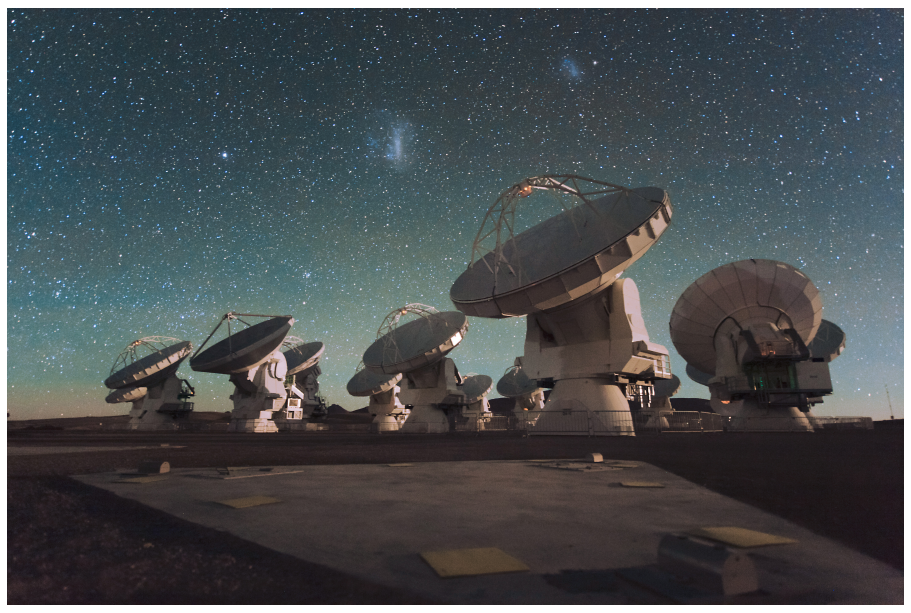

The night sky over ALMA, located on the Chajnantor Plateau in the Chilean Andes. Image: ESO/C. Malin

Unlike most radio telescopes, the ALMA antennas are located at a roughly 5000 km altitude on the Llano de Chajnantor in northern Chile, one of the driest locations on Earth. The atmospheric transparency and stability at sub-mm wavelengths are essential for ALMA, and combined with the large number of antennas and wide bandwidth, make the ALMA interferometer the most sensitive mm observatory on Earth.

To enable the integration of phased ALMA in VLBI networks, the ALMA beamformer was developed within the ALMA Phasing project (Matthews et al. 2018) and new calibration strategies designed (Goddi et al. 2019). The aggregated collection area of phased ALMA, when all 50 12-m antennas are considered, is equivalent to a single dish with $\sim$84 m diameter, boosting the signal-to-noise ratio of VLBI baselines to ALMA. Starting from Cycle 4, ALMA VLBI observing mode has been offered in conjunction with the GMVA (86 GHz) and EHT (230 GHz, and possibly 345 GHz in the future; see A.1.7).

The plan for future developments of ALMA has been published in 2018, in the ALMA Development Roadmap [1].

ALMA Array operations are the responsibility of the Joint ALMA Observatory (JAO), while the telescope itself is located at the Array Operations Site operated from the Operations Support Facility. The JAO has a central office in Santiago (Chile). The interface between the observatory and the global astronomical community is through the ALMA Regional Centers (ARCs) spread across East Asia, Europe and North-America. The proposal submission deadline is once a year, in April.

Aside from phased-ALMA, the ALMA data are an excellent complement to VLBI data. ALMA can reach tens of mas angular resolution, comparable to the short baselines in the EVN. Contrasting with VLBI, ALMA can observed thermal spectral lines and thermal continuum emission. For example, 6.7 GHz methanol and water masers have been combined with ALMA $CH_3CN$ and $CH_3OH$ maps to study the disk and jet system around a 10 $M_\odot$ protostar (Moscadelli et al. 2019).

[1] http://www.eso.org/sci/facilities/alma/announcements/20180712-alma-development-roadmap.pdf

## A.1.12 Space VLBI missions

The radio astronomy technique of VLBI involves simultaneous observations of the same radio source by multiple widely separated telescopes. Data collected from each of these telescopes is correlated to produce a resultant radio image. In ‘traditional’ VLBI, radio telescopes are distributed across the globe. The longer the distance between the observing telescopes – the baseline – the higher the resolution of the eventual image. Consequently, one of the limiting factors of traditional VLBI is the baseline that can be achieved between telescopes on Earth.

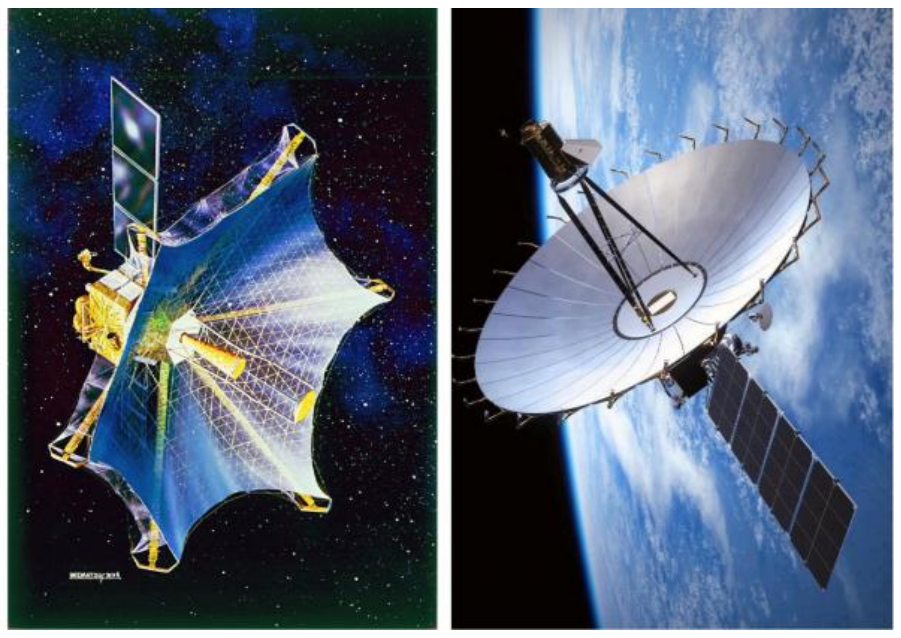

Left: (*HALCA*) of the *VSOP* programme (JAXA). Right: *Spektr-R* spacecraft (Lavochkin Scientific and Production Association).

In 1997, the Institute of Space and Astronautical Science (ISAS) in Japan led an international collaboration, which placed the space-borne radio telescope HALCA in orbit. This was part of the VLBI Space Observatory Programme (*VSOP*) - the first dedicated Space VLBI (SVLBI) mission - which operated until 2003. In 2011, the next dedicated space VLBI mission *RadioAstron*, led by the Astro Space Center of Lebedev Physical Institute and Lavochkin Science and Production Association in Russia, advanced space VLBI by offering baselines comparable with the distance to the Moon. In 2019, *RadioAstron* completed its in-orbit operations. The *VSOP* and *RadioAstron* missions constituted the first generation of dedicated space-based VLBI instruments.

While the comprehensive lessons learned from the first demonstration experiment and two dedicated space VLBI missions are still awaiting thorough attention, several preliminary conclusions are summarised in Gurvits (2020b). A broad picture of the current state of affairs and prospects of high-resolution space-borne radio astronomy are presented in twenty papers (and references therein) in the Special Issue of Advances in Space Research (Gurvits, 2020a). Of special interest to the overall development in high-resolution radio astronomy is a highly synergistic to the EVN and global VLBI the concept of TeraHertz Exploration and Zooming-in for Astrophysics (THEZA, Gurvits et al. 2019).

## A.2 The current technological framework for the EVN and prospects for its development

The last EVN Vision Document titled “The future of the European VLBI Network” (EVN2015) was released in 2007 and contained a number of recommendations related to technological improvements of the array and the correlator, based on requirements that were derived from the science goals. These recommendations mainly focused on wider bandwidths (and thus higher bitrates), more observing frequencies, more participating telescopes, RFI mitigation, the inclusion of the *e*-MERLIN baselines, higher speed real-time observations, and active preparation for operations in the SKA era. A number of these have been realised through the past decade. In this section we will focus on the technological developments achieved since 2007 and on the prospects for the forthcoming decade.

### A.2.1 Current status

To define our future goals we will first describe the current status of the technological areas, referred to in the previous EVN science document.

- New **telescopes** have become members of the EVN since 2007. The previous document listed 7 potential candidates for the EVN of which Latvia (32 m), Yebes (40m), SRT (64 m) and Kunming (40 m) are already fully operational at the observing frequencies for which they have receivers. Moreover, the three Russian 30 m Quasar VLBI Network telescopes, Svetloe, Zelenchukskaya and Badary, joined the EVN, as did the Korean VLBI Network (KVN), composed of three 21m radio telescopes (Yonsei, Ulsan and Tamna). On the other hand, the phased-up WSRT was lost to the EVN with the installation of the APERTIF front ends in most of its dishes, leaving just a single 25m dish for VLBI observations.
- Very high-bandwidth (more than 13 GHz) **receiver systems** for ~GHz frequencies are currently under development. This development is being funded by RadioNet, as part of the Work Package BRAND. Related developments have taken place within VGOS (VLBI Global Observing System). However, no wide bandwidth receiver below 15 GHz has been built yet.
- Several **back-ends** have been developed within the EVN in the past years. Although most of the stations operate DBBC2s there have been back-end developments by IAA (Russia) and ShAO (China). The Korean VLBI Network (KVN) and Jet Propulsion Laboratory (JPL) are also using new back-ends, while Haystack Observatory has continued its development in line with three different versions of RDBEs. On the other hand the DBBC3, the European back-end, is already a reality and some early versions have been delivered to some institutes and observatories. This world-wide activity has enlarged the knowledge of creating FPGA-based systems, has provided higher bandwidths and data rates in the EVN, and increased its operational reliability. Compatibility, however, may well become an issue.
- **Multi-frequency receivers** at 22, 43, 86 and 129 GHz are deployed in the KVN (South Korea). This multi-frequency receiver which allows observing four different frequencies simultaneously has proven extremely efficient for VLBI studies at frequencies above 20 GHz. Currently the 40 m Yebes radio telescope is the only telescope within the EVN equipped with such a receiver and, together with the KVN telescopes, capable of 22/43 GHz simultaneous observations. A compact prototype has already been developed to fit in telescopes with tight space constraints (Han et al. 2017).
- Internet connections at **multiples of 10 and 100 Gbps** have become standard in current research networks (see the GÉANT map topology of January 2018). There are still differences amongst countries within Europe and the highest speeds are not generally available outside of Europe. Some of the observatories within the EVN are still connected at multiples of **1 Gbps**.

- **Disk shipping** has become obsolete in a large part of the EVN, greatly simplifying observing and scheduling logistics. The current scheme consists of using local storage at the stations able to contain at least one complete EVN session, and deploying a similar amount of storage at the JIVE correlator, allowing the processing of one more EVN session. The observations are automatically transferred to the correlator via e-shipping.
- ***e*-MERLIN** telescopes have been successfully included in regular EVN operations. This inclusion provides short baselines that increase the fidelity of the images obtained with the EVN.
- The design of the **SKA** comes with high performance challenges which will lead to new technological approaches which might be applicable to VLBI data processing.
- The **VDIF data format** has been adopted widely making global observations far easier than previously. Scheduling and correlation have benefited from this adoption.
- **Software correlation** has delivered a number of important advantages like flexiblity and scalability, as well as many new features like multiple phase centres and mixed bandwidth observing. It has also enabled individual stations to make tests and check the quality of their data in near real time. Distributed correlation is possible in principle but not used in practice. Currently, the two major correlator software packages are DifX and SFXC. These have been compared extensively, both to the previous Mark4 hardware correlator and to each other, and found to produce identical results. These correlators mostly run on modestly sized CPU clusters, which can be easily extended for added computing power. GPU correlation has been investigated, but while GPU correlators are well suited for large-N networks, it is not entirely clear yet whether this holds for VLBI as well.

In summary, over the past years correlation capabilities and data transport have developed well, aided by several international projects that have provided funds. Further improving the sensitivity of the EVN array with the current telescopes is still possible, by increasing the observing bandwidth which is currently limited to 256 MHz.

### A.2.2 New telescopes: sensitivity and fidelity

Sensitivity is improved by increasing the number of radio telescopes, preferably with large collecting areas, and/or by increasing the observing bandwidth. Fidelity is achieved by increasing the number and variety of baselines to improve the coverage of the *uv*-plane.

#### Extending the EVN

A number of telescopes might join the EVN in the coming years. Three types of telescopes are considered.

- Newly built or to-be-built radio telescopes: FAST 500 m (China), QTT 110 m (China), NARIT 40m (Thailand), MeerKAT (South Africa) and the proposed 30-40 m radio telescope in the United Arab Emirates. The latter project is still in a conceptual phase.
- Refurbished telecommunication antennas: Goonhilly 26 m (UK), Usuda 64 m (Japan), Sao Miguel 32 m (Portugal), the Hellenic telescope 30 m (Greece), Ghana 32m (Ghana), Xi'An 40 m (China) and ROT 54 m (Armenia).
- Geodetic 13.2 m telescopes: two at Ny Ålesund, two at Wettzell, one at RAEGE Santa María and one at RAEGE Gran Canaria. The inclusion of these telescopes however will not be straightforward due to limited time availability.

Including and validating new telescopes, in particular the refurbished telecommunication dishes,

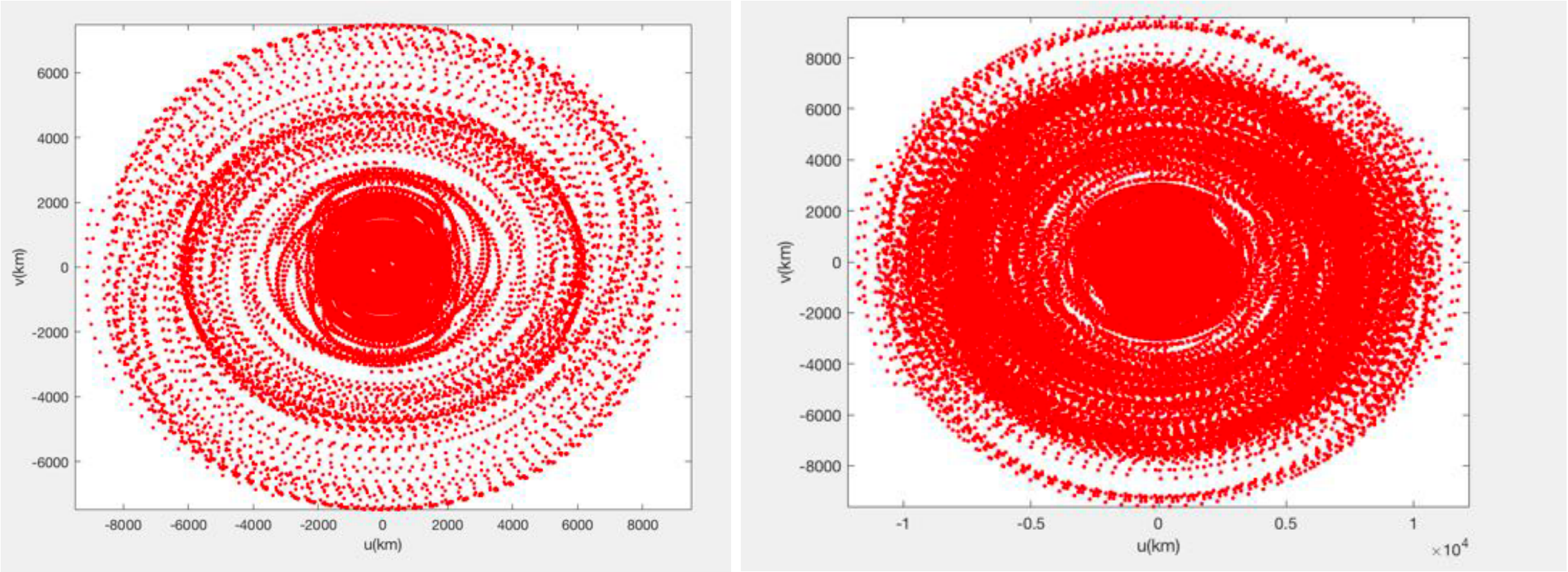

Figure A.1: Left: *uv*-coverage of EVN and *e*-MERLIN with 18 stations total. Right: global VLBI *uv*-coverage with 30 stations world-wide. In both cases for wide bandwidth continuum observations via using Multi-Frequency Synthesis remaining gaps in *uv*-coverage can be robustly filled.

may require the temporary deployment of VLBI back-ends at those sites. The EVN might consider providing such equipment for this purpose.

#### Extending the e-EVN

The real-time capacity of the EVN can be extended, by adding more telescopes observing at higher aggregate bit rates, going from the current 2 Gbps to 4 and 8 Gbps. Real-time VLBI has important advantages above recorded VLBI, enabling rapid response to transient events, as well as faster turn-over of scientific results. To increase its scientific relevance it will be crucial to add the *e*-MERLIN short baselines, the intermediate Quasar VLBI Network baselines filling the gap between the European and Chinese parts of the EVN, and the great sensitivity of the Sardinia and FAST telescopes. Although the transport of recorded data has been greatly simplified through the advent of FlexBuff recorders, the independence of recording media when doing e-VLBI still offers a tremendous advantage, especially at higher data rates.

Real-time data rates of 8 Gbps and above will obviously necessitate higher connectivity to JIVE, and a further upgrade of correlator capacity at JIVE. This could be in the form of upgrading the SFXC hardware, possibly including GPUs. Special observing modes require recording data during real-time sessions at JIVE, and recorrelation at a later time. This is done occasionally, but could be routinely available. Having sufficient recording capacity in the EVN would of course reduce this problem to doing regular e-VLBI in parallel to regular recorded VLBI.

### A.2.3 EVN-light and fast response

The concept of EVN-light, a sub-array of EVN stations, has been considered for many years, but has never been pursued. Such an array could be deployed in the gaps between the regular EVN sessions and would be very useful, for example, for monitoring campaigns (e.g. astrometry, transients) and other long-term projects.

The capability to quickly respond to transient events is becoming more and more important as new instruments come online. Although, depending on the type of trigger, as a rule as many telescopes as possible will be requested, the presence of an operational EVN-light would make follow-up observations at short notice far easier.

### A.2.4 (Much) Wider bandwidths

Wider bandwidth at the telescopes is one the most important goals to be accomplished by the EVN. It will provide both higher sensitivity and increased frequency coverage. Ideally all EVN telescopes would be equipped with low system temperature receivers between 1.4 GHz and 24 GHz; their number should be as low as possible to reduce maintenance efforts and minimise the space needed in the receiver cabins.

#### Spectral studies

Ultra-wideband receivers will revolutionise our ability to carry out spectral imaging, since it will become possible to obtain information about the intensity of the radio emission at all the frequencies covered by the full bandwidth simultaneously. This will not only be much more efficient and reliable than the earlier practice of combining a few narrow bands, but it will provide much more complete information about the spectra measured, thereby providing much more complete and reliable information about the emission and absorption mechanisms operating in a given region and the corresponding conditions in the radio source. The continuous frequency coverage will enable the identification of particular features in radio spectra and trends displayed by the curvature of the spectrum, rather than just a crude description using a spectral index derived between two fairly widely spaced frequencies.

#### Polarisation studies - Faraday Rotation mapping

The frequency dependence of the degree of polarisation and the polarisation angle carries a wealth of information about the conditions in the radio source and in its immediate vicinity. Faraday rotation of the linearly polarised waves – as they travel through an ionised medium – is a unique source of information about the magnetic fields and electron densities in thermal plasma in the immediate vicinity of the radio emitting regions. Ultra-wideband receivers will make EVN Faraday rotation studies possible with much more complete and extensive frequency coverage than can be currently carried out with any VLBI system. The technique of "Faraday RM synthesis" will allow us to separate inhomogeneities within the source and along the line of sight, that independently contribute to the Faraday rotation observed. This will have a major impact to the science cases described later in the document.

#### Possible realisation of ultra-wide bands

For geodetic VLBI, which needs wide instantaneous bandwidths to determine the group delay, a system has been developed that samples $4 \times 1$ GHz bands within a range of 2–14 GHz. The EHT currently operates at high frequencies producing a full 64 Gbps per telescope. The technology for wide bandwidth observing clearly is available right now.

The EVN BRAND project aims to develop and build a prototype broad-band digital receiver, with a frequency range from 1.5 GHz to 15.5 GHz, covering the C, M and X band, and opening up new unexplored bands.

Ultra-wide frequency coverage brings with it a set of problems that will need to be tackled. For example disk space availability, lower feed efficiency, correlation capacity, higher connectivity demands and saturation due to RFI. On the other hand, the greatly increased correlated bandwidth will compensate for the lower efficiency, while also providing the option to observe at several widely spaced frequencies. Cost for maintenance and energy for cooling will also be reduced by replacing multiple narrow-band receivers with one wide-band receiver.

The coverage for frequencies above 22 GHz may also be considered at telescopes with a reasonable aperture efficiency and low atmospheric opacity. As mentioned above there is already a de facto standard defined by the KVN which consists of a tri-band receiver simultaneously covering 22, 43 and 86 GHz. The lower frequencies help in determining the noise phase caused by the atmosphere and allow us to increase the integration time at higher frequencies.

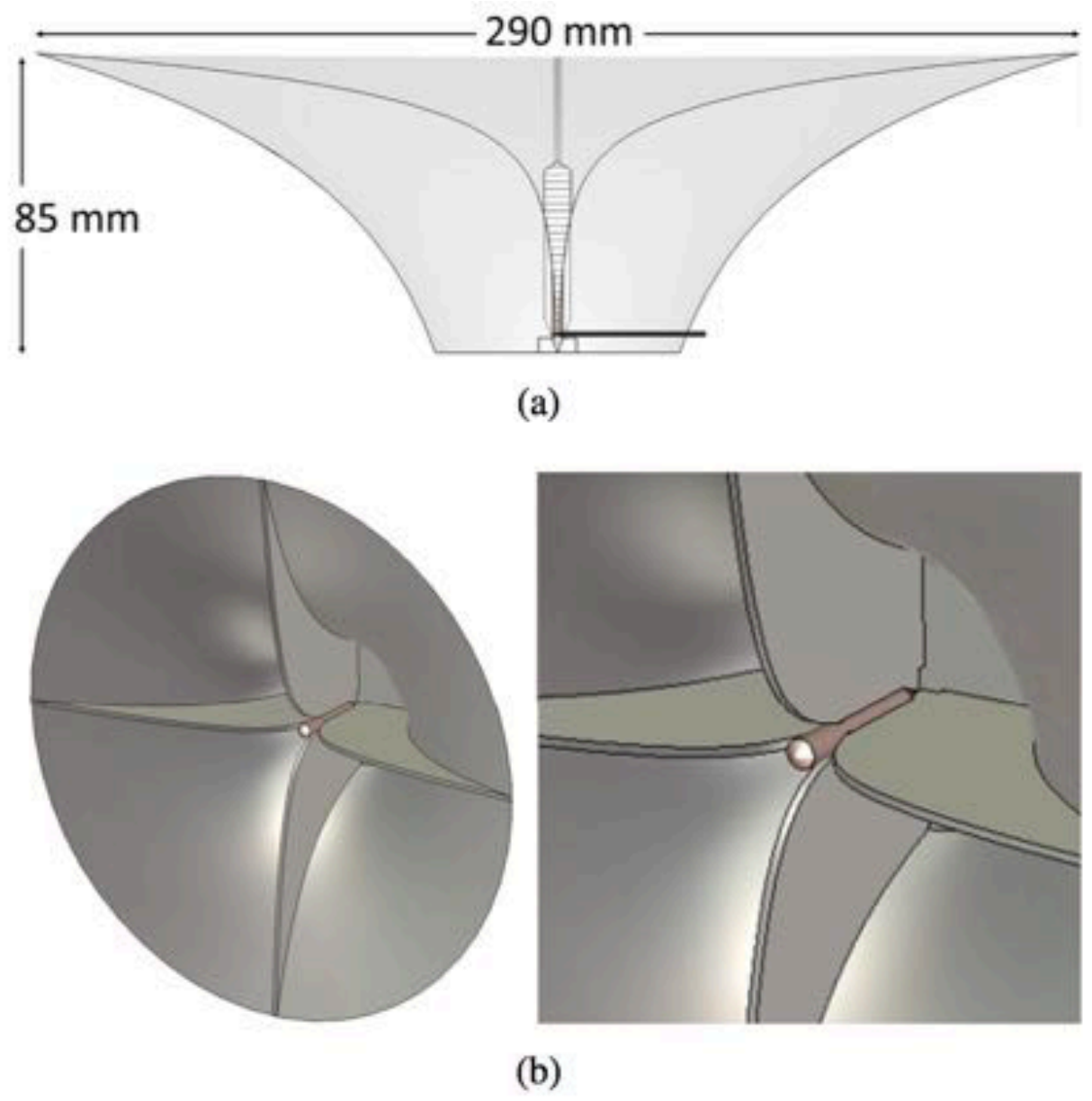

Figure A.2: Feed design for prime focus Effelsberg 100m (Flygare, Pantaleev & Olvhammar 2018). Quad-Ridge Flared Horn (QRFH) illustrated in a) cross-section and b) perspective view with zoom in on the dielectric load at the centre.

Whether the EVN decides to go with a 10:1 system like BRAND, or, like the ngVLA is considering, a suite of wide-band receivers in the 4:1 regime, will no doubt depend on performance, cost and specific science cases. However, it is clear that wide-band observing will be essential for the future of the EVN.

### A.2.5 Back-ends: higher rates and flexibility

Future back-ends are likely to be very flexible. There are already some developments towards a 'universal back-end', capable of doing VLBI, continuum, spectral and pulsar observations. These back-ends rely on FPGAs and ADC (Analogue to Digital Converters) with many bits and very high sampling rates. Different observing modes can be implemented by loading different types of firmware. As an example, currently, the Xilinx RFSoC provides an FPGA that includes 8 ADCs with 12 bits and 4 Gsamples per second.

This 'universal back-end' allows for the implementation of algorithms, like band separation, removal of RFI, conversion of linear to circular polarisation (although this can be done off-line as well), packaging and sampling of data into VDIF packets. VDIF packets can be stored for later processing or processed on the fly by other equipment connected via Ethernet.

Alternatively, one could try to separate the sampling from the rest of the processing, sample as much RF or IF as possible and, possibly after coarse channelisation and the distribution of the data stream to several computers, package the sampled data into VDIF packets. A fast ethernet switch would then connect the VDIF output streams to storage, or directly to various CPU/GPU clusters for different types of processing. In this way VLBI would simply become one of several modes of observing that one could switch to in a matter of minutes.

### A.2.6 Radio Frequency Interference

Radio Frequency Interference (RFI) is a severe global problem which will worsen in the future and which will require action from the EVN to mitigate its effects. Its impact on astronomical observations is a major concern for all observatories and may jeopardise the investments to achieve better sensitivity and wider frequency coverage. The RFI from space satellite services has become one of the most severe threats (e.g. Iridium) and requires an urgent global and coordinated response from the radio astronomy community. The plans of SpaceX on the deployment of a constellation of 42000 satellites orbiting the Earth[2] and the first launch of 60 satellites of the *Starlink* constellation in May 2019 highlights such urgency. Around 2000 more satellites are expected in the next 7 years and, according to the documentation sent to the US Federal Communications Commission, they will use nine frequency bands between 10 and 30 GHz jeopardising the usage of wide-band receivers compatible with SKA band 5 and higher frequencies.

Several strategies can be pursued to overcome the effects of RFI:

- Continuous monitoring at the EVN stations. Small antennas covering the frequency interval between 1 and 18 GHz and capable of moving in azimuth and elevation would be a necessary investment to identify the source and frequency of the RFI.
- The installation of High Temperature Superconductor filters prior to the cryogenic LNAs to avoid the saturation of wide band amplifiers.
- An increase of the number of bits to digitise the recorded signal to avoid saturation. Traditionally VLBI has used 2 bits but the presence of strong RFI signals advises the usage of 8 or 10 bits.
- Legal measurements to prohibit the usage of the electromagnetic spectrum in the vicinity of the observatories.
- A global policy defended by the CRAF.
- RFI flagging prior to correlation using advanced algorithms.

### A.2.7 Scheduling, monitoring and the Field System

VLBI depends to a high degree on the seamless interaction of many elements, spread around the globe. To make this technique possible at all, many specialised tools and methods have been developed over the years. While still functional, some of these tools are outdated, out of support and consequently very hard to upgrade to the new instruments and observational modes that are constantly being developed.

SCHED is the scheduling programme used by the EVN and several other VLBI networks. It was written entirely in Fortran, in the eighties. Although well written, structured and maintained, the person responsible for its maintenance has recently retired. Both the fact that Fortran as a language is not suited for this type of functionality, and that very few software engineers nowadays are comfortable programming in Fortran, make keeping SCHED up-to-date increasingly problematic.

To address this, a re-factoring of the SCHED software has started. In the new version, named pySCHED, all the existing functionality can still be accessed, albeit through a Python interface, while new functionality, written in Python, can be easily added. Some fairly essential new features have already been implemented, such as "VEX"2 support and the basic scheduling of DBBC recording. While still at an early stage, this development, making use of a language that has become the de facto standard for scientific programming, should ensure that VLBI scheduling will be able to keep up with future developments.

[2] https://mc03.manuscriptcentral.com/csb?URL_MASK=5159d9cca5bb40a99ad8bc1739cb141e, http://licensing.fcc.gov/myibfs/download.do?attachment_key=1158350

Although the advent of real-time VLBI has made it far easier to detect problems both at the stations and at the correlator, in a timely fashion, recorded VLBI is still very much needed for many observations. Consequently, monitoring of the network and all its elements remains a crucial part of VLBI observations.

As many VLBI networks are heterogeneous arrays, consisting of widely different telescopes and telescope control systems, designing one single monitoring system is not a trivial task, and more so if this monitoring is coupled to remote control, like in the case of some geodetic stations. Nevertheless, over the years quite a few different monitoring systems have been designed and deployed.

An effort is underway to produce a central, web-based monitoring system, usable for both astronomical and geodetic VLBI. As a first step, an extensive evaluation and rating has been undertaken of the currently existing monitoring systems, their suitability, ease of use, etc. Based on this evaluation a system is being designed that takes advantage of existing monitoring systems at various telescopes and is capable of accessing local databases. This will be followed by the actual integration of existing software into a central infrastructure. The end product should greatly enhance the ability to monitor the performance of the network, and with it, the quality of its data product.

The Field System (FS) is a key element at most VLBI telescopes, controlling the telescope itself, the backend and the recording system. As is the case for SCHED, it is legacy code, written in Fortran and C and developed more than 30 years ago. Its maintenance and development are currently done by NVI Inc, contracted to NASA. Developments are driven by requests from the geodetic and the astronomical communities and some members of these communities actively take part in the debugging of the code. On one specific occasion the EVN has funded the implementation of support for the two main recording modes of the DBBC2.

To ensure its sustainability and prevent a possible fragmentation of the code driven by local needs, the FS could be upgraded and modernised. This should be discussed and agreed upon by NASA, NVI, and technical representatives of the VLBI community as a whole. Modularisation could help to open up the code to contributions from developers from other institutes, and ease the inclusion of different VLBI back-ends. A common basic command syntax for VLBI back-ends and recorders would also greatly facilitate their deployment at different telescopes.

### A.2.8 Correlator

Software correlators have seen a tremendous development over the past decade. The JIVE-developed SFXC has gone from first fringes in 2007 to becoming a powerful, highly versatile, flexible and multi-functional backbone of the EVN network. Moore's law has easily kept up with increasing bit rates, and to this date (2020) the correlator capacity has never been a bottleneck in EVN operations.

Over the years, other solutions have been considered and developed, mainly driven by the assumption that data rates would grow at such a rate that CPU-based software correlators would quickly be overwhelmed, an assumption that did not materialise. However, with the advent of ultra-wide-band receivers this issue may have to be reconsidered.

One of these solutions was the JUC, the JIVE UniBoard Correlator, a Field Programmable Gate Array (FPGA)-based computing platform. Although powerful and energy-efficient, such platforms tend to take a very long time to develop and debug, while the implementation of firmware modifications equally is a very lengthy process. Simply put, they are highly suitable for the processing of large identical data streams, with identical data formats and rates. Many VLBI arrays are by their very nature inhomogeneous, deploying different equipment which moreover is constantly being upgraded. This, combined with their lack of flexibility, severely limits the usefulness of FPGAs in VLBI correlation.

Correlation on Graphics Processing Units (GPU) has been successfully implemented, an example being the LOFAR COBALT correlator. For VLBI arrays however, consisting of relatively few array elements (10 to 30 telescopes), GPUs do not seem to bring a substantial improvement. In 2018, during a hackathon in Perth, engineers from CSIRO, Swinburne and JIVE collaborated to port parts of the DIFX software correlator to a GPU platform, aided by a number of GPU experts. The final result was that the gain in correlation speed compared to a CPU of similar price, was roughly a factor of two. The main bottleneck for VLBI correlation seems to be the memory access between GPU and its memory, causing only 10 to 20% of compute resources to be actually used. It should be noted that GPUs are in constant development, and new versions may be better suited to the VLBI use case. A hybrid approach using both CPUs and GPUs may be the way of the future; it might be possible to run large-bandwidth continuum correlation on GPUs while reserving the more flexible CPUs for non-standard, complex and/or innovative correlation jobs.

Cloud correlation has finally been suggested as a solution to cope with the large data streams that geodetical and astronomical VLBI are, or will start, producing in the near future. Recently, this has been investigated by Gill et al. (2019). In this case, too, a hybrid approach may be optimal, for example by using HPC/cloud resources for occasional high-volume standard correlation jobs and an in-house correlator for experimental work. Whether such a solution will ever be economically/logistically feasible is not entirely clear, but it is certainly worth investigating in the coming years.

### A.2.9 Future improvements

In the following section, we list the improvements that are technically feasible on time scales of 5 and 10 years.

#### Five years

- Wide-band receivers on at least a sub-set of the EVN. What frequency range will be covered is unclear, and will no doubt partly depend on the performance and cost of BRAND, partly on specific science cases. Aiming for maximum sensitivity by using the full band will of course bring its own set of problems, in terms of availability of media, connectivity, correlation capacity. However, there will be fewer receivers to maintain, and multiple discontinuous bands can be observed simultaneously.
- At higher frequencies, multi-band receivers like the KVN quasi-optics system on a sub-set of the EVN. These are large systems that are not suitable for prime focus, but a compact prototype has been developed recently (Han et al. 2017).
- DBBC3 backends at all stations.
- RFI mitigation at the stations (in-built feature of DBBC3).
- Multiple 100G connectivity to JIVE.
- Inclusion of Quasar VLBI Network telescopes in real-time operations.
- Inclusion of Ghana, Xi'An, MeerKAT telescopes.
- Regular correlation of geodetic observations at JIVE.
- (semi-) Automated follow-up of transient events by sub-set of EVN stations.

- EVN-light observations, e.g. more regular observing sessions within a subset of EVN stations.
- Upgraded scheduling and common monitoring tools.

### Ten years

- 'Real' global VLBI, seamless cooperation of different networks.
- Global VLBI with the SKA.
- Deployment of 'Software Defined Radio' back-ends, to some degree doing away with the need for different hardware back-ends for VLBI, pulsar astronomy, etc. Such a system would isolate the digitisation step from the rest of the back-end, sampling the RF signal, or IF, and packaging the samples into VDIF packets. These VDIF packets can be stored for later processing or processed on the fly by any suitable piece of equipment, connected via Ethernet. Such a system is currently being considered at the VLBA.
- True wide-field VLBI, enabling high-resolution blind surveys.
- Upgraded software control tools for telescopes and backends.

## A.3 Spectral lines of interest to radio astronomy (∼<100 GHz)

| **Atom/molecule** | Line name | Rest frequency [GHz] |
|---|---|---|
| H I | neutral hydrogen | 1.420405752 |
| OH | hydroxyl radical | 1.6122310 |
| OH | hydroxyl radical | 1.6654018 |
| OH | hydroxyl radical | 1.6673590 |
| OH | hydroxyl radical | 1.7205300 |
| CH | methylidyne | 3.263794 |
| CH | methylidyne | 3.335481 |
| CH | methylidyne | 3.349193 |
| OH | hydroxyl radical | 4.660242 |
| OH | hydroxyl radical | 4.765562 |
| $H_2CO$ | ortho-formaldehyde | 4.829660 |
| OH | hydroxyl radical | 6.030747 |
| OH | hydroxyl radical | 6.035092 |
| $CH_3OH$ | methanol | 6.668518 |
| $HC_3N$ | cyanoacetylene | 9.098332 |
| $CH_3OH$ | methanol | 9.936202 |
| $CH_3OH$ | methanol | 12.178593 |
| OH | hydroxyl radical | 13.4414173 |
| $H_2CO$ | ortho-formaldehyde | 14.488490 |
| $C_3H_2$ | ortho-cyclopropenylidene | 18.343145 |
| $NH_3$ | para-ammonia | 18.499390 |
| $NH_3$ | ortho-ammonia | 19.757538 |
| $CH_3OH$ | methanol | 19.967415 |
| $NH_3$ | para-ammonia | 20.804830 |
| $NH_3$ | ortho-ammonia | 21.070739 |
| $H_2O$ | ortho-water | 22.23507985 |
| $NH_3$ | para-ammonia | 22.653022 |
| $CH_3OH$ | methanol | 23.121024 |
| $NH_3$ | para-ammonia | 23.694506 |
| $NH_3$ | para-ammonia | 23.722634 |
| $NH_3$ | ortho-ammonia | 23.870130 |
| $CH_3OH$ | methanol | 24.9287 |
| $CH_3OH$ | methanol | 24.933468 |
| $CH_3OH$ | methanol | 24.95908 |
| $CH_3OH$ | methanol | 25.018123 |
| $NH_3$ | ortho-ammonia | 25.056025 |

With help from Katharina Immer. Useful resources: `https://www.narrabri.atnf.csiro.au/observing/spectral.html` compiled by Bob Sault and Robin Wark. Also see Lovas, Snyder & Johnson “Recommended rest frequencies for observed interstellar molecular transitions” (1979), Wilson, Rohlfs, & Hüttemeister “Tools of radio astronomy” (1986), and Malcolm Gray “Maser sources in astrophysics” (2012). More molecules/transitions are available in the Lovas search engine: `https://physics.nist.gov/cgi-bin/micro/table5/start.pl`

| **Atom/molecule** | Line name | Rest frequency [GHz] |
|---|---|---|
| $CH_3OH$ | methanol | 25.124873 |
| $CH_3OH$ | methanol | 25.294411 |
| $CH_3OH$ | methanol | 25.54143 |
| $CH_3OH$ | methanol | 25.87818 |
| $CH_3OH$ | methanol | 26.84727 |
| $CH_3OH$ | methanol | 27.47258 |
| $CH_3OH$ | methanol | 28.16952 |
| $CH_3OH$ | methanol | 28.90585 |
| $CH_3OH$ | methanol | 28.9699 |
| $CH_3OH$ | methanol | 29.63711 |
| $CH_3OH$ | methanol | 30.30808 |
| $NH_3$ | ortho-ammonia | 31.424943 |
| $CH_3OH$ | methanol | 36.16924 |
| $CH_3OH$ | methanol | 37.70372 |
| $CH_3OH$ | methanol | 38.2935 |
| $CH_3OH$ | methanol | 38.4526 |
| SiO | silicon monoxide | 42.820570 |
| SiO | silicon monoxide | 43.122090 |
| SiO | silicon monoxide | 43.423853 |
| $CH_3OH$ | methanol | 44.06949 |
| CS | carbon monosulfide | 48.990955 |
| $DCO^+$ | deuterated formylium | 72.039331 |
| DCN | deuterated hydrogen cyanide | 72.413484 |
| $CH_3OH$ | methanol | 84.521169 |
| SiO | silicon monoxide | 85.640456 |
| SiO | silicon monoxide | 86.243442 |
| $CH_3OH$ | methanol | 86.615602 |
| $H^{13}CO^+$ | formylium | 86.754294 |
| SiO | silicon monoxide | 86.846998 |
| $CH_3OH$ | methanol | 86.9029479 |
| HCN | hydrogen cyanide | 88.631847 |
| $HCO^+$ | formylium | 89.188518 |
| HNC | hydrogen isocyanide | 90.663574 |
| $N_2H^+$ | diazenylium | 93.173809 |
| $CH_3OH$ | methanol | 95.16944 |
| $H_2O$ | para-water | 96.26116 |
| CS | carbon monosulfide | 97.980968 |
| $CH_3OH$ | methanol | 107.01385 |
| $CH_3OH$ | methanol | 108.893963 |
| $C^{18}O$ | carbon monoxide | 109.782182 |
| $^{13}CO$ | carbon monoxide | 110.20137 |
| CO * | carbon monoxide | 115.271203 |

* redshifted CO for $z > 1.67$ falls in the Q-band (43 GHz).

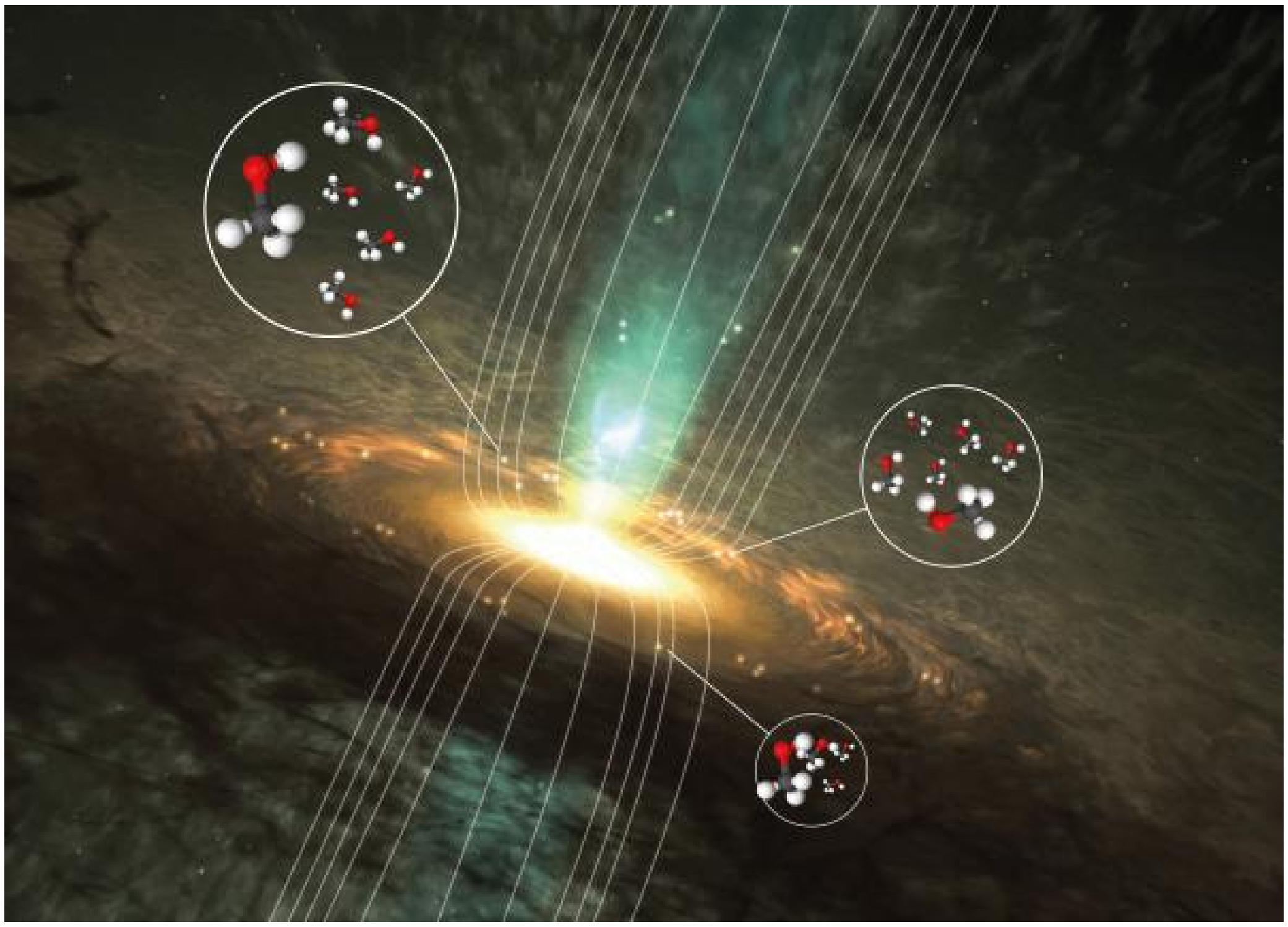

Figure A.3: Magnetic fields play an important role in the places where most massive stars are born. This illustration shows the surroundings of a forming massive star, and the bright regions where radio signals from methanol can be found. The bright spots represent methanol masers - natural lasers that are common in the dense environments where massive stars form - and the curved lines represent the magnetic field. Thanks to new calculations by astrochemists, astronomers can now start to investigate magnetic fields in space by measuring the radio signals from methanol molecules in these bright sources. (JIVE press release: `http://www.jive.eu/astrochemists-reveal-magnetic-secrets-methanol`) Credit: Wolfgang Steffen/Chalmers/Boy Lankhaar (molecules: Wikimedia Commons/Ben Mills).

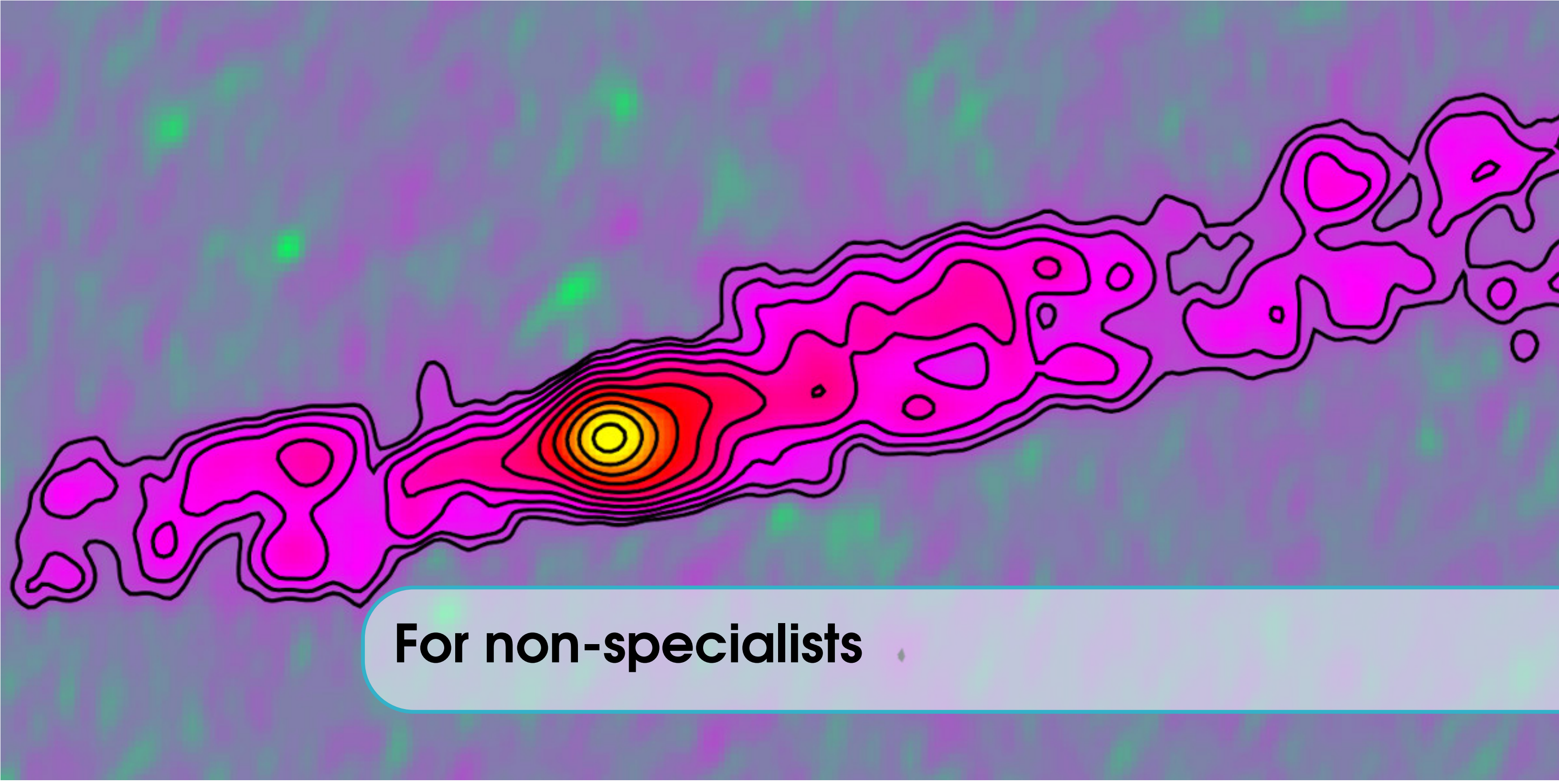

# For non-specialists

**"VEX"** VLBI schedules are written in vex-format files. These are used to control antennas for observations.. 174

***uv*-coordinates** VLBI interferometers sample the data in the *uv*-plane, which is in a Fourier relation with the sky brightness distribution. A higher number of telescopes and longer observing time mean more dense sampling of the acrshortuv-plane, i.e. better *uv*-coverage – also depending on the distribution of the antennas. This results in higher fidelity and more sensitive images of the radio sky. 10

**e-EVN** Is a subset of the EVN telescopes participating in real-time e-VLBI operations. All telescopes are in principle e-VLBI capable, albeit different data rate limits may apply. 6

**e-VLBI** Is real-time data transfer and correlation through high-speed networks using a software correlator. The EVN and JIVE have played a pioneering role in the development of the e-VLBI technique. 148

**EVN** The European VLBI Network is the most sensitive standalone VLBI array to date. It includes some of the greatest radio antennas in Europe, Africa, Asia and the Caribbeans. 3

**EVN-light** Is a concept of using part of the EVN array for VLBI observations outside of regular disc recording sessions. 15, 170

**JIVE** The Joint Institute for VLBI ERIC was formerly known as the Joint Institute for VLBI in Europe, and generally referred to as JIVE. It is a European Research Infrastructure Consortium (formally registered as JIV-ERIC) based in Dwingeloo, Netherlands. It operates the central data processor of the EVN. 3

**SFXC** The EVN software correlator at JIVE. 92

**SKA-VLBI** Is a concept to include phased-up phase-I. SKA telescopes in VLBI observations. The EVN will have joint sky coverage with SKA1-MID. 141, 152

**SKA1-MID** Is the planned, phase-I. mid-frequency telescope of the Square Kilometre Array in South Africa. 5

**Space VLBI** The resolving power of a VLBI array depends on the largest distances between its elements. The longer these 'baselines' are, the finer details the images will have. In space VLBI one (or in the future, possibly more) of the telescopes is in space, to overcome the natural limitation to the maximum baselines a VLBI array can have on Earth.. 15, 167

**VLBI** Very Long Baseline Interferometry is a technique that combines radio telescopes at great distances, to produce the highest angular resolution images in astronomy. 3

---

VLBI image of the archetypal radio galaxy Cygnus A at the frequency of 86 GHz. The high spatial resolution gives insight into the transverse width profile at the onset of the two-sided flow. The jet base appears wide with a minimum width of 227 Schwarzschild radius RS. This is significantly larger than the radius of the innermost stable circular orbit (ISCO), suggesting that the emission is produced by a mildly relativistic, parabolically expanding outer disk wind. The existence of a central and faster jet, possibly driven by the rotation of the inner regions of the accretion disk or by the spinning black hole and invisible due to Doppler deboosting, is suggested by the kinematic properties and by the observed limb brightening of the flow. Credit: Boccardi et al. (2016), *A&A*, **588**, L9, reproduced with permission ©ESO.

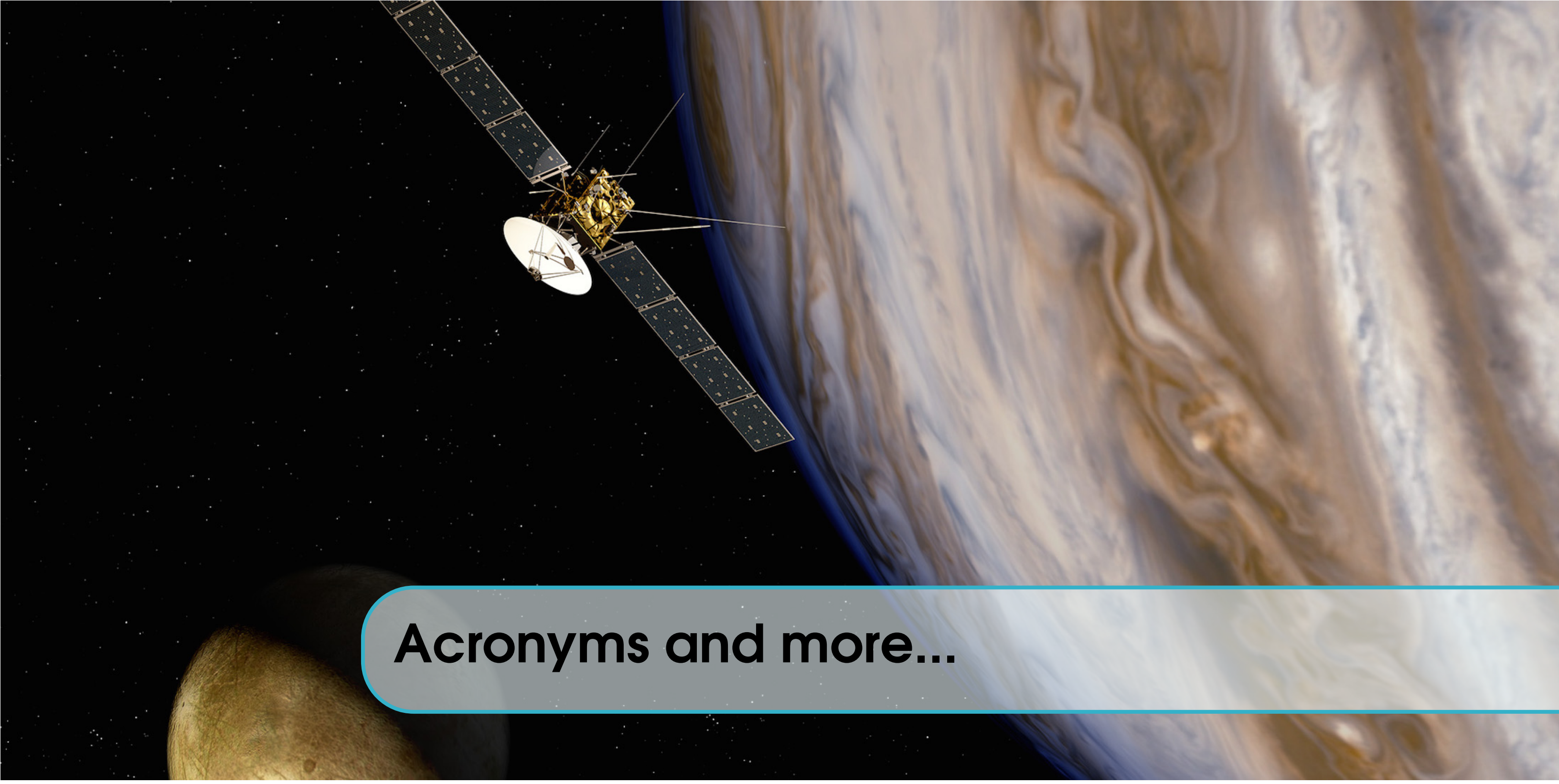

# Acronyms and more...

***e*-MERGE** *e*-MERlin Galaxy Evolution survey. 38, 39

***e*-MERLIN** *enhanced* Multi-Element Remotely Linked Interferometer Network. 5, 7, 11, 15, 37–40, 42, 50, 51, 96, 97, 105, 112–114, 122, 124–126, 140, 141, 148, 157, 168–170

*uv* *uv*-coordinates/plane/coverage. 10, 25, 26, 46, 52, 61, 70, 72, 78, 82, 91–93, 97, 125–127, 142, 148, 149, 152, 165, 169, 170, 183

*Athena* Advanced Telescope for High-ENergy Astrophysics. 7

*Chandra* Chandra X-ray Observatory. 37, 38, 69

*Euclid* Euclid satellite. 7, 17, 18

*Fermi* Fermi Gamma-ray Space Telescope. 63, 64, 86, 142, 146

*Gaia* Gaia mission. 10, 79, 80, 95, 105, 122, 123, 128, 132–134, 142, 152

*HALCA* Highly Advanced Laboratory for Communication and Astronomy. 167

*HST* Hubble Space Telescope. 21, 69, 85, 142

*IKAROS* Interplanetary Kite-craft Accelerated by Radiation Of the Sun. 135

*IXPE* X-ray Polarimetry Explorer. 78

*JUICE* Jupiter Icy Satellites Explorer. 136

*JWST* James Webb Space Telescope. 7, 72, 142

*Kepler* Kepler Space Telescope. 96

*LISA* Laser Interferometer Space Antenna. 9, 68, 144, 147, 159

*MSX* Midcourse Space Experiment. 124

*Planck* Planck satellite. 17

*RadioAstron* Space VLBI mission. 58, 59, 136, 155, 167

*SELENE* Selenological and Engineering Explorer. 135

*Spektr-R* Space VLBI telescope – *RadioAstron* mission. 167

*Starlink* Satellite constellation being constructed by American company SpaceX to provide satellite Internet access. 173

*Swift* Neil Gehrels Swift Observatory; formerly Swift Gamma-Ray Burst Mission. 64, 84, 86, 142

*VSOP* VLBI Space Observatory Programme. 167

*XIPE* X-ray Imaging Polarimetry Explorer. 78

*eROSITA* X-ray instrument on Spektr-RG. 52, 142

**ADC** Analog-to-digital Converter. 172
**AGB** Asymptotic Giant Branch. 112, 114, 120, 122–126
**AGN** Active Galactic Nuclei. 6, 9, 13, 14, 23, 24, 26, 27, 29, 36–40, 43, 44, 46–52, 55–60, 62–76, 78, 85, 92, 104, 131–133, 141, 146, 148, 149, 151, 156, 157
**AIPS** Astronomical Image Processing System. 126
**ALMA** Atacama Large Millimeter/submillimeter Array. 7, 15, 51, 52, 103–105, 112–114, 121, 123, 124, 126, 127, 142, 159, 161, 166
**APERTIF** APERture Tile In Focus. 7, 50, 52, 93, 168
**ASASSN** All-Sky Automated Survey for Supernovae. 84
**ASKAP** Australian Square Kilometre Array Pathfinder. 50, 90, 93, 111, 113, 149, 161
**ASTRON** Netherlands Institute for Radio Astronomy. 11
**ATCA** Australia Telescope Compact Array. 78, 160, 161
**ATNF** Australia Telescope National Facility, Marsfield, Australia. 160
**AU** Astronomical Unit. 98, 104, 113, 116–120, 127, 145
**AVN** African VLBI Network. 108, 125, 127, 140, 145, 149, 157

**BAaDE** Bulge Asymmetries and Dynamical Evolution. 123, 124
**BAO** Baryon Acoustic Oscillation. 18, 20, 26, 29
**BAT** *Swift* Burst Alert Telescope. 64
**BeSSel** Bar and Spiral Structure Legacy Survey. 123
**BH** Black Hole. 71, 77–80, 88, 144, 147, 154
**BKG** Federal Agency for Cartography and Geodesy, Frankfurt am Main, Germany. 11
**BL-Ic** Broad-line type Ic supernova. 83, 84
**BLR** Broad Line Region. 43
**BNS** Binary Neutron Star. 88
**BRAND** BRoad bAND EVN. 31, 61, 62, 125, 168, 171, 172, 175

**CASA** Common Astronomy Software Applications. 126, 150, 151
**CBD** Consortium Board of Directors. 155, 156
**CCSN** Core-collapse Supernova. 82
**CDM** Cold Dark Matter. 17, 19, 23, 66
**$CH_3OH$** methanol. 103, 117–119, 151, 177, 178
**CHIME** Canadian Hydrogen Intensity Mapping Experiment. 7, 90, 93, 142
**CLASS** Cosmic Lens All-Sky Survey. 21, 22, 24
**CMB** Cosmic Microwawe Background. 17, 18, 27, 29
**CME** Coronal Mass Ejection. 108, 109
**CO** Carbon monixide. 51
**COBRaS** Cygnus OB2 Radio Survey. 112, 114
**COSMOS** Cosmic Evolution Survey. 67
**CPU** Central Processing Unit. 169, 172, 174, 175
**CRAF** Committee on Radio Astronomical Frequencies. 173
**CSE** Circumstellar Envelope. 120, 121, 124, 126
**CSIRO** Commonwealth Scientific and Industrial Research Organisation, Canberra, Australia. 160, 175
**CSO** Compact Symmetric Object. 66, 151

**CTA** Cherenkov Telescope Array. 7, 142, 145, 146, 152
**CVN** Chinese VLBI Network. 15, 158

**DBBC** Digital Base Band Converter. 161, 168, 174, 175
**DeltaDOR** Delta-Differential One-Way Ranging. 135
**DES** Dark Energy Survey. 17
**DiFX** Distributed FX-style software correlator. 159–161
**DM** Dispersion Measure. 89
**DR2** Gaia Data Release 2. 79
**DSA-10** Deep Synoptic Array. 90
**DSN** Deep Space Network. 135

**E-CDFS** Extended Chandra Deep Field-South. 36
**EAVN** East Asian VLBI Network. 15, 62, 139, 158–160
**EC-H2020** European Commission Horizon 2020 programme. 3
**ECM** Electron Cyclotron Maser. 109
**EGP** Extrasolar Giant Planet. 111
**EHT** Event Horizon Telescope. 10, 58, 59, 62, 139, 159, 162, 166, 171
**ELT** Extremely Large Telescope. 7, 94, 142, 144, 145, 152
**EM** Electromagnetic. 68, 144, 147
**EMU** Evolutionary Map of the Universe. 113
**ERIC** European Research Infrastructure Consortium. 156
**ERIS** European Radio Interferometry School. 140
**ESA** European Space Agency. 17, 122, 135–137, 144, 152, 180
**ESTRACK** European Space Tracking. 135
**EVN** European VLBI Network. 1, 3, 5–8, 10, 14, 15, 17, 19, 20, 25, 26, 31, 38–41, 44, 46, 49–52, 57, 62, 65–67, 69, 71, 72, 77, 81, 82, 84–86, 88, 90, 92–98, 101, 105, 107–110, 113, 114, 117–127, 134–137, 139–142, 144–152, 155–160, 166–176, 181

**FAST** Five-hundred-meter Aperture Spherical radio Telescope. 93–95, 97, 148, 158, 159, 169, 170
**FIRST** Faint Images of the Radio Sky at Twenty-Centimeters. 63, 64
**FoV** Field of View. 90, 92, 93, 96
**FPGA** Field-programmable Gate Array. 168, 172, 174, 175
**FR** Fanaroff-Riley. 68
**FRB** Fast Radio Burst. 89–93, 139, 141, 147, 152, 156, 162
**FUV** Far ultraviolet light. 36
**FWHM** Full Width at Half Maximum. 38, 61, 88, 144

**GÉANT** Pan-European data network for research and education. 168
**Gbps** Gigabit per second. 10, 95, 148, 160, 161, 168–171
**GBT** Robert C. Byrd Green Bank Telescope. 17, 20, 97, 161
**GLT** Greenland Telescope Project. 161
**GMRT** Giant Metrewave Radio Telescope. 164, 165
**GMVA** Global Millimeter VLBI Array. 58, 59, 61, 62, 139, 155, 159, 161, 166
**GOBELINS** Gould's Belt Distances Survey. 106, 107
**GOODS-N** Great Observatories Origins Deep Survey - North. 38, 147
**GPU** Graphics Processing Unit. 161, 169, 170, 172, 175

**GRB** Gamma Ray Burst. 35, 82–84, 86–88, 90
**GUT** Grand Unified Theory. 147
**GW** Gravitational Wave. 28, 30, 68, 70, 86, 94, 142, 144, 156, 159, 162

**$H_2O$** water. 7, 46, 103, 116, 117, 120, 121, 126, 177, 178
**HBA** LOFAR High Band Antenna. 163
**HDF** Hubble Deep Field. 147
**HERA** Hydrogen Epoch of Reionisation Array. 2nd generation instrument that combines efforts from MWA and PAPER. 7
**HMSFR** High-mass Star-forming Region. 116–118, 120, 122, 123, 125, 126
**HPC** High Performance Computing. 175
**HSA** High Sensitivity Array. 46, 159, 161

**IAA** Institute of Applied Astronomy, St. Petersburg, Russia. 11, 100, 168
**IAU** International Astronomical Union. 131
**IceCube** IceCube neutrino detector. 146, 147, 153, 180
**ICRF** International Celestial Reference Frame. 131–134, 152
**ICT** Information and Communication Technology. 140
**IFRS** Infra-red Faint Radio Sources. 39
**IGM** Intergalactic Medium. 89, 93
**IGN** Instituto Geográfico Nacional, Spain. 11
**ILT** International LOFAR Telescope. 5, 140
**IMBH** Intermediate Mass Black Hole. 71, 72, 144
**INAF** Italian National Institute for Astrophysics. 3
**INDIGO** Indian Initiative in Gravitational-wave Observations. 144
**IPTA** International Pulsar Timing Array. 93
**IR** Infrared light. 35, 36, 38, 43, 44, 85, 86, 107, 113, 121
**IRA** Istituto di Radioastronomia, Bologna, Italy. 3, 11
**ISAS** Institute of Space and Astronautical Science. 167
**ISM** Interstellar Medium. 8, 27, 28, 40, 41, 47, 51, 80, 82, 83, 95, 97, 105, 115, 120
**IVIA** Iniciativa VLBI IberoAmericana. 140
**IVS** International VLBI Service. 139, 155, 159, 160

**JIVE** Joint Institute for VLBI ERIC. 5, 10, 11, 15, 24, 93, 97, 120, 136, 137, 141, 151, 155–157, 169, 170, 174, 175
**JPL** Jet Propulsion Laboratory. 55, 137, 168
**JUMPING** Joining up Users for Maximising the Profile, the Innovation and Necessary Globalisation of JIVE. 3, 141
**JVAS** Jodrell/VLA Astrometric Survey. 21, 23
**JVLA** Jansky Very Large Array. 15, 37–39, 87, 90, 105, 108, 109, 124, 157, 159, 162
**JVN** Japanese VLBI Network. 158

**KIC** Kepler Input Catalog. 97
**KIDS** KIlo-Degree Survey. 17
**KPNO** Kitt Peak National Observatory. 147
**KVN** Korean VLBI Network. 11, 52, 62, 63, 126, 158, 160, 161, 168, 172, 175

**LAT** *Fermi* Large Area Telescope. 63, 64

**LBA** LOFAR Low Band Antenna. 163, 164
**LBA** Long Baseline Array. 15, 52, 107, 139, 160, 161
**LC** Light-curve. 30
**LEAP** Large European Array for Pulsars. 95
**LIGO** Laser Interferometer Gravitational-Wave Observatory. 7, 28, 86, 88, 144
**LNA** Low-noise Amplifier. 173
**LO** Local Oscillator. 163
**LOCATe** LOCalising Astronomical Transients with e-VLBI. 90
**LOFAR** Low-Frequency Array. 5, 7, 10, 15, 52, 111, 139–141, 149, 163, 164, 175
**LSR** Local Standard of Rest. 106
**LSST** Large Synoptic Survey Telescope / Vera C. Rubin Observatory. 7, 17, 79, 142

**MALS** MeerKAT Absorption Line Survey. 50
**mas** milliarcsecond. 18, 152
**MCP** Megamaser Cosmology Project. 27, 29
**MeerKAT** Karoo Telescope Array. 7, 50, 68, 72, 82, 94, 95, 143, 148, 149, 169, 175
**MERLIN** Multi-Element Radio Linked Interferometer Network. 21, 22, 25, 39
**MPIfR** Max Plank Institute for Radio Astronomy, Bonn, Germany. 11, 156
**MRO** Metsähovi Radio Observatory, Aalto University, Finland. 11
**MSP** Millisecond Pulsar. 93
**MWA** Murchison Widefield Array. 164

**NAIC** National Astronomy and Ionospheric Center. 11
**NARIT** National Astronomical Research Institute of Thailand. 169
**NASA** National Aeronautics and Space Administration. 3, 55, 98, 135, 174
**NGC** New General Catalogue of Nebulae and Clusters of Stars. 28, 40, 46, 48, 52, 68, 86, 116, 117
**ngVLA** Next Generation Very Large Array. 7, 15, 94, 95, 140, 162, 172
**NIRSpec** Near-Infrared Spectrograph. 72
**NOEMA** NOrthern Extended Millimeter Array; formerly Plateau de Bure observatory. 7, 52, 126
**NS** Neutron Star. 77–80, 88, 93, 94, 147
**NVSS** NRAO VLA Sky Survey. 165

**O3** LIGO/Virgo O3 observing run. 88
**OB** Early type (O, B) stars. 112, 114, 115
**OD** Orbit Determination. 135, 136
**OH** hydroxyl radical. 44, 103, 120, 121, 123, 125, 177
**OSO** Onsala Space Observatory, Onsala, Sweden. 11

**PAF** Phased Array Feed. 93, 149
**Pan-STARRS1** Panoramic Survey Telescope & Rapid Response System 1. 67
**PC** Programme Committee. 155, 156
**pc** parsec. 10
**Pierre Auger Observatory** Pierre Auger Observatory. 146
**PMN** Parkes-MIT-NRAO catalog. 24, 25
**PMS** Pre-main Sequence. 107
**PRIDE** Planetary Radio Interferometry and Doppler Experiment. 135–137
**PTA** Pulsar Timing Array. 94

**PTF** Palomar Transient Factory. 84
**PWN** Pulsar Wind Nebula. 145

**QRFH** Quad-Ridge Flared Horn. 172
**QTT** Qitai 110m Radio Telescope. 159, 169

**RAEGE** Rede Atlântica de Estações Geodinâmicas e Espaciais. 169
**RDBE** Roach Digital Backend. 168
**RFC** Radio Fundamental Catalog. 29
**RFI** Radio Frequency Interference. 9, 49–52, 89, 90, 92, 93, 96, 97, 165, 168, 171–173, 175
**RFSoC** Radio Frequency System-on-Chip. 172
**RM** Rotation Measure. 92, 171
**rms** root mean square. 21, 26, 42, 46, 71, 94, 113, 114, 125
**RSG** Red Supergiant. 112

**SARAO** South African Radio Astronomy Observatory, Cape Town, South Africa. 11, 143
**SCORPIO** Stellar Continuum Originating from Radio Physics In Our galaxy. 113
**SDSS** Sloan Digital Sky Survey. 18, 41, 52, 63, 64, 96
**SED** Spectral Energy Distribution. 38, 64, 88
**SETI** Search for Extraterrestrial Intelligence. 9, 10, 14, 90, 96, 97
**SFG** Star-forming Galaxies. 37, 38, 43
**SFPR** Source Frequency Phase-referencing technique. 62
**SFR** Star Formation Rate. 38
**SFRD** Star Formation Rate Density. 35, 36
**SFXC** Software FX Correlator – EVN software correlator at JIVE. 92, 97, 120, 136, 151, 156, 169, 170, 174
**ShAO** Shanghai Astronomical Observatory. 11, 168
**SHARP** Searching for H I Absorption with APERTIF. 49, 50, 52
**SiO** Silicium Oxide. 103, 117, 120–124, 126, 149
**SKA** Square Kilometre Array. 5–7, 10, 15, 24, 26, 27, 29, 46, 47, 49–52, 68, 72, 82, 85, 90, 92–95, 105, 111, 126, 127, 134, 140–145, 148–152, 155, 156, 160, 164, 168, 169, 173, 176
**SMA** Submillimeter Array. 118
**SMBH** Supermassive Black Hole. 6, 14, 39, 40, 43, 44, 55, 63–71, 73, 75, 84, 85, 147
**SMBHB** Supermassive Black Hole Binary. 94
**SN** Supernova. 20, 26, 43, 80–84, 120
**SNR** Signal-to-Noise Ratio. 62
**SpaceX** Space Exploration Technologies Corporation. Private American aerospace manufacturer and space transportation services company. 173
**SRT** Sardinia Radio Telescope. 97, 168
**Super-Kamiokande** Super-Kamiokande neutrino detector. 147

**TAC** Time Allocation Committee. 160
**TCfA** Toruń Centre for Astronomy, Toruń, Poland. 11
**TDE** Tidal Disruption Event. 84–86, 144
**THEZA** TeraHertz Exploration and Zooming-in for Astrophysics. 144, 154, 167
**TOG** Technical and Operations Group. 155, 156
**ToO** Target of Opportunity. 155, 156

**UCD** Ultracool Dwarf. 109
**UGC** Uppsala General Catalogue of Galaxies. 31
**uGMRT** upgraded Giant Metrewave Radio Telescope. 7, 15, 164, 165
**UHECR** Ultra-High Energy Cosmic Rays. 146
**UHF** Ultra High Frequency. 10
**ULIRG** Ultra-luminous Infrared Galaxies. 38, 42
**ULX** Ultra-luminous X-ray sources. 40
**UT** Universal Time. 79
**UV** Ultraviolet light. 35, 38, 44, 84, 85, 112, 115
**UWB-H** Ultra Wideband High receiver at Parkes. 161
**UWB-L** Ultra Wideband Low receiver at Parkes. 161

**VDIF** VLBI Data Interchange Format. 157, 161, 169, 172, 176
**VERA** The VLBI Exploration of Radio Astrometry. 123–126, 158
**VGOS** VLBI Global Observing System. 168
**VHE** Very High Energy. 145, 146
**VIRAC** Ventspils International Radio Astronomy Centre, Ventspils, Latvia. 11
**Virgo** Virgo interferometer, gravitational wave detector. 7, 28, 86, 88, 144
**VLA** Very Large Array. 7, 22, 37, 39, 44, 52, 71, 78, 80, 81, 88, 113, 114, 118, 124, 126, 162–164
**VLASS** VLA Sky Survey. 162
**VLBA** Very Long Baseline Array. 15, 17, 20, 27, 46, 47, 52, 62, 65, 79, 88, 109, 110, 118, 120, 123–126, 139, 145, 156, 159–161, 176
**VLBI** Very Long Baseline Interferometry. 3, 5–10, 13–15, 17–32, 37–44, 46–52, 55–63, 65–68, 70–72, 77–80, 82–85, 88–98, 103–109, 111, 112, 114–118, 120–127, 131–137, 139–146, 148–152, 154–176
**VO** Virtual Observatory. 151, 152

**WALLABY** ASKAP H I All-Sky Survey. 50
**WCR** Wind Collision Region. 114, 115
**WD** White Dwarf. 80, 144
**WDM** Warm Dark Matter. 19, 23
**WEAVE** next-generation wide-field survey facility for the William Herschel Telescope. 52
**Wi-Fi** The Standard for Wireless Fidelity. 90
**WMAP** Wilkinson Microwave Anisotropy Probe. 29
**WR** Wolf-Rayet type stars. 112, 114, 115
**WSRT** Westerbork Synthesis Radio Telescope. 7, 41, 93, 147, 168

**XAO** Xinjiang Astronomical Observatory, Urumqi, China. 11
**XRB** X-ray Binaries. 40, 57, 77–80

**YSO** Young Stellar Object. 116–119

**ZAMS** Zero Age Main Sequence. 118
**ZTF** Zwicky Transient Facility. 79

---

The European Space Agency JUICE mission will visit the icy moons of Jupiter. Credit: ESA.